\documentclass[12pt,a4]{article}
\pdfoutput=1
\usepackage[utf8]{inputenc}
\usepackage[labelfont=bf,textfont=it]{caption}
\usepackage{a4wide}
\usepackage{xspace,setspace}
\usepackage[labelfont=bf,textfont=it]{caption}
\usepackage{amsmath,amssymb,slashed,bbm}
\usepackage{braket}
\usepackage{booktabs}
\usepackage{graphicx}
\usepackage{url}
\usepackage[breaklinks=true]{hyperref}
\hypersetup{
     colorlinks   	= true,
     citecolor    	= LightSeaGreen,
     linkcolor 	= black,
     menucolor	= black,
}
\usepackage{hhline,multirow}
\usepackage{enumitem}
\allowdisplaybreaks
\usepackage{cite}
\usepackage{subcaption}
\usepackage{adjustbox}
\usepackage{placeins}
\usepackage{footmisc}
\usepackage{enumitem}
\usepackage{array}
\usepackage{bm}

\usepackage[titles]{tocloft}

\numberwithin{equation}{section}
\numberwithin{figure}{section}
\numberwithin{table}{section}

\makeatletter
\@addtoreset{footnote}{section}
\makeatother

\usepackage[dvipsnames,svgnames,x11names,hyperref]{xcolor}
\definecolor{comment}{rgb}{0,0.3,0}
\definecolor{identifier}{rgb}{0.0,0,0.3}
\usepackage{listings}
\newcommand{\be}{\begin{equation}}
\newcommand{\ee}{\end{equation}}

\newcommand{\dzero}{\ensuremath{\text{D${\slashed{0}}$}}\xspace}
\newcommand{\Afb}{\ensuremath{A_\mathrm{FB}}\xspace}
\newcommand{\Ac}{\ensuremath{A_\mathrm{C}}\xspace}

\newcommand{\U}[2]{\ensuremath{\text{U{(#1)}}_{#2}}}
\newcommand{\su}[2]{\ensuremath{\text{SU{(#1)}}_{#2}}}
\newcommand{\smgg}{\ensuremath{\text{SU(3)}_C \times \text{SU(2)}_L \times \text{U(1)}_Y} }
\newcommand{\cp}{\ensuremath{\mathcal{CP}}\xspace}
\newcommand{\x}{\ensuremath{\mathcal{\times}}\xspace}
\renewcommand{\d}{\ensuremath{\text{d}}\xspace}
\newcommand{\gev}{\ensuremath{\text{G}\mspace{0.2mu}\text{e}\mspace{-1mu}\text{V}}\xspace}
\newcommand{\tev}{\ensuremath{\text{T}\mspace{0.2mu}\text{e}\mspace{-1mu}\text{V}}\xspace}
\newcommand{\fb}{\ensuremath{\text{fb}}\xspace}
\newcommand{\pb}{\ensuremath{\text{pb}}\xspace}
\newcommand{\ifb}{\ensuremath{\text{fb}^{-1}}\xspace}
\newcommand{\iab}{\ensuremath{\text{ab}^{-1}}\xspace}
\newcommand{\sqrts}{\ensuremath{\sqrt{s}}\xspace}
\newcommand{\ttbar}{\ensuremath{t\bar{t}}\xspace}
\newcommand{\ttz}{\ensuremath{t\bar{t}Z}\xspace}
\newcommand{\ep}{\ensuremath{e^+e^-}\xspace}
\newcommand{\pp}{\ensuremath{pp\xspace}}

\newcommand{\lag}[1]{\ensuremath{\mathcal{L}_\mathrm{#1}}}
\newcommand{\ord}[1]{\ensuremath{\mathcal{O}(\mathrm{#1})}}
\newcommand{\amp}[1]{\ensuremath{\mathcal{A}_\mathrm{#1}}}
\newcommand{\m}[1]{\ensuremath{\mathcal{M}_{#1}}}
\newcommand{\D}[1]{\ensuremath{D=#1}}
\newcommand{\summ}[1]{\ensuremath{\sum_{\substack{#1}} }}
\newcommand{\op}[2][]{\ensuremath{\mathcal{O}_{#2}^{#1}}}
\newcommand{\co}[2][]{\ensuremath{c_{#2}^{#1}}}
\newcommand{\cb}[2][]{\ensuremath{\bar{c}_{#2}^{#1}}}
\newcommand{\qqquad}{\qquad \qquad}
\newcommand{\qqqquad}{\qquad \qquad \qquad}

\newcommand{\csection}[1]{\begin{center}\section*{#1}\end{center}}

\newenvironment{mydescription}{%
   
   \begin{description}[leftmargin=0.25cm, style=sameline]%
}{%
   \end{description}%
}

\begin{document}

\title{ \huge{\textbf{Top Quark Physics in \\the Large Hadron Collider era}} \\
\normalsize ~\\ ~\\ ~\\
\includegraphics[width=0.4\textwidth]{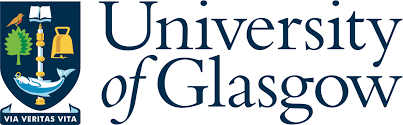} ~\\ ~\\
}

\author{{\sc{\LARGE{Michael Russell}}} ~\\ ~\\ ~\\
  \normalsize ~\\ 
\multicolumn{1}{p{.5\textwidth}}{\centering\emph{Particle Physics Theory Group, School of Physics \& Astronomy, University of Glasgow}} ~\\ ~\\ ~\\
}

\date{September 2017}
\maketitle
\thispagestyle{empty}

\vfill
\begin{center}
\begin{minipage}{0.7\textwidth}\centering
\large{A thesis submitted in fulfilment of the requirements for the degree of Doctor of Philosophy}
\end{minipage}
\end{center}

\newpage
\csection{Abstract}
We explore various aspects of top quark phenomenology at the Large Hadron Collider and proposed future machines. After summarising the role of the top quark in the Standard Model (and some of its well-known extensions), we discuss the formulation of the Standard Model as a low energy effective theory. We isolate the sector of this effective theory that pertains to the top quark and that can be probed with top observables at hadron colliders, and present a global fit of this sector to currently available data from the LHC and Tevatron. Various directions for future improvement are sketched, including analysing the potential of boosted observables and future colliders, and we highlight the importance of using complementary information from different colliders. Interpretational issues related to the validity of the effective field theory formulation are elucidated throughout. Finally, we present an application of artificial neural network algorithms to identifying highly-boosted top quark events at the LHC, and comment on further refinements of our analysis that can be made.

\newpage
\csection{Acknowledgements}
First and foremost I must thank my supervisors, Chris White and Christoph Englert, for their endless support, inspiration and encouragement throughout my PhD. They always gave me enough freedom to mature as a researcher, whilst providing the occasional necessary nudge to keep me on the right track. I also have to thank David Miller for his friendly advice and irreverent sense of humour, and for helping me settle into the group in Glasgow, and Christine Davies, for financial support for several research trips, and for creating a wonderful place to do physics.

This thesis would not have been written without the foundations laid down by Liam Moore, so I have to thank him for the countless hours he has put into the project, and for being a mate. Of the other students in Glasgow, I also have to mention Karl Nordstr\"om for many useful conversations, both about the work presented here and often tangential (but always illuminating) topics, and his occasional computational wizardry.
In Heidelberg, special mention must go to Torben Schell, for his patience in showing me some of the many ropes of jet substructure, and to Tilman Plehn for stimulating collaboration, and for helping me start the next chapter in my life as a physicist. I thank the students and postdocs in both of these places for making such a friendly working environment.

Throughout my PhD I have been fortunate to work with several excellent experimentalists; James Ferrando and Andy Buckley in Glasgow, and Gregor Kasieczka in Z\"urich. I thank them for all they have taught me about collider physics (and especially Andy for coding assistance in the rocky early days of {\sc{TopFitter}}), and I hope to collaborate with them again in the future. Though I never had the opportunity to work with Sarah Boutle or Chris Pollard directly, their unwavering fondness for a pint helped keep me sane after many a long day.  

Finally, I have to acknowledge the huge intellectual (and general) debt I owe to my parents. Though they do not share my passion for physics, they have always supported me along the way, and politely endured far too many wearisome physics rants to mention. This thesis is dedicated to them. Thanks also to Daniel and Lucy for trying to keep me in the real world.

\newpage
\csection{Declaration}
I declare that this thesis is the result of my own original work and has not been previously presented for a degree. In cases where the work of others is presented, appropriate citations are used. Chapters 1 and 2 serve as an introduction to the research topics presented in the rest of the thesis. Chapters 3 to 5 are a result of my own original work, in collaboration with the authors listed below.

Work appearing in this thesis was presented in the following publications~\cite{Buckley:2015nca,Buckley:2015lku,Englert:2016aei,Englert:2017dev,Kasieczka:2017nvn}:

\begin{itemize}

\item A.~Buckley, C.~Englert, J.~Ferrando, D.~J. Miller, L.~Moore, M.~Russell, and
  C.~D. White, ``{Global fit of top quark effective theory to data},''
  \href{http://dx.doi.org/10.1103/PhysRevD.92.091501}{{\em Phys. Rev.}
  {\bfseries D92} no.~9, (2015) 091501},
\href{http://arxiv.org/abs/1506.08845}{{\ttfamily arXiv:1506.08845 [hep-ph]}}.

\item A.~Buckley, C.~Englert, J.~Ferrando, D.~J. Miller, L.~Moore, M.~Russell, and
  C.~D. White, ``{Constraining top quark effective theory in the LHC Run II
  era},'' \href{http://dx.doi.org/10.1007/JHEP04(2016)015}{{\em JHEP}
  {\bfseries 04} (2016) 015},
\href{http://arxiv.org/abs/1512.03360}{{\ttfamily arXiv:1512.03360 [hep-ph]}}.

\item C.~Englert, L.~Moore, K.~Nordstr\"{o}m, and M.~Russell, ``{Giving top quark
  effective operators a boost},''
  \href{http://dx.doi.org/10.1016/j.physletb.2016.10.021}{{\em Phys. Lett.}
  {\bfseries B763} (2016) 9--15},
\href{http://arxiv.org/abs/1607.04304}{{\ttfamily arXiv:1607.04304 [hep-ph]}}.

\item A.~Buckley, C.~Englert, J.~Ferrando, D.~J. Miller, L.~Moore, K.~Nordstr\"om,
  M.~Russell, and C.~D. White, ``{Results from TopFitter},'' in {\em {9th
  International Workshop on the CKM Unitarity Triangle (CKM 2016) Mumbai,
  India, November 28-December 3, 2016}}.
\newblock 2016.
\newblock \href{http://arxiv.org/abs/1612.02294}{{\ttfamily arXiv:1612.02294
  [hep-ph]}}.
\newblock
\newblock

\item G.~Kasieczka, T.~Plehn, M.~Russell, and T.~Schell, ``{Deep-learning Top Taggers
  or The End of QCD?},'' \href{http://dx.doi.org/10.1007/JHEP05(2017)006}{{\em  
  JHEP} {\bfseries 05} (2017) 006},
\href{http://arxiv.org/abs/1701.08784}{{\ttfamily arXiv:1701.08784 [hep-ph]}}.

\item C.~Englert and M.~Russell, ``{Top quark electroweak couplings at future lepton
  colliders},'' \href{http://dx.doi.org/10.1140/epjc/s10052-017-5095-z}{{\em    
  Eur. Phys. J.} {\bfseries C77} no.~8, (2017) 535},
\href{http://arxiv.org/abs/1704.01782}{{\ttfamily arXiv:1704.01782 [hep-ph]}}.

\end{itemize}

My specific contributions to these chapters are as follows. In chapter 3, I set up the EFT fitting framework, and used the {\sc{Mathematica}} model developed by Liam Moore to simulate events, construct theory observables and derive confidence limits on the operators considered. In chapter 4, I wrote the boosted analysis and derived the various projected limits on the operators considered. All numerical results in those chapters were derived by myself. In chapter 5, my contributions involved generating events, writing the analysis code to construct the fat jets and subsequent images used, formulating the boosted decision tree used as a comparison, and helping to design the neural network architecture used. All figures in this thesis were generated by the author, with the following exceptions: Fig.~\ref{fig:residuals} (A. Buckley), Figs.~\ref{fig:pols}-\ref{fig:stu} (C. Englert), Fig.~\ref{fig:img_mass_step} and Figs.~\ref{fig:arc_best} -\ref{fig:deep_layers_and_pcc} (T. Schell).

%

\newpage
\null
\newpage

\begingroup
\setstretch{1.0}
\tableofcontents
\endgroup

\newpage
\section*{A historical introduction}
\addcontentsline{toc}{section}{A historical introduction}

The origins of what is now known as the `Standard Model' of particle physics can be traced to the late 1940s. The attention of theoretical physicists was centred on how to consistently embed the postulates of quantum mechanics; the laws describing (sub)atomic particles, within the framework of special relativity; the laws of motion for objects with very large velocities. Their efforts culminated in the development of quantum electrodynamics; the quantum theory of the electromagnetic field. The problems associated with the infinities arising in calculations had been brought under control by the development of renormalisation, which allowed properties of the electron and photon to be calculated to high precision, showing extraordinary agreement with experiment.

It had already been known for some time, however, that electrodynamics could not be the full story. Firstly, it was immediately obvious that the stability of the atomic nucleus would not withstand the electrostatic repulsion between the positively-charged protons, therefore an additional force must have been present to stabilise the nucleus. This force had to be strong, and extremely short-ranged (no further than the typical size of a nucleus) so was dubbed the \emph{strong interaction}. Moreover, the observation of certain types of radioactive decay, which necessitated the existence of a new, extremely light particle (what is now called the neutrino), could not be accommodated with the known facts about electromagnetism and the strong force. These interactions did not allow for processes which changed electric charge, which nuclear $\beta$-decay plainly did. Due to the relatively long lifetimes associated with these decay processes, the force responsible was called the \emph{weak interaction}.

The first attempt to write down a theory of the strong nuclear force was made by Yukawa in 1935~\cite{Yukawa:1935xg}. He proposed that the force binding together the nucleus was due to an interaction between protons and neutrons mediated by a scalar particle he dubbed the `pion', which he calculated should have a mass of around one tenth of the proton mass. Tentative discoveries of these pions were made in 1947 in photographic emulsion recordings of cosmic ray showers~\cite{Lattes:1947mw}. The problem with the theory was that it could not predict anything precisely. The strong force is (by definition) strong, and all the known calculational tools of the day relied on treating the interactions as small perturbations, and the particles as almost non-interacting. These approximations failed spectacularly when applied to the strong force. 

Besides, the cosmic ray observations posed an additional problem: in Yukawa's model the protons, neutrons and mediating pions were considered fundamental; that is, not containing any substructure. However, experiments made using the more advanced bubble chamber discovered a slew of new particles - similar in properties to the pion and proton, but with different masses. By Yukawa's token, each of these new particles were just as fundamental as the proton or the pion. By the mid 1950s, however, dozens of such particles had been discovered, none of which had, or could have, been predicted, and with no underlying theory to relate them. Fundamental physics in this era was in a state of excited disarray. 

Motivated purely by the observed properties of nuclear $\beta$-decay, and by the discovery of the neutron by Chadwick two years previously~\cite{Chadwick:1932ma}, Fermi wrote down the first model of the weak interaction in 1934~\cite{Fermi:1934hr}, which modelled $\beta$-decay as a contact interaction in which a neutron decays into a proton by emitting an electron and a neutrino. It was extremely successful in predicting observables in $\beta$-decay such as the electron energy spectrum, but soon it was realised that the theory was non-renormalisable: the infinities that had plagued early calculations in QED cropped up again, but unlike in QED, they could not be removed. Therefore it was abandoned as a fundamental theory.

An improvement came in the form of the intermediate vector boson model~\cite{Feynman:1958ty,Sudarshan:1958vf,Schwinger:1957em,Lee:1957jf,Bludman:1958zz,LeiteLopes:1958tva}, where, rather than a contact interaction, the decay was described as mediated by the exchange of vector bosons, completely analogously to the pions that mediated the strong force in Yukawa's theory, and the photons of QED. This immediately raised the problem that, unlike the photon and pion, these vector bosons had not been discovered, and had to be extremely heavy to give the correct radioactive decay rates. More startlingly, new tests of the properties of $\beta$-decay showed that the weak interaction, unlike all the other known forces, violated parity symmetry~\cite{Lee:1956qn,Wu:1957my}, i.e. it was able to distinguish between left and right. Any new theory of the weak interaction would have to be radically different in structure to accommodate these facts. 

Despite these puzzles, progress was made in the 1950s and early 1960s on two fronts. The first was the quark model of Gell-Mann and Zweig~\cite{Zweig:1964jf,GellMann:1964nj}. In an effort to classify the myriad of new particles emerging from the bubble chamber experiments, they postulated that rather than being fundamental, these particles were composed of smaller particles (Gell-Mann coined them `quarks', in a literary homage to James Joyce's \emph{Finnegan's Wake}). Requiring only 3 \emph{flavours} of quark (which were dubbed the \emph{up}, \emph{down} and \emph{strange}) as elements of the global symmetry group SU(3), that is, location-independent transformations on the quark fields by 3$\times$3 unitary matrices, the model was able to accommodate the observed mass spectrum of many of the observed mesons and baryons, and predicted new ones, several of which were duly discovered. Still, for most physicists, these quarks were no more than an idealisation, a useful bookkeeping device for classifying the bubble chamber results, and few took their existence seriously as fundamental particles. 

The other major development was made in 1954 by Yang and Mills~\cite{Yang:1954ek} who made the observation that the electromagnetic interaction could be described as resulting from a U(1) \emph{gauge} symmetry, a type of local symmetry where the fields receive a location-dependent phase transformation but the full theory is left invariant. They observed that since the proton and neutron were almost equal in mass, it was instructive to model them as elements of a 2D symmetry group, hence they modelled the proton-neutron nuclear force as originating from a local SU(2) symmetry, which they dubbed \emph{isospin}. The immediate difference from QED was that the gauge boson of this force would be self-interacting, unlike the photon. This was perfectly allowed by the experimental facets of the strong interaction, and offered intriguing insights into the possibility of constructing a quantum theory of gravity. 

The bugbear of the gauge theories suggested by Yang and Mills was how to accommodate mass. If gauge symmetry was to be an exact symmetry of the weak and strong interactions, the mediating particles of this symmetry; the gauge bosons, would have to be massless and so (it was presumed), these forces would have to be long-ranged, just like in electromagnetism, which they clearly were not. On the other hand, adding in mass terms for the gauge bosons violated the gauge symmetry explicitly, defying the point of introducing it in the first place, and, as was later shown~\cite{Veltman:1968ki,Veltman:1970nh}, just as in the case of the intermediate vector boson models, led to unacceptable physical behaviour when extrapolated to higher energies. Despite their mathematical beauty, the application of gauge theories to particle physics was stymied in the late 1950s and early 1960s by the apparent incompatibility between the symmetry patterns of the theory and the basic observation of particle masses.

The missing piece of the puzzle emerged from a completely different area of physics: superconductivity. When a conductor is cooled below a certain ultra-cool temperature, it displays the bizarre property of having almost no electrical resistance. Anderson noted~\cite{Anderson:1963pc} that this effect could be explained by the photons which transmit the electric forces inside the bulk of the superconductor effectively gaining a mass, which would break the long range electromagnetic interactions and allow currents to flow with effectively zero resistance. The U(1) electromagnetic symmetry still remained in the conductor, but it was \emph{spontaneously}, rather than explicitly, broken when it entered the superconducting phase. Anderson speculated that this phenomena might have important consequences for the application of gauge theories to elementary particle physics.

Ideas about spontaneously broken symmetries were already being tested in particle physics, but in the wrong way. It had been suggested by several authors that the known approximate symmetries of the strong interactions, such as the proton-nucleon isospin and the symmetries of Gell-Mann's quark model, could have originated from the spontaneous breaking of some exact symmetry of the system, perhaps at a higher energy. This idea suffered an apparently fatal blow when it was proved by Goldstone, Salam and Weinberg~\cite{Goldstone:1961eq,Goldstone:1962es} that any spontaneously broken symmetry would necessarily lead to the appearance of massless, interacting scalar bosons. These scalars would have easily been observed experimentally long ago and they had not, and, ignorant of the developments in superconductivity, most particle theorists regarded Goldstone's theorem as the death knell for this idea.

The importance of Anderson's observations was immediately appreciated by Higgs, however. He had been trying to find a loophole in Goldstone's theorem, and showed that if the symmetry of the system was not global but local~\cite{Higgs:1964ia}(as in the gauge theories of Yang and Mills), the unwanted massless scalars that resulted from the symmetry breaking would be absorbed by the gauge bosons, giving them a mass. In this way, gauge symmetry could be preserved whilst still giving masses to the gauge bosons, just as the U(1) electromagnetic gauge symmetry was preserved while the photons inside the superconductor gained an effective mass. The massive Yang-Mills problem had, in principle, been solved. He speculated that this mechanism could be applied to a gauge theory of the weak interaction, and would allow the vector bosons mediating the interaction to gain the mass they needed. Crucially, he also predicted the appearance of a new, massive scalar boson, which now bears his name~\cite{Higgs:1964pj}. The same ideas were published at almost exactly the same time by Brout and Englert~\cite{Englert:1964et}, and by Guralnik, Hagen and Kibble~\cite{Guralnik:1964eu}, who were attempting to give a mass to the pion in a gauge theory of the strong interaction.

The ideas of Brout, Englert, Guralnik, Hagen, Higgs and Kibble were put to use by Weinberg~\cite{Weinberg:1967tq} and Salam~\cite{Salam:1968rm} in 1967. They wrote down a gauge group with an SU(2)$\times$U(1) symmetry, as had been suggested in studies by Glashow~\cite{Glashow:1961tr} and Salam and Ward in 1961~\cite{Salam:1964ry}. Unlike the earlier papers, however, which contained explicit gauge boson mass terms, they applied the Higgs mechanism to it, and showed that the charged vector bosons mediating the weak interaction (the $W$ bosons) gained a very large mass, at least forty times the mass of the proton. Echoing Higgs' conclusions, they predicted the appearance of a massive scalar. They also predicted the appearance of a heavier still, electrically neutral vector boson. Since this was the last new particle required by the model, it was called the $Z$. After the symmetry breaking, an unbroken subgroup remained, this was identified as electromagnetism, with a massless photon. Hence, their theory unified the weak and electromagnetic interactions into one single model. This `electroweak' theory, along with the Higgs mechanism, forms one of the two pillars of what we now call the Standard Model of elementary particles.

Still, the spectre of renormalisability loomed over the electroweak theory. Though renormalisability had been demonstrated in electromagnetism by Feynman, Schwinger, Tomonaga and Dyson twenty years earlier~\cite{Feynman:1948ur,Feynman:1949zx,Feynman:1950ir,Schwinger:1948yk,Schwinger:1948yj,Schwinger:1949ra,Tomonaga:1946zz,Koba:1947rzy,Dyson:1949ha,Dyson:1952tj,Dyson:1949bp}, little progress had been made in tackling the problem for the more sophisticated Yang-Mills theories. The question then remained of whether the symmetry breaking mechanism spoiled the renormalisability of the electroweak theory, in which case it would have been little improvement over its Fermi and intermediate vector boson model predecessors.  Most theorists thought the answer to this question was yes, and so the unified electroweak theory received little attention at first. In a series of \emph{tour de force} calculations~\cite{tHooft:1972tcz}, it was shown by `t Hooft and Veltman that the gauge theories of Yang and Mills were, in fact, renormalisable. First this was demonstrated in the massless case~\cite{tHooft:1971akt}, and then in the more involved case where the vector bosons get their masses from the Higgs mechanism~\cite{tHooft:1971qjg}. After this work was published, interest in the Weinberg-Salam-Glashow model exploded, and a dedicated experimental program for uncovering the precise gauge structure of the electroweak interaction took off.

The other half of the Standard Model is the strong interaction. In contrast to the agitation that engulfed the theory community in the 1950s and early 1960s, particle accelerators evolved rapidly during this era. These developments allowed the quark model of Gell-Mann to be put to a crucial test in experiments at SLAC and MIT in 1967~\cite{Bloom:1969kc,Breidenbach:1969kd}. The experiments drew analogy with the famous Rutherford experiment of 1912, in which a beam of $\alpha$-particles were fired at a strip of gold foil, and the rare collisions in which the $\alpha$-particles rebounded from the foil provided evidence for a positively charged nucleus within the centre of the atom. The SLAC-MIT experiments instead fired a high-energy beam of electrons into a fixed proton target, to probe the putative inner structure of the proton. The results were unequivocal. High scattering rates at large angles were observed, containing events with the detection of the scattered electron and large numbers of hadrons. The sole explanation for these events was that the electron shattered the proton into its intermediate pieces, which interacted some time later to re-form into into the various types of hadrons observed in the final state.

These results gave strong weight to the existence of quarks as fundamental particles, but that relied on a strange presumption: while the scattering rates at large angles were consistent with electromagnetic interactions between the electrons and the proton's sub-components, they could only be explained if the strong interaction between the proton's inner parts was much weaker at high-energy, i.e. that when the incoming electron approached the proton, it saw the proton constituents as almost non-interacting. Then when the constituents became separated again after the collision, the strong interaction between them switched on again, binding them into the hadrons that were observed in the final state. These ideas were put on a firm mathematical footing in the parton model of Feynman and Bjorken~\cite{Bjorken:1968dy,Feynman:1969ej}. Still, the behaviour of the strong force: weak at short distances and strong at long distances (like the restoring force on a stretched elastic band) was at odds with all the known forces at the time. Electromagnetism and gravity both become weaker as the separation between the interacting objects is increased. 

A few years later, in 1973, Gross, Politzer and Wilczek~\cite{Gross:1973id,Politzer:1973fx} discovered a class of theory which exhibited precisely this property, which is known as \emph{asymptotic freedom}. They were exactly the same theories used to construct the electroweak interaction: Yang-Mills gauge theories. The strong interaction was modelled by an SU(3) gauge symmetry between the quarks. Their models assumed that the quarks had, as well as electric charge, a property of `strong' charge, which came to be known as colour charge. The SU(3) structure presumed there were 3 types of such colour.  The theory thus became known as quantum chromodynamics, in analogy with QED 30 years previously. The gauge bosons of this interaction were ultimately responsible for binding together the quarks into hadrons such as the proton, so they were dubbed gluons. 

The puzzle remained of how the strong interaction remained short-ranged, and how come the massless gluons had not been observed themselves. It was initially presumed that the SU(3) symmetry was broken so that the gluons gained mass, as in the case of the $W$s and $Z$s, but were too heavy to observe. It was soon realised that the gluons could indeed be massless, but the same phenomena which bound the quarks together was responsible for keeping the gluons confined with the hadrons. This phenomena of \emph{confinement} has intriguing consequences for how we view mass; most of the mass of hadrons such as the proton and neutron (and, by extension, most of the mass of the observable Universe) originates not from the mass of their constituents, but in the binding energy between the constituents. An analytic proof of confinement in Yang-Mills theory has not been rigorously obtained, and the Clay Mathematical Foundation continues to offer a \$1 million prize for a first-principles solution. Nonetheless, numerical studies on the lattice have demonstrated that confinement is indeed a property of QCD.

The Standard Model was beginning to take shape, but several observations were still at odds with its early predictions. Firstly, the observed rates of certain types of \emph{strangeness violating} weak decay processes were much lower than expected from the electroweak model. Secondly, it had been known since the 1950s that the weak interactions violated the so-called $CP$-symmetry, effectively a symmetry between matter and antimatter, by a small amount, but the electroweak model contained no terms which violated $CP$. The first problem was solved by Glashow, Ilopoulos and Maiani~\cite{Glashow:1970gm}, who showed that the existence of a fourth flavour of quark (they dubbed it the \emph{charm}) was able to suppress the rates by much more than the na\"ive prediction. It was quickly realised that this fourth quark could lead to many new different kinds of meson. The simplest of these would be a bound state of a charm quark and antiquark. This was promptly discovered in 1974~\cite{Augustin:1974xw,Aubert:1974js}, with a production rate and mass in excellent agreement with the Standard Model predictions. It was then realised by Kobayashi and Maskawa~\cite{Kobayashi:1973fv}, building on earlier work by Cabbibo~\cite{Cabibbo:1963yz},, that $CP$-violation could be obtained by adding in a 3rd generation (a fifth and sixth flavour) of quarks, these were called the \emph{top} and the \emph{bottom}. The bottom quark was discovered in 1978~\cite{Herb:1977ek,Innes:1977ae}, the much heavier top quark in 1995~\cite{Abe:1995hr,Abachi:1995iq}. For theoretical reasons mainly pertaining to the cancellation of anomalies, it was also presumed that a 3rd generation of leptons would exist, a heavier extension of the electron and muon. The charged lepton of this generation: the $\tau$, was discovered in experiments between 1974 and 1977~\cite{Perl:1975bf}, its neutrino was finally discovered in 2000~\cite{Kodama:2000mp}. 

\begin{table}[t]
\begin{center}
\begin{tabular}{|c|c|c|} \hline
\multicolumn{2}{|c|}{Fermions} & Bosons \\ \hline
Quarks & \( \left( \begin{array}{c} \mathrm{u} \\ \mathrm{d} \end{array} \right) \,\,\, \left( \begin{array}{c} \mathrm{c} \\ \mathrm{s} \end{array} \right) \,\,\, \left( \begin{array}{c} \mathrm{t} \\ \mathrm{b} \end{array} \right) \) & \( \begin{array}{c} \gamma \\ g \end{array} \) \\ \cline{1-2}
Leptons & \( \left( \begin{array}{c} \mathrm{e^-} \\ \nu_{e} \end{array} \right) \,\,\, \left( \begin{array}{c} \mu^- \\ \nu_{\mu} \end{array} \right) \,\,\, \left( \begin{array}{c} \tau^- \\ \nu_{\tau} \end{array} \right) \) & \( \begin{array}{c}\ W^{\pm} ~ Z^{0} \\ H \end{array} \) \\ \hline
\end{tabular}
\end{center}
\caption[Particle content of the Standard Model.]{Particle content of the Standard Model of particle physics. The mass of each matter generation increases from left to right. The fermions form doublets which differ in electric charge by 1: for the quarks this is of the form $(q_u,q_d) = (+2/3e, -1/3e)$, for the leptons this is $(q_l,q_{\nu})=(-1e,0)$.}
\label{table:sm}
\end{table}

These discoveries complete what we now call the Standard Model of particle physics: The matter content consists of six quarks and six leptons, in three generations of increasing mass. The force carries are the gauge bosons: the photon of electromagnetism, the $W$ and $Z$ of the weak interaction, and the gluon mediating the strong interaction. Underpinning all of it is the Higgs boson, which breaks the electroweak symmetry and gives mass to the $W$ and $Z$, and also to the quarks and charged leptons. This is summarised in table \ref{table:sm}. The model is strikingly minimal; with just a handful of particles it can explain all the observable matter content in the Universe, and its interactions (other than gravity). The main concern is the \emph{ad hoc} nature of its structure. It was largely cobbled together to fit experiment, and all the parameters relating to the masses of the fermions, the mixing between different generations, and the relative strengths of the weak, electromagnetic and strong interactions, as well as the mass of the Higgs boson, are not predicted by it, and have to be determined by experiment. 

Whatever aesthetic qualms one may have about its structure, however, the successes of the Standard Model as a physical theory describing Nature have been nothing short of astounding. The first coup of the Glashow-Weinberg-Salam model was the discovery of the neutral currents in 1973~\cite{Hasert:1973ff,Hasert:1973cr}, lending strong indirect evidence for the existence of the $Z$. The $W$ and $Z$ bosons were discovered outright at CERN in 1983~\cite{Arnison:1983rp,Arnison:1983mk,Banner:1983jy,Bagnaia:1983zx}, four years after Glashow, Weinberg and Salam were awarded the Nobel Prize in Physics for their electroweak theory. The gluon was discovered in three-jet events at the PETRA experiment in 1978~\cite{Brandelik:1979bd}. The predictions of the Standard Model continued to be tested throughout the 1980s in collider experiments across the world. These efforts culminated in the precision electroweak measurements at LEP and SLC~\cite{ALEPH:2005ab}, which probed the SU(3)$\times$SU(2)$\times$U(1) gauge structure to per-mille level accuracy, providing incontrovertible evidence that the Standard Model is an excellent description of Nature up to energies around 100 GeV. The last outstanding piece of the theory; the Higgs boson, was discovered in 2012~\cite{Aad:2012tfa,Chatrchyan:2012xdj}, 48 years after it was first hypothesised, and a detailed program for the precise measurements of its properties is now well underway~\cite{deFlorian:2016spz}.

Despite these triumphs, ever since the inception of the Standard Model, physicists have been looking for evidence for new physics which will take us beyond the current SM paradigm. After the unification of the weak and electromagnetic interactions into one simple model, it was natural to ask if the strong interaction could be unified with the electroweak in a single gauge group. More ambitious still were attempts to include gravitational interactions in such a framework, a so-called theory of everything. Early attempts at these models pointed out that the unification would happen at a very large energy scale, inaccessible to any conceivable future collider experiment. However, several pieces of indirect evidence point to new physics just above the electroweak scale, well within reach of current colliders.

With a centre-of-mass energy of 14 TeV, the LHC is best poised to answer the question of whether new physics beyond the SM resides at the TeV energy scale. However, there are a large number of well-motivated scenarios, and their experimental signatures are often very similar. Given the huge catalogue of measurements published by the LHC, and the possibility of different manifestations of new physics hiding in many of them, it is best to ask not ``Does my new physics model explain this particular measurement better than the Standard Model alone?" but ``Which consistent theory best describes \emph{all} the data?". This has led to renewed interest in being able to describe the data in a model-independent way. Effective field theory provides such a description.

Since its discovery in 1995, the top quark continues to mystify physicists with its properties. As it is the only fermion with a mass around the electoweak scale ($m_t$ = 173 \gev), and as the precise mechanism which breaks the electroweak symmetry is unexplained in the Standard Model, the top quark usually plays a special role in theories of physics beyond the Standard Model.  The top quark sector is thus one of the many well-motivated places to look for the effects of potential new physics, as only now are its properties beginning to be scrutinised with high precision. The language of effective field theory provides a powerful, systematic way of doing this. This is the primary topic of this thesis.

The thesis is structured as follows. In chapter 1, I will discuss the unique role of the top quark in the Standard Model. In chapter 2, I will outline some of the main hints for physics beyond the Standard Model and some well-studied new physics models, and their relevance for top quark phenomenology. I will then describe the formulation of the Standard Model as a low energy effective theory where all the ultraviolet degrees of freedom have been integrated out, and the sector of this effective theory that can be studied with top measurements from hadron colliders. In chapter 3 I will discuss a global fit of the top quark sector of the Standard Model EFT to data from the LHC. Chapter 4 is concerned with refinements of the analysis of chapter 3, such as how the increase in LHC energy from 8 to 13 \tev can be best exploited; how `boosted observables' that draw on high-momentum transfer final states can improve the fit results, and how proposed future lepton colliders can complement results extracted from hadron collider measurements. In chapter 5 we move away from effective theory, and study how the performance of certain algorithms for reconstructing `boosted' final states may be augmented by recent developments in machine learning, before summarising the conclusions of this thesis.

The motivations for this work are thus threefold: 1. With the abundant data from the LHC, top properties can be examined with precision for the first time. 2. The top quark continues to be a sensible place to search for new physics. 3. Effective field theories are a powerful tool for confronting new physics models with data in a systematic way. It is worth remembering that the story of the Standard Model began with an effective theory when Fermi wrote down his model of nuclear $\beta$-decay in 1935. Perhaps it would be fitting if the story ended with one as well.

\newpage
\section{The top quark in the Standard Model}

\subsection{Introduction}
The top quark was discovered in 1995 by the CDF~\cite{Abe:1995hr} and \dzero~\cite{Abachi:1995iq} experiments at the Tevatron. Still, only recently have its couplings begun to be measured with sub-10\% level accuracy, thanks to the much higher production rates at the LHC and the large integrated luminosity collected over the total lifetime of the Tevatron. The role that the top quark might play in specific realisations of electroweak symmetry breaking is just beginning to be tested. Before we can turn to these questions, however, we must summarise the unique role of the top quark within the Standard Model. This is the subject of this chapter.

This introductory chapter is structured as follows. In section \ref{sec:sm} I discuss the building blocks of the Standard Model of particle physics: the unified \su2L\x\U1Y electroweak theory; the \su3C gauge theory of the strong interaction known as quantum chromodynamics and the role of spontaneously broken local gauge symmetry via the Higgs mechanism, before discussing the free parameters of the Standard Model. In section \ref{sec:colliders} I discuss some generalities about hadron collider phenomenology, including the main theoretical uncertainties that crop up in scattering calculations. In section \ref{sec:topphys} I discuss the main production mechanisms for top quarks at hadron colliders, and some properties of top production and decay, before summarising in section \ref{sec:conc_ch1}.
 
\subsection{The Standard Model of Particle Physics}
\label{sec:sm}

The Standard Model has three main ingredients. Firstly, there is quantum chromodynamics (QCD)~\cite{Gross:1973id,Politzer:1973fx,Fritzsch:1973pi}: the theory of the strong interaction between `coloured' quarks and gluons (the mediators of this interaction), described by a gauge group with a local \su3C symmetry. Secondly, the electroweak theory described by the model of Glashow, Salam and Weinberg~\cite{Glashow:1961tr,Salam:1968rm,Weinberg:1967tq}, which unifies the electromagnetic and weak interactions of quarks and leptons under the gauge group \su2L\x \U1Y; its charges are weak isospin $L$ and weak hypercharge $Y$. Finally, there is the celebrated Higgs mechanism~\cite{Higgs:1964pj,Englert:1964et,Guralnik:1964eu}: a complex scalar field doublet (with four degrees of freedom) whose potential acquires a non-zero minimum which spontaneously breaks the electroweak symmetry into a U(1) group describing QED; its charge is the familiar electromagnetic coupling. Three of the four degrees of freedom form the longitudinal polarisation states of the $W^{\pm}$ and $Z^0$ bosons, giving mass to these particles and thus being responsible for the phenomena of nuclear $\beta$-decay and other weak processes. The remaining one forms a massive scalar particle: the Higgs boson. The Higgs mechanism is also responsible for giving mass to the quarks and charged leptons through a Yukawa-type interaction. 

\subsubsection{Before electroweak symmetry breaking}
The Standard Model before electroweak symmetry breaking has two types of field:
\begin{mydescription}

\item[Matter fields $\psi$:] 
Since the weak interactions are known to violate parity, the matter fields are constructed out of left-handed and right-handed (\textit{chiral}) fermions.
\begin{equation}
\psi = \psi_L + \psi_R
\end{equation}
where
\begin{equation}
\begin{split}
\psi_L &= P_L \psi \hspace{10pt} \text{with} \hspace{10pt} P_L=\frac{(1-\gamma_5)}{2} \\
\psi_R &= P_R \psi \hspace{10pt} \text{with} \hspace{10pt} P_R=\frac{(1+\gamma_5)}{2}.
\end{split}
\end{equation}
The operators $P_{L,R}$ project out the chiral states of each fermion, whose kinetic terms in the Dirac Lagrangian \lag{Dirac} can thus be decomposed:
\begin{equation}
\bar{\psi}\gamma^\mu\partial_\mu \psi = \bar{\psi}_L\gamma^\mu\partial_\mu \psi_L + \bar{\psi}_R\gamma^\mu\partial_\mu \psi_R
\end{equation}
 i.e. massless fermions decouple into chiral components. For the SM we have three generations of left-handed and right-handed spin-$\frac{1}{2}$ fermions which can be categorised into quarks and leptons. To reproduce the chiral structure of the weak interaction, left-handed fermions are in weak-isospin doublets, and right-handed fermions fall into weak-isospin singlets. 
\begin{equation}
\begin{split}
Q_1  &= \left(\begin{array}{c} u \\ d \end{array}\right)_L \quad , \quad u_{R_{1}} = u_R \quad , \quad d_{R_{1}} = d_R \quad , \quad  L_1  = \left(\begin{array}{c} \nu_e \\ e^- \end{array}\right)_L \quad , \quad e_{R_{1}} = e_R^- \\
Q_2  &= \left(\begin{array}{c} c \\ s \end{array}\right)_L \quad , \quad  u_{R_{2}} = c_R \quad , \quad d_{R_{2}} =  s_R \quad , \quad  L_2  = \left(\begin{array}{c} \nu_\mu \\ \mu^-\end{array}\right)_L \quad , \quad e_{R_{2}} =  \mu_R^-  \\
Q_3  &= \left(\begin{array}{c} t \\ b \end{array}\right)_L \quad , \quad u_{R_{3}} =  t_R \quad , \quad d_{R_{3}} =  b_R \quad , \quad  L_3  = \left(\begin{array}{c} \nu_\tau \\ \tau^-\end{array}\right)_L \quad , \quad e_{R_{3}} =  \tau_R^-  .
\end{split}
\end{equation}
Members of each doublet have 3rd component of weak isospin $I^3_f = \pm \frac{1}{2}$, which is related to \U1Y hypercharge  $Y_f$ and electric charge  $Q_f$ by 
\begin{equation}
Y_f = Q_f - I^3_f,
\end{equation}
where
\begin{equation}
\quad Y_{L_{i}} = -\frac{1}{2}, \quad Y_{e_{R}} =  -1, \quad Y_{Q_{i}} = +\frac{1}
{6}, \quad Y_{u_{R_{i}}} =  +\frac{2}{3},  \quad Y_{d_{R_{i}}} =  -\frac{1}{3} .
\end{equation}
These hypercharge assignments ensure the fermions have the correct electric charge: isodoublets  differ in electric charge by 1: for the quarks this is of the form $(q_u,q_d) = (+2/3e, -1/3e)$, for the leptons this is $(q_l,q_{\nu})=(-1e,0)$. The quark fields are charged under \su3C, i.e. each quark appears as a triplet of 3 colours, whereas the leptons are singlets. This important feature ensures that the anomaly cancellation condition
\begin{equation}
\sum_{\substack{f}} Y_f= 0  ,
\end{equation}
where the sum runs over all fermions in a generation, is satisfied. Hence gauge-invariance is not spoiled by radiative corrections and the theory remains renormalisable. 

\item[Gauge fields $V_\mu$:] These are the spin-1 bosons that mediate the electroweak and strong interactions. The \su2L symmetry of the electroweak sector gives rise to 3 vector fields $W_{\mu}^{1,2,3}$ corresponding to the generators $T^I$ ($I=1,2,3$), expressed in terms of the Pauli matrices $\tau^I$ as\footnote{Throughout this thesis, weak isospin indices are denoted by lowercase Roman letters $\{i,j,k\ldots\} \in \{1,2\}$, while \su3C and \su2L adjoint indices are denoted by uppercase Roman: $\{A,B,C\ldots\} \in \{1\ldots8\}$ and $\{I,J,K\ldots\} \in \{1,2,3\}$.}
\begin{equation}
T^I = \frac{1}{2}\tau^I \quad; \quad \tau^1 = \left(\begin{array}{cc}0 & 1 \\ 1 & 0 \end{array} \right) \quad , \quad \tau^2 = \left(\begin{array}{cc}0 & -i \\ i & 0 \end{array} \right) \quad , \quad \tau^3 = \left(\begin{array}{cc}1 & 0 \\ 0 & -1 \end{array} \right) ,
\end{equation}
which satisfy the commutation relations
\begin{equation}
[T^I,T^J] = i\epsilon^{IJK}T^K,
\end{equation}
where $\epsilon^{IJK}$ is the antisymmetric tensor. The \U1Y symmetry corresponds to a vector field $B_{\mu}$, which has the unique generator $Y$. The strong sector has an \su3C symmetry, corresponding to 8 gluon fields $G_{\mu}^{1,...,8}$, expressed in terms of the $3\times3$ Gell-Mann matrices $T^A$ which we do not write explicitly here and which satisfy
\begin{equation}
[T^A,T^B] = if^{ABC}T^C \hspace{10pt} \text{and} \hspace{10pt} \text{Tr}[T^AT^B] = \frac{1}{2}\delta^{AB},
\end{equation}
where $f^{ABC}$ denote the \su3C structure constants. From these fields one may construct gauge invariant field strength tensors
\begin{equation}
\begin{split}
G^A_{\mu \nu} &= \partial_{\mu} G^A_{\nu} - \partial_{\nu} G^A_{\mu} + g_sf^{ABC}G^B_{\mu}G^C_{\nu} \\
W^I_{\mu \nu} &= \partial_{\mu} W^I_{\nu} - \partial_{\nu} W^I_{\mu} + g\epsilon^{IJK}W^J_{\mu} W^K_{\nu} \\
B_{\mu \nu} &= \partial_{\mu} B_{\nu} - \partial_{\nu} B_{\mu} ,
\end{split}
\end{equation}
 where $g_s$ and $g$ respectively denote the \su3C and \su2L coupling constants. The \U1Y coupling is denoted as $g'$. 
 
\end{mydescription}  

\noindent 
To couple the matter fields to the gauge fields, we replace the ordinary derivative $\partial_\mu$ with the gauge covariant derivative $D_\mu$:
\begin{equation}
D_\mu = \partial_\mu + ig_sT^AG^A_\mu + igT^IW^I_\mu + ig'\frac{Y}{2}B_\mu .
\end{equation}
which leads to matter-gauge couplings of the form
\begin{equation}
\text{fermion-gauge couplings} \hspace{10pt}:\hspace{10pt} g_i\bar{\psi}V_\mu\gamma^\mu \psi 
\end{equation}
In addition to matter-gauge interactions, the non-Abelian nature of the SM leads to self-interactions among the gauge bosons, which we can generically class into 3-point and 4-point couplings:
\begin{equation}
\begin{split}
\text{3-point couplings} &\hspace{10pt}:\hspace{10pt} -g_i\text{Tr}(\partial_{\mu} V_{\nu} - \partial_{\nu} V_{\mu})[V_\mu,V_\nu] \\
\text{4-point couplings} &\hspace{10pt}:\hspace{10pt} g_i^2\text{Tr}[V_\mu,V_\nu]^2
\end{split}
\end{equation}
where $g_i \in \{g_s,g,g'\}$. The Standard Model Lagrangian at this point consists only of kinetic terms for massless fermions and gauge bosons: 
\begin{equation}
\lag{SM} = \lag{gauge} + \lag{fermion}  ,
\end{equation}
where
\begin{equation}
\begin{split}
\lag{gauge} &= -\frac{1}{4}G^A_{\mu\nu}G^{A,\mu\nu} -\frac{1}{4}W^I_{\mu\nu}W^{I,\mu\nu}-\frac{1}{4}B_{\mu\nu}B^{\mu\nu} \\
\lag{fermion} &= i\bar{L}_iD_\mu\gamma^{\mu} L_{i} + i\bar{e}_{Ri}D_\mu\gamma^{\mu} e_{Ri} +  i\bar{Q}_iD_\mu\gamma^{\mu} Q_{i} +  i\bar{u}_{Ri}D_\mu\gamma^{\mu} u_{Ri} +  i\bar{d}_{Ri}D_\mu\gamma^{\mu} d_{Ri} .
\label{eqn:lsm_nomass}
\end{split}
\end{equation}
It is manifestly invariant (by construction) under local \su3C\x\su2L\x \U1Y gauge transformations. For instance, under an \su2L transformation,
\begin{equation}
\begin{split}
e_{Ri} \to e'_{Ri} &= e_{Ri} \\
L_i \to L'_i &= e^{-i\omega^I(x)T^I}L_i.
\end{split}
\end{equation}
So we see that the \su2L singlets $\psi_R$ are trivially \su2L invariant and therefore do not couple to the corresponding gauge fields $W_{\mu}^{1,2,3}$. So far, the theory is self-consistent. When we try to include particle masses, however, we run into two problems:

\begin{enumerate}
\item \textbf{Fermion Masses:} Explicit fermion masses take the form $\lag{mass}= -m\bar{\psi}{\psi}$, which when decomposed into chiral components become:
\begin{equation}
  m\bar{\psi}{\psi} = m\bar{\psi}\left(\frac{(1-\gamma_5)}{2}+\frac{(1+\gamma_5)}{2}\right)\psi = m(\bar{\psi}_R\psi_L+\bar{\psi}_L\psi_R)
\end{equation}
which is not \su2L invariant, as it mixes left-handed and right-handed fermion components.

\item \textbf{Gauge boson masses:} The observed short-range of the weak interaction $\sim$ 0.1 fm tells us that the vector bosons mediating to the weak interaction have masses of order $\sim$ 10 GeV. However, when we include explicit mass terms in the Lagrangian, it is easy to see they are not gauge invariant. Using the simple $\text{U(1)}$ case of QED with a massive photon as an example:
\begin{equation}
\frac{1}{2}M^2_A A_\mu A^\mu \to \frac{1}{2}M^2_A (A_\mu -\frac{1}{e}\partial_\mu \alpha) (A_\mu -\frac{1}{e}\partial^\mu \alpha) \neq \frac{1}{2}M^2_A A_\mu A^\mu
\end{equation}

\end{enumerate}

\begin{figure}[t!]
\begin{center}
\includegraphics[width=0.375\textwidth]{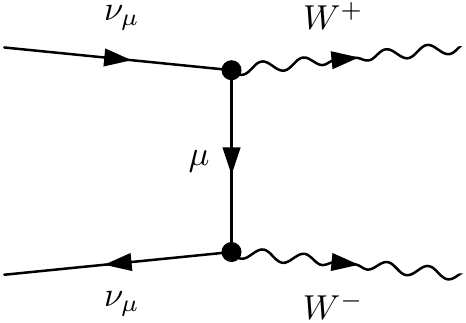}
 \hspace*{0.1\textwidth}
\includegraphics[width=0.375\textwidth]{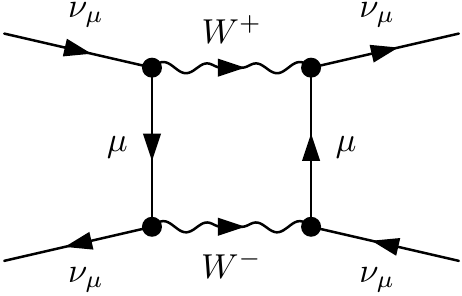}
\end{center}
\caption[Neutrino scattering in the intermediate vector boson model.]{Left: The tree-level process $\nu_{\mu}\nu_{\mu}\to W^+ W^-$ in the intermediate vector boson model, which violates unitarity in the high-energy limit. Right: One-loop neutrino scattering in the IVB model; a non-renormalisable interaction. }
\label{fig:neutrino_scat}
\end{figure}

\noindent
To appreciate the problems caused by explicit breaking of gauge invariance, consider the propagator for a generic massive vector boson.
\begin{equation}
\frac{i}{p^2-M^2}\left(-g^{\mu\nu}+\frac{p^\mu p^\nu}{M^2}\right)
\end{equation}
and the weak-interaction process $\nu_\mu \bar{\nu}_\mu \to W^+ W^-$, the leading order Feynman diagram for which is sketched on the left-hand side of Fig. \ref{fig:neutrino_scat}. 

Although this process would be rather difficult to implement experimentally, its cross-section can be straightforwardly calculated. In the high-energy limit, the $p^\mu p^\nu/M^2$ term, corresponding to the longitudinal $W$ polarisation states, will dominate contributions to the cross-section. In fact, one finds for this process
\begin{equation}
\sigma \sim G^2_F E^2
\label{eqn:nunuww}
\end{equation}
where $G_F$ is a coupling constant which must have dimensions of (energy)$^{-2}$: i.e. the cross-section grows quadratically with energy. We can decompose the scattering amplitude $A$ for this process into partial waves $f_\ell$ of orbital angular momentum $\ell$:
 \begin{equation}
 A = 16\pi\sum\limits_{\ell=0}^\infty (2\ell+1)P_\ell(\cos\theta) f_\ell
 \end{equation}
 where $P_\ell$ are the Legendre polynomials and $\theta$ is the scattering angle. Noting that for 2 $\to$ 2 processes with massless external legs the cross-section is given by $d\sigma/d\Omega = |A|^2/64\pi^2s$, with $d\Omega = 2\pi d\cos\theta$, the total cross-section is then
  \begin{equation}
 \begin{split}
 \sigma &= \frac{8\pi}{s}\sum\limits_{\ell=0}^\infty\sum\limits_{\ell'=0}^\infty (2\ell+1)(2\ell'+1)f_\ell f_{\ell^\prime}\int\limits_{-1}^{1} d\cos\theta P_\ell(\cos\theta) P_{\ell'}(\cos\theta) \\
 	 &= \frac{16\pi}{s}\sum\limits_{\ell=0}^\infty (2\ell+1)|f_\ell|^2
 \end{split}
 \end{equation}
where the orthogonality condition $\int d\cos\theta P_\ell P_{\ell'}=2\delta_{\ell\ell'}/2\ell+1$ was used. From the optical theorem (a simple consequence of unitarity), $\sigma$ is equal to the imaginary part of the forward ($\theta=0$) scattering amplitude~\cite{Schwartz:2013pla}, so that, at each order in the partial wave expansion, unitarity requires:
\begin{equation} 
\begin{split}
|f_\ell|^2 = \text{Im}(f_\ell) &\Rightarrow [\text{Re}(f_\ell)]^2+[\text{Im}(f_\ell)]^2 = \text{Im}(f_\ell) \\
                         &\Rightarrow [\text{Re}(f_\ell)]^2+ [\text{Im}(f_\ell)-\frac{1}{2}]^2 = \frac{1}{4}
\end{split}
\end{equation}
which is just the equation of a circle in the $[\text{Re}(f_\ell),\text{Im}(f_\ell)]$ plane, of radius $\frac{1}{2}$ centred at $[0,\frac{1}{2}]$. Hence $|\text{Im}(f_\ell)| \leq 1$, and the cross-section in each partial wave projection has the unitarity bound
\begin{equation}
\sigma \leq \frac{16\pi (2\ell+1)}{s}.
\label{eqn:unitarity}
\end{equation}
Comparing Eq.~\eqref{eqn:unitarity} with Eq.~\eqref{eqn:nunuww}, we see that unitarity is violated at some finite energy. Plugging the numbers in we find this is around $E \sim$ 1 TeV~\cite{Lee:1977eg,Lee:1977yc,Djouadi:2005gi}, indicating that beyond this energy the theory is perturbatively not well-defined. 

Since this is only a perturbative statement, one might well argue that the theory may still be consistent if strong dynamics take over in this regime. However, we could instead consider the case where the $W$ bosons appear as virtual particles, e.g. in the one-loop process $\nu_\mu \bar{\nu}_\mu \to \nu_\mu \bar{\nu_\mu}$, as depicted on the right-hand side of Fig. \ref{fig:neutrino_scat}, in which the longitudinal states $W_L$ lead to quadratically divergent loop-momenta. Renormalising this divergence would require the inclusion of a counterterm corresponding to a four-neutrino vertex. However, no such vertex exists in the theory. Hence, the theory is non-renormalisable, and cannot be expected to make predictions for arbitrarily high-energies. 

To summarise, it seems there is a fundamental conflict between constructing renormalisable gauge theories for particle physics, and allowing particles in those theories to have mass. If there was a way to generate mass \textit{dynamically}, i.e. not through explicit mass terms but through a gauge-invariant interaction between fields, perhaps the gauge principle can be saved. The Higgs mechanism provides such an interaction.

\subsubsection{The Higgs mechanism}
\begin{figure}[t!]
\begin{center}
\includegraphics[width=\textwidth,height=7cm]{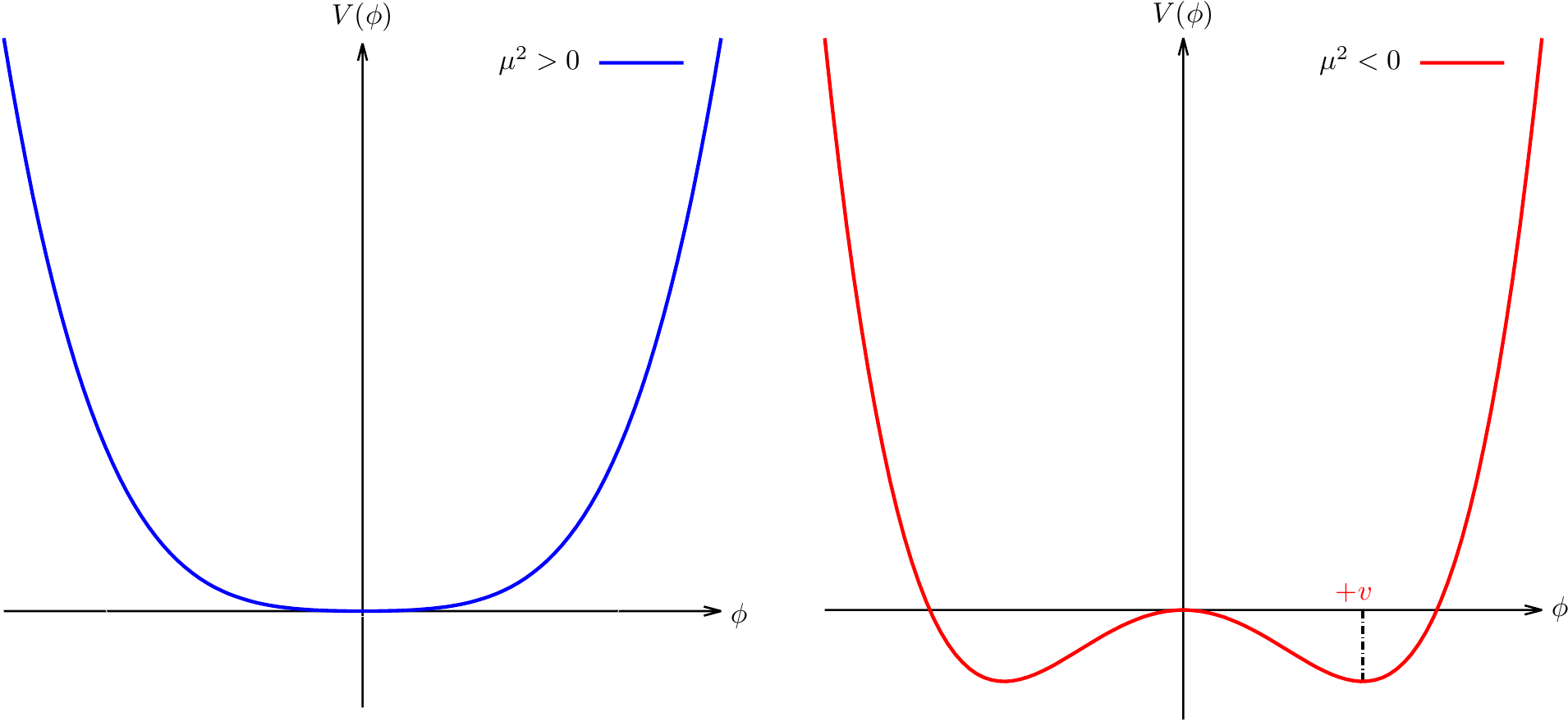}
\end{center}
\caption[The Higgs potential]{The potential $V(\phi)$ of Eq.~\eqref{eqn:phi4} in the cases $\mu^2>0$ (left) and $\mu^2<0$ (right). }
\label{fig:phi_4}
\end{figure}

As a warmup, we consider the example of a real scalar field $\phi$ with the Lagrangian
\begin{equation}
\lag{} = \frac{1}{2}\partial_\mu\phi\partial^\mu\phi - V(\phi) \hspace{10pt}\text{where}\hspace{10pt} V(\phi) = \frac{1}{2}\mu^2\phi^2+\frac{1}{4}\lambda\phi^4
\label{eqn:phi4}
\end{equation}
\lag{} is invariant under reflections $\phi \to -\phi$. For \lag{} to describe any physical system, $\lambda$ must be positive-semidefinite, otherwise the potential is unbounded from below. $\mu^2$ can take positive or negative values, however. For $\mu^2>0$ the minimum of the potential (in quantum field theoretic terms, its \textit{vacuum expectation value} $\langle 0 | \phi | 0 \rangle$) is located at the origin $\phi_0 = 0$.  In this case \lag{} is just the Lagrangian of a spin-zero particle of mass $\mu$, as shown in the left-hand side of Fig. \ref{fig:phi_4}.
However, when $\mu^2<0$ this no longer represents the Lagrangian of a particle of mass $\mu$. The minima of the potential are now at 
\begin{equation}
\langle 0 | \phi | 0 \rangle = \phi_0 = \pm \sqrt{-\frac{\mu^2}{\lambda}} \equiv \pm v
\end{equation}
The field has picked up a non-zero vacuum expectation value $v$, as highlighted in the right-hand side of Fig \ref{fig:phi_4}. To extract the interactions of this theory, we expand the field around $\phi = v$. Defining $\phi = v + \sigma$, \lag{} is, up to constant terms
\begin{equation}
\lag{} = \frac{1}{2}\partial_\mu\sigma\partial^\mu\sigma + \mu^2\sigma^2 - \sqrt{-\mu^2\lambda}\sigma^3-\frac{1}{4}\lambda^4
\end{equation}

The theory now describes a new scalar field of mass $m^2_\sigma = -2\mu^2$, with trilinear and quartic self-interactions. The $\sigma^3$ term breaks the original reflection symmetry; that is, a symmetry of the Lagrangian is no longer a symmetry of the vacuum, it has been \textit{spontaneously broken}.

The next simplest example of spontaneously broken symmetry is that of four scalar fields (equivalently a complex doublet of scalars) with Lagrangian
\begin{equation}
\lag{} = \frac{1}{2}\partial_\mu\phi_i\partial^\mu\phi_i - \frac{1}{2}\mu^2\phi_i\phi_i+\frac{1}{4}\lambda(\phi_i\phi_i)^2
\end{equation}
which is invariant under the transformation $\phi_i = R_{ij}\phi_j$ where $R_{ij}$ are 4-dimensional orthogonal matrices, i.e. transformations under the rotation group in four dimensions, O(4). Setting $\mu^2<0$ and expanding around the minima at $\phi_i$ = $(0,0,0,v)$, where $v^2=\frac{\mu^2}{\lambda}$, \lag{} becomes
\begin{equation}
\begin{split}
\lag{} &= \frac{1}{2}\partial_\mu\sigma\partial^\mu\sigma + \mu^2\sigma^2 - \sqrt{-\mu^2\lambda}\sigma^3-\frac{1}{4}\lambda^4 \\
&+   \frac{1}{2}\partial_\mu\pi_i\partial^\mu\pi_i - \frac{1}{4}\lambda(\pi_i\pi_i)^2 -\lambda v \pi_i\pi_i\sigma -\frac{1}{2}\pi_i\pi_i\sigma^2 ,
\end{split}
\end{equation}
where $i$ now runs from 1 to 3, and $\sigma = \phi_4 - v$, $\pi_i = \phi_i$. Again, a massive $\sigma$ boson with mass $m^2_\sigma = -2\mu^2$ has appeared, but so have three massless pions, among which there is a residual O(3) symmetry. 
This is an example of a general property of spontaneously broken continuous symmetries known as Goldstone's theorem~\cite{Goldstone:1962es}, which can be stated as follows:

\textit{For a continuous symmetry group $\mathbb{G}$ spontaneously broken down to a subgroup $\mathbb{H}$, the number of broken generators is equal to the number of massless scalars that appear in the theory.} 

The O($N$) group has $N(N-1)/2$ generators, so O($N-1$) has $(N-1)(N-2)/2$ and $N-1$ Goldstone bosons appear (the above example is the case of $N=4$). 

The above example applied to global symmetries, but if the mechanism is extendable to local (gauge) symmetries, it would provide a viable way of giving mass to the vector bosons of the weak interaction.
We begin with the case of an Abelian U(1) symmetry, with the Lagrangian
\begin{equation}
\lag{} = -\frac{1}{4}F_{\mu\nu}F^{\mu\nu}+ (D_{\mu}\phi)^*(D^{\mu}\phi) - V(\phi) \hspace{10pt}\text{where}\hspace{10pt} V(\phi) = \mu^2\phi^*\phi + \lambda(\phi^*\phi)^2,
\end{equation}
where $D_{\mu} = \partial_\mu+igA_{\mu}$ is the usual covariant derivative. This is invariant under local U(1) transformations:
\begin{equation}
\phi(x) \to e^{i\theta(x)}\phi(x) , \hspace{10pt} A_{\mu} \to A_{\mu}+\frac{1}{g}\partial_\mu\theta(x) .
\end{equation}
The case $\mu^2>0$ corresponds to scalar QED: interactions between a charged scalar of mass $\mu$ and a massless vector boson, with an additional four-point scalar self-interaction. For $\mu^2<0$, $\phi$ as usual obtains a non-zero vev, and the potential is minimised at 

\begin{equation}
\langle 0 | \phi | 0 \rangle = \sqrt{-\frac{\mu^2}{2\lambda}} \equiv \frac{v}{\sqrt{2}} .
\end{equation}
Expanding the potential around the vev,
\begin{equation}
\phi(x) = \frac{1}{\sqrt{2}}(v+\phi_1(x)+i\phi_2(x)) \equiv \frac{1}{\sqrt{2}}(v+H(x))e^{i\chi(x)/v} ,
\label{eqn:phi_comp}
\end{equation}
the Lagrangian describing the vacuum state is now
\begin{equation}
\begin{split}
\lag{} &= -\frac{1}{4}F_{\mu\nu}F^{\mu\nu}+(\partial_\mu-igA_{\mu})\phi^*(\partial^\mu+igA^{\mu})\phi - \mu^2\phi^*\phi - \lambda(\phi^*\phi)^2 \\
		&= -\frac{1}{4}F_{\mu\nu}F^{\mu\nu}+\frac{1}{2}(\partial_\mu\phi_1)^2 + \frac{1}{2}(\partial_\mu\phi_2)^2 +\mu^2\phi^2_1+\frac{1}{2}g^2v^2A_{\mu}A^{\mu} +gvA_{\mu}\partial^{\mu}\chi \\
		&+ \text{(interaction terms)}.
\end{split}
\end{equation}		
The photon has obtained a mass $M_A^2=g^2v^2$, the scalar particle $\phi_1$ has a mass $M_{\phi_1}^2=-2\mu^2$. The $\phi_2$ has apparently emerged as the Goldstone boson of this symmetry breaking.

However, \lag{} now contains the bilinear term $gvA_{\mu}\partial^{\mu}\chi$, which neither corresponds to an interaction or a field strength. The symmetry breaking has also apparently created an extra degree of freedom. Before, there were four: two in the massless photon and two in the complex field $\phi$. Now there appear to be five: three for the massive photon, and one each for $\phi_1$ and $\phi_2$. The resolution of this paradox lies in the fact that we are free to make a gauge transformation:
\begin{equation} 
\phi(x) \to e^{-i\chi(x)/v}\phi(x), \hspace{10pt} A_{\mu} \to A_{\mu}+\frac{1}{gv}\partial_\mu\chi(x)
\end{equation}
which removes all $\chi(x)$ terms from the Lagrangian. Counting degrees of freedom, we see the massless photon has absorbed the Goldstone boson, and gained mass: it has a longitudinal polarisation state. The U(1) symmetry has been spontaneously broken, leading to a massive vector boson and the appearance of a massive scalar boson. This is the Higgs mechanism.

\subsubsection{The Higgs mechanism in the Standard Model}
To apply the Higgs mechanism to the Standard Model, we need to generate mass for the $W^\pm$ and $Z^0$ bosons, whilst keeping the photon massless. So the \su2L\x \U1Y electroweak symmetry should be broken to a U(1) subgroup describing electromagnetism. This means that at least 3 degrees of freedom are needed.  We also want to introduce a gauge-invariant interaction that gives masses to fermions without mixing chiral components. The simplest object that satisfies these criteria is an SU(2) doublet of scalar fields $\phi$
\begin{equation}
\Phi = \left(\begin{array}{c}\phi^+ \\ \phi^0 \end{array}\right) ,
\end{equation}
where the superscript denotes the electric charge in each component. We add the usual $\phi^4$ Lagrangian \lag{Higgs} to the SM Lagrangian in Eq.~\eqref{eqn:lsm_nomass}
\begin{equation}
\lag{Higgs} =  (D_{\mu}\Phi)^\dagger(D^{\mu}\Phi) - V(\Phi) \hspace{10pt}\text{where}\hspace{10pt} V(\Phi) = \mu^2\Phi^\dagger\Phi + \lambda(\Phi^\dagger\Phi)^2 .
\end{equation}
$V(\Phi)$ gets a minimum at $\Phi^\dagger\Phi = \mu^2/2\lambda$, which we take to be in the neutral direction to preserve $ \text{U(1)}_{e.m}$
\begin{equation}
\langle \Phi \rangle_0 = \langle 0 | \Phi | 0 \rangle = \frac{1}{\sqrt{2}}\left(\begin{array}{c} 0  \\ v \end{array}\right) ,
\end{equation}
with $v=\sqrt{-\frac{\mu^2}{\lambda}}$. Expanding around the vev as before:
\begin{equation}
\Phi = \frac{e^{i\theta^I(x)\tau^I/v}} {\sqrt{2}}\left(\begin{array}{c} 0  \\ v + h(x) \end{array}\right) ,
\end{equation}
and expanding out the covariant derivative term in \lag{Higgs}, we have
%
\begin{align}
(D_{\mu}\Phi)^\dagger(D^{\mu}\Phi) &=\left| \left( \partial_\mu + ig\frac{\tau^I}{2}W^I_\mu + ig'\frac{Y}{2}B_\mu \right) \right|^2 \nonumber \\
&= \frac{1}{2} \left| \left( \begin{array}{cc} \partial_\mu + \frac{i}{2}(gW^3_\mu + g'\frac{Y}{2}B_\mu) & i\frac{g}{2}(W^1_\mu - iW^2_\mu) \\ i\frac{g}{2}(W^1_\mu + iW^2_\mu)  &  \partial_\mu - \frac{i}{2}(gW^3_\mu - g'\frac{Y}{2}B_\mu) \end{array} \right) \left(\begin{array}{c} 0  \\ v + h \end{array}\right) \right |^2 \nonumber \\
&=\frac{1}{2}(\partial_\mu h)^2+\frac{1}{8}(v+h)^2|W^1_\mu+iW^2_\mu|^2+\frac{1}{8}(v+h)^2|gW^3_\mu-g'B_\mu|^2 \nonumber \\
&+ \text{(interaction terms)}.
\label{eqn:covdevsq}
\end{align}
Eq.~\eqref{eqn:covdevsq} shows that there are terms mixing the fields $W^3_\mu$ and $B_\mu$. The physical bosons must be superpositions of these fields such that there are no mixing terms. The physical fields can be obtained by performing the rotation
\begin{equation}
\left(\begin{array}{c} Z_\mu \\ A_\mu \end{array}\right) = \left(\begin{array}{cc} \cos\theta_W & -\sin\theta_W \\ \sin\theta_W & \cos\theta_W\end{array}\right) \left(\begin{array}{c} W^3_\mu \\ B_\mu \end{array}\right),
\end{equation}
where the weak mixing/Weinberg angle 
\begin{equation}
\tan\theta_W \equiv \frac{g'}{g},
\end{equation}
has been introduced. With this, Eq.~\eqref{eqn:covdevsq} becomes 
\begin{equation}
(D_{\mu}\Phi)^\dagger(D^{\mu}\Phi) = \frac{1}{2}(\partial_\mu h)^2 + \frac{g^2v^2}{4}W^+_\mu W^{-\mu} + \frac{g^2v^2}{8\cos^2\theta_W}Z_{\mu}Z^{\mu} + 0A_{\mu}A^{\mu},
\end{equation}
where $W^\pm = (W^1\mp W^2)/\sqrt{2}$. The $W$ and $Z$ bosons have acquired masses
\begin{equation}
M_W = \frac{1}{2}gv, \quad M_Z = \frac{1}{2} \frac{gv}{\cos\theta_W}
\end{equation}
i.e. there is a mass relation
\begin{equation}
M_Z = \rho M_W\cos\theta.  
\end{equation}
The parameter $\rho$ has been introduced: at tree-level $\rho = 1$ but radiative quantum effects give corrections to this relation. The $\text{SU(2)}$ gauge structure of the electroweak theory ensures that these corrections are small, however; a feature known as \textit{custodial symmetry}~\cite{Sikivie:1980hm}. Different choices of representations for the Higgs field (e.g. an $\text{SU(2)}$ triplet) would not protect the $\rho$ parameter from large corrections.
The linear combination $A$ has remained massless, so is to be identified with the photon. To see that a  $\text{U(1)}$ subgroup remains unbroken, consider the symmetry associated with the generator 
\begin{equation}
Q \equiv  T^3 +\frac{Y}{2} = \left(\begin{array}{cc} 1 & 0 \\ 0 & 0 \end{array}\right)
\end{equation}
where we have included the explicit representation of $Y$ as a 2$\times$2 identity matrix. Then
\begin{equation}
Q | 0 \rangle \sim \left( \begin{array}{cc} 1 & 0 \\ 0 & 0 \end{array} \right)  \left(\begin{array}{c} 0  \\ v + h \end{array}\right) = 0
\end{equation}
i.e. the symmetry associated with this generator is unbroken by the vacuum, so the corresponding field $gW^3+g^\prime B \equiv A$ is massless. We can similarly expand the potential terms around the vacuum:
\begin{equation}
\begin{split}
V(\Phi) &= \frac{\mu^2}{2}(0, v+h)  \left(\begin{array}{c} 0  \\ v + h \end{array}\right) +\frac{\lambda}{4} \left|(0, v+h)  \left(\begin{array}{c} 0  \\ v + h \end{array}\right)\right|^2 \\
&= -\lambda^2v^2-\lambda v h^3 -\frac{\lambda}{4}h^4 + \text{(constants)}.
\end{split}
\end{equation}
So the scalar particle has gained a mass $m^2_h = -2\mu^2 = 2\lambda v^2$, and has trilinear and quartic self-interactions. This is the Higgs boson.
Next we turn to the issue of generating masses for the fermions. This too can be done in a gauge invariant way through a Yukawa-type interaction $\phi\bar{\psi}\psi$
\begin{equation}
\lag{Yukawa} = -y_e\bar{L}\Phi e_R  - y_d\bar{Q}\Phi d_R - y_u\bar{Q}\tilde{\Phi}u_R + \mathit{h.c. } 
\end{equation}
where $\Tilde{\Phi} = i\tau_2\Phi^*$ is used instead of $\Phi$ for the up quark because the vev is in the lower component of the Higgs doublet. Upon spontaneous symmetry breaking, we have, e.g. for the electron 
\begin{equation}
\lag{Yukawa} = -\frac{1}{\sqrt{2}}y_e\begin{array}{cc} (\bar{\nu}_L & \bar{e}_L) \end{array} \left(\begin{array}{c} 0  \\ v + h \end{array}\right)e_R + \textit{h.c.} \Rightarrow -\frac{y_e v}{\sqrt{2}}\bar{e}{e} + \text{interaction term,}
\end{equation}
and similarly for the up and down quarks. To summarise, using an $\text{SU(2)}$ doublet $\Phi$ we have generated masses for both the $W$ and $Z$ vector bosons and the fermions. The \su2L\x \U1Y symmetry is no longer apparent in the vacuum; it has been spontaneously broken down to an unbroken U(1) subgroup, identified as electromagnetism. The color SU(3) symmetry is also unbroken, so has been omitted in this section. Because gauge invariance has not been explicitly broken, the Standard Model remains renormalisable~\cite{tHooft:1971akt,tHooft:1972tcz} and unitary~\cite{LlewellynSmith:1973yud,Bell:1973ex} up to high energies. The Standard Model can thus be summarised by the following Lagrangian\footnote{The full Lagrangian also contains gauge-fixing and Fadeev-Popov `ghost' terms to eliminate redundant gauge field configurations. For brevity these are not included here.}.
\begin{equation}
\lag{SM} = \lag{gauge}+\lag{fermion}+\lag{Yukawa}+\lag{Higgs}
\end{equation}
where
\begin{equation}
\begin{split}
\lag{gauge} &=  -\frac{1}{4}G^A_{\mu\nu}G^{A,\mu\nu} -\frac{1}{4}W^I_{\mu\nu}W^{I,\mu\nu}-\frac{1}{4}B_{\mu\nu}B^{\mu\nu} \\
\lag{fermion} &= i\bar{L}_iD_\mu\gamma^{\mu} L_{i} + i\bar{e}_{Ri}D_\mu\gamma^{\mu} e_{Ri} +  i\bar{Q}_iD_\mu\gamma^{\mu} Q_{i} \\
& \hspace{10pt} +i\bar{u}_{Ri}D_\mu\gamma^{\mu} u_{Ri} +  i\bar{d}_{Ri}D_\mu\gamma^{\mu} d_{Ri} \\
\lag{Yukawa} &= -y_e\bar{L}\Phi e_R  - y_d\bar{Q}\Phi d_R - y_u\bar{Q}\tilde{\Phi}u_R + \mathit{h.c. }  \\
\lag{Higgs} &=  (D_{\mu}\Phi)^\dagger(D^{\mu}\Phi) -  \mu^2\Phi^\dagger\Phi - \lambda(\Phi^\dagger\Phi)^2.
\end{split}
\end{equation}
\subsubsection{The parameters of the Standard Model}
For one generation of fermions, the free parameters in the Standard Model are:
\begin{itemize}
\item The three gauge couplings $\{g_s,g,g'\}$
\item The two parameters in the Higgs potential $V(\phi)$: $\mu$ and $\lambda$
\item The three Yukawa coupling constants $\{y_u,y_d,y_e\}$
\end{itemize}
Although these are the `fundamental' parameters, they are typically expressed in terms of the more directly measurable quantities:
%
\begin{align}
\tan\theta_W &= \frac{g'}{g}  \nonumber \\
e &= g\sin\theta_W  \nonumber \\
m_H &= \sqrt{2}{\mu}  = \sqrt{2\lambda} v \nonumber \\
M_W &= \frac{g\mu}{2\sqrt{\lambda}} = \frac{gv}{2} \nonumber \\
m_f &= \frac{y_f\mu}{\sqrt{2\lambda}} = \frac{y_f}{\sqrt{2}} v .
\end{align}
%
Once these parameters have been measured precisely, predictions for $M_Z$ and $G_F$ (the Fermi coupling) can be made. Thus the interaction strengths of the entire electroweak sector of the Standard Model are fixed by seven parameters (the strong interaction is determined by 1: $g_s$) . 

Adding additional generations brings some complications, however. For instance, the presence of a second and third generation of quarks leads to the Yukawa couplings
\begin{equation}
-[y_d]_{ij}\bar{Q}_i\Phi d_{Rj} - [y_u]_{ij}\bar{Q}_i\Phi u_{Rj} +\textit{h.c.}
\end{equation}
where $i,j$ are generation/flavour indices. The Yukawa couplings are now 3 $\times$ 3 matrices, and off-diagonal terms are perfectly allowed by gauge-invariance. This would mix quarks of different flavour. To obtain the physical particles we diagonalise the mass matrix and extract the terms bilinear in each field, just as we did to extract the physical $Z$ and $A$ fields. This can be done by performing a unitary rotation on each quark field. However, this means that we must also rotate the quark kinetic terms, so the off-diagonal structure has merely been transferred to the fermion-gauge couplings. To relate the weak eigenstates to the mass eigenstates, the convention is to define the up-type quarks as in the mass-eigenstate basis to begin with, then to relate the down-type quark weak eigenstates $q'$ to the mass eigenstates $q$ through a unitary rotation

\begin{equation}
\begin{array}{ccc} (d' & s' & b') \end{array} = \mathbf{V} \begin{array}{ccc} (d & s & b) \end{array}
\end{equation}
where $\mathbf{V}$ is the $3 \times 3$ Cabibbo-Kobayashi-Maskawa (CKM) matrix~\cite{Cabibbo:1963yz,Kobayashi:1973fv}, whose values are~\cite{Agashe:2014kda}:
\begin{equation}
\begin{split}
\mathbf{V_{CKM}} = &\left(\begin{array}{ccc} 	V_{ud} & V_{us} & V_{ub} \\
									V_{cd} & V_{cs} & V_{cb} \\
									V_{td} & V_{ts} & V_{tb} \end{array} \right)  \\
			     =	&\left(\begin{array}{ccc} 	0.97427 \pm 0.00014 & 0.22536 \pm 0.00061 & 0.00355 \pm 0.00015 \\
									0.22522 \pm 0.00061 & 0.97343 \pm 0.00015 & 0.0414 \pm 0.0012 \\
									0.00886 \pm 0.00033 & 0.0405 \pm 0.0012 & 0.99914 \pm 0.00005 \end{array} \right)
\end{split}									
\label{eqn:ckm}
\end{equation}									
To count the parameters of this matrix, we first note that a general unitary $3 \times 3$ matrix has nine independent parameters. With six quarks we can absorb five relative phases into the quark field strengths $q\to e^{i\theta}q$, which leaves four independent parameters: three mixing angles (akin to the Euler rotation angles) and a residual complex phase. The off-diagonal terms in the CKM matrix are subleading, and a well-known parametrisation of the CKM matrix which mimics this structure is due to Wolfenstein~\cite{Wolfenstein:1983yz}, which can be approximated as:

\begin{equation}
\mathbf{V_{CKM}} = \left(\begin{array}{ccc} 1 - \lambda^2/2 & \lambda & A\lambda^3(\rho-i\eta) \\
						          	   -\lambda        & 1 - \lambda^2/2 & A\lambda^2 \\
								A\lambda^3(1-\rho-i\eta) & 1-A\lambda^2 & 1 \end{array} \right) + \mathcal{O}(\lambda^4)
\label{eqn:wolf}
\end{equation}

where the complete expression involving  $\{\bar{\rho},\bar{\eta}\} = \{\rho,\eta\}(1-\lambda^2/2+\mathcal{O}{(\lambda^4)})$ has not been displayed here. For massless neutrinos, there is no analogous mixing in the lepton sector: the weak eigenstates are the same as the mass eigenstates, a property of the Standard Model known as \textit{lepton universality}.

Hence for the Standard Model with three generations, we have the following free parameters, 18 in total\footnote{Adding in neutrino masses would bring in 7 more parameters: 3 masses and 4 PMNS~\cite{Pontecorvo:1957qd,Maki:1962mu,Pontecorvo:1967fh} mixing angles. For the purposes of this thesis neutrinos can be considered massless.}.
\begin{itemize}
\item The 8 parameters mentioned above.
\item Three extra Yukawa couplings for each additional generation: six in total.
\item Four parameters in the CKM matrix:  $\{A,\bar{\rho},\lambda,\bar{\eta}\}$.
\end{itemize}
The most up-to-date values of these parameters\footnote{There is also a 19th parameter: the QCD $\theta$-term, which does not lead to physical effects in perturbation theory, but can be generated by non-perturbative instanton effects. This will be discussed in some detail in chapter 2.}, expressed through more conveniently measurable quantities, are shown in Table \ref{table:smparam}~\cite{Agashe:2014kda}.
\begin{table}[h!]
\begin{center}
\begin{tabular}{| c | c || c | c | } \hline 
\multicolumn{1}{ | c |} {Parameter} & \multicolumn{1}{ | c ||} {Value} & \multicolumn{1}{ | c |} {Parameter} & \multicolumn{1}{ | c | }{Value} \\ \hline \hline
$\alpha$ &  1/137.035 999 074(44) & $m_e$ & 0.510 998 928(11) MeV \\
$\alpha_s$ & 0.1185(6) & $m_\mu$ & 105.6583715(35) MeV \\
$G_F$ & 1.166 378 7(6)  $\times$ 10$^{-5}$  GeV$^{-2}$ & $m_\tau$ & 1776.82(16) MeV \\
$m_h$ & 125.7(4) GeV & $m_u$ & 2.3$^{+0.6}_{-0.5}$ MeV \\
$M_W$ & 80.385(15) GeV & $m_d$ & 4.8$^{+0.5}_{-0.3}$ MeV \\
$A$ & 0.814$^{+0.023}_{-0.024}$ & $m_c$ & 1.275(25) GeV \\
$\lambda$ & 0.22537(61) & $m_s$ & 95(5) MeV \\
$\bar{\rho}$ & 0.117(21) & $m_t$ & 173.21 $\pm$ 0.51 $\pm$ 0.71 GeV \\
$\bar{\eta}$ & 0.353(13) & $m_b$ & 4.18(3) GeV \\ \hline
\end{tabular}
\end{center}
\caption{\label{table:smparam} 
The 18 free parameters of the Standard Model with massless neutrinos.}
\end{table}

 A few remarks are in order:

\begin{itemize}

\item The coupling constants $\alpha$ are related to their respective gauge coupling parameters by
\begin{equation}
\alpha = \frac{e^2}{4\pi} \hspace{10pt}, \hspace{10pt} \alpha_s = \frac{g_s^2}{4\pi}
\end{equation}
Coupling constant is something of a misnomer, however. The renormalisation of the couplings by higher-order corrections ensures the values of these parameters depend on the scale at which they are resolved. This will be discussed in more detail later. In Table \ref{table:smparam} $\alpha$ is quoted at the scale $Q^2=0$, whereas $\alpha_s$ is quoted at $Q^2=M^2_Z$. 

\item In a similar way, the quark masses $m_i = \frac{y_iv}{\sqrt{2}}$ are renormalised by QCD effects (QED renormalisation of lepton masses is negligible), and the values quoted refer to the `running masses' in the $\overline{MS}$ renormalisation scheme, each evaluated at the scale $\mu$ = 2 \gev, with the exception of the top quark.

\item The top quark presents an additional ambiguity: the measured value quoted above is obtained from fitting Monte Carlo templates with different input values for the top quark mass. This was formerly interpreted as equal to the top quark \textit{pole} mass: the renormalised mass corresponding to the pole in the top propagator. This analogy is flawed, however, due to subtleties in the showering and hadronisation of partons in the Monte Carlo. The ambiguity of definition here introduces a theoretical uncertainty of $\sim$ 1 GeV in additional to the statistical and systematic errors quoted above. This issue is discussed in more detail in the next section.

\end{itemize}

The predictions of the Standard Model have been tested in numerous fixed-target and collider experiments over the last forty years. The rich phenomenology of the strong interaction has been extensively studied in electron-positron collisions and deep-inelastic scattering in electron-proton events at HERA~\cite{Aaron:2009aa}. In addition, the precision measurements carried out at LEP, the Tevatron and elsewhere~\cite{ALEPH:2005ab,Schael:2013ita} have probed the electroweak couplings to sub-percent level accuracy. The 2012 discovery~\cite{Aad:2012tfa,Chatrchyan:2012xdj} of a Higgs boson at a mass of $\sim$ 125 GeV by the ATLAS and CMS experiments has filled in the last piece of the SM picture, and studies of the Higgs sector to a similar precision are now underway~\cite{deFlorian:2016spz}. 

The subject of this thesis concerns the properties of the top quark, and how they may be probed at hadron colliders, so the next section reviews some general features of hadron collider machines such as the LHC and Tevatron, before reviewing the physics of the top quark that may be studied with them.

\subsection{Hadron collider physics}
\label{sec:colliders}

\subsubsection{Scattering theory}
\begin{figure}[t!]
\begin{center}
\includegraphics[width=8cm,height=5.5cm]{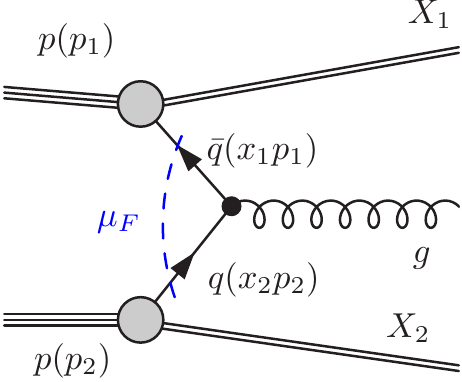}
\end{center}
\vspace{-20pt}
\caption[Factorisation of the hard process from the parton densities.]{Schematic diagram for the process $pp\to g X$, where $X$ denotes any other final state products of the collision. The factorisation scale $\mu_F$ separates the perturbative `hard' process from the non-perturbative parton densities of the incoming hadrons.  }
\label{fig:hadronic_xsec}
\end{figure}

The starting point for calculating scattering amplitudes in quantum field theory is the S-matrix, which can be split into a trivial `free-propagation piece' and a scattering piece $T$.
\begin{equation}
S = \mathbbm{1} + iT = \mathbbm{1} + i\delta^4(p_f-p_i)\mathcal{M}_{fi}
\end{equation}
The delta function appears in every scattering amplitude to enforce momentum conservation, so can be factored out of any scattering amplitude to define the matrix-element  $\mathcal{M}_{fi}$. From this, and Fermi's golden rule, the cross-section for producing a general final state $X$ from initial state particles $a_1$ and $a_2$ of momenta $p_1$ and $p_2$ is
\begin{equation}
\sigma(a_1(p_1)a_2(p_2) \to X) = \frac{1}{\Phi} \int d\Pi_n |\mathcal{M}_{fi}|^2
\end{equation} 
where
\begin{equation}
\Phi = |v_{a_1}-v_{a_2}|(2E_{a_1})(2E_{a_2})
\end{equation}
is the flux factor, defined in terms of the relative velocities of the incoming beams in the lab frame, $v_{a_i}$, and 
\begin{equation}
\int d\Pi_n = \int \prod\limits_{i=1}^n\frac{d^3k_i}{(2\pi)^32E_{k_i}} = \int d\text{LIPS}(X)
\end{equation}
is the volume of the n-body final-state Lorentz invariant phase-space. Cross-sections can be calculated order-by-order in perturbation theory, provided that coupling constants are not too large so that higher-order terms in the perturbation series can be neglected. The initial states relevant for hadron collisions are the constituents of the hadron: quarks and gluons. The perturbative cross-section is calculated completely in terms of quarks and gluons (collectively known as `partons') and is related to the full hadronic cross-section by
\begin{equation}
\sigma(p(k_1)p(k_2)\to X) = \sum_{\substack{i,j} }\int\limits^{1}_{0} dx_1 dx_2f_i(x_1,\mu_F^2)f_j(x_2,\mu_F^2)\hat{\sigma}_{ij}(x_1,x_2,s,\alpha_S(\mu_R,Q^2)).
\label{eqn:hadronicxsec}
\end{equation}
Here $\hat{\sigma}_{ij\to X}$ is the partonic cross-section for final state $X$ from partons $i$ and $j$, where $i,j \in \{q,g\}$. $f_i$ are the parton density functions (pdfs): the probability of finding a parton $i$ in the proton with fraction $x$ of the total proton momentum.  To obtain the full hadronic cross-section, we calculate the partonic cross-section for momenta $x_1p_1$ and $x_2p_2$, then integrate these over the full range of $x$ for each proton, then sum over all allowed partonic subprocesses.  This is illustrated in Fig. \ref{fig:hadronic_xsec}. 

The parameters $\mu_{R,F}$ are arbitrary scales necessarily introduced in fixed-order perturbation theory. Calculations for hadron colliders are plagued by theoretical uncertainties, which can be broadly classed into three categories: scale dependence, pdf uncertainties and finite accuracies for Standard Model parameters such as $\alpha_s$, which enter as inputs into the calculation. Here we briefly discuss each of these in turn.

\subsubsection{Scale uncertainties}
It is well-known that quantities calculated beyond the leading Born approximation in quantum field theory often feature ultraviolet divergences. These arise from quantum fluctuations with unconstrained high-momenta. For a certain class of quantum field theories, it is possible to remove these divergences by defining the theory at some renormalisation scale $\mu_R$ which separates the low energy field theory from the unknown short-distance physics and allows one to make low-energy predictions regardless of the underlying degrees of freedom. Although separated out, the degrees of freedom at different scales have the effect of introducing a \emph{scale dependence} of the coupling constants and masses of the theory. In QED, for instance, the effective electromagnetic coupling runs from $\alpha = 1/137$ at the scale $\mu_R = 0$ to  $\alpha(\mu_R = M_Z) \sim 1/129$.

The renormalisation scale is an arbitrary parameter and so predictions for physical quantities should be independent of $\mu_R$. The renormalisation group equations define precisely how the renormalised couplings should vary with scale such that order-by-order in perturbation theory, measurable quantities are independent of $\mu_R$. Truncating the perturbative expansion at a fixed order, however, means that the cancellation of $\mu_R$ in physical quantities is incomplete, i.e. there is a residual dependence on $\mu_R$ proportional to the next order in the perturbative expansion. 

It is not immediately clear which value of $\mu_R$ should be chosen for a process, but it should be a characteristic energy scale entering the process that absorbs the large logarithms $\log(s/\mu_R^2)$ arising from separate scales involved in the process. To estimate the size of unknown higher-order corrections, one typically varies the scale over the range $\mu/2 \leq \mu_R \leq 2\mu$, using the variation in the prediction as the scale uncertainty. For most processes where NNLO corrections have been calculated, they have been found to lie in the scale uncertainty band of the NLO estimate, which vindicates this rather \textit{ad hoc} procedure. Counter-examples exist, however, where the NLO and NNLO scale uncertainty bands do not overlap, such as in cases where widely different scales enter\footnote{One example is associated Higgs production $pp\to HV$~\cite{Brein:2003wg}.}, and to truly quantify higher-order effects there is no substitute for doing the actual calculation.

\begin{figure}[t!]
\begin{center}
\includegraphics[width=\textwidth,height=7cm]{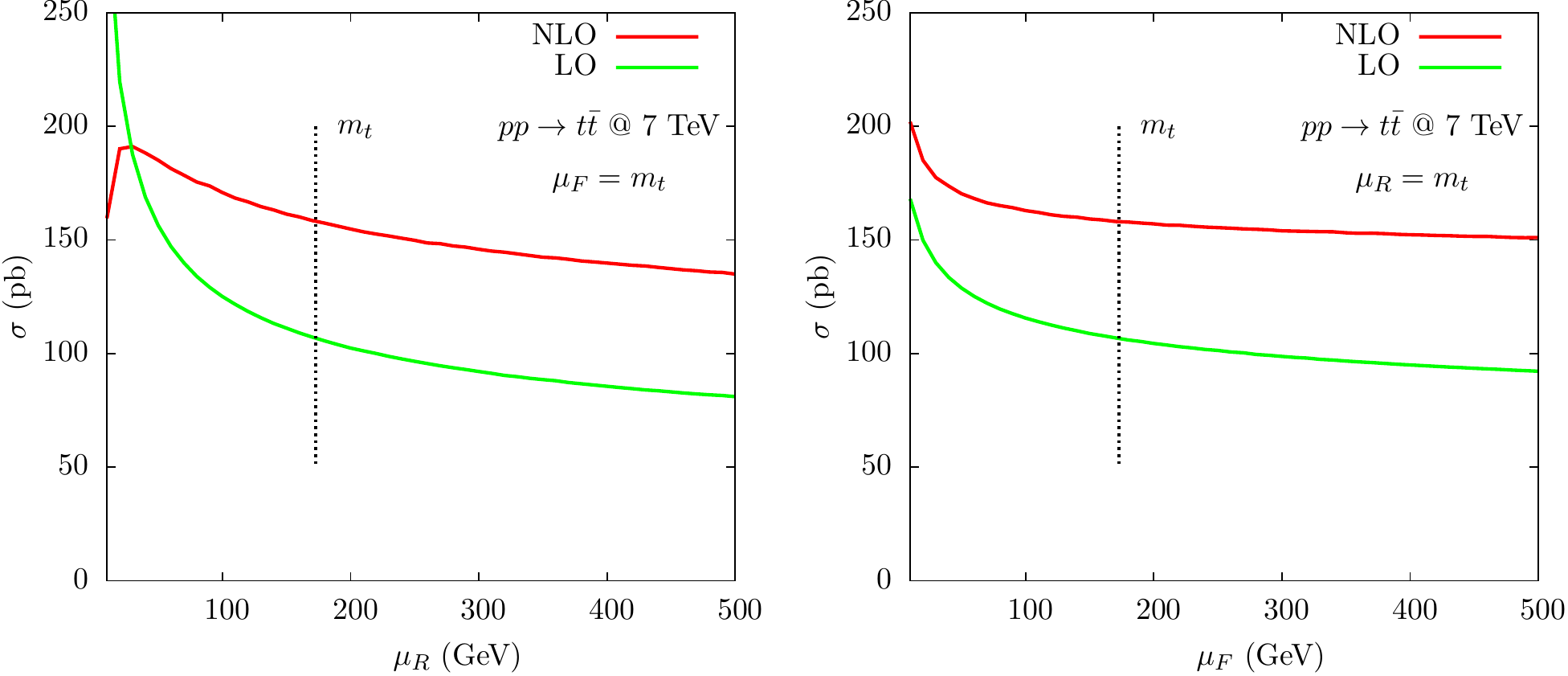}
\end{center}
\vspace{-20pt}
\caption[Renormalisation and factorisation scale dependence of the \ttbar cross-section.]{Renormalisation (left) and factorisation (right) scale dependence of the $pp \to t\bar{t}$ cross-section at 7 TeV, using the MSTW2008 pdf sets.  }
\label{fig:scales}
\end{figure}

One must also choose the factorisation scale for a process. This defines at what energy we separate the hard (high-momentum) scattering process cross-section, which is calculated in perturbation theory, from the parton density functions (PDFs) for the incoming protons, which are extracted from data. 

The factorisation scale is also not a physical quantity, it is a definition of what energy scale corresponds to the partonic process and what falls into the definition of the incoming protons. Any initial state radiation with energy $E < \mu_F$ is absorbed into the hadron. Again, there is no `correct' scale, one simply chooses a value typical of the process and varies over [1/2,2] to estimate the uncertainty. For most of our predictions (save a few special cases) we set a common central scale $\mu_R = \mu_F = \mu = m_t $ and vary both independently over [$\mu$/2,2$\mu$]. The dependence of the total $t\bar{t}$ production cross-section at the 7 TeV LHC, at leading and next-to-leading order, on these scales is sketched in Fig. \ref{fig:scales}, showing that the range [$\mu$/2,2$\mu$] captures most of the scale dependence.

\subsubsection{The parton densities}

The inner structure of the proton is determined by quantum chromodynamics in the strongly coupled, low-momentum transfer regime where perturbative techniques are not valid, so the parton densities $f_i$ are not calculable from first principles. The choice of pdf set introduces an additional theoretical uncertainty and several such sets are available. Of course, predictions should be independent of the set used: the structure of the proton at a certain energy scale is a universal physical property. In practice, however, the different approaches each group uses to extract parton densities from data introduce systematic uncertainties leading to different results.

Due to the vastly different methodologies used by the main pdf groups, and the different input measurements used in their fits, it is often not possible to compare their results in an unbiased way. Instead the discrepancies resulting from different pdf choices are resolved in the most conservative way, by calculating predictions for each of the main pdf groups: CT14~\cite{Dulat:2015mca}, MMHT~\cite{Harland-Lang:2014zoa} and NNPDF~\cite{Ball:2014uwa}, and taking the maximum range as an additional (`pdf') uncertainty. This prescription is the recommendation of the PDF4LHC~\cite{pdf2} working group, and is the one adopted throughout this thesis, unless otherwise stated.

\begin{figure}[t!]
\begin{center}
\includegraphics[width=\textwidth]{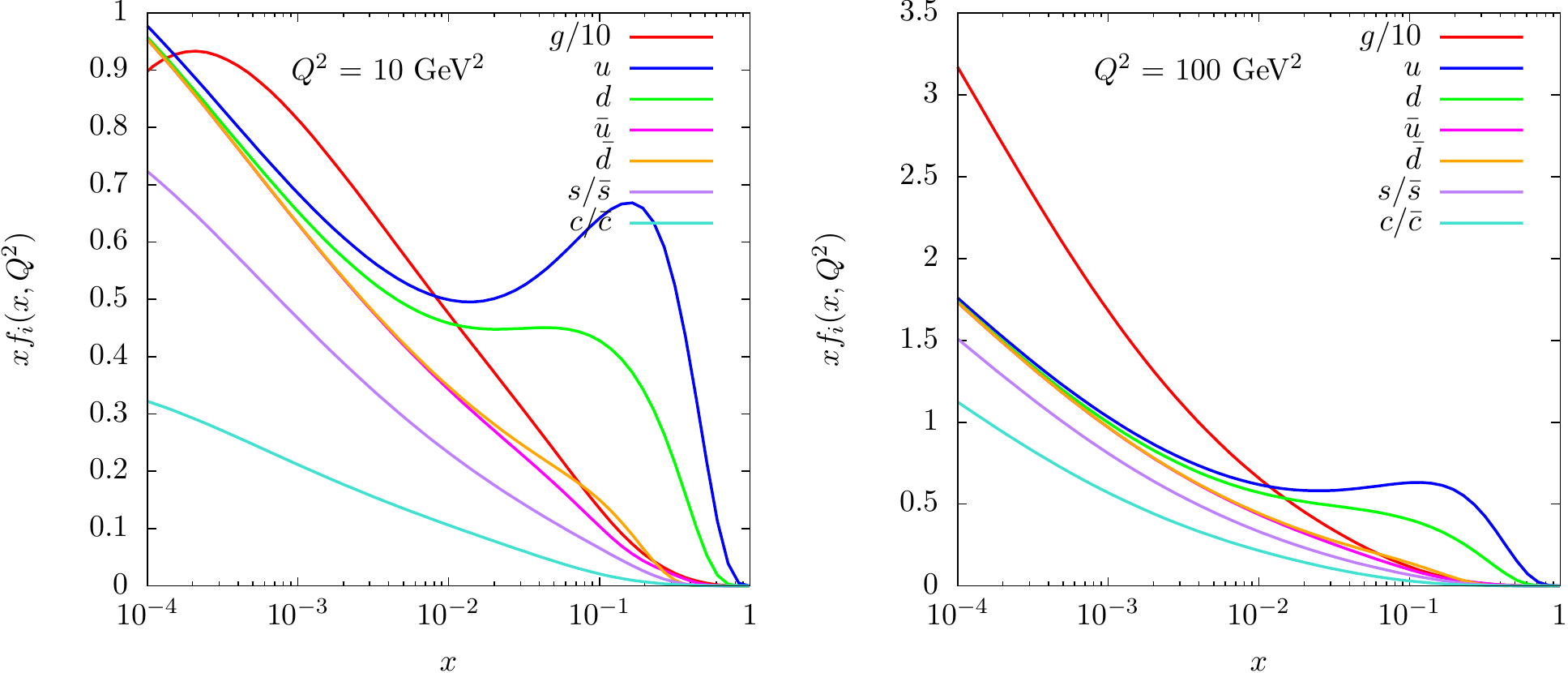}
\end{center}
\vspace{-20pt}
\caption[Parton densities of the proton at low and high momentum transfer.]{Parton distribution functions for the proton from the MSTW2008 NNLO fit~\cite{Martin:2009iq}, at low (left) and high (right) momentum transfer. Uncertainties are not shown. }
\label{fig:pdfs}
\end{figure}

\subsubsection{Standard Model parameters}
In addition to the theoretical uncertainties arising from scale and pdf choices, an additional source arises from the finite precision with which the SM parameters entering the calculation have been measured. Of the 18 parameters in Table \ref{table:smparam}, the two most relevant for top quark production are the strong coupling constant $\alpha_s$ and the top quark mass $m_t$. The former is known to sub per-mille accuracy, having been extracted mainly from high-precision $e^+e^-$ experiments at LEP and SLC, as well as through deep inelastic scattering measurements. Its value, quoted at the $Z$-pole, can be calculated at any energy scale using the QCD $\beta$-function, which has recently been calculated to five-loop accuracy~\cite{Baikov:2016tgj,Luthe:2017ttc,Herzog:2017ohr}. It is therefore one of the most precisely measured quantities of the Standard Model, and the inclusion or omission of its experimental uncertainty rarely has a substantial effect on perturbative QCD predictions.

The top quark mass, however, presents additional challenges, as mentioned above. Experimentally, the top quark mass has been measured to sub percent-level accuracy, This is typically achieved by directly reconstructing the top quark from its decay products: a $b$-tagged jet and either a charged lepton and missing transverse energy, or an additional pair of jets. Kinematic distributions of these decay products are constructed, e.g. the reconstructed top mass $m_t$, and Monte-Carlo predictions (`templates') with different values of $m^{MC}_t$ are fit to the data. The best-fit value is then defined as the top quark mass. This is a well-defined statistical procedure. However, ambiguity arises when relating $m^{MC}_t$ to a renormalised mass in quantum field theory. 

The top quark mass is renormalised by self-energy corrections. The UV divergent pieces of these corrections are absorbed unambiguously into the running of the mass. However, different treatments of the finite corrections admit different definitions of what is meant by a `mass' in quantum field theory. The most intuitive is the \textit{pole} mass, the mass corresponding to the pole in the propagator, where \textit{all} divergent and finite corrections are absorbed into the mass. Owing to non-perturbativity, however, loop corrections with momenta $\lesssim$ 1 GeV (the QCD hadronisation scale) cannot be calculated, which defines a maximum precision on the pole mass definition. A Monte Carlo generator never runs into such problems. The MC top mass is defined as the pole in the hard matrix element. When this is interfaced to the parton shower, which generates successive parton splittings at increasingly low momenta, self-energy corrections are ignored, so they must be viewed as already included in the definition of $m^{MC}_t$. However, when the typical parton momenta in the shower reaches $\mathcal{O}$(1 GeV), showering stops and the hadronisation model takes over. There is thus a fundamental precision of $\sim$ 1 GeV with which we can relate $m^{pole}_t$ to the experimentally measured $m^{MC}_t$, and this uncertainty should be included in any calculations involving $m_t$ (see Ref.~\cite{Hoang:2014oea} for a recent review of these issues). 

\subsection{Top quark physics at hadron colliders}
\label{sec:topphys}

The top quark couples directly to all of the Standard Model gauge and Higgs bosons. The interaction with gluons is described by a vectorial fermion-gauge coupling $\bar{\psi}\psi A_\mu $ 
\begin{equation}
\vcenter{\hbox{ \includegraphics[width=0.225\textwidth]{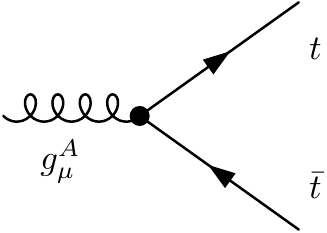} }} \qquad = -ig_sT^A\gamma^\mu ,
\label{eqn:ttgvertex}
\end{equation}
as is the coupling to photons,
\begin{equation}
\vcenter{\hbox{ \includegraphics[width=0.225\textwidth]{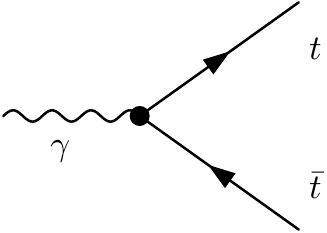} }} \qquad = -i\frac{2}{3}e\gamma^\mu .
\end{equation}

Due to the $V-A$ structure of the charged weak currents, only the left-handed top couples to the $W^\pm$, with coupling
\begin{equation}
\vcenter{\hbox{ \includegraphics[width=0.225\textwidth]{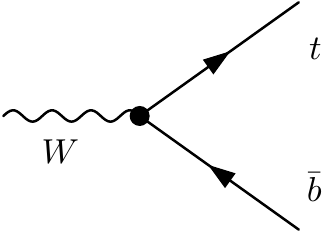} }} \qquad = ig\gamma^\mu(1-\gamma^5)V_{tb},
\end{equation}
where the value of $V_{tb}$ is given in Eq.~\eqref{eqn:ckm}. The top couples to the $Z$ with unequal left and right-handed components, given by
\begin{equation}
\vcenter{\hbox{ \includegraphics[width=0.225\textwidth]{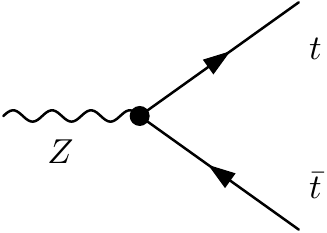} }} \qquad = \frac{ig}{2\cos\theta_W}\gamma^\mu(v_t-a_t\gamma^5),
\end{equation}
where $v_t = T_{3t} - 2Q_t\sin^2\theta_W \simeq 0.19$ and $a_t = T_{3t} = 1/2$. Finally, it couples to the Higgs boson with a Yukawa-type interaction $\bar{\psi}\psi\phi$,
\begin{equation}
\vcenter{\hbox{ \includegraphics[width=0.225\textwidth]{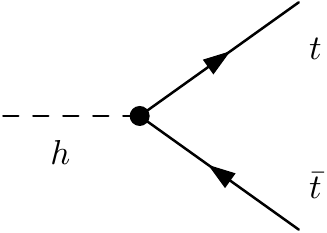} }} \qquad =  y_t = \frac{\sqrt{2} m_t}{v} .
\label{eqn:tthvertex}
\end{equation}
All of these couplings are flavour-conserving, with the exception of the charged-current interaction with the $W^\pm$. Since we are interested in top quark production at hadron colliders, the QCD triple gluon vertex will also be relevant for our discussion. Its Feynman rule is 
\begin{equation}
\vcenter{\hbox{ \includegraphics[width=0.275\textwidth]{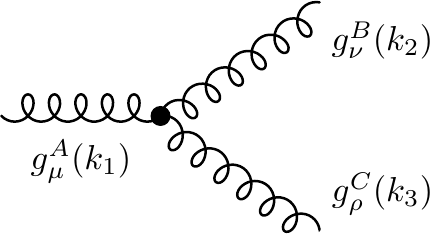} }}  = g_sf^{ABC}[g_{\mu\nu}(k_1-k_2)_\rho  + g_{\nu\rho}(k_2-k_3)_\mu + g_{\rho\mu}(k_3-k_1)_\nu ] , \quad
\label{eqn:gggvertex}
\end{equation}
where all momenta are defined as towards the vertex.

The structure of the top quark couplings is identical to those of the other quarks, but the top enjoys properties unique amongst the quarks, namely its large coupling to the Higgs boson ($y_t \simeq 1$ in the SM) which suggests it plays a special role in electroweak symmetry breaking, and its large coupling to $b$-quarks ($V_{tb}$ has been measured to be very close to 1), an observation which is unexplained in the SM. For these reasons and others, the top is often viewed as a possible window to physics beyond the Standard Model (indeed, this is the subject of this thesis). However, before turning our attention to BSM physics, we conclude this chapter with a discussion of the main production mechanisms for top quarks at hadron colliders.

\subsubsection{Top pair production}
\begin{figure}[t!]
\begin{center}
\includegraphics[width=5cm,height=2.5cm]{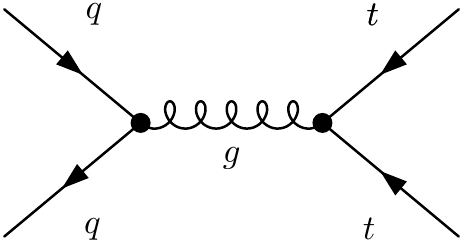}
\end{center}
\caption{The leading order Feynman diagram for $q\bar{q}\to \ttbar$ in the SM.}
\label{fig:qqttbarsm}
\end{figure}

\begin{figure}[t!]
\begin{center}
\includegraphics[width=15cm,height=4cm]{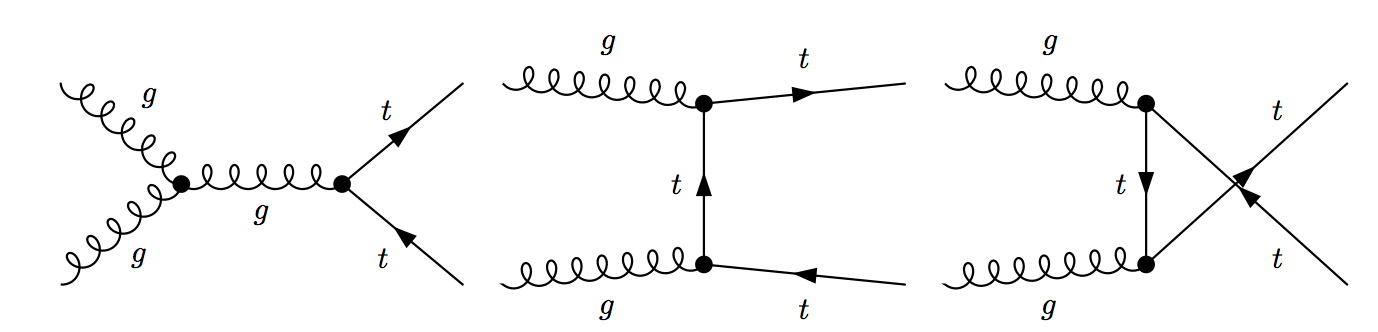}
\end{center}
\caption[Leading order Feynman diagrams for $gg\to \ttbar$ in the SM.]{The leading order Feynman diagrams for $gg\to \ttbar$ in the SM. Diagrams with these topologies are generically labelled (from left to right): $s,t$ and $u$-channel diagrams.}
\label{fig:ggttbarsm}
\end{figure}

By far the dominant production mechanism for top quarks in hadron collisions is top pair production $pp/p\bar{p}\to t\bar{t}$. The main contributions to this process come from QCD; production through intermediate Z bosons are negligible because the $t\bar{t}$ threshold is far from the $Z$ pole, while QED contributions are parametrically suppressed by $(\alpha/\alpha_s)^2$. At leading-order in $\alpha_s$, the partonic subprocesses $q\bar{q}\to t\bar{t}$ and $gg\to t\bar{t}$ both contribute. For the former, the partonic cross-section is, averaging (summing) over initial (final) state spins and colours:
\begin{equation}
\sigma_{q\bar{q}\to t\bar{t}} = \frac{g^4_s}{108\pi s} \beta(3-\beta^2),
\end{equation}

where  $\beta = \sqrt{1-4m^2_t/s}$ is the velocity of the top quark in the centre of mass frame (generically referred to as the `threshold variable'). The leading-order Feynman\footnote{Our convention for Feynman diagrams is to represent the flow of charge through diagrams with arrows, in keeping with the Feynman-Stuckleberg interpretation of antimatter as matter under a time reversal transformation.} diagram for the $q\bar q$ channel is sketched in Fig. \ref{fig:qqttbarsm}. 
For the $gg$ channel, we have:
\begin{equation}
\sigma_{gg\to t\bar{t}} = \frac{g^4_s}{768\pi s}\left(31\beta^3 -59\beta + (33 - 59\beta + (33-18\beta^2+\beta^4))\log\frac{1+\beta}{1-\beta}\right).
\end{equation}
Feynman diagrams for this process are sketched in Fig. \ref{fig:ggttbarsm}.

To obtain the full hadron-level cross-section, we convolute these expressions with the parton densities, as in Eq.~\eqref{eqn:hadronicxsec}. Displaying a closed-form expression for the hadron-level cross-section would thus require functional forms for the parton densities $f_g(x;Q^2)$ and $f_{q}(x;Q^2)$ used to fit the data. Here we simply discuss the numerical results, obtained from numerical tables of the pdf data.

The relative contributions of the partonic subprocesses are determined by the nature of the incoming hadrons. At the Tevatron $p\bar{p}$ collider, antiquarks exist as valence quarks in the initial state, so $q\bar{q}\to t\bar{t}$ is the dominant subprocess: it contributes around 85\% of the total cross-section, the remainder is made from gluon-fusion. At a centre of mass energy of $\sqrt{s} = 1.96$ TeV, the leading order cross-section is calculated to be around 7 pb, for $\mu_R = \mu_F = m_t$ and using the CTEQ6l1 parton sets. At the LHC, antiquarks only appear as sea quarks, whilst the large kinematic reach means the proton is resolved down to much smaller momentum fraction $x_{min} \sim 10^{-5}$. In this regime the gluon luminosity becomes dominant, so the $gg$ channel contributes up to 90\% of the total cross-section. At a centre of mass energy of 7 TeV, the leading-order cross-section is around 100 pb ~\cite{Alwall:2014hca}. 

\subsubsection*{Higher-order corrections}
Understanding the effects of higher-order radiative corrections is necessary for obtaining precise Standard Model cross-section predictions. The size of (as yet) uncalculated higher-order effects can be estimated by noting the change in the cross-section with respect to scale variations. Leading-order estimates are typically correct within a factor of two, i.e. they provide a good ballpark estimate, but, owing to the fact that they include information about appropriate scale choices that should absorb the large logarithms that occur at higher orders, next-to-leading order (NLO) and often higher still corrections must be included for truly accurate estimates. They can be approximately included by defining a $K$-factor  
\begin{equation}
K = \frac{\sigma(pp \to X)_{(N)NLO}}{\sigma(pp \to X)_{LO}},
\end{equation}

The higher-order estimate is then simply calculated by multiplying (`reweighting') the leading-order estimate by the $K$-factor. For top-pair production, the current `state-of-the-art' SM prediction is the full next-to-next-to leading order estimate, which includes the resummation of terms involving soft gluon emissions to next-to-next-to-leading logarithmic accuracy (shorthand NNLO+NNLL) leading to the following values~\cite{Czakon:2012zr,Czakon:2012pz,Czakon:2013goa}:
\begin{equation}
\begin{split}
\sigma(pp \to t\bar{t} + X) = 172.0\hspace{5pt}^{+4.4}_{-5.8}\hspace{5pt}\text{(scale)} \hspace{5pt}^{+4.7}_{-4.8}\hspace{5pt}  \text{(pdf)} \hspace{10pt}  &\sqrt{s}=7\hspace{5pt} \text{TeV} \\
\sigma(pp \to t\bar{t} + X) = 245.8\hspace{5pt}^{+6.2}_{-8.4}\hspace{5pt}\text{(scale)} \hspace{5pt}^{+6.2}_{-6.4}\hspace{5pt}  \text{(pdf)} \hspace{10pt}  &\sqrt{s}=8\hspace{5pt} \text{TeV} \\
\sigma(p\bar{p} \to t\bar{t} + X) = 7.164\hspace{5pt}^{+0.11}_{-0.20}\hspace{5pt}\text{(scale)} \hspace{5pt}^{+0.17}_{-0.12}\hspace{5pt}  \text{(pdf)} \hspace{10pt}  &\sqrt{s}=1.96\hspace{5pt} \text{TeV}
\end{split}
\end{equation}
As well as the total cross-section, it is useful to study the dependence of the cross-section on kinematic observables that can be measured at colliders. The most commonly studied variables are briefly outlined:

\begin{itemize}

\item The invariant mass, defined as
\begin{equation}
m^{2} =\left(\sum_{\substack{i}}E_{i}\right)^{2}-\left(\sum_{\substack{i}}\boldsymbol{p_{i}}\right)^{2}
\label{eqn:mass}
\end{equation}
where the sum is over all final state particles $i$. Final state invariant mass distributions are the classic way of searching for new particles. A peak in the $t\bar{t}$ invariant mass distribution at high $m_{t\bar{t}}$ would be an unambiguous signal of a new resonance decaying to top quarks. 

\item A related kinematic quantity is the transverse momentum $p_T$ of the top; large-$p_T$ events correspond to events in the high-energy region, where possible new physics effects are most likely to lie.

\item The distribution of particles throughout the geometry of the detector is usually specified in terms of the rapidity $y$, defined as 
\begin{equation}
y = \frac{1}{2} \ln \left (\frac{E+p_z}{E-p_z}\right).
\end{equation}
This is typically used as a geometrical proxy for polar-angle $\theta$\footnote{Most collider experiments use a spherical co-ordinate system, where $\theta$ is the angle between the beam ($z$-axis) and the particle track, and $\phi$ is the azimuthal angle between the track and the vertical, i.e. looking down the beam.}  as, unlike $\theta$, it is additive under Lorentz boosts in the $z$-direction.

\end{itemize}

\begin{figure}[t!]
\begin{center}
\includegraphics[width=\textwidth,height=15cm]{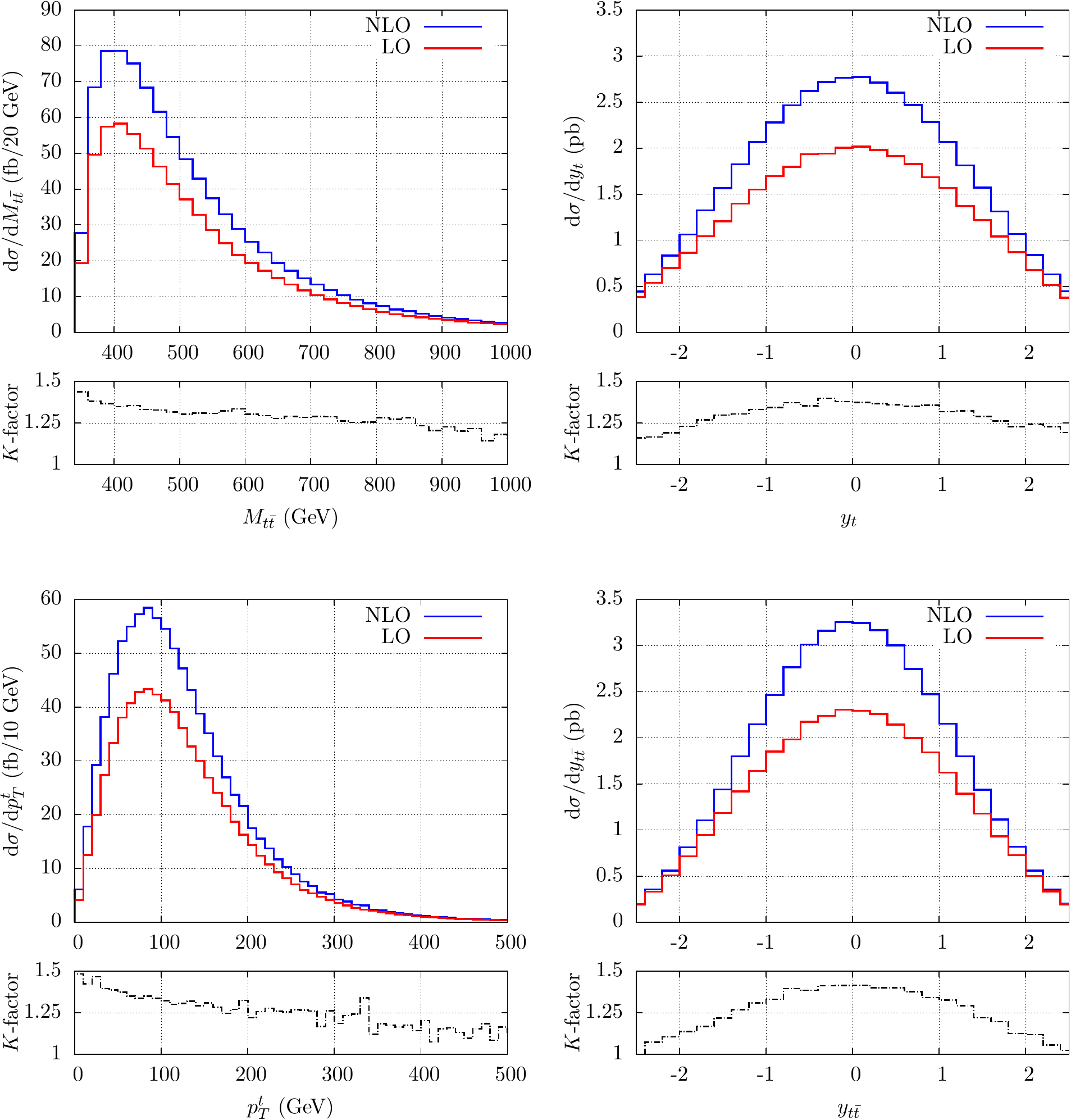}
\end{center}
\vspace{-20pt}
\caption[Kinematic distributions in \ttbar production at LO and NLO.]{Cross-section distributions in $pp \to t\bar{t}$ collisions at the LHC at NLO and LO, with associated bin-by-bin $K$-factors, as calculated with {\sc{Mcfm}}~\cite{Campbell:2010ff}. Here $y_{t\bar{t}} \equiv y_t - y_{\bar{t}}$.}
\label{fig:kfac}
\end{figure}

Top quark differential distributions have been calculated at NLO and are now fully automated in various Monte Carlo event generator programs~\cite{Campbell:2014kua,Frixione:2010wd,Campbell:2010ff,Gleisberg:2008ta}. Full phase-space results (at parton level) for top quark differential distributions are now available at NNLO QCD~\cite{Czakon:2015owf,Czakon:2016ckf}, however they are not yet implemented in a Monte Carlo simulation such that they can be interfaced to a parton shower and implemented in a realistic experimental cutflow. To illustrate the importance of NLO corrections, in Fig. \ref{fig:kfac} we plot kinematic distributions in $\sigma$ at LO and NLO. Uncertainties related to scales and pdfs have not been shown, the point is merely to illustrate that NLO corrections are large (nearing 50\% in some bins) which highlights the need to include them.

\subsubsection{Charge asymmetries}
An important probe of the Standard Model in top pair production is through charge asymmetries~\cite{Bernreuther:2012sx,Aguilar-Saavedra:2014kpa,AguilarSaavedra:2011vw,Kamenik:2011wt,Kuhn:2011ri}. The most well-known of these is the so-called `forward-backward' asymmetry in proton-antiproton collisions, which is most conveniently expressed as a difference between the number of top pairs in the forward direction (parallel with the incoming proton) and the backward direction (antiparallel with the incoming proton):
\begin{figure}[!t]
\begin{center}
\includegraphics[width=\textwidth,height=5cm]{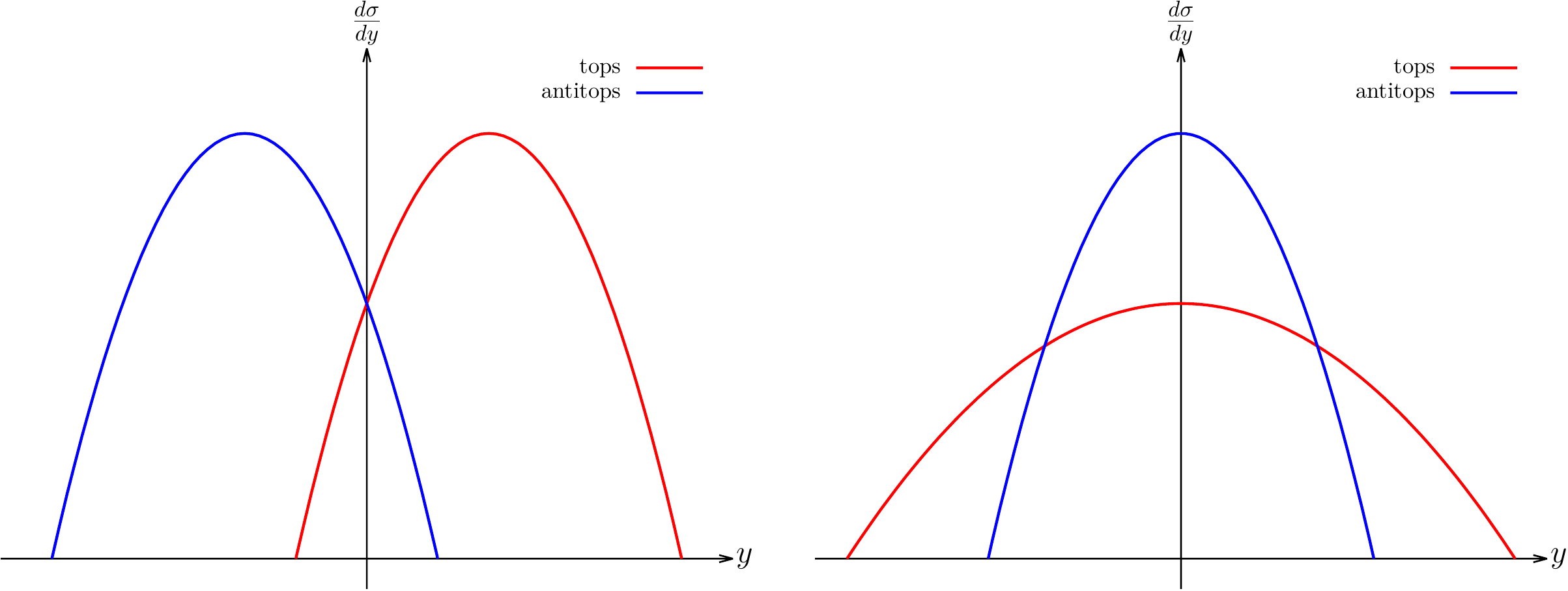}
\end{center}
\caption[Schematic of LHC and Tevatron \ttbar asymmetries.]{Exaggerated schematic of the origin of the asymmetries $A_{FB}$ at the Tevatron (left), and $A_C$ at the LHC (right).}
\label{fig:asymmetries}
\end{figure}

\begin{equation}
A_{FB} = \frac{N(\Delta y > 0)-N(\Delta y <0)}{N(\Delta y > 0)+N(\Delta y <0)},
\label{eqn:afb}
\end{equation}
where $\Delta y = y_t - y_{\bar{t}}$. An asymmetry arises in the subprocess $q\bar{q}\to t\bar{t}$ due to terms which are odd under the interchange $t\leftrightarrow\bar{t}$ (with initial quarks fixed), specifically from the interference between the tree-level diagram for $q\bar{q}\to t\bar{t}$ and the 1-loop `box' diagram, and interference between the real emission contributions for $q\bar{q}\to t\bar{t}g$. Thus, the asymmetry originates at next-to-leading order in QCD. The SM prediction at NNLO QCD is \Afb = 7.24+1.04-0.67~\cite{Czakon:2014xsa}, where the errors are from scale variation.

A different, but related, asymmetry can be defined at the LHC, where the charge symmetric initial state does not define a `forward-backward' direction. Instead, a \textit{central} charge asymmetry $A_C$ can be defined
\begin{equation}
A_{C} = \frac{N(\Delta |y| > 0)-N(\Delta |y| <0)}{N(\Delta |y| > 0)+N(\Delta |y| <0)},
\label{eqn:ac}
\end{equation}
where $\Delta |y| = |y_t| - |y_{\bar{t}}|$. This definition makes use of the fact that in $q\bar{q}\to t\bar{t}$ the quark in the initial state is almost always a valence quark and is likely to carry more longitudinal momentum than the antiquark, which is always a sea quark. The net result is that tops, being more correlated with the direction of the initial state quarks, tend to be produced at larger absolute rapidities than antitops. However, at LHC energies, gluons dominate the beam composition, so the $gg\to t\bar{t}$ channel, for which $A_C = 0$, dominates the cross-section. This means \Ac is much more diluted than \Afb. Its SM prediction is \Ac = 0.0123 $\pm$ 0.0005~\cite{Bernreuther:2012sx}, which includes NLO QCD and electroweak corrections. The asymmetries at the Tevatron and the LHC are visualised in Fig. \ref{fig:asymmetries}.

\subsubsection{Single top production}
\begin{figure}[!t]
\begin{center}
\includegraphics[width=\textwidth,height=3.5cm]{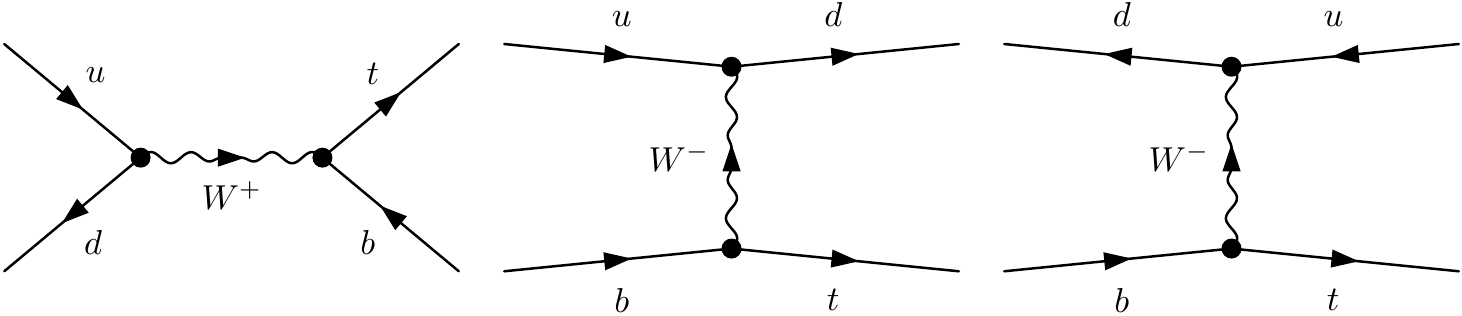}
\end{center}
\caption[Feynman diagrams for electroweak single top production in the SM.]{The leading order Feynman diagrams for electroweak single top production in the SM. The corresponding antitop diagrams are constructed by reversing the fermion arrows. }
\label{fig:stchan}
\end{figure}

\begin{figure}[t!]
\begin{center}
\includegraphics[width=10cm,height=3.5cm]{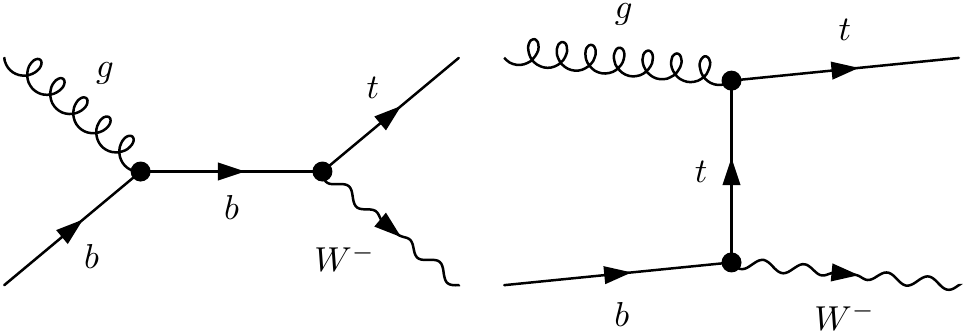}
\end{center}
\caption[Feynman diagrams for $Wt$ associated production in the SM.]{The leading order Feynman diagrams for $Wt$-associated production in the SM. The corresponding antitop diagrams are constructed by reversing the fermion arrows. }
\label{fig:wtprod}
\end{figure}

 The next-most-dominant way of producing top quarks at hadron colliders is the single-top process, which can be sub-categorised into the purely electroweak processes $q\bar{q}' \to t\bar{b}$ and $qb \to tq'$, mediated by $W$ bosons in the $s$~\cite{Cao:2004ap,Campbell:2004ch,Kidonakis:2010tc} and $t$-channel~\cite{Kidonakis:2011wy,Campbell:2009ss,Falgari:2010sf,Cao:2005pq,Brucherseifer:2014ama}, and the electroweak+QCD process $gb \to tW$; referred to as $Wt$-associated production~\cite{Willenbrock:1986cr,Tait:1999cf,Zhu:2001hw,Campbell:2005bb,Frixione:2008yi,Cao:2008af,Re:2010bp}.  Feynman diagrams for both cases are shown in Figs. \ref{fig:stchan} and \ref{fig:wtprod}.

The $s$-channel cross-section has a relatively large rate at the Tevatron, but at the LHC it is much rarer than its $t$-channel counterpart, because it is initiated by antiquarks and so is suppressed by the initial parton densities. The signature for $s$-channel production is a pair of $b$-tagged quarks, one originating from the primary vertex and one from the decay of the top quark, a high $p_T$ lepton, and missing transverse energy, corresponding to a neutrino from the leptonic top quark decay. It remains a challenging channel to reconstruct, however, due to its small event rate and large backgrounds, namely from top pair and $W$+jets. The leading order partonic cross-section for $s$-channel top production is

\begin{equation}
\sigma_{u\bar{d}\to t\bar{b}} = \frac{|V_{ud}|^2|V_{tb}|^2g^4(s-m^2_t)^2(2s+m^2_t)}{384\pi s^2(s-M^2_W)^2}.
\end{equation}
 
In $t$-channel production, in order to produce a top quark, the spacelike $W$ must be highly off-shell, and so there is a large momentum transfer between the outgoing partons, hence the light quark tends to recoil against the heavy top, leading to an untagged jet in the forward region of the detector. Moreover, the exchange of a color singlet between the two outgoing partons means there is relatively little QCD radiation in the region between them, leading to suppressed central jet activity between the top quark decay products and the jet from the light quark, known as a  \textit{rapidity gap}. Though this defines a very clear experimental signature, at the theoretical level there exists some ambiguity in the parton-level definition of this process. One may choose to define the incoming $b$-quark as originating directly from the incoming proton, using a so-called 5-flavour scheme for the proton pdf, leading to the $2\to2$ topology as shown in Fig. \ref{fig:stchan}. Alternatively, one may treat the $b$-quark as the product of the collinear splitting of a gluon ($g\to b\bar{b}$) in the initial state, leading to a $2\to3$ event topology.

Formally, these two treatments should lead to the same cross-section prediction, but differ when truncated at fixed-order in perturbation theory, in particular due to the accuracy at which the logarithms originating from the gluon splitting are resummed, and the treatment of these splittings in the evolution of the pdfs.
The leading-order parton level cross-section for $t$-channel production in the 5-flavour scheme, in both the $ub\to dt$ and $d\bar{b}\to\bar{u}t$ channels are:
 \begin{equation}
 \begin{split}
 \sigma_{ub \to dt} &=  \frac{|V_{ud}|^2|V_{tb}|^2g^4(s-m^2_t)^2}{64\pi s M^2_W(s-m^2_t +M^2_W)} \\
 \sigma_{d\bar{b}\to\bar{u}t} &= \frac{|V_{ud}|^2|V_{tb}|^2g^4\left((s-m^2_t)^2(2s+m^2_t)-M^2_W(2s+2M^2_W-m^2_t)\log\frac{s+M^2_W-m^2_t}{M^2_W}\right)}{4\pi s^2 M^2_W} .
 \end{split}
 \end{equation}
 Finally, for $Wt$-associated production, the cross-section in the 5-flavour scheme takes the form
 \begin{equation}
 \begin{split}
 \sigma_{gb\to Wt} =  &\frac{|V_{ud}|^2|V_{tb}|^2g^2g^2_s}{384s^3M^2_W}\bigg(-3((m^2_t - 2M^2_W)s+7(m^2_t-M^2_W)(m^2_t+2M^2_W))\lambda^{1/2}(s,m^2_t,M^2_W)  \\
 & +2(m^2_t+2M^2_W)(s^2+2(m^2_t-M^2_W)+2(m^2_t-M^2_W))\log\Big(\frac{s+m^2_t+M^2_W+\lambda^{1/2}}{s+m^2_t+M^2_W-\lambda^{1/2}}\Big)\bigg)
 \end{split}
 \end{equation}
 where $\lambda(x,y,z) = x^2+y^2+z^2 - 2xy -2xz -2yz$ is the K\"all\'en function~\cite{Kallen:1964lxa}.
 
The pure electroweak single top production processes typically have cross-sections an order of magnitude smaller than for top pair production. Although the available phase space for producing one top instead of two is much larger, the matrix elements are parametrically suppressed by the strength of the electroweak coupling relative to the strong coupling. For the same reason, the cross-sections are more stable against higher-order corrections, and $K$-factors for $s$ and $t$-channel production are more flat in differential distributions and scale choices, typically at the 10-20\% level. The most up-to-date calculations for electroweak single-top production are at approximate NNLO, although there are different definitions of this term. One calculation calculates in the so-called structure function approximation, where only factorisable vertex corrections are considered~\cite{Brucherseifer:2014ama}. The remaining terms are colour suppressed $\sim 1/N_c^2$ and kinematically subdominant. Another approach is to expand the resummed leading-order cross-section to $\mathcal{O}(\alpha_s^2)$~\cite{Kidonakis:2011wy}. Both calculations are in general agreement. The latter yields for the $t$-channel:
\begin{equation}
\begin{split}
\sigma(pp \to tq + X) &= 65.7\hspace{5pt}^{+1.9}_{-1.9} \hspace{10pt}  \sqrt{s}=7\hspace{5pt} \text{TeV} \\
\sigma(pp \to tq + X) &= 87.1\hspace{5pt}^{+0.24}_{-0.24} \hspace{10pt}  \sqrt{s}=8\hspace{5pt} \text{TeV} \\
\sigma(p\bar{p} \to tq + X) &= 2.06\hspace{5pt}^{+0.13}_{-0.13} \hspace{10pt}  \sqrt{s}=1.96\hspace{5pt} \text{TeV} .
\end{split}
\end{equation}
where the uncertainties quoted have added scale and pdfs in quadrature, and for the $s$-channel:
\begin{equation}
\begin{split}
\sigma(pp \to tb + X) &= 4.5\hspace{5pt}^{+0.2}_{-0.2} \hspace{10pt}  \sqrt{s}=7\hspace{5pt} \text{TeV} \\
\sigma(pp \to tb + X) &= 5.5\hspace{5pt}^{+0.2}_{-0.2} \hspace{10pt}  \sqrt{s}=8\hspace{5pt} \text{TeV} \\
\sigma(p\bar{p} \to tb + X) &= 1.03\hspace{5pt}^{+0.05}_{-0.05} \hspace{10pt}  \sqrt{s}=1.96\hspace{5pt} \text{TeV} .
\end{split}
\end{equation}
These values are summed over the top and antitop channels. At the Tevatron, owing to its charge symmetric initial states, both channels contribute equally, while at the LHC the relative top/antitop contributions are 65\% to 35\% for $t$-channel, and 69\% to 31\% for $s$-channel.

Since the $Wt$ process is QCD initiated, it is expected to receive sizeable corrections from higher-order terms. However, an ambiguity arises  when one tries to define an NLO estimate for this process. Generically, NLO corrections result from both virtual `loop' corrections, and emission of real particles. The latter type in $Wt$ production include diagrams of the form shown in Fig. \ref{fig:wtint}:
\begin{figure}[t!]
\begin{center}
\includegraphics[width=10cm,height=3.5cm]{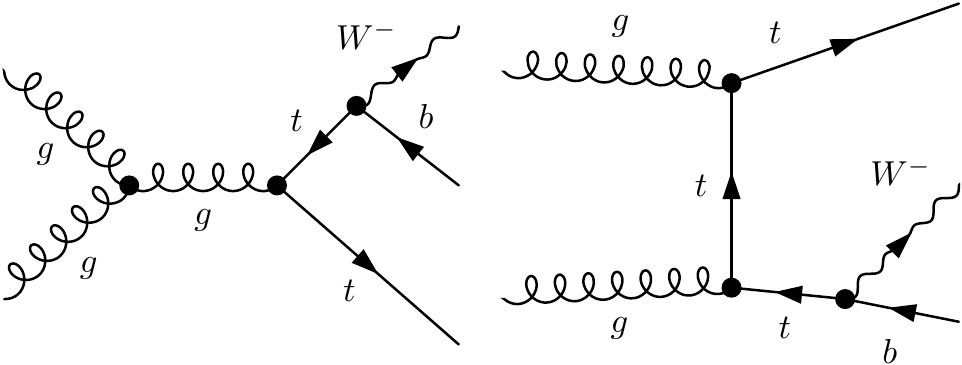}
\end{center}
\caption[Interference between NLO $Wt$ associated production and \ttbar production.]{Real emission contributions to $Wt$-associated production at NLO. These also correspond to diagrams for $t\bar{t}$ production, where the antitop has decayed. }
\label{fig:wtint}
\end{figure}
which are also present in resonant top-pair production, with one top quark decay $t \to Wb$. When this intermediate top quark goes on-shell, the $Wt$ cross-section becomes of the order of the $t\bar{t}$ one, which is an order of magnitude larger. In this regime, such a large $K$-factor means a perturbative definition of the $Wt$ process is ill-defined. The question is then, do such diagrams belong to $Wt$ or $t\bar{t}$, and can one make an NLO definition of $Wt$ production that avoids the interference with resonant top-pair production? This problem has been studied in some detail~\cite{Campbell:2005bb,White:2009yt,Belyaev:1998dn,Kersevan:2006fq,Kauer:2001sp,Demartin:2016axk}. 

Two well-known prescriptions for removing the effects of top pair production in $Wt$ at NLO are \textit{diagram removal} and \textit{diagram subtraction}. The former removes the diagrams of the form of Fig. \ref{fig:wtint} at the amplitude level, so they do not enter the calculation. The latter subtracts their contributions from the final cross-section. The difference between the NLO $Wt$ cross-section predictions from these two methods thus provides a measure of the interference effect of the diagrams of Fig. \ref{fig:wtint}~\cite{Frixione:2008yi,White:2009yt}. Reasonable choices of experimental cuts can be made to minimise this interference (an obvious choice, for instance, would involve an invariant mass cut close to the $t\bar{t}$ threshold), so that $Wt$ and $t\bar{t}$ can, for all practical purposes, be considered as separate processes\footnote{For an alternative viewpoint, see e.g. Ref.~\cite{Kauer:2001sp}.}.
 
However one decides to define the $Wt$-associated production process, the cross-section is too small at the Tevatron to be of any phenomenological relevance. At the LHC, however, the approximate NNLO cross-section, defined from the NNLO expansion of the NNLL-resummed cross-section, is:
\begin{equation}
\begin{split}
\sigma(pp \to Wt + X) &= 15.5\hspace{5pt}^{+1.2}_{-1.2} \hspace{10pt}  \sqrt{s}=7\hspace{5pt} \text{TeV} \\
\sigma(pp \to Wt + X) &= 22.1\hspace{5pt}^{+1.2}_{-1.2} \hspace{10pt}  \sqrt{s}=8\hspace{5pt} \text{TeV} .
\end{split}
\end{equation}
Since the process is initiated by a gluon and a $b$-quark, and $f_b(x) = f_{\bar{b}}(x)$ in the proton, the top/antitop contributions are equal.  

\subsubsection{Higher-order processes}
As the LHC probes kinematic regions inaccessible to previous colliders such as the Tevatron, new event topologies with a higher multiplicity of hard partons become increasingly commonplace. Of special interest for top physics are processes where a top quark pair is produced in association with an additional particle in the hard process, dubbed \textit{higher-order} because they already have a 2 $\to$ 3 topology at tree-level. These processes probe directly the top couplings of Eqs.~\eqref{eqn:ttgvertex}-\eqref{eqn:tthvertex}, allowing for a model-independent way of constraining new top interactions. For instance, a measurement of top pairs in association with a $Z$ boson directly probes the $t\bar{t}Z$~\cite{Lazopoulos:2008de,Kardos:2011na} coupling, allowing contact to be made with precision LEP observables. In principle this could be extracted from the simple $pp\to Z \to t\bar{t} $ process, but this signal is drowned out by the much larger QCD $t\bar{t}$ rate. 

Top pairs produced in association with Higgs bosons ($t\bar{t}H$)~\cite{Dawson:2003zu,Beenakker:2002nc} are of particular interest in this regard, because they allow for a model-independent extraction of the top quark Yukawa coupling, thus offering discriminating power between models where much of the top quark mass is generated from a non-SM mechanism. These processes have small rates, typically $\mathcal{O}$(100 fb), and are thus experimentally challenging. Nonetheless, significant progress has been made towards their discovery in LHC Run I~\cite{Khachatryan:2015ila,Aad:2016zqi}, and the high statistics forecast for the LHC lifetime suggest they can ultimately be measured with similar precision to the leading order processes discussed above. 

\subsubsection{Top quark decay}
The unique properties of the top quark stem largely from the characteristics of its decay. The top quark is the only quark in the Standard Model with a mass larger than that of the W boson. Hence it can decay directly through the process $t\to Wb$. The mass difference $m_t -M_W \sim$ 90 GeV means that the allowed phase-space for the decay is large, and so the top quark decays before the strong interaction can bind it into hadrons. At next-to-leading order in $\alpha_s$, the top decay width is given by~\cite{Jezabek:1988iv} 

\begin{equation}
\Gamma_t = \frac{G_Fm^3_t}{8\pi\sqrt{2}}\left(1-\frac{M^2_W}{m^2_t}\right)\left(1+2\frac{M^2_W}{m^2_t}\right)\left[1-\frac{2\alpha_s}{3\pi}\left(\frac{2\pi^2}{3}-\frac{5}{2}\right)\right].
\end{equation}

where terms of order $m_b^2/m_t^2, \alpha_s^2 $ and $(\alpha_s/\pi)M_W^2/m_t^2$ have been neglected, and it is assumed $|V_{tb}|^2 =1$. For $m_t$ = 173.3 GeV this gives a value of $\Gamma_t  \sim$ 1.3 GeV, corresponding to a top quark lifetime of $0.5\times 10^{-24}$ s. Despite the large phase space available for the decay, the top still satisfies the narrow width approximation $\Gamma \ll m $. This means that one can make the replacement of the propagator

\begin{equation}
\frac{1}{((s-m^2)^2+(m\Gamma)^2)} \to \frac{\pi}{m\Gamma}\delta(s-m^2),
\end{equation}

in the squared matrix element. This substantially simplifies the calculation of decay amplitudes. Nonetheless, the narrow width approximation should be treated with care, as it is not valid for observables whose main contributions originate from regions of phase space where the top is far off-shell. In addition, the raw top width is a difficult quantity to measure at hadron colliders, without making assumptions (such as a SM-like cross-section). Other observables relating to top decay can be much more precisely measured. 

For instance, the fraction of events in which the top decays to $W$-bosons with a given helicity: left-handed,
right-handed or zero-helicity, can be expressed in terms of helicity fractions, which for leading order with a finite $b$-quark mass are
\begin{equation}
\begin{split}
F_0 &=\frac{ (1-y^2)^2- x^2(1+y^2)}{(1-y^2)^2+x^2(1-2x^2+y^2)} \\
F_L &= \frac{x^2(1-x^2+y^2)+\sqrt{\lambda}}{(1-y^2)^2+ x^2(1-2x^2 + y^2)} \\
F_R &=  \frac{x^2(1-x^2+y^2) -\sqrt{\lambda}}{(1-y^2)^2+ x^2(1-2x^2 + y^2)},
\label{eqn:helfrac}
\end{split}
\end{equation}
where $x=M_W/m_t$, $y=m_b/m_t$ and $\lambda = 1 + x^4 + y^4 - 2x^2y^2 - 2x^2 - 2y^2 $. A desirable feature of these quantities is that they are relatively stable against higher order corrections, so the associated scale uncertainties are small. The Standard Model NNLO estimates for these are: $\{F_0,F_L,F_R \} = \{0.687 \pm 0.005, 0.311 \pm 0.005, 0.0017 \pm 0.0001 \}$~\cite{AguilarSaavedra:2006fy,Czarnecki:2010gb}, i.e. the uncertainties are at the per mille level. The large enhancement of decays to longitudinal $W$ bosons results from the Goldstone boson equivalence theorem, which states that in the limit $s \gg M_W$, $W$ scattering is dominated by the longitudinal components, so $W$s may be approximated by Goldstone scalars in calculations.

Another unique feature of the top quark is that its decay width is much larger than the QCD spin decorrelation width $\Lambda^2_{QCD}/m_t \sim $ 0.1 MeV, therefore the correlation between the spins of tops and their decay products is completely preserved, and spin correlations in, for example, top pair production can be measured directly through spins of the decay leptons (selecting dilepton events). 

\subsection{Summary}
\label{sec:conc_ch1}
To summarise, the top quark plays a special role among the Standard Model fermions, and, owing to its unique experimental properties, offers a valuable hadron collider testing ground for many SM predictions. However, the real interest in the top quark stems from its role in potential TeV scale new physics, perhaps within reach of the LHC. This is the subject of the next chapter.

\newpage

\newpage
\section{The top quark beyond the Standard Model}
\subsection{Introduction}

Despite the vast list of experimentally verified predictions of the SM over the last forty years, it still paints a somewhat unsatisfying picture of Nature. In order for it to be predictive, it requires fixing the values of 18\footnote{Excluding the QCD $\theta$-parameter, which will be discussed later in this chapter.} arbitrary parameters from experiment. These parameters span several orders of magnitude, with no apparent pattern between them. Any fundamental theory worth its salt ought to be able to predict the values of these numbers, or at least relate them in terms of a smaller subset of more fundamental parameters.

Beside the aesthetic issue of the large number of free parameters, the Standard Model also has some deep structural problems that have motivated new physics model building for the last few decades. Owing to its unique properties among the SM fermions, the top quark has played a special role in most of these scenarios. Indeed, the potential for using measurements of top quark couplings to place bounds on the effects of new physics is the main topic of this thesis.

This chapter is structured as follows: In section \ref{sec:motivations} I will outline the main motivations for physics beyond the Standard Model (BSM). In section \ref{sec:models} I discuss a few of the most popular scenarios of BSM physics, and the role the top quark plays in them. In section \ref{sec:eft} I discuss some generalities of using effective field theories to parameterise the effects of heavy degrees of freedom on low energy observables, before moving onto discussing the formulation of the Standard Model as an effective theory in section \ref{sec:smeft}, and the parts of that EFT that are relevant for top quark physics in section \ref{sec:topeft}. Conclusions are presented in section \ref{sec:conc_ch2}.

\subsection{Motivations for physics beyond the Standard Model}
\label{sec:motivations}

\subsubsection{The hierarchy problem}
Perhaps the best-known motivation for physics beyond the SM is the hierarchy problem: the vast difference between the electroweak scale and the Planck scale where quantum gravity becomes important: $v/M_{Pl} \sim 10^{-16}$. This large mass hierarchy is not a specific problem of the Glashow-Salam-Weinberg model, but a general feature of theories containing fundamental scalars. To see this, we return to the simple case of a real scalar field in four dimensions. 
\begin{equation}
\mathcal{L} = \frac{1}{2}\partial_\mu\phi\partial^\mu\phi - V(\phi) \hspace{10pt}\text{where}\hspace{10pt} V(\phi) = \frac{1}{2}\mu^2\phi^2+\frac{1}{4}\lambda\phi^4
\end{equation}

At tree-level the potential will simply correspond to the classical potential $V_{\text{tree}} = V(\phi)$. However, radiative corrections will modify this relation, and the \textit{true} potential; that is, the one that the vacuum expectation value seeks to minimise, is the so-called \textit{effective potential} $V_{\text{eff}}$, which will receive radiative corrections.

The radiative corrections originate from the effects of virtual particle emission and absorption on the interaction energy, so in principle includes all one-particle irreducible diagrams with any number of external legs $n>2$\footnote{Vacuum energy graphs with $n=0$ just shift the energy by a constant, diagrams with $n=1$ can be absorbed into a shift of the field~\cite{Peskin:1995ev}.}. The potential involves only non-derivative terms in the Lagrangian, so the momenta of the external legs can be taken to be zero without loss of generality, i.e. calculating the radiative corrections to the scalar potential amounts to summing up all 1PI diagrams with zero external momentum (for a more formal proof of this statement, see e.g. Refs~\cite{Coleman:1973jx,Sher:1988mj}). This can be done order-by-order in perturbation theory.
\begin{figure}[!t]
\begin{equation*}
\vcenter{\hbox{ \includegraphics[width=0.22\textwidth]{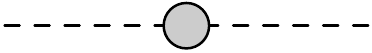} }} \qquad  = -i\mathcal{M}^2 \qqquad
\vcenter{\hbox{ \includegraphics[width=0.25\textwidth]{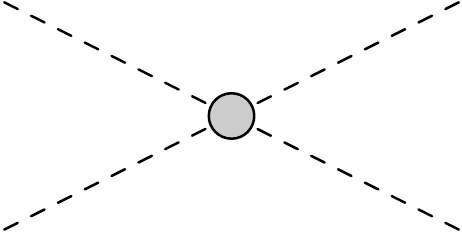} }} \qquad  = \lambda
\end{equation*}
\caption{Feynman rules for the 1-loop mass and coupling constants defined in Eqs.~\eqref{eqn:1loopcoup}-\eqref{eqn:1loopmass}.} 
\label{fig:scalarfeynrules}
\end{figure}
To isolate the quantum corrections, it is useful to split $\phi$ into a classical `background' or external field $\phi_c$, corresponding to the field in the tree potential, and quantum corrections $\delta\phi$.  
\begin{equation}
\phi = \phi_c + \delta\phi .
\end{equation}
Since we will encounter divergent loop momenta, we must define renormalisation conditions for the couplings of the theory to absorb them. Conventionally, one defines the renormalised mass of the scalar field $\phi$ as the `pole' mass, i.e. the negative of the inverse propagator at zero momenta
\begin{equation}
\mathcal{M}^2 \equiv -\frac{\delta^2\mathcal{L}}{\delta\phi_c^2} \bigg |_{\phi_c=0} ,
\end{equation}
while the renormalised coupling is defined from the 4-point function at zero momentum:
\begin{equation}
\lambda \equiv -\frac{\delta^4\mathcal{L}}{\delta\phi_c^4} \bigg |_{\phi_c=0}.
\label{eqn:1loopcoup}
\end{equation}
In the spontaneously broken theory, the subtraction point is shifted from $\phi_c = 0$ to $\phi_c = \braket{\phi}$ so that
\begin{equation}
\mathcal{M}^2 \equiv -\frac{\delta^2\mathcal{L}}{\delta\phi_c^2} \bigg |_{\phi_c= \braket{\phi}} \quad = \mu^2+\frac{\lambda}{2} \braket{\phi}^2.
\label{eqn:1loopmass}
\end{equation}
This means that the renormalised masses and couplings will in general be functions of $\braket{\phi}$. The corresponding Feynman rules are shown in Fig. \ref{fig:scalarfeynrules}.

As mentioned above, the all-order effective potential is given by the sum of all 1PI vacuum diagrams with zero external momentum. At one-loop then, calculating the contributions to $V_{\text{eff}}$ amounts to summing up the vacuum `bubble' diagrams of the form shown below~\cite{Sher:1988mj}, 

\begin{figure}[!h]
\begin{equation*}
\begin{split}
&V_{\text{eff}}^{\text{1-loop}}(\phi_c) = \\
&\vcenter{\hbox{\includegraphics[width=0.2\textwidth]{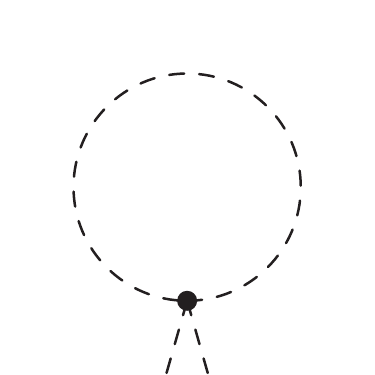} }}+ \vcenter{\hbox{\includegraphics[width=0.2\textwidth]{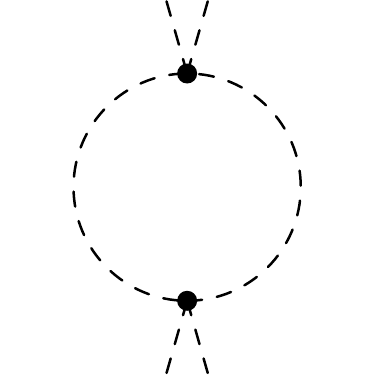} }}+\vcenter{\hbox{\includegraphics[width=0.2\textwidth]{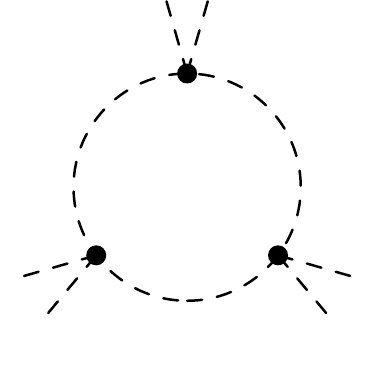} }}+\quad\vcenter{\hbox{\includegraphics[width=0.2\textwidth]{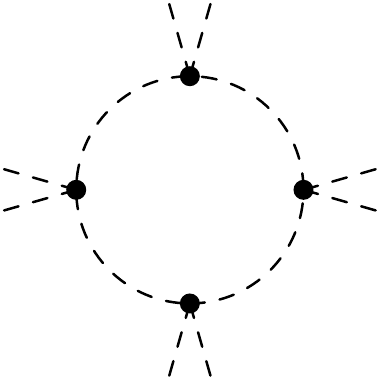} }}+\ldots
\end{split}
\end{equation*}
\end{figure}

leading to a geometric series which can be resummed, giving (up to constant terms) the well-known one-loop Coleman-Weinberg effective potential~\cite{Coleman:1973jx}:

\begin{equation}
\begin{split}
V(\phi_c) &= V_{\text{tree}}+ V^{(1)}(\phi_c) \\
V^{(1)}(\phi_c) &= \frac{\Lambda^2}{32\pi^2} \mu^2 + \frac{\lambda \phi_c^2 \Lambda^2}{32\pi^2} + \frac{(\mu^2+\lambda\phi_c^2)^2}{64\pi^2} \left[\log\frac{(\mu^2+\lambda\phi_c^2)}{\Lambda^2}-\frac{1}{2}\right].
\label{eqn:colwein}
\end{split}
\end{equation}

Clearly, this expression is divergent: the mass term has a divergence proportional to the UV cutoff $\Lambda^2$, and the quartic coupling has a logarithmic divergence. These may be absorbed into counterterms $\delta_\mu$ and $\delta_\lambda$ specified by the renormalisation conditions of Eqs.~\eqref{eqn:1loopcoup} and~\eqref{eqn:1loopmass}, so that the full potential at 1-loop order is then
\begin{equation}
V(\phi_c) = \mu^2_{\text{renorm}}\phi_c^2 + \frac{\lambda_{\text{renorm}}}{4} \phi_c^4 =  V_{\text{tree}}+ V^{(1)}(\phi_c) + V_{\text{c.t.}}
\end{equation}
This, however, means that the renormalized (physical) mass will receive corrections of the form
\begin{equation}
\mu^2_{\text{renorm}} = \mu^2_0 + \frac{\lambda_0 \Lambda^2}{32\pi^2} + \frac{\lambda_0\mu^2_0}{64\pi^2}\biggl[\log\left(\frac{\mu^2_0+\frac{\lambda\braket{\phi}^2}{2}}{\Lambda^2}\right)^2-\frac{1}{2} \biggr] - \delta_\mu . 
\end{equation}
So in order to keep the renormalised mass of the same order as the bare mass $m_0 = \sqrt{-2\mu^2}$, we require a cancellation between the quadratically divergent term $\Lambda$ and the counterterm $\delta_\mu$ (the logarithmic term remains of the same order since it is multiplied by $\mu^2_0$). If the physical mass is to be much smaller than the cutoff $\Lambda$, we must assume a miraculous cancellation between contributions below the cutoff and the unknown UV degrees of freedom, parameterised by the counterterms, above the cutoff~\cite{Gildener:1976ai,Dimopoulos:1981zb}.

This analysis can be applied to the $\mu^2$ term in the Higgs potential of the Standard Model. Here the cutoff $\Lambda$ denotes the generic scale at which the Standard Model is no longer valid, it could for instance represent the mass of a new heavy scalar. Since it can be subtracted off in mass renormalisation, it should not affect low-energy physics. To ensure this, however, requires an extraordinarily precise fine-tuning of parameters. Suppose, for instance, that the SM were valid all the way up to the Planck scale. To keep the renormalized mass at the 100 GeV scale one would need to arrange for the cancellation between the `bare' mass, describing the low-energy theory, and the counterterms, describing unknown high-energy degrees of freedom. This cancellation would have to be precise to 16 orders of magnitude, and hold through several orders in perturbation theory. Even if one started without a tree level mass, and generated it radiatively, i.e. by just considering the quartic term in the tree-level potential, the large mass corrections would still be present, because the scalar mass renormalisation is additive, not multiplicative\footnote{Although other regularisation schemes such as dimensional regularisation `hide' the UV divergences by lacking a UV cutoff $\Lambda$, this does not mean the hierarchy problem is an artefact of using a momentum-dependent regulator: it is a statement that parameters at one scale are sensitive to parameters at a vastly different scale, whether one expresses this scale in terms of a cutoff or not.}. 

We know of no other situation in physics where degrees of freedom separated by so many orders of magnitude would conspire to produce the phenomena that we observe. To calculate the Bragg diffraction angles on a crystal, for instance, one does not need to know the mass of the $Z$ boson. In keeping with this separation of scales principle, it seems that the natural mass for a fundamental scalar in a theory is close to the cutoff of that theory. What mechanism is it, then, that keeps the Higgs so light? This is the hierarchy problem, and has been the main driving force for physics beyond the Standard Model for the last forty years. 

One could restate the argument in a different way. The hierarchy problem is \textit{not} the fact that there is a large difference between the electroweak scale and the Planck scale. One does not complain, for instance, about the large hierarchy ($\sim 10^6$) between the electron mass and the electroweak scale. This is because the electron mass term in the Standard Model originates from its chiral Yukawa coupling to the Higgs field:

\begin{equation}
\bar{L}\varphi e_R \to y_e (\nu_L, e_L) H e_R.
\end{equation}

This is the only term in the SM Lagrangian that breaks the electron's chiral symmetry. Consequently, any radiative corrections that break chiral symmetry can only be proportional to positive powers of $y_e$. Setting $y_e = 0$ thus enlarges the symmetry group of the Standard Model. Approximate symmetries like this have physical consequences such as (approximately) conserved currents, so there is a `natural' reason for $m_e$ to take such a small value in relation to other relevant scales. $y_e$ is an example of a `technically natural' parameter~\cite{tHooft:1979rat}. The Higgs mass term $\mu^2(H^\dagger H)$, on the other hand, is not technically natural; since it is already invariant under any chiral transformation $H\to e^{i\theta\gamma_5} H$, so it has no natural reason for being so much smaller than na\"ive power counting would suggest.

\subsubsection{Vacuum stability}
Arguments for Naturalness of the Higgs mass as evidence for the need for new physics are convincing, but not incontrovertible. The Standard Model is a renormalisable field theory, which means that it is in principle a valid description of Nature from the electroweak scale all the way up to the Planck scale, where the degrees of freedom of quantum gravity will become important. It is thus possible, that there is no new physics in the region in between. Extrapolating the Standard Model across this many orders of magnitude, however, leads to an interesting implication for cosmology. To show this, we consider the 1-loop renormalisation group equations for the following Standard Model parameters, with $n_f = 6$ flavours of quark: 

\begin{itemize}
\item The hypercharge coupling $g'$:
\begin{equation}
\mu \frac{dg'}{d\mu} = \frac{41}{6} \frac{g'^3}{16\pi^2}
\label{eqn:rge_gprime}
\end{equation}

\item The SU(2) gauge coupling $g$:
\begin{equation}
\mu \frac{dg}{d\mu} = -\frac{19}{6} \frac{g^3}{16\pi^2}
\label{eqn:rge_g}
\end{equation}

\item The strong coupling constant $g_s$:
\begin{equation}
\mu \frac{dg_s}{d\mu} = -7 \frac{g_s^3}{16\pi^2}
\label{eqn:rge_gs}
\end{equation}

\item The Higgs quartic coupling $\lambda$:
\begin{equation}
\mu \frac{d\lambda}{d\mu} = \frac{1}{16\pi^2}\left( \frac{3g'^4}{8}+\frac{3g^2g'^2}{4}+\frac{9g^4}{8}-6y_t^4-\lambda(3g'^2+9g^2-12y_t^2)+24\lambda^2\right)
\end{equation}

\item The top quark Yukawa coupling $y_t$:
\begin{equation}
\mu \frac{dy_t}{d\mu} = \frac{y_t}{16\pi^2}\left( \frac{9y_t^2}{2}-\frac{17g'^2}{12}- \frac{9g^2}{2}-8g_s^3 \right).
\end{equation}

\end{itemize}

\begin{figure}[!t]
\begin{center}
\includegraphics[width=\textwidth]{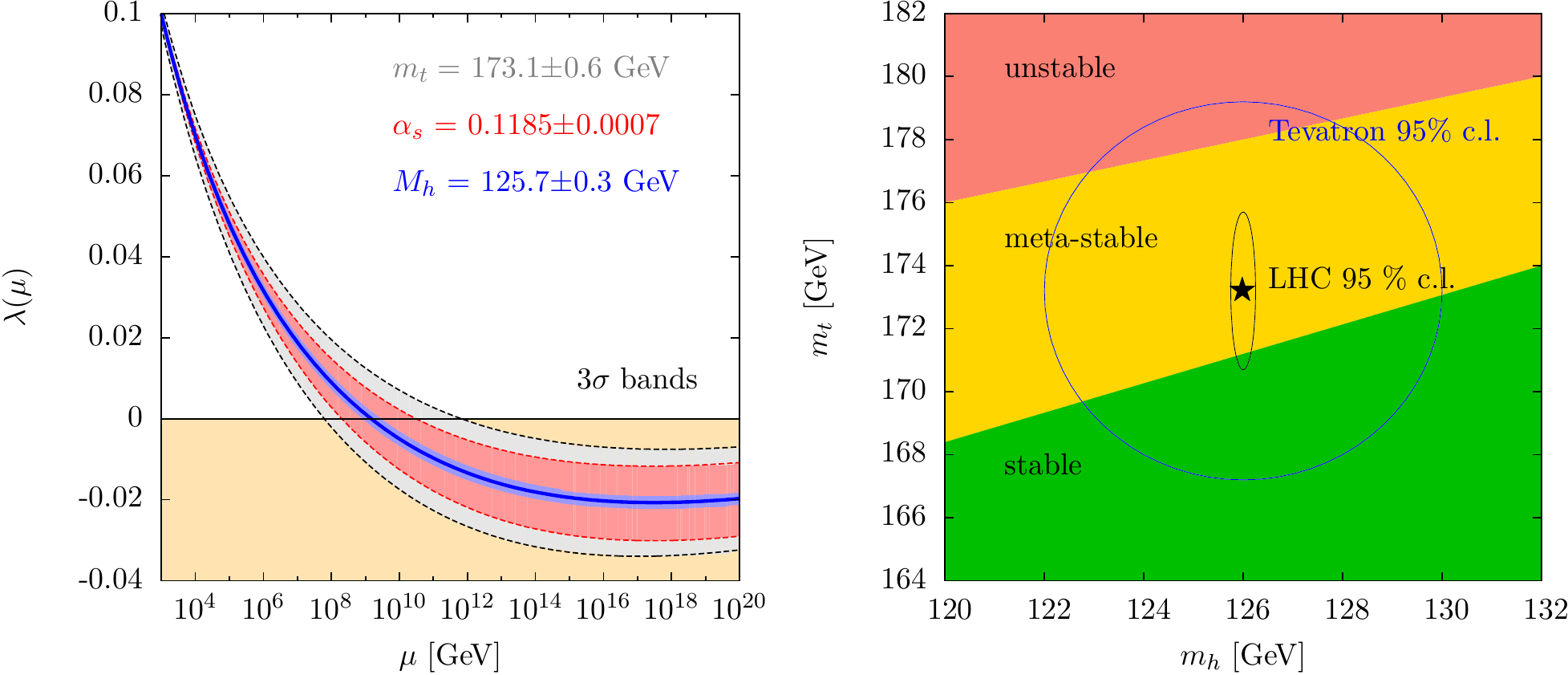}
\caption[Renormalisation group evolution of the Higgs self-coupling and regions of stability, instability and metastability in the $m_t$-$M_h$ plane.]{Left: One-loop renormalisation group evolution of the SM Higgs self-coupling $\lambda$. Also shown are the 3$\sigma$ bands for the three most dominant sources of uncertainty: the measured value of the top quark mass $m_t$, the strong coupling constant $\alpha_s$, and the Higgs mass $M_h$. Right: Corresponding regions of stability, meta-stability and instability (as described in the text) of the electroweak vacuum, as a function of the Higgs and top quark masses, overlaid with the most recent best fit contours for both from the Tevatron and LHC.} 
\label{fig:stability}
\end{center}
\end{figure}

The evolution of the Higgs self-coupling is clearly most sensitive to the top quark Yukawa (all of its other fermionic couplings can be safely neglected). It is also sensitive to the large running of the strong coupling constant (indirectly through $y_t$) and to itself, $\lambda$. Substituting in as boundary conditions the values of SM couplings at the electroweak scale $\mu = v$, taken from Tab 1.1, and using $\lambda = m_h^2/2v^2$, we can straightforwardly solve for $\lambda(\mu)$. The running of the Higgs self-coupling is plotted on the left of Fig. \ref{fig:stability}.

The $y_t^4$ term tends to drive the Higgs potential negative at large renormalisation scales. The precise scale at which this happens is extremely sensitive to the values of the electroweak scale inputs, but it is clear that $\lambda < 0$ at some scale $\mu = \Lambda < M_{Planck}$. It is unclear how this should be interpreted. As mentioned in chapter 1, the absolute stability of the Higgs potential requires that $\lambda > 0$, otherwise the potential will be unbounded from below. In the SM alone, there are no terms which can rescue the boundedness of the potential: it is negative definite for $\mu > \Lambda$. However, one can reasonably assume that couplings between the Higgs sector and Planck scale physics, which would manifest in the Higgs potential as higher-dimensional operators such as $(\varphi^\dagger\varphi)^3$, will restore the boundedness of the potential. Still, this means that the electroweak vacuum is not a \textit{true vacuum}: there is another vacuum at a much higher scale (perhaps at $M_{Planck}$) for which it is energetically favourable for the Universe to tunnel into. Based on whatever numerical value the tunnelling rate $\xi$ takes\footnote{An expression for the tunnelling/vacuum decay rate in terms of $\lambda(\mu)$ can be found in e.g. Ref.~\cite{Espinosa:2007qp}.} , one can draw three possible consequences for the fate of the electroweak vacuum~\cite{Coleman:1977py,Callan:1977pt}:

\begin{itemize}

\item \underline{$\xi$ = 0:}\\ The tunnelling rate is exactly zero, and $v$ is actually the true minimum of the Higgs potential. This means the Higgs potential is absolutely stable. It is apparently disfavoured by data, unless the top mass and Higgs mass are respectively somewhat smaller and larger than their current measurements suggest.

\item \underline{$\xi < 1/T_{Universe}$ :}\\ The tunnelling rate is non-zero, but with a decay lifetime larger than the current age of the Universe, which would explain why the tunnelling has not yet taken place.

\item \underline{$\xi > 1/T_{Universe}$ :}\\ The tunnelling rate is faster than the Hubble rate, meaning the Universe should have undergone a phase transition from the electroweak vacuum to the true one, sometime between the Big Bang and today.

\end{itemize}

One can express these different possibilities in terms of $m_t$ and $M_h$, as shown on the right of Fig. \ref{fig:stability}. The current measurements place us squarely in the metastable region. This fact alone is unremarkable (although if we were in the unstable region we ought to have a good explanation). However, it is a peculiar outcome for the evolution of the Universe. Shortly after the Big Bang, the Universe was in a state of very large free thermal energy. At some point the Universe must have then cooled enough to undergo a phase transition to the electroweak vacuum. It is then unknown what caused the Universe to choose the less energetically favourable of the two vacua, or what stabilised this false vacuum against quantum tunnelling and thermal fluctuations, which would have been much more likely in the early Universe (i.e. when $\xi$ was much larger than $1/T_{Universe}$)\footnote{The state-of-the-art calculation of vacuum stability in the SM is at NNLO~\cite{EliasMiro:2011aa,Degrassi:2012ry,Espinosa:2015qea,Ellis:2009tp}, rather than the leading-order one presented here, and a more comprehensive discussion of the issue can be found in those papers.}. 

If one believes in Naturalness as a guiding principle, one should take this problem as seriously as the hierarchy problem. This apparent paradox may be taken as indirect evidence that new physics lies between $v$ and $M_{Planck}$ which stabilises the potential, because the preceding argument was only valid if there was no such new physics. At any rate, it shows that despite satisfying the requirement of renormalisability, the Standard Model alone paints a rather unsatisfactory picture of physics between $v$ and $M_{Planck}$.
\newpage

\subsubsection{Gauge coupling unification}

\begin{figure}[!t]
\begin{center}
\includegraphics[width=\textwidth]{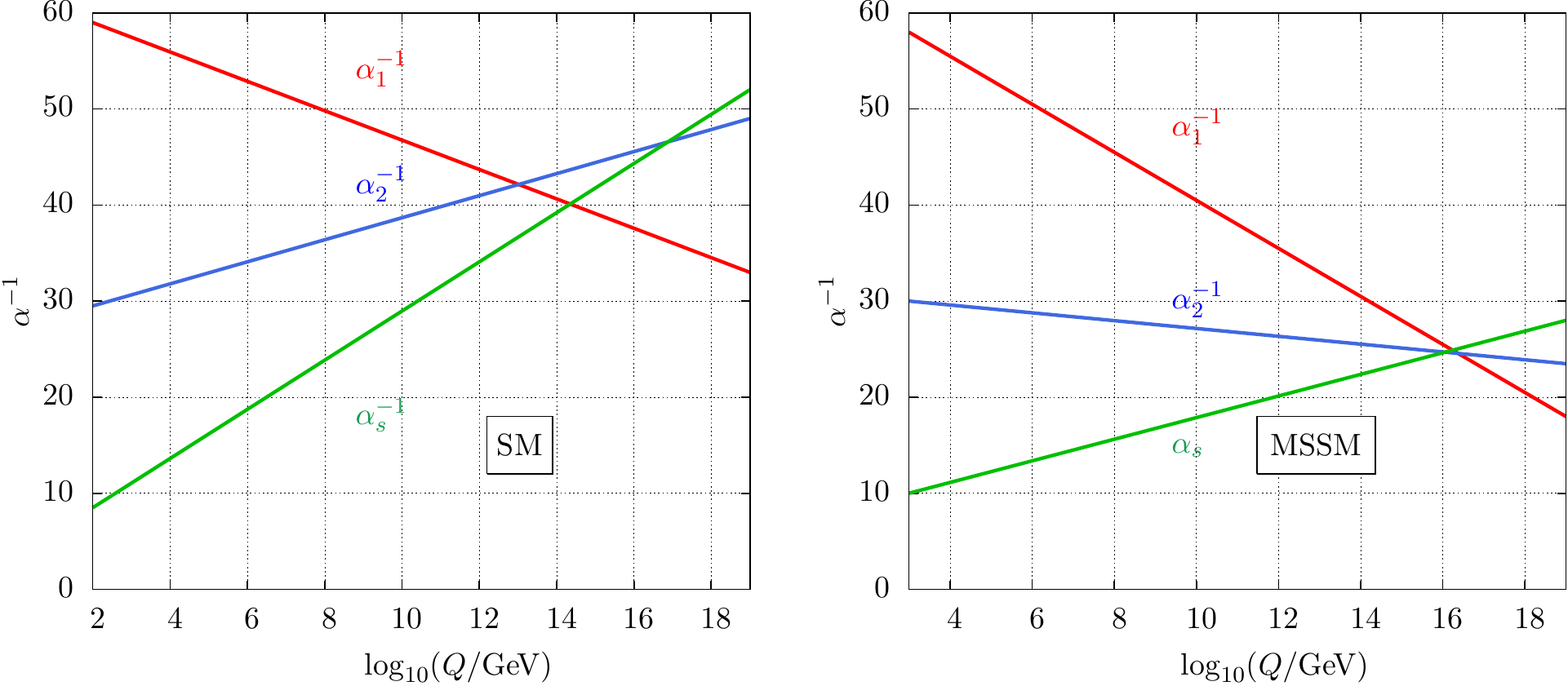}
\caption[Coupling constant unification in the MSSM.]{Left: One-loop renormalisation group evolution of the SM gauge couplings. Right: Their renormalisation group flow in the Minimal Supersymmetric extension of the Standard Model (MSSM), showing unification at a scale $Q \sim 10^{16}$ GeV.} 
\label{fig:unification}
\end{center}
\end{figure} 

The extraordinary accuracy of the electroweak theory in describing physics up to $v$ has led to speculation that there may be an even larger unification scenario, in which the electroweak and strong interactions are unified under a single gauge group, with new degrees of freedom occupying that unification scale. Consider instead Eqs.~\eqref{eqn:rge_gprime}-\eqref{eqn:rge_gs}. Although the coupling constants have very different values at the electroweak scale, they run in different directions, so perhaps unify at a single energy scale. The one-loop renormalisation group flow of $g',g$ and $g_s$ in the Standard Model is shown on the left of Fig. \ref{fig:unification}.

It can be seen that the couplings do not meet, and so no unification takes place. On the other hand, adding in new degrees of freedom between $v$ and $M_{Planck}$ will also alter the running of the couplings. One of the most well-studied candidates for new physics is surely supersymmetry. In the minimal supersymmetric extension of the MSSM, where every SM particle is supplemented with a superpartner of opposite spin (i.e. there is only one copy of the supersymmetry algebra), the couplings unify\footnote{In fact, the couplings do not exactly unify in the MSSM, but exact unification can be achieved by adding appropriate threshold corrections at the unification scale, see e.g. Ref.~\cite{Carena:1993ag}.} at a scale of around $Q \sim 10^{16}$ GeV~\cite{Georgi:1974yf,Dimopoulos:1981yj,Langacker:1980js}. This is a striking result, and is often considered one of the main motivations for supersymmetry as a candidate for new physics at the TeV scale.

\subsubsection{Strong CP violation}
Although there were 18 parameters in the Standard Model listed in the previous chapter, in principle one could write down a 19th (\cp-odd) term that could lead to experimental effects in the strong interaction. The fact that these effects have not been observed is known as the \emph{strong \cp  problem}\footnote{See Ref.~\cite{Dine:2000cj} for a pedagogical review.}. The origin of this problem lies in the fact that one can write down another gauge invariant field strength kinetic term:
\begin{equation}
\lag{}= \frac{\theta}{16\pi^2} F_{\mu\nu}\tilde F^{\mu\nu} \hspace{10pt} \text{where} \hspace{10pt} \tilde F_{\mu\nu} =  \epsilon_{\mu\nu\rho\sigma}  F^{\rho\sigma}.
\end{equation}
Since this term can be written as a total derivative;
\begin{equation}
\lag{}  = \partial_\mu K^\mu \hspace{10pt} \text{where} \hspace{10pt}  K_\mu = \frac{1}{2} \epsilon_{\mu\nu\rho\sigma} A^\nu F^{\sigma \rho} ,
\end{equation}
the usual argument is that it only contributes a surface term to the action, so based on the boundary condition that the fields should go to zero (or more concretely, that the $K_\mu$ term goes to zero faster than the surface element diverges) in the limit that $r \to \infty$, this term can be neglected as it has no physical consequences. While this is certainly true in the abelian QED case, it does not hold in QCD. There, the total derivative has the form
\begin{equation}
K_\mu = \frac{1}{2} \epsilon_{\mu\nu\rho\sigma}( A^a_\nu G^a_{\sigma\rho} -\frac{2}{3}f_{abc}A^a_\nu A^b_\sigma A^c_\rho).
\end{equation}
Hence, even if the field strength tensor $G^a_{\mu\nu}$ goes to zero rapidly enough, the non-Abelian $A^3$ term above means that $K_\mu$ might still not vanish at spatial infinity. In fact, there are field configurations that do not vanish. These $F\tilde F$ terms do not appear in a perturbative expansion of \lag{QCD}, however, they do have physical effects. To show this, we consider \lag{QCD} with only two massless quarks:
\begin{equation}
\lag{QCD} = i\bar{Q}D_\mu\gamma^{\mu} Q +  i\bar{u}_{R}D_\mu\gamma^{\mu} u_{R} +  i\bar{d}_{R}D_\mu\gamma^{\mu} d_{R} .
\end{equation}
Given that $m_{\{u,d\}} \ll \Lambda_{QCD}$, this is a fair approximation at low energy.  At energies above $\Lambda_{QCD}$, the Lagrangian has a global `chiral' \su2L\x\su2R symmetry. Approaching $\Lambda_{QCD}$, this is spontaneously broken by QCD condensates $\braket{\bar qq}$ to a vectorial subgroup \su2V. By Goldstone's theorem, there are three massless scalars associated with this symmetry breaking. These are identified with the pions. In fact the residual vectorial symmetry is explicitly broken by a small amount by electromagnetic interactions and the small $u-d$ mass splitting, which gives the pions a small mass (\ord{\Lambda_{QCD}}, though this is not calculable from first principles). The Lagrangian also has a \U1A symmetry which is broken by QCD condensation. There is no Goldstone boson associated with this symmetry, however. 

Na\"ively, this spontaneous breaking of the axial symmetry should result in the appearance of a pseudoscalar $0^-$ meson with mass of order the pion mass, but no such particle exists in the meson spectrum of QCD. The next such candidate is the $\eta$ meson, but it is too heavy, as Weinberg showed that this particle can have a mass no greater than $\sqrt{3}m_\pi$~\cite{Weinberg:1975ui}. This was referred to as the $U(1)$ problem of QCD.

The resolution of this problem, due to `t Hooft~\cite{tHooft:1976snw,tHooft:1986ooh}, was that non-perturbative gauge configurations known as \emph{instantons} also contribute to the QCD vacuum, so that the full action is given by
\begin{equation}
S_{\text{QCD vacuum}} = \int d^4x \lag{QCD} + \frac{\theta g_s^2}{32\pi^2}\int d^4x \tilde{G}^a_{\mu\nu} G^{a, \mu\nu} .
\end{equation}
That $m_\eta \gg m_\pi$ requires $\theta \neq 0$. However, other observables are sensitive to, and place strong bounds on, $\theta$. It generates a contribution to the neutron electric dipole $d_n$ for instance, $d_n \sim e\theta m_q/M_N^2$. Current bounds require $d_n \lesssim 3\times10^{-26} e$ cm~\cite{Peccei:2006as}. This translates into the bound $\theta \lesssim 10^{-9}$. The question is then, why is the dimensionless parameter $\theta$ so small but apparently nonzero? There is no additional symmetry enhancement when $\theta$ is taken to zero, so this is not a technically natural small parameter. This is the strong-\cp problem, and cannot be resolved within the Standard Model alone, therefore new physics, such as axions~\cite{Weinberg:1977ma,Peccei:1977ur,Wilczek:1977pj,Peccei:1977hh}, is required to explain it.

\subsection{The role of the top quark in specific BSM scenarios}
\label{sec:models}

\subsubsection{Low energy supersymmetry}
The most widely known solution to the hierarchy problem is supersymmetry. In general, supersymmetry is a postulate that the theory exhibits a symmetry under the transformation $Q\ket{fermion}=\ket{boson}$, $Q\ket{boson}=\ket{fermion}$, i.e. there is a unique transformation which maps each boson in the theory into a corresponding fermion and vice versa, thus keeping the overall theory invariant. 

Clearly, the Standard Model does not exhibit this symmetry. This is linked to the problem of quadratic divergences in the Higgs mass. To see this, one can compute the explicit corrections to the Higgs mass due to loops of Standard Model particles. Focusing explicitly on the fermionic  corrections (although the same argument applies for the $W$ and $Z$ bosons), due to $N_f$ flavours of fermion, one finds~\cite{Djouadi:2005gj}

\begin{figure}[!t]
\begin{center}
\includegraphics[width=\textwidth]{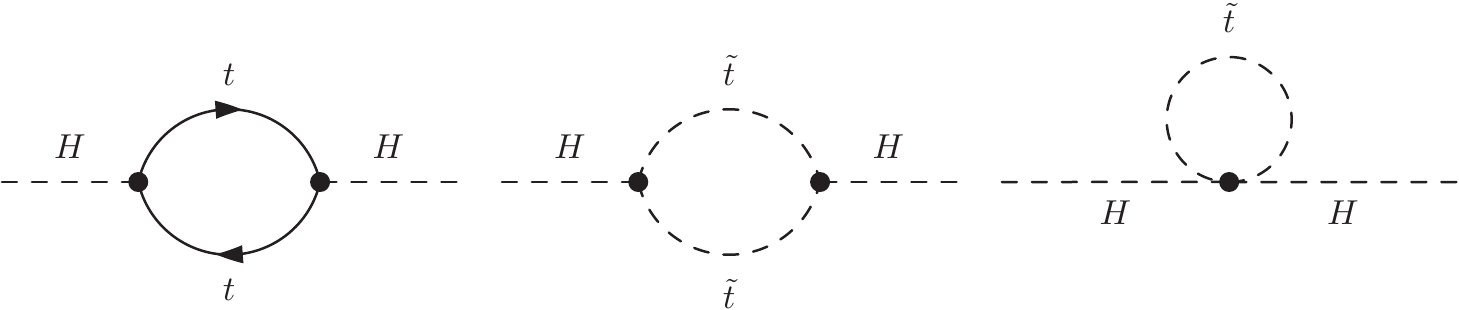}
\caption[Cancellation of quadratic corrections to the Higgs mass in the MSSM.]{One-loop diagrams contributing to the renormalisation of the Higgs mass in the Minimal Supersymmetric Standard Model: the correction due to a top quark loop (left) and corrections due to a loop of stop quarks $\tilde t$. } 
\label{fig:stops}
\end{center}
\end{figure} 

%
\begin{equation}
m_H^2 = m^2_{H,tree} + N_f^2\frac{\lambda_f^2}{8\pi^2}\left(-\Lambda^2+6m_f^2\log\frac{\Lambda}{m_f}-2m_f^2\right).
\end{equation}
In a supersymmetric theory, for each fermion loop there would be a contribution due to a corresponding scalar particle in the two diagrams on the right of Fig. \ref{fig:stops}, giving the mass correction
\begin{equation}
m_H^2 = m^2_{H,tree} + N_s^2\frac{\lambda_s}{16\pi^2}\left(-\Lambda^2+2m_s^2\log\frac{\Lambda}{m_s}\right)
-N_s^2\frac{\lambda_s}{16\pi^2}v^2\left(-1+2\log\frac{\Lambda}{m_s}\right).
\end{equation}
Counting degrees of freedom, each fermion (a two-component Weyl spinor) must have two scalar `partners', so $N_s = 2 N_f$. If one makes the additional assumption that $\lambda_s =\lambda_f^2$, then upon adding these two contributions together, one finds the quadratic divergences in $m_H^2$ cancel entirely, leaving a logarithmic contribution to the Higgs mass
\begin{equation}
m_H^2 = m^2_{H,tree}+N_f^2\frac{\lambda_f^2}{4\pi^2}\left[(m_f^2-m_s^2)\log\frac{\Lambda}{m_s}+3m_f^2\log\frac{m_s}{m_f}\right].
\end{equation}
If supersymmetry is exact, then $m_s=m_f$ and the corrections to the Higgs mass are exactly zero, to all orders in perturbation theory. This can be understood from symmetry grounds: a process involving virtual particles without their corresponding superpartners will violate supersymmetry and reintroduce divergences. 

In practice, SUSY must be violated by some amount, otherwise the superpartners would have the same masses as their standard model counterparts, and would surely have been observed already. Therefore one must introduce SUSY breaking parameters to break the $m_s=m_f$ relation by some amount, though this amount cannot be too large otherwise the hierarchy problem will reappear. If SUSY is broken at the TeV scale, then it should show rich phenomenology at colliders such as the LHC. The top quark would play a central role here, both directly and indirectly. 

\begin{figure}[!t]
\begin{center}
\includegraphics[width=\textwidth]{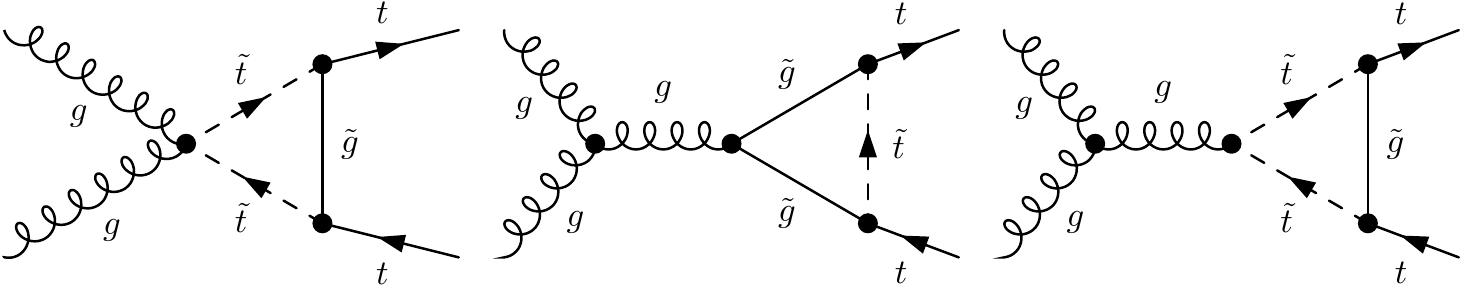}
\caption[Top pair production in the MSSM.]{Example one-loop diagrams contributing to top pair production in the $gg$ channel, owing to the effects of virtual SUSY particles, namely the gluing $g$ and the stop $\tilde t$. } 
\label{fig:ggstops}
\end{center}
\end{figure}

To emphasise the latter, consider top pair production in the $gg$ channel, with loop effects of scalar top (\emph{stop}) quarks, as shown in Fig. \ref{fig:ggstops}. If the top pair production cross-section were very precisely measured, it could be used to place indirect bounds on the mass of the stop quark, and thus the scale of supersymmetry breaking. Indirect stop contributions show up elsewhere, such as in the Higgs decay to $\gamma\gamma$, which is part mediated by a top loop~\cite{Drozd:2015kva}. This shows that the properties of the top quark are a valuable testing ground for the effects of new physics such as supersymmetry.  

\subsubsection{Little Higgs}
The Higgs mass naturalness problem was solved in supersymmetry by adding extra degrees of freedom to cancel off the quadratic divergences, but more fundamentally this is due to extending the Poincar\'e algebra of spacetime symmetries. In general the Higgs mass can be made `technically natural' by enlarging the SM symmetry group, such that quadratic corrections to $\mu^2$ are forbidden by the extra symmetries. The `Little Higgs' family~\cite{Schmaltz:2005ky} of models use this idea, by having the \smgg gauge group of the SM emerge from a spontaneously broken \emph{global symmetry}. The gauge couplings of the SM break the residual symmetry explicitly by a small amount, and the Higgs emerges as one of the pseudo-Nambu-Goldstone bosons of this symmetry breaking. The pattern of the symmetry breaking ensures that the corrections to the Higgs mass are at most logarithmically sensitive to the cutoff $\Lambda$. 

\begin{figure}[!t]
\begin{center}
\includegraphics[width=\textwidth]{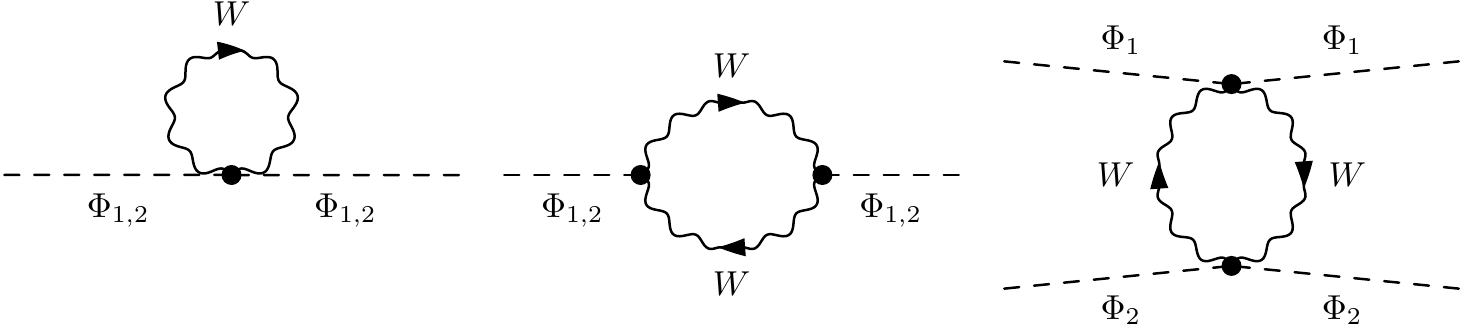}
\caption[1-loop diagrams in the little Higgs model.]{Quadratically (left) and logarithmically (centre) divergent one-loop diagrams from Eq.~\eqref{eqn:littlehiggs} that do not renormalise the Higgs mass, and logarithmically divergent 1-loop diagrams that do renormalise the Higgs mass (right). } 
\label{fig:littlehiggs}
\end{center}
\end{figure}

To illustrate this more concretely, we consider an extension of the minimal example of a global symmetry breaking SU(3)$\to$SU(2) by the vev of a complex triplet field $\braket{\Phi^T} = \braket{(\phi_1,\phi_2,\phi_3)} = (0,0,f)$. Instead, we consider two complex fields $\Phi_1$ and $\Phi_2$, each with its own set of 5 Nambu-Goldstone bosons, i.e.  the symmetry breaking pattern is [SU(3)$\to$SU(2)]$^2$. To parameterise this symmetry breaking, $\Phi_1$ and $\Phi_2$ can be written as
\begin{equation}
\Phi_1 = e^{i\vec\pi_1/f_1}\begin{pmatrix} 0 \\ 0 \\ f_1 \end{pmatrix}, \hspace{10pt} \Phi_2 = e^{i\vec\pi_2/f_2}\begin{pmatrix} 0 \\ 0 \\ f_2 \end{pmatrix} .
\end{equation}
For convenience we assume that the two vevs $f_1$ and $f_2$ are aligned. The Nambu-Goldstone bosons $\vec\pi = \pi^a T^a$ are given by the generators of SU(3) that are not also generators of SU(2). We can write explicitly, for each set $\vec\pi$
\begin{equation}
\pi^aT^a = \frac{1}{\sqrt{2}} \left(\begin{array}{cr} \begin{matrix} 0 & 0 \\ 0 & 0 \end{matrix} & H \\ H & 0 \end{array} \right) + \frac{\eta}{2}\begin{pmatrix} 1 & 0 & 0 \\ 0 & 1 & 0 \\ 0 & 0 &-2 \end{pmatrix} .
\end{equation}
Four of the Goldstones are in the complex doublet $H$, the remaining one is in the singlet $\eta$, which can be ignored for our purposes. The factors in front of the Goldstone matrices ensure that the $H$ and $\eta$ kinetic terms are canonically normalised. As well as the Goldstones, there are also massive radial excitations $r$, which are assumed to be heavy so that they are integrated out. It is also assumed for simplicity that there are degenerate symmetry breaking scales $f_1 = f_2 = f$. The Lagrangian for this toy model is then
\begin{equation}
\lag{LH} = |D_\mu\Phi_1^2|^2+|D_\mu\Phi_2^2|^2.
\label{eqn:littlehiggs}
\end{equation}
Upon expanding out the Lagrangian, one generates the 1-loop correlation functions corresponding to the two diagrams on the left hand side of Fig. \ref{fig:littlehiggs}:
\begin{equation}
\amp{1-loop} \sim \frac{g^2\Lambda^2}{16\pi^2}(\Phi_1^\dagger\Phi_1+\Phi_2^\dagger\Phi_2)= \frac{g^2\Lambda^2}{16\pi^2}(2f^2).
\end{equation}
%
%
In addition, one generates the term corresponding to the diagram on the right of Fig.  \ref{fig:littlehiggs}.
\begin{equation}
\amp{1-loop} \sim \frac{g^4}{16\pi^2}\log\left(\frac{\Lambda^2}{\mu^2}\right)|\Phi_1^\dagger\Phi_2|^2 = \frac{g^4}{16\pi^2}\log\left(\frac{\Lambda^2}{\mu^2}\right)(f^2-2H^\dagger H+\frac{(H^\dagger H)^2}{f^2}+\ldots).
\end{equation}
Hence, the 1-loop corrections to the Higgs mass term are at most logarithmically divergent. To see why this has happened, let us focus on the gauge part of \lag{LH}:
\begin{equation}
\lag{LH} = |g_1^2A_\mu\Phi_1|^2+|g_2^2A_\mu\Phi_2|^2.
\end{equation}
This explicitly breaks the SU(3)$\times$SU(3) global symmetry to a gauged diagonal subgroup: so only one of the SU(3)$\to$SU(2) breaking mechanisms is exact. The other is explicitly broken by a small amount, giving its Goldstone bosons a small mass, one of which we then take to be the Higgs. Setting either $g_1$ or $g_2$ to zero restores the full [SU(3)]$^2$ symmetry. In the case of $g_2 = 0$, for example, we have two independent symmetries
\begin{equation}
\Phi_1 \to U_1\Phi_1, \hspace{10pt} A_\mu\to U_1A_\mu U_1^\dagger, \hspace{10pt}\Phi_2 \to U_2\Phi_2,
\end{equation}
whereas in the case of $g_1 = 0$ we have the symmetries
\begin{equation}
\Phi_1 \to U_1\Phi_1, \hspace{10pt} A_\mu\to U_2A_\mu U_2^\dagger, \hspace{10pt}\Phi_2 \to U_2\Phi_2.
\end{equation}
So when either of the gauge couplings is set to zero, $\pi$ is an exact Nambu-Goldstone boson, so the corrections to its mass can only be proportional to $g_1g_2$, i.e. the exact symmetry can only be \emph{collectively} broken by the two fields. There are no quadratically divergent diagrams involving $g_1$ and $g_2$ at 1-loop, however, so the Higgs mass is stabilised at this order by this \emph{collective symmetry breaking}~\cite{Kaplan:2003uc,Schmaltz:2004de}.

The same trick for the gauge loops can be played for the top quark loops. To ensure that there is collective symmetry, one enlarges the quark doublets into triplets; $Q_L \to \Psi_L = (t_L, b_L, T_L)$, by adding an extra fermionic partner for each generation. One finds again that the quadratic divergences due to the top quark loop are cancelled by the top partner $T$. It is this extra top partner that has an impact on top quark phenomenology. It will in general mix with the top quark, so it can be produced via $Wb$ fusion, Depending on its quantum numbers, it may decay via $T\to th$, $T\to tZ$, $T\to bW$~\cite{Han:2003wu}, which would lead to large enhancements of cross sections for top quarks associated with electroweak and Higgs bosons.

\subsubsection{Warped extra dimensions}
An alternative approach to explaining the large hierarchy between the electroweak scale and the Planck mass is through warped extra dimensions. The most studied scenario is the Randall-Sundrum (RS) model~\cite{Randall:1999ee,Randall:1999vf}, where the Standard Model field content resides on the 4-dimensional boundary of a 5-dimensional bulk, in which the gravitational degrees of freedom propagate. The extra dimension is compactified onto a $S^1/\mathbb{Z}_2$ orbifold (a circle with an additional $\mathbb{Z}_2$ symmetry $\phi = -\phi$) of radius $r_c$,  and the two boundaries of the 5D bulk are taken to be at the points $\phi=0$ and $\phi=\pi$, where a 4-dimensional field theory resides, such that there is a \emph{visible} boundary or brane and a \emph{hidden} one. The full 5D metric $G_{MN}(x^\mu,\phi)$ is then related to the 4D metrics by the boundary conditions:
\begin{equation}
g^{vis}_{\mu\nu} = G_{\mu\nu}(x^\mu,\phi = \pi), \hspace{10pt} g^{hid}_{\mu\nu} = G_{\mu\nu}(x^\mu,\phi = 0).
\end{equation}
The Einstein-Hilbert action for the 5D theory is then given by
\begin{equation}
S = S_{grav}+S_{vis}+S_{hid} = \int d^4x \int\limits_{-\pi}^{\pi}d\phi \sqrt{-G}(\Lambda+2M^3R)+ \int d^4x ( \sqrt{-g_{vis}}\lag{vis}+  \sqrt{-g_{hid}}\lag{hid}),
\end{equation}
where $\Lambda$ is a cosmological constant, $M$ is a universal mass scale extracted to give the field $\phi$ the same units as in the 4D theory and $R$ is the Ricci scalar. A solution to the Einstein field equations for the above action is
\begin{equation}
ds^2= e^{-2kr_c|\phi|}\eta_{\mu\nu}dx^\mu dx^\nu +r_c^2d\phi^2 .
\end{equation}
This non-factorisable metric describes flat 4D spacetime modified by an exponential \emph{warp factor} $2kr_c|\phi|$. The parameter $k$ is a scale relating the `observed' 4D Planck scale to the Planck scale in the bulk. The radius $r_c$ describes the compactification of the 5th dimension, and may be taken to be near the Planck length, if one views the model as originating from a string/M-theory UV completion. To see how this is relevant for low energy physics, we can expand the metric about its local fluctuations $g_{\mu\nu}(x) = \eta_{\mu\nu}+h_{\mu\nu}(x)$, substitute in a Higgs field into \lag{vis} in the action, and perform a wave-function renormalisation so that the Higgs kinetic term is canonically normalised, and one finds that any mass or vev $m_0$ in the fundamental 5D theory is related to the mass/vev in the visible theory by 
\begin{equation}
m_{vis} = e^{-kr_c\pi}m_0 .
\label{eqn:wed}
\end{equation}
So a hierarchy of order $v/M_{Pl}\sim10^{-16}$ in the visible theory translates into a hierarchy of size $kr_c\sim 12$ in the fundamental theory, thus it is much more natural for the bulk curvature $k$ to live near the Planck scale $1/r_c$. This is a compelling solution of the hierarchy problem, because it shows that weak scale masses can be determined by parameters not far from the Planck scale, but in a natural way. Why is this relevant for top quark physics?

The main phenomenological prediction of warped extra dimension models is that, because the extra dimension has periodic (Dirichlet) boundary conditions, the fields that are allowed to propagate in the bulk will have an infinite tower of \textit{Kaluza-Klein} modes of mass $m_n$, analogous to standing wave modes on a closed string\footnote{Kaluza-Klein excitations are also present in other theories of extra dimensions, such as the ADD models~\cite{ArkaniHamed:1998rs,ArkaniHamed:1998nn} and universal extra dimensions~\cite{Dienes:1998vh,Appelquist:2000nn}.}, which can be coupled to the SM fields that also propagate in the bulk~\cite{Davoudiasl:1999tf,Gherghetta:2000qt}, and which, according to Eq.~\eqref{eqn:wed} can have masses near the electroweak scale. Moreover, the KK modes will couple most strongly to fields that are localised near the $\phi = \pi$ IR brane, and most weakly coupled to fields localised on the UV brane. 

To keep a hierarchy between $v$ and $M^{4D}_{Pl}$, the Higgs field must be localised on the IR brane ($\phi = \pi$), cf Eq.~\eqref{eqn:wed}. However, the large top Yukawa means there must be a strong overlap between the top and Higgs wave-functions, so the top most also be located close to the IR~\cite{Davoudiasl:1999jd,Pomarol:1999ad}. Hence, the top is expected to couple strongly to the new Kaluza-Klein modes of the 5th dimension, which motivates the search for heavy resonances of spin-1 (the KK modes of the $\gamma$, $g$, $W$ and $Z$) and spin-2 (the KK modes of the massless graviton), decaying into $t\bar{t}$ pairs. 

The preceding section provides a strong argument for processes involving top quarks as a well-motivated place to look for BSM physics at colliders, albeit in the limited context of specific models. The remainder of this chapter will turn to a more model-independent formulation of the effects of new particles and couplings, which views the Standard Model as the first part of an effective field theory, and discusses the sector of this effective theory that may be probed at hadron colliders.

\subsection{Principles of effective field theory}
\label{sec:eft}
\subsubsection{The Euler-Heisenberg Lagrangian}
As a warmup, we consider an example from electromagnetism. Imagining a different Universe where the electron was much heavier and had not yet been detected directly, so that the `full' Lagrangian for electromagnetism is just that of the free Maxwell theory. 

\begin{equation}
\lag{} = -\frac{1}{4}F_{\mu\nu}F^{\mu\nu}.
\end{equation}

If the electromagnetic process of four-photon scattering had been experimentally observed, it could not be described by this Lagrangian. The lowest-order Lagrangian that could describe this process would have to have four field-strength tensors, since the Maxwell theory is abelian. It should also respect Lorentz symmetry and U(1) gauge symmetry, and can be written as
\begin{equation}
\lag{4\gamma} = A(F_{\mu\nu}F^{\mu\nu})^2+B(F_{\mu\nu}\tilde{F}^{\mu\nu})^2 .
\end{equation}
Since the Lagrangian must have mass dimension 4 (we work in natural units), the  coefficients $A$ and $B$ must have mass dimension -4. We can then write
\begin{equation}
\lag{4\gamma} \equiv  \frac{1}{\Lambda^4}\left(c_1(F_{\mu\nu}F^{\mu\nu})^2+c_2(F_{\mu\nu}\tilde{F}^{\mu\nu})^2\right) .
\end{equation}
The coefficients $c_1$ and $c_2$ are dimensionless, and $\Lambda$ is some generic mass scale. From this Lagrangian we know that the scattering cross-section $\sigma$ for $\gamma\gamma\to\gamma\gamma$ scales as $1/\Lambda^8$. However, since $\sigma$ is just an area, it has units of $[mass]^{-2}$, so there must be some other mass scale in the problem. The only other mass scale in the problem, however, is the frequency of the incoming photons $\omega$\footnote{For simplicity we assume them to be of the same order of magnitude.}, so we can say that the cross-section scales as
%
\begin{figure*}[!t]
\begin{center}
 \includegraphics[width=0.4\textwidth]{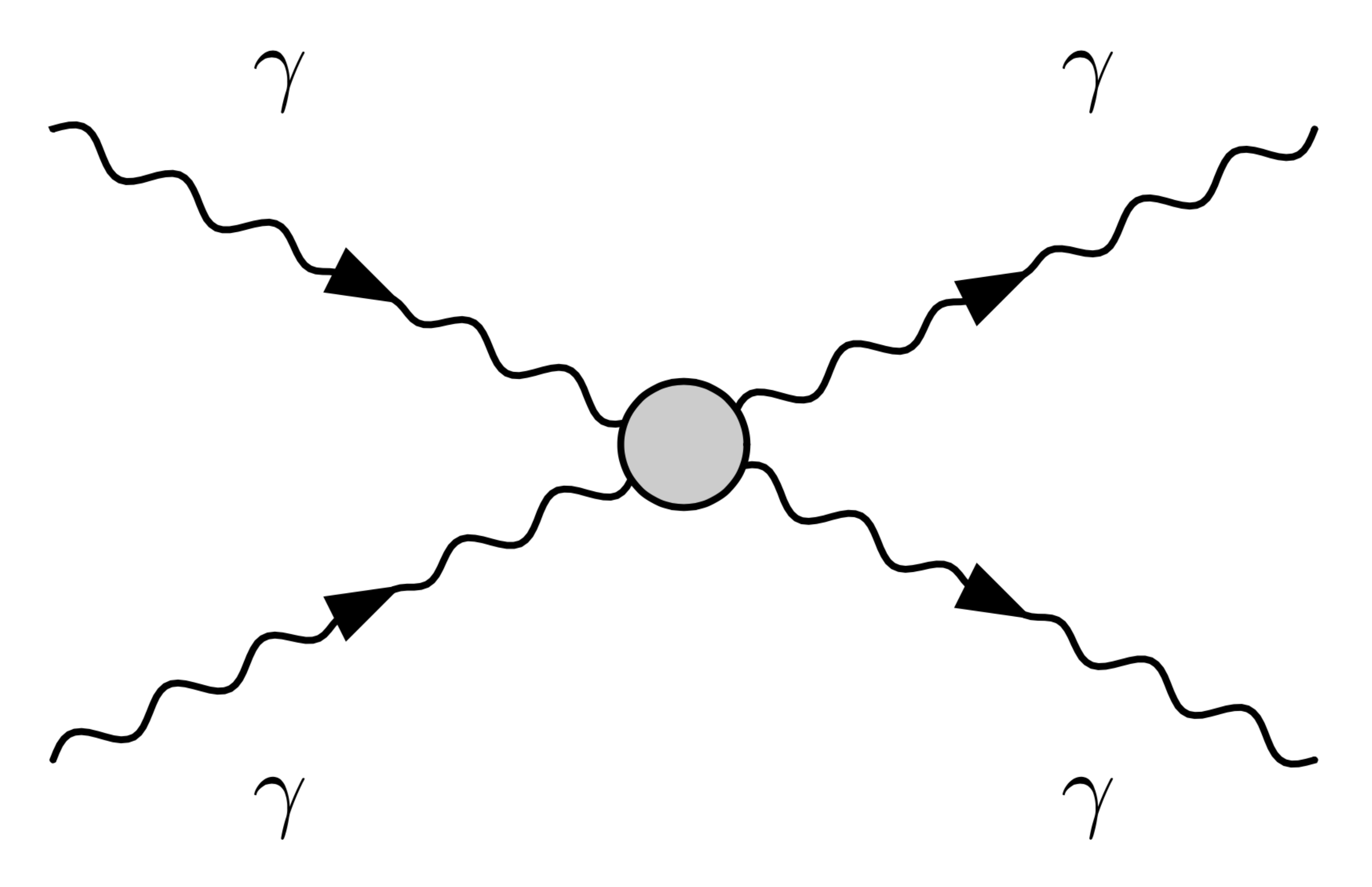}
  \hspace{2cm}
  \includegraphics[width=0.4\textwidth]{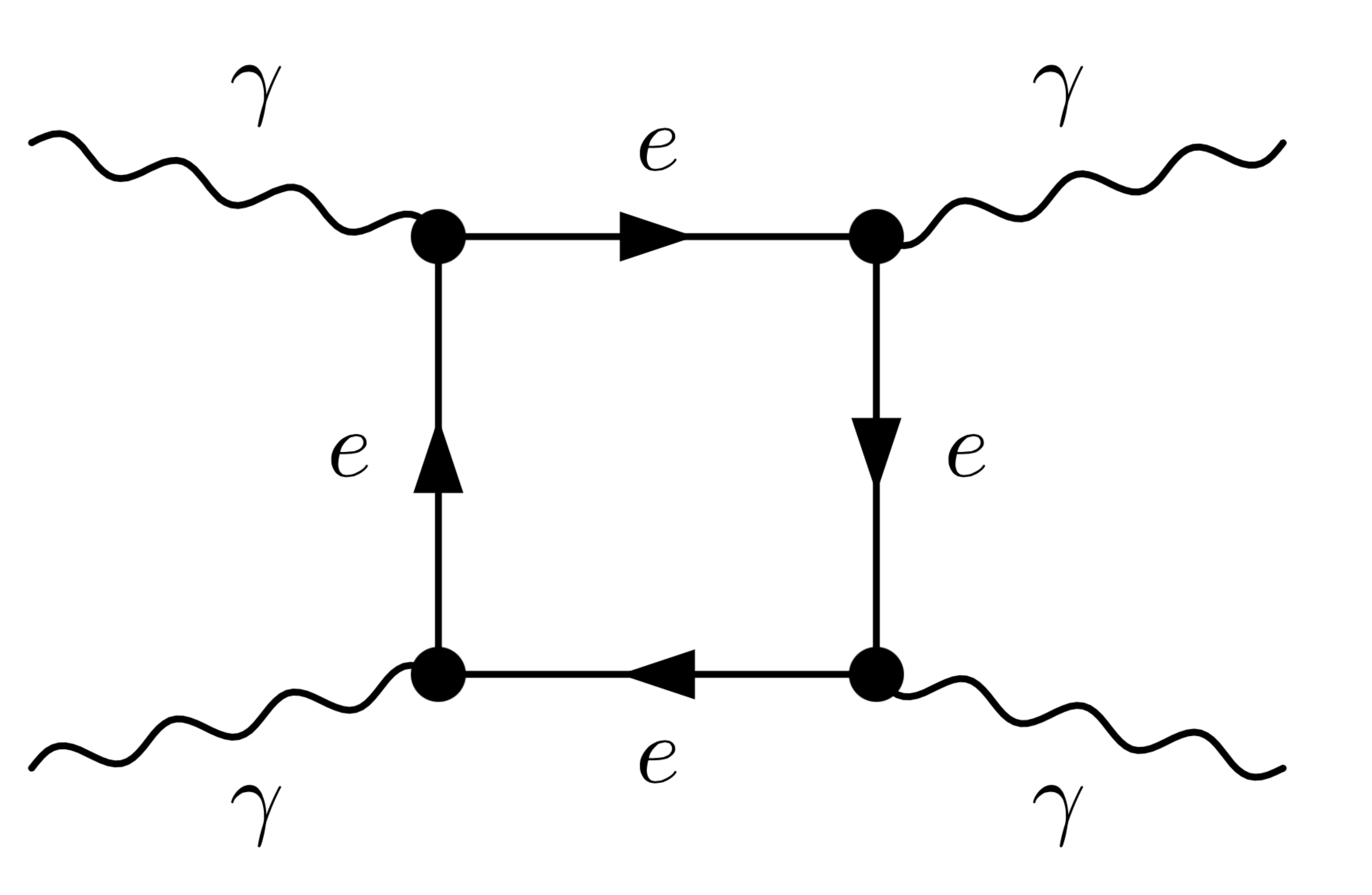}
 \caption[Four photon scattering in the Euler-Heisenberg EFT and in quantum electrodynamics.]{$\gamma\gamma\to\gamma\gamma$ scattering at tree-level in the Euler-Heisenberg effective theory (left) and at one-loop level in quantum electrodynamics (right).}
\label{fig:4gamma}
\end{center}
\end{figure*}  
%
\begin{equation}
\sigma \sim \frac{\omega^6}{\Lambda^8} + \ldots
\end{equation}
The power of effective field theory is manifest: by general arguments of symmetry and making no assumptions about the nature of the underlying interaction, one can make powerful deductions about the scaling behaviour of scattering processes in quantum field theory. 

There is a caveat: the ellipsis denotes higher-order corrections due to neglected operators of dimension $D > 8$, the leading term of which will be \ord{\omega^8/\Lambda^{10}}. There are an infinite number of such operators, so these corrections must be small in order for their omission to be valid. In other words, the validity of the effective theory requires that $\omega \ll \Lambda$. But in the pure effective theory, $\Lambda$ is a free parameter, so there is no \emph{a priori} guarantee that this is true. In reality, of course there is a full theory of electromagnetism: quantum electrodynamics, so one can compute the scattering amplitude in terms of its parameters: the fine-structure constant $\alpha$ and the electron mass $m_e$, corresponding to the Feynman diagram on the right of Fig. \ref{fig:4gamma}, and match them onto the EFT parameters. One finds~\cite{Manohar:1995xr}
\begin{equation}
\Lambda  = \frac{m_e}{\sqrt{\alpha}} , \hspace{10pt} c_1 = \frac{1}{90}  , \hspace{10pt} c_2 = \frac{7}{90}.
\end{equation}
Therefore, provided the photon scattering frequencies $\omega \ll m_e/\sqrt{\alpha} \sim \ord{MeV}$, this treatment is valid, and the loop diagram of Fig. \ref{fig:4gamma} does not have to be calculated.

\subsubsection{Fermi theory of weak decay}
The textbook example of an effective field theory is the Fermi low-energy theory of weak decay. One can arrive at the theory by starting from the full electroweak model and integrating out the heavy degrees of freedom, i.e. the $W$ and $Z$ boson. In the electroweak theory, nuclear $\beta$ decay is mediated by the transition $d\to uW\to ul\nu$. This has the transition amplitude~\cite{Burgess:2007pt}
\begin{equation}
\mathcal{A}_{\text{full}} = \left(\frac{ig}{\sqrt{2}}\right)^2(\bar u \gamma_\mu P_L d)(\bar l \gamma_\nu P_L \nu)\left(\frac{-ig^{\mu\nu}}{p^2-M_W^2}\right) ,
\end{equation}
where $P_L$ is the left-handed projection operator $(1-\gamma_5)/2$. The low energy continuum is obtained by Taylor expanding the $W$ boson propagator in the limit $p^2 \ll M_W^2$.
\begin{equation}
\left(\frac{-ig^{\mu\nu}}{p^2-M_W^2}\right) = \frac{-ig^{\mu\nu}}{M_W^2}\left(1+\frac{p^2}{M_W^2}-\frac{p^4}{M_W^4}+\ldots\right) .
\end{equation}
Hence the higher order terms decouple rapidly in this limit, so that the amplitude in the low energy theory is given by the first term in the series, multiplied by a \D6 four-fermion contact term:
\begin{equation}
\mathcal{A}_{\text{EFT}} = \left(\frac{g^2}{2M_W^2}\right)(\bar u \gamma_\mu P_L d)(\bar l \gamma^\mu P_L \nu).
\end{equation}
In fact this is the same $\beta$ decay amplitude that was written down by Fermi in his contact interaction model, though he wrote in terms of the nucleon wavefunctions $u\to p$ and $d\to n$, and omitted the projection operator $P_L$ as parity violation had not yet been observed. He also parametrised it in terms of an overall dimensionful coupling $G_F$, which we can obtain an expression for by matching the electroweak parameters to the parameters of the Fermi theory, giving
\begin{equation}
\frac{G_F}{\sqrt{2}} = \frac{g^2}{8M_W^2},
\end{equation}
with the value of $G_F$ given in the first chapter.  Provided the condition $p^2 \ll M^2_W$ is justified, one can calculate to a good approximation all weak scattering processes, such as muon decay and meson mixing, without knowing the details of the gauge structure of the underlying electroweak theory. In fact, if one makes assumptions about the underlying couplings, one can even predict the regime of validity of the EFT. Setting $g = 1$ and using the measured value of $G_F$, and neglecting \ord{1} coefficients above, for example, one finds $M_W \approx 290$ \gev, not far from its actual value of 80 \gev, and very close to the electroweak scale $v \sim$ 246 \gev where the EFT would no longer be valid. This highlights another strength of EFT: using the measurements of the low energy parameters, one can infer details about some of the high-energy ones, by making broad assumptions about the perturbativity of the underlying theory. This \emph{matching} procedure will be returned to later in the thesis. 

\subsubsection{Renormalisation group treatment}
Here we perform a more systematic analysis of how the dimensionality of an operator determines at which scale it becomes relevant, by means of the renormalisation group. To isolate the behaviour of operators of a certain dimensionality when the renormalisation scale is varied, let us return to our prototypical example of a scalar field in four dimensions with a quartic self-interaction. This time, we supplement the Lagrangian with an infinite series of higher-dimensional (i.e. $D > 4$) operators~\cite{Kaplan:1995uv}.
\begin{equation}
\lag{E} = \frac{1}{2}(\partial\phi)^2 + \frac{1}{2}m^2\phi^2 +\frac{\lambda}{4!}\phi^4 + \summ{n}\left( \frac{c_n}{\Lambda^{2n}}\phi^{4+2n} + \frac{d_n}{\Lambda^{2n}}(\partial\phi)^2\phi^{2+2n} + \ldots \right),
\label{eqn:eftlag}
\end{equation}
where the Lagrangian has been rotated to imaginary time in order to perform the path integral, defining a `Euclidean action'. The ellipsis denotes all operators with higher derivatives. Demanding that the effective theory preserve the original $\phi\to -\phi$ symmetry eliminates terms of odd mass dimension. The scale $\Lambda$ has been introduced to keep the coupling constants $c_i$ and $d_i$ dimensionless. The kinetic and mass terms both have dimension \D2, the quartic coupling has \D4. The question is then: how does the dimensionality of an operator influence its renormalisation scaling behaviour?

To isolate field configurations that are most relevant at a certain momentum/length, we can perform the path integral
\begin{equation}
\int D \phi e ^{-S_E} \hspace{10pt} \text{where} \hspace{10pt} S_E = \int d^4 x \lag{E}.
\end{equation}
We can consider a field configuration $\tilde{\phi}$ in the path integral with amplitude $\phi_k$, and wavenumber $k_\mu$, i.e. a `wavelet' that is confined to a spacetime volume $L^4 = (2\pi/k)^4$. In momentum space, the Euclidean action can be trivially obtained by Fourier transforming  Eq.~\eqref{eqn:eftlag}.  Then the action is given by
\begin{equation}
S_E = (2\pi)^4\left[\frac{\hat{\phi}^2_k}{2}+ \frac{m^2}{k^2}\hat{\phi}^2_k+ \frac{\lambda}{4!}\hat{\phi}^4_k
+ \summ{n}\left( c_n \left(\frac{k^2}{\Lambda^{2n}}\right)^2\hat{\phi}^{4+2n}_k + d_n \left(\frac{k^2}{\Lambda^{2n}}\right)^2\hat{\phi}^{4+2n}_k+\ldots \right) \right] ,
\end{equation}
where $\hat{\phi}_k \equiv \phi_k/k$. The contribution of this single mode to the path integral is
\begin{equation}
\int d \hat{\phi}_k e^{-S_E}.
\end{equation}
Clearly it will be dominated by values of $\hat{\phi}_k$ for which $S_E \lesssim 1$. As the amplitude $\hat{\phi}_k$ becomes large, then the kinetic term $(2\pi)^4\hat{\phi}^2_k/2$ will dominate the action, i.e the path integral will get its dominant contribution for $\phi_k \sim k/(2\pi^2)$.

When $k$ is decreased, the higher-dimensional terms proportional to $c_i$ and $d_i$ get smaller and smaller; they are called \textit{irrelevant} operators; their effects become increasingly decoupled as we move from the ultraviolet to the infrared. The mass term, on the other hand, becomes increasingly dominant; it is a \textit{relevant} operator. The quartic operator is neither relevant nor irrelevant, its effects are of apparently equal strength for small and large $k$; it is referred to as a \textit{marginal} operator.

Another way of deriving the scaling properties of these operators (the method originally employed by Wilson~\cite{Wilson:1971bg,Wilson:1971dh,Wilson:1973jj}) is to consider a random field configuration $\phi(x)$, and look at how its corresponding action changes when we perform a passive transformation $\phi(x)\to \phi(\xi x)$, i.e. when we move across different length scales. If we just consider a plane wave, for instance, then the transformation is given by $\phi(\xi x) = e^{i\xi k \cdot x}$, so that the limit $\xi \to \infty$ corresponds to shorter wavelengths $k' = \xi k$. Then the action becomes

\begin{equation}
\begin{split}
S_E(\phi(\xi x); \Lambda,m^2,\lambda,c_n,d_n )  =  &\int d^4 x \frac{1}{2}(\partial\phi(\xi x))^2 + \frac{1}{2}m^2\phi(\xi x)^2 + \frac{\lambda}{4!}\phi(\xi x)^4  \\
& + \summ{n}c_n \frac{\phi^{4+2n}(\xi x )}{\Lambda^{2n}}  + \summ{n}d_n \frac{(\partial \phi(\xi x))^2 \phi^{2n}(\xi x )}{\Lambda^{2n}} \\
= &\int d^4 x' \frac{1}{2}(\partial' \phi' (x'))^2 + \frac{1}{2}m^2\xi^{-2}\phi '(x')^2+\frac{\lambda}{4!}\phi '(x')^4 \\
& + \summ{n}\left({c_n}\xi^{2n}\frac{\phi'^{4+2n}(x')}{\Lambda^{2n}} + d_n\xi^{2n} \frac{(\partial'\phi'(x'))^2\phi'^{2n}(x')}{\Lambda^{2n}} \right) ,
\end{split}
\end{equation}

where $\phi'(x) = \xi^{-1}\phi(x)$, and $x' = \xi x$. But since we integrate over $x$ and $x'$, we can just compare the integrands directly. Relabelling the dummy variable $x' \to x$ shows that our transformation returns the original action, but with rescaled fields and couplings.
\begin{equation}
S_E(\phi(\xi x); \Lambda,m^2,\lambda,c_n,d_n) = S_E(\xi^{-1}\phi(x);\xi^{-2}m^2,\lambda,c_n\xi^{2n},d_n\xi^{2n}) ,
\end{equation}
so the rescaled fields and couplings are:
\begin{equation}
\phi \to \xi^{-1} \phi, \hspace{10pt} m^2\to \xi^{-2}m^2,  \hspace{10pt} \lambda \to \lambda,  \hspace{10pt} c_n \to \xi^{2n} c_n, \hspace{10pt}  d_n \to \xi^{2n} d_n.
\end{equation}
In the infrared limit $\xi\to 0$,  it can be seen that the mass term becomes increasingly important, the higher-dimensional operators $c_n$ and $d_n$ become increasingly irrelevant, and the quartic and kinetic terms stay marginal. This result follows purely from dimensional analysis, and did not rely on any unique symmetry properties of scalar fields. We can then make the general statement that there are three types of operator scaling behaviour in a four-dimensional quantum field theory.

\begin{itemize}

\item{\textbf{Relevant operators:}} Operators \op[(d_i)]{i} with dimension $d_i < 4$. Their effects become increasingly large at low energies, and increasingly decoupled at high energies.

\item{\textbf{Marginal operators:}} Operators \op[(d_i)]{i} with dimension $d_i = 4$. Their effects are na\"ively the same across all scales, i.e. they appear to be conformal. However they may scale logarithmically, and a full radiative calculation is needed to obtain their scaling behaviour.

\item{\textbf{Irrelevant operators:}} Operators \op[(d_i)]{i} with dimension $d_i > 4$. Their effects become increasingly large at high energies, and increasingly decoupled in the infrared.

\end{itemize}

It was once considered a miracle that the Standard Model contained only marginal and relevant (and thus renormalisable) operators. From a modern perspective, we know that there is no miracle and this follows completely from the renormalisation group: the Standard  Model is a theory of low energy physics (compared to the Planck scale), therefore the higher-dimensional operators would not be expected, because they only `switch on' as we move towards the cutoff $\Lambda$ for the Standard Model. If there exist heavy new degrees of freedom, then at some scale between the electroweak and Planck scales their effects on electroweak scale observables can be described generally by supplementing the Standard Model Lagrangian with operators of dimension $>$ 4. These operators are built completely out of Standard Model fields since the underlying new heavy degrees of freedom have been integrated out. This is the Standard Model effective field theory (SMEFT), and it thus provides a completely model-independent way of searching for the effects of unknown heavy new physics on electroweak scale observables. In the next section we will discuss the SMEFT in detail.

\subsection{The Standard Model effective field theory}
\label{sec:smeft}
To derive the effective Lagrangian for the Standard Model~\cite{Burges:1983zg,Leung:1984ni,Buchmuller:1985jz}, all we have to do is write down the expansion in powers of $c_n/\Lambda^{n}$, as in Eq.~\eqref{eqn:eftlag}, but with the full SM field content, rather than just one scalar with a quartic interaction. This time we do not have a $\mathbb{Z}_2$ symmetry to respect, so odd powers of $\Lambda^{-1}$ are allowed. The effective Lagrangian can then be written
\begin{equation}
\lag{\text{eff}} = \lag{\text{SM}}+\summ{i}\frac{\co[(5)]{i}\op[(5)]{i}}{\Lambda}+\summ{i}\frac{\co[(6)]{i}\op[(6)]{i}}{\Lambda^2} + \ldots, 
\end{equation}
where the sum is over the full operator set at each mass dimension and the ellipsis denotes all operators at  $D\geq7$.  The full set of operators can be derived systematically, simply by using dimensional analysis to write down the list of operators of a given dimension that respect the symmetry constraints, i.e. Lorentz invariance and the full \smgg SM gauge symmetry. Then, care must be taken to ensure there are no redundancies in the operator set; that is, each operator generates a unique contribution to the $S$-matrix that cannot be expressed in terms of other operators. We will derive the dimension-five SM Lagrangian as an illustrative example.

\subsubsection{\D5}
To recap, the SM Lagrangian is composed entirely of spin-$\frac{1}{2}$ fermion fields of dimension 3/2, scalar and vector fields of dimension 1, and field strength tensors of dimension 2, as well as various covariant derivative operators. At dimension 5, then, na\"ively there are several types of operator that can be constructed, but more careful analysis shows that most are forbidden for symmetry reasons:

\begin{itemize}

\item Clearly, no fermion-only operators are allowed, because 5 is not a multiple of $\frac{3}{2}$. Scalar only operators are also forbidden, because the Higgs only appears in doublets, so there must be an even number of scalars in the term.

\item An operator with two fermions and two scalars is allowed dimensionally. Two combinations of scalars would be allowed, either $(\varphi^\dagger\varphi)$ or $(\varphi \varphi)$. The first case requires that the two fermions must also combine to hypercharge zero, so they must be Hermitian conjugates of the same fermion multiplet: $\bar\psi \psi$, which vanishes for chiral fermions. The second case is allowed, provided the scalars multiply to give an SU(2) triplet (the singlet product of two equal doublets is zero). Then the two fermions must also form a triplet to dot this into a scalar. Each fermion must then be an SU(2) doublet. The term can then be written as 
\begin{equation}
\lag{5} = \epsilon_{ij} \bar{L_i}\varphi_j\epsilon_{kl}L_k\varphi_l + h.c.
\label{eqn:weinbergop}
\end{equation}
where $i,j \in \{1,2\}$ etc. denote weak isospin indices. 
\end{itemize}
In fact, this is the only allowed dimension-five operator~\cite{Weinberg:1979sa} in the Standard Model effective theory. No analogous operator may be formed with quark fields, since replacing $L\to Q$ does not give a colour singlet. Operators with two vector fields and two fermions are forbidden because the fermion bilinear must be a hypercharge zero SU(2) singlet $\bar\psi \psi$, and so vanishes for chiral fermions. Other combinations involving vector fields cannot be constructed on dimensional grounds. Expanding Eq.~\eqref{eqn:weinbergop} after electroweak symmetry breaking generates a Majorana-like mass term for the neutrino $m_\nu\bar\nu_L\nu_L^C$, and mixing between the neutrino flavour eigenstates. It is therefore required to be non-zero by neutrino phenomenology. However, the mass terms are proportional to $m_\nu \sim v^2/\Lambda$, which points to $\Lambda \sim 10^{13} $ GeV based on current neutrino mass limits. It is therefore not accessible at collider energies, and not explored in the remainder of this thesis.

\subsubsection{\D6}
At dimension-six, many more operators are allowed\footnote{The elimination of all the redundancies of the original operator set written down in Ref. ~\cite{Buchmuller:1985jz} to the basis of Ref.~\cite{Grzadkowski:2010es} was partially done in several intermediate papers~\cite{GrosseKnetter:1993td,Simma:1993ky,Wudka:1994ny,Arzt:1994gp,AguilarSaavedra:2008zc,AguilarSaavedra:2009mx, AguilarSaavedra:2010zi,Fox:2007in}.}. Deriving the full, non-redundant dimension-six operator set is somewhat more involved, so it will not be fully reproduced here, we will merely comment on some of its features. All the operators are built out of the same objects: field strength tensors of dimension two (shorthanded as $X$), Higgs doublets of dimension one (denoted $\varphi$), fermion fields $\psi$ of dimension $\frac{3}{2}$ and various covariant derivatives $D$ of dimension one. By simple power-counting, we can denote the operators as belonging to one of three classes:

\begin{table}[t!]
\begin{center}
\def\arraystretch{1.5}
\begin{tabular}{| c | c || c | c || c | c |} \hline 
\multicolumn{2}{ | c ||} {$X^3$} & \multicolumn{2}{ | c ||}{$\varphi^6$ and $\varphi^4D^2$} & \multicolumn{2}{ | c |} {$\psi^2\varphi^3$}  \\ \hline \hline
\op{G} & $f_{ABC} G_\mu^{A,\nu}G_\nu^{B,\rho}G_\rho^{C,\mu}$ &  \op{\varphi} & $(\varphi^\dagger\varphi)^3 $ & \op{e\varphi} & $(\varphi^\dagger\varphi)(\bar L e\varphi)$ \\
\op{\tilde G} & $f_{ABC}\tilde G_\mu^{A,\nu}G_\nu^{B,\rho}G_\rho^{C,\mu}$ &  \op{\varphi\Box} & $(\varphi^\dagger\varphi)\Box (\varphi^\dagger\varphi) $ & \op{u\varphi} & $(\varphi^\dagger\varphi)(\bar Q u\tilde \varphi)$ \\
\op{W} & $ \epsilon^{IJK}W_\mu^{I,\nu}W_\nu^{J,\rho}W_\rho^{K,\mu}$ & \op{\varphi D} & $(\varphi^\dagger D^\mu \varphi)^\star (\varphi^\dagger D_\mu \varphi) $ & \op{d\varphi} & $(\varphi^\dagger\varphi)(\bar Q d\varphi)$ \\
\op{\tilde W} & $ \epsilon^{IJK}\tilde W_\mu^{I,\nu}W_\nu^{J,\rho}W_\rho^{K,\mu}$ & & & &  \\ \hline \hline
\multicolumn{2}{ | c ||} {$X^2\varphi^2$} & \multicolumn{2}{ | c ||}{$\psi^2 X \varphi$} & \multicolumn{2}{ | c |} {$\psi^2\varphi^2X$}  \\ \hline \hline
\op{\varphi G} & $\varphi^\dagger\varphi G_{\mu\nu}^AG^{A,\mu\nu}$ & \op{eW} & $(\bar L\sigma^{\mu\nu}e)\tau^I\varphi W^I_{\mu\nu} $ & \op[(1)]{\varphi l} & $(\varphi^\dagger i\overleftrightarrow{D_\mu} \varphi)(\bar L \gamma^\mu L)$ \\ 
\op{\varphi \tilde G} & $\varphi^\dagger\varphi \tilde G_{\mu\nu}^AG^{A,\mu\nu}$ & \op{eB}  & $(\bar L\sigma^{\mu\nu}e)\varphi B_{\mu\nu} $ & \op[(3)]{\varphi l} & $(\varphi^\dagger i\overleftrightarrow{D^I_\mu} \varphi)(\bar L \tau^I \gamma^\mu L)$ \\ 
\op{\varphi W} & $\varphi^\dagger\varphi W_{\mu\nu}^IW^{I,\mu\nu}$ & \op{uG} & $ (\bar Q\sigma^{\mu\nu}T^A u)\tilde\varphi G^A_{\mu\nu} $ & \op{\varphi e} & $(\varphi^\dagger i\overleftrightarrow{D_\mu} \varphi)(\bar e \gamma^\mu e)$ \\ 
\op{\varphi \tilde W} & $\varphi^\dagger\varphi \tilde W_{\mu\nu}^IW^{I,\mu\nu}$ & \op{uW} & $(\bar Q\sigma^{\mu\nu}u)\tau^I\tilde \varphi W^I_{\mu\nu} $ & \op[(1)]{\varphi q} & $(\varphi^\dagger i\overleftrightarrow{D_\mu} \varphi)(\bar Q\gamma^\mu Q)$ \\ 
\op{\varphi B} & $\varphi^\dagger\varphi B_{\mu\nu}B^{\mu\nu}$ & \op{uB} & $(\bar Q\sigma^{\mu\nu}u)\tilde \varphi B_{\mu\nu} $ & \op[(3)]{\varphi q} & $(\varphi^\dagger i\overleftrightarrow{D^I_\mu} \varphi)(\bar Q \tau^I \gamma^\mu Q)$  \\ 
\op{\varphi \tilde B} & $\varphi^\dagger\varphi \tilde B_{\mu\nu}B^{\mu\nu}$ & \op{dG} & $ (\bar Q\sigma^{\mu\nu}T^A d)\varphi G^A_{\mu\nu} $ &  \op{\varphi u}  & $(\varphi^\dagger i\overleftrightarrow{D_\mu} \varphi)(\bar u \gamma^\mu u)$ \\ 
\op{\varphi WB} & $\varphi^\dagger\tau^I\varphi W_{\mu\nu}^IB^{\mu\nu} $ & \op{dW}  & $(\bar Q\sigma^{\mu\nu}d)\tau^I \varphi W^I_{\mu\nu} $  &  \op{\varphi d}  & $(\varphi^\dagger i\overleftrightarrow{D_\mu} \varphi)(\bar d \gamma^\mu d)$ \\ 
\op{\varphi \tilde WB} & $\varphi^\dagger\tau^I\varphi \tilde W_{\mu\nu}^IB^{\mu\nu} $ & \op{dB} & $(\bar Q\sigma^{\mu\nu}d)\varphi B_{\mu\nu} $ &  \op{\varphi ud} & $(\tilde\varphi^\dagger i\overleftrightarrow{D_\mu} \varphi)(\bar u \gamma^\mu d)$ \\ \hline
\end{tabular}
\end{center}
\caption[Bosonic and single-fermion current \D6 operators in the Warsaw basis.]
{The non-redundant bosonic and single fermionic-current \D6 operators in the `Warsaw basis' described here. For readability we do not explicitly display the fermion generation indices, but where relevant they are denoted by an extra superscript. For example $\op[23]{e\varphi} = (\varphi^\dagger\varphi)(\bar\mu,\bar\nu_\mu)\tau \varphi $.\label{tab:bosonicops} }
\end{table}

\begin{table}[t!]
\begin{center}
\def\arraystretch{1.5}
\begin{tabular}{| c | c || c | c || c | c |} \hline 
\multicolumn{2}{ | c ||} {$(\bar{L}L)(\bar{L}L)$} & \multicolumn{2}{ | c ||}{$(\bar{R}R)(\bar{R}R)$} & \multicolumn{2}{ | c |} {$(\bar{L}L)(\bar{R}R)$}  \\ \hline \hline
\op{ll}  & $(\bar L \gamma_\mu L) (\bar L \gamma^\mu L)$ &  \op{ee} & $(\bar e \gamma_\mu e) (\bar e \gamma^\mu e)$ & \op{le} & $(\bar L \gamma_\mu L)(\bar e \gamma^\mu e)$ \\ 
 \op[(1)]{qq} & $(\bar Q \gamma_\mu Q) (\bar Q \gamma^\mu Q)$ & \op{uu} & $(\bar u \gamma_\mu u) (\bar u \gamma^\mu u)$ & \op{lu} & ($\bar L \gamma_\mu L) (\bar u \gamma^\mu u)$ \\ 

 \op[(3)]{qq} & $(\bar Q \gamma_\mu \tau^I Q) (\bar Q \gamma^\mu \tau^I Q)$ & \op{dd} & $(\bar d \gamma_\mu d) (\bar d \gamma^\mu d)$ & \op{ld} & ($\bar L \gamma_\mu L) (\bar d \gamma^\mu d)$ \\ 
  \op[(1)]{lq} & $(\bar L \gamma_\mu L) (\bar Q \gamma^\mu Q)$ &  \op{eu} & $(\bar e \gamma_\mu e) (\bar u \gamma^\mu u)$ & \op{qe} & $(\bar Q \gamma_\mu Q)(\bar e \gamma^\mu e)$ \\ 
   \op[(3)]{lq} & $(\bar L \gamma_\mu \tau^I L) (\bar Q \gamma^\mu \tau^I Q)$ &  \op{ed} & $(\bar e \gamma_\mu e) (\bar d \gamma^\mu d)$ & \op[(1)]{qu} & $(\bar Q \gamma_\mu Q)(\bar u \gamma^\mu u)$  \\
   & & \op[(1)]{ud} &  $(\bar u \gamma_\mu u) (\bar d \gamma^\mu d)$ & \op[(8)]{qu} & $(\bar Q \gamma_\mu T^A Q)(\bar u \gamma^\mu T^A u)$  \\
    & & \op[(8)]{ud} &  $(\bar u \gamma_\mu T^A u) (\bar d \gamma^\mu T^A d)$ & \op[(1)]{qd} & $(\bar Q \gamma_\mu Q)(\bar d \gamma^\mu d)$  \\
   & & & & \op[(8)]{qd} & $(\bar Q \gamma_\mu T^A Q)(\bar d \gamma^\mu T^A d)$ \\
 \hline \hline 
\multicolumn{2}{ | c ||} {$(\bar{L}R)(\bar{R}L)$ and $(\bar{L}R)(\bar{L}R)$} & \multicolumn{4}{ | c |}{$B$-violating}  \\ 
\hline \hline
\op{ledq} & $(\bar L e )(\bar d Q)$ & \op{duq} & \multicolumn{3}{| c |}{$\epsilon^{\alpha\beta\gamma}\epsilon_{jk}[(d^\alpha)^T C u^\beta][(Q^{\gamma j})^T C L^k]$} \\
\op[(1)]{quqd} & $(\bar Q u)\epsilon_{jk}(\bar Q d) $ & \op{qqu} & \multicolumn{3}{| c |}{$\epsilon^{\alpha\beta\gamma}\epsilon_{jk} [(Q^{\alpha j})^TCQ^{\beta k}] [(u^\gamma)^TCe] $} \\
\op[(8)]{quqd} & $(\bar Q T^A u)\epsilon_{jk}(\bar Q T^A d)$ & \op[(1)]{qqq} & \multicolumn{3}{| c |}{$\epsilon^{\alpha\beta\gamma}\epsilon_{jk}\epsilon_{mn}  [(Q^{\alpha j})^TCQ^{\beta k}] [(Q^{\gamma m})^TC L^n] $} \\
\op[(1)]{lequ} & $(\bar L^j e)\epsilon_{jk}(\bar Q^k u)$ &  \op[(3)]{qqq} & \multicolumn{3}{| c |}{$\epsilon^{\alpha\beta\gamma}(\tau^I\epsilon)_{jk}(\tau^I\epsilon)_{mn}  [(Q^{\alpha j})^TCQ^{\beta k}] [(Q^{\gamma m})^TC L^n]  $} \\
\op[(3)]{lequ} & $(\bar L^j \sigma_{\mu\nu} e)\epsilon_{jk} (\bar Q^k \sigma^{\mu\nu} u)$ & \op{duu} & \multicolumn{3}{| c |}{$\epsilon^{\alpha\beta\gamma}[(d^\alpha)^T C u^\beta][(u^\gamma)^TCe] $} \\  
\hline  
\end{tabular}
\end{center}
\caption[Four-fermion \D6 operators in the Warsaw basis.]{\label{tab:4fops} 
The non-redundant four-fermion \D6 operators in the `Warsaw basis'. For readability we do not explicitly display the fermion generation indices, but where relevant they are denoted by an extra superscript $\op[prst]{}\sim \bar \psi_p \psi_r\bar \psi_s \psi_t$. For example $\op[1231]{ll} = (\bar e \gamma_\mu \mu)(\bar \tau\gamma^\mu e)$ .}
\end{table}

\begin{mydescription} 


\item [Bosonic operators:] These contain no fermion fields. There must be an even number of Higgs doublets, and an even number of covariant derivative operators, to ensure that all Lorentz indices are contracted. The allowed combinations are then $X^3$, $X^2\varphi^2$, $X^2D^2$, $X\varphi^4$, $XD^4$, $X\varphi^2D^2$, $\varphi^6$, $\varphi^4D^2$ and $\varphi^4D^2$. We can eliminate several of these classes. Firstly, Lorentz symmetry forbids $X\varphi^4$ terms, which are not Lorentz contracted. All $XD^4$ terms can also be moved to $X^2D^2$ terms by use of the identity $[D_\mu,D_\nu] \sim X_{\mu\nu}$. As for the remaining classes:

\underline{$\varphi^2D^4$:} By the equations of motion:
\begin{equation}
\begin{split}
(D_\mu D^\mu \varphi)^j &= m^2\varphi^j - \lambda(\varphi^\dagger\varphi)\varphi^j - \bar e y^\dagger_e L^j + \epsilon_{jk}\bar Q^k y_u u - \bar d y_d^\dagger Q^j , \\
(D^\rho G_{\rho \mu})^A &= g_s(\bar Q \gamma_\mu T^A Q +  \bar u \gamma_\mu T^A u +  \bar d \gamma_\mu T^A d), \\
(D^\rho W_{\rho\mu})^I &= \frac{g}{2}\left(\varphi^\dagger i \overleftrightarrow D_\mu^I\varphi + \bar L \gamma_\mu \tau^I L + \bar Q \gamma_\mu \tau^I Q\right), \\
\partial^\rho B_{\rho\mu} &= g^\prime Y_\varphi \varphi^\dagger i \overleftrightarrow D_\mu \varphi + g^\prime \summ{i} Y_\psi \bar \psi \gamma_\mu \psi .
\end{split}
\label{eqn:bosoneom}
\end{equation}
these can be moved to operators in the bosonic classes $\varphi^2XD^2$ and $\varphi^4D^2$, as well as the single-fermionic current operator class $\psi^2\phi D^2$. 

\underline{$\varphi^2XD^2$:} The identity $[D_\mu,D_\nu] \sim X_{\mu\nu}$ again moves some operators to the $\varphi^2X^2$ class. Also, using the equations of motion for the gauge field and the Bianchi identity $D_{[\rho}X_{\mu\nu]} = 0$, the remaining operators of this class are moved either to the bosonic class $\varphi^4D^2$ or the fermionic class $\psi^2\phi D^2$.

\underline{$X^2D^2$:} All operators of this class can be reduced to operators in the class $X^3$,  $\varphi^2XD^2$, or $\psi^2XD$, or made to vanish by the equations of motion.

The only surviving bosonic operator classes are then $X^3, X^2\varphi^2, \varphi^6$ and $\varphi^4D^2$. The non-redundant operators in each of these classes are shown in columns 1, 2 and 4 of Tab. \ref{tab:bosonicops}.

\item [Single-fermionic current operators:] The classes allowed here are: $\psi^2D^3$, $\psi^2\varphi D^2$, $\psi^2XD$, $\psi^2\phi^3$, $\psi^2X\varphi$ and $\psi^2\phi^2D$. We can make use of the following equations of motion for the fermion currents.
\begin{equation}
i \slashed D l = y_e e\varphi, \hspace{10pt} i \slashed D e = y^\dagger_e \varphi^\dagger l, \hspace{10pt} i \slashed D Q = y_u u \tilde \varphi + y_d d \varphi, \hspace{10pt}  i\slashed D_\mu = y_u^\dagger \tilde \varphi^\dagger Q, \hspace{10pt} i \slashed D d = y_d^\dagger \varphi^\dagger Q .
\end{equation}
\underline{$\psi^2D^3$:} By reordering derivatives, we can use the equations of motion to reduce these operators to operators of the class $\psi^2\varphi D^2$.

\underline{$\psi^2\varphi D^2$:} All operators in this class can be reduced to (up to total derivatives) operators in the single fermionic current classes $\psi^2\phi^3$ and $\psi^2X\varphi$ or four-fermion operators $\psi^4$, plus operators that vanish by the equations of motion.

\underline{$\psi^2XD$:} Using the equations of motion for the gauge field and the Bianchi identities, one finds that all operators in this class can be reduced to operators in the classes $\psi^2X\varphi$ and $\psi^2\phi^2D$, and four-fermion operators $\psi^4$, plus total derivatives.

The remaining non-redundant operators of the single-fermionic current operator classes are thus all in the $\psi^2\varphi^3$, $\psi^2X\varphi$ and $\psi^2\varphi^2D$ subclasses. They are displayed in columns 3, 5 and 6 of  Tab. \ref{tab:bosonicops}.

\item[Four-fermion operators:] Although all operators in this class are all of the simple form $\psi^4$, they constitute by far the most numerous, though they can be straightforwardly classified. Noting that they are generically constructed out of left-handed fields $L$ and right-handed fields $R$, they can be constructed out of products of hypercharge zero currents $(\bar{L}L)(\bar{L}L)$, $(\bar{R}R)(\bar{R}R)$ and $(\bar{L}L)(\bar{R}R)$, and a few others of the form $(\bar{L}R)(\bar{R}L)$ and $(\bar{L}R)(\bar{L}R)$, as well as four baryon-number violating operators. Though the equations of motion cannot be used to whittle down this operator set, the Fierz identity, 
\begin{equation}
(\bar\psi_L\gamma_\mu \psi_L) (\bar\chi_L\gamma_\mu \chi_L)  = (\bar \psi_L\gamma_\mu \chi_L) (\bar\chi_L\gamma_\mu \psi_L)
\end{equation}
as well as the identity for the SU(N) generators
\begin{equation}
T^A_{\mu\nu} T^A_{\alpha\beta} = \frac{1}{2} \delta_{\mu\alpha} \delta_{\nu\beta} -\frac{1}{2N}  \delta_{\mu\nu} \delta_{\alpha\beta} ,
\end{equation}
can be used. The complete non-redundant set of four-fermion operators are shown in Tab. \ref{tab:4fops}.

\end{mydescription}
The 64 operators of those tables complete what is commonly referred to as the `Warsaw basis' of the \D6 Standard Model effective theory. This is the operator basis used throughout this thesis. Other bases for the \D6 operator set are also commonly used~\cite{Hagiwara:1993ck,Giudice:2007fh,Artoisenet:2013puc,Gupta:2014rxa}, and it is merely an exercise in linear algebra to translate between them\footnote{In fact there are tools which automate this procedure entirely~\cite{Falkowski:2015wza}.}. Excluding the five $B$-violating operators, whose effects must be strongly suppressed to respect proton decay bounds, we have 59 independent operators. In fact, one can relax the flavour assumptions and allow all possible flavour combinations to be an independent operator. This increases the operator set to 2499 operators. In order to make an analysis tractable, it is typically assumed that the operators obey minimal flavour violation, so that 59 $B$-conserving operators form a complete set. 

The operator set at \D7 and \D8 has also been computed, and there now exist tools for computing the operator set to arbitrarily high dimension, though not all the redundancies are automatically eliminated. The \D7 operator set all violate lepton number, so are typically not interesting for LHC energies, where lepton number conservation has been demonstrated to an extremely high degree. There are 993 \emph{structures} at \D8~\cite{Henning:2015alf}, even assuming minimal flavour violation (not all of these correspond to a unique operator, however). In order to avoid the proliferation of large numbers of operators, for phenomenological purposes one typically cuts off the expansion at \D6. Since higher-order terms will be proportional to higher-powers of $\Lambda$, then provided there is a large enough separation between the low-energy theory and the cutoff, then this truncation is allowed. The \D6 truncation is typically referred to as the Standard Model Effective Field Theory (SMEFT).

\subsection{The top quark sector of the Standard Model effective theory}
\label{sec:topeft}

In order to access the sector of the SMEFT that is relevant for top quark physics at hadron colliders, it is necessary to compute the Feynman rules of the operators in Tables \ref{tab:bosonicops} and \ref{tab:4fops}, and calculate which of them lead to modifications of the processes and observables listed in the first chapter. We will take each of these processes in turn, but first we make some general comments about the modifications of a collider observable due to a \D6 operator.

We will focus first on cross-sections, though the same arguments will apply to decay observables as well. In general, the Lorentz-invariant matrix element \m{} is related to the differential cross-section in some observable $X$ by
\begin{equation}
\frac{\d\sigma}{\d X} = \int \d\Pi_{\text{LIPS}} \delta^{(4)}(X-X^\prime) |\m{}|^2, 
\end{equation}
where the Lorentz-invariant phase space element $\d\Pi_{\text{LIPS}} \sim \d X^\prime$. $X$ may be a one particle inclusive quantity such as the $p_T$ of one of the final state particles, or $N$-particle inclusive, such as the final state invariant mass. In the presence of \D6 operators, the matrix element is modified to $\m{\text{full}} = \m{\text{SM}}+\m{\text{D6}}$, so that the expression for the squared matrix element is
\begin{equation}
|\m{\text{full}}|^2 = |\m{\text{SM}}|^2+2\Re\m{\text{SM}}^*\m{\text{D6}}+|\m{\text{D6}}|^2.
\label{eqn:d6me}
\end{equation}
The linear term $2\Re\m{\text{SM}}^*\m{\text{D6}}$ is proportional to $1/\Lambda^2$, and is generated by interference between Standard Model and new physics amplitudes, while the quadratic term is generated solely by new physics contributions, and is proportional to $1/\Lambda^4$. Since there is no dependence on the matrix element in the final state phase space, we can schematically write the cross-section as
\begin{equation}
\d\sigma_{\text{full}} \sim \d\sigma_{\text{SM}}+\co{i}\d\sigma_{\text{D6}}+\co[2]{i} \d\sigma_{\text{D6}^2},
\label{eqn:d6xsec}
\end{equation}
where we have displayed the \D6 Wilson coefficients explicitly. 

Since the matrix element receives contributions from terms at different orders in $\Lambda$, one might worry that the corresponding cross-section is not properly defined in $\Lambda$. Namely, the quadratic terms $\m{\text{D6}}$ and the interference between \D8 operators and the Standard Model $\sim 2\Re\m{\text{SM}}^*\m{\text{D8}}$ are formally both \ord{1/\Lambda^4}, however we only consider the former and not the latter. This issue ties in with a broader discussion of the validity of the EFT description of an observable, and we will return to it throughout this thesis. For now it suffices to assume that the higher-order interference terms can be neglected.

\subsubsection{Top pair production}
\begin{figure}[!t]
\begin{center}
\includegraphics[width=\textwidth]{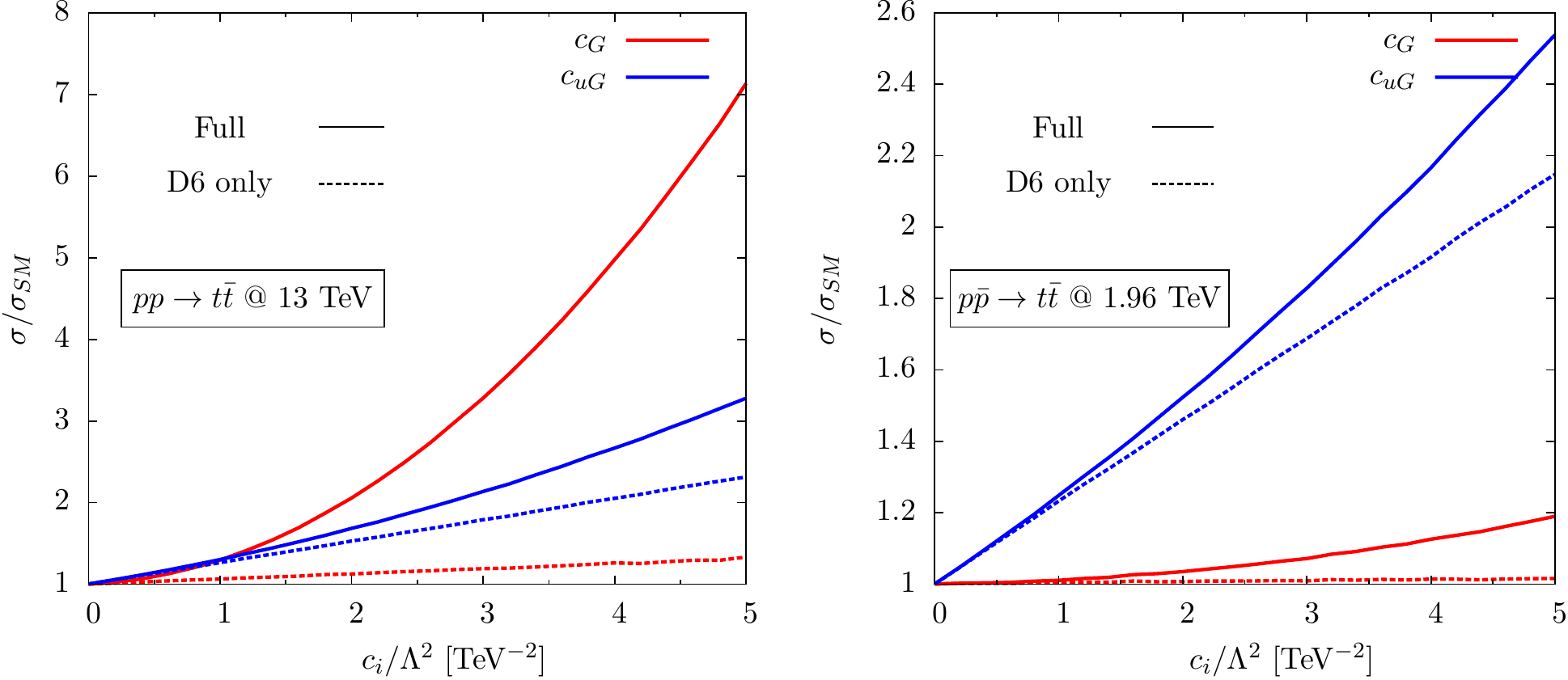}
\caption[\ttbar production cross-section with the operators \op{G} and \op{uG}.]{Ratio of the full cross-section for $t\bar t$ production at the 13 \tev LHC (left) and Tevatron (right) for the operators \op{G} and \op[33]{uG} to the Standard model prediction, as a function of the operator Wilson coefficient. The dashed lines show the effects of the interference term only, the solid lines show the interference and quadratic terms.}
\label{fig:xsec_vs_c1c3}
\end{center}
\end{figure}  
As in the case of the Standard Model, we can (at leading-order in $\alpha_s$) split up top-quark pair production into the $gg$ and $q\bar{q}$ channels. For the former, any new physics which couples to the top quark directly will modify the top-gluon vertex. The only operator that does this directly is the so-called chromomagnetic moment operator \op[33]{uG}$\equiv$\op{tG}, where the superscript denotes the generation index explicitly. Its interference with the SM $gg\to t\bar t$ amplitude gives the term~\cite{Degrande:2010kt}\footnote{We want to focus on the interference terms, so we do not show the quadratic terms explicitly, though we will compute their numerical contributions to observables.}
\begin{equation} 
2\Re\m{uG}^{33 *}\m{\text{SM}} = \frac{g_s^3}{2\sqrt{2}} \frac{vm_t\co[33]{uG}}{\Lambda^2}\left(\frac{1}{6\tau_1\tau_2}-\frac{3}{8} \right) ,
\end{equation}
where $\tau_{1,2}$ are functions of the Mandelstam invariants $\tau_1 = (m_t^2-t)/s$, $\tau_2 = (m_t^2-u)/s$ and $\rho = 4m_t^2/s$ is the threshold variable. 

In fact, this is the only operator that modifies the $gg\to t\bar t$ production cross-section by directly coupling to the top. We can indirectly modify the cross-section, however, by modifying the triple-gluon vertex in the initial state with the operator \op{G}. This leads to the interference matrix element~\cite{Cho:1993eu,Cho:1994yu}
\begin{equation} 
2\Re\m{G}^*\m{\text{SM}} = \frac{9}{8}\frac{\co{G}g_s^3}{\Lambda^2} \frac{m_t^2(\tau_1-\tau_2)^2}{\tau_1\tau_2}.
\end{equation}
The partonic differential cross-sections are folded with the incoming parton densities to give the proton-(anti)proton cross-sections. To get an idea of the strength of the operators, we plot the ratio of the full cross-section to the SM only value, broken up into the interference and quadratic pieces, as a function of the dimensional Wilson coefficient $ \tilde {\co{i}} = \co{i}/\Lambda^2$ for \op[33]{uG} and \op{G}. The results are shown in Fig. \ref{fig:xsec_vs_c1c3}, for both the 13 \tev LHC and Tevatron.

\begin{figure}[!t]
\begin{center}
\includegraphics[width=\textwidth]{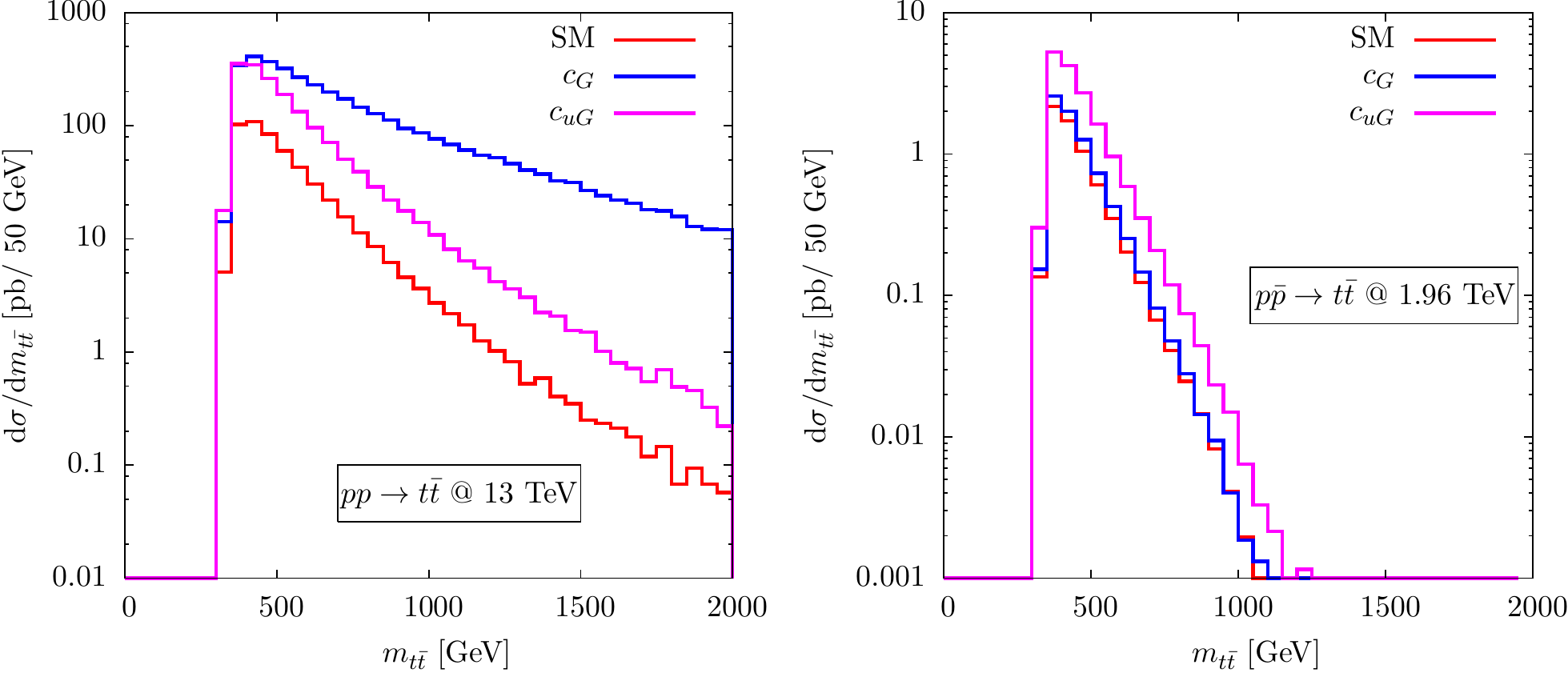}
\caption[\ttbar invariant mass distributions with the operators \op{G} and \op{uG}.]{Top quark pair invariant mass distributions at the 13 TeV LHC (left) and the Tevatron (right). The red curves are for  the SM only, while the blue and magenta curves are for (respectively) the coefficients $\co{G}/\Lambda^2$ and $\co[33]{uG}/\Lambda^2$ set to a value of 5 \tev$^{-2}$.}
\label{fig:mtt_vs_cs}
\end{center}
\end{figure}  

We see that the effects of the operator \op{G} come almost entirely from its quadratic term, and its interference is very small, therefore in a \D6 framework, constraints on its coefficient should be taken with caution. It is also clear that \op{G} is much stronger at the LHC than at the Tevatron, due to the much higher gluon densities in the proton beam at LHC energies. For the chromomagnetic moment \op[33]{uG}, the interference term dominates up to very large values of its Wilson coefficient. Its effects are also typically much stronger at the Tevatron, because its larger contribution is from the $q\bar q$ channel, which dominates here since both quark and antiquark are valence and dominate over gluons at the typical $x$ values probed at the Tevatron.

We can also show the effects of these two operators on kinematic distributions. In Fig. \ref{fig:mtt_vs_cs} we plot the invariant mass distributions at a point in the parameter space for each operator, namely $\co{i}/\Lambda^2$ = 5 \tev$^{-2}$, again at the 13 \tev LHC and Tevatron. We see again that in the case of the LHC, the operator \op{G} has the much stronger effect, whereas the operator \op[33]{uG} dominates at the Tevatron. The operator \op[33]{uG}, in both cases, modifies the SM distribution by an overall normalisation factor, whereas the operator \op{G} has a much stronger effect in the tail of the distribution at the LHC, corresponding to the region where the gluon pdf becomes increasingly dominant.

\begin{figure}[!t]
\begin{center}
\includegraphics[width=\textwidth]{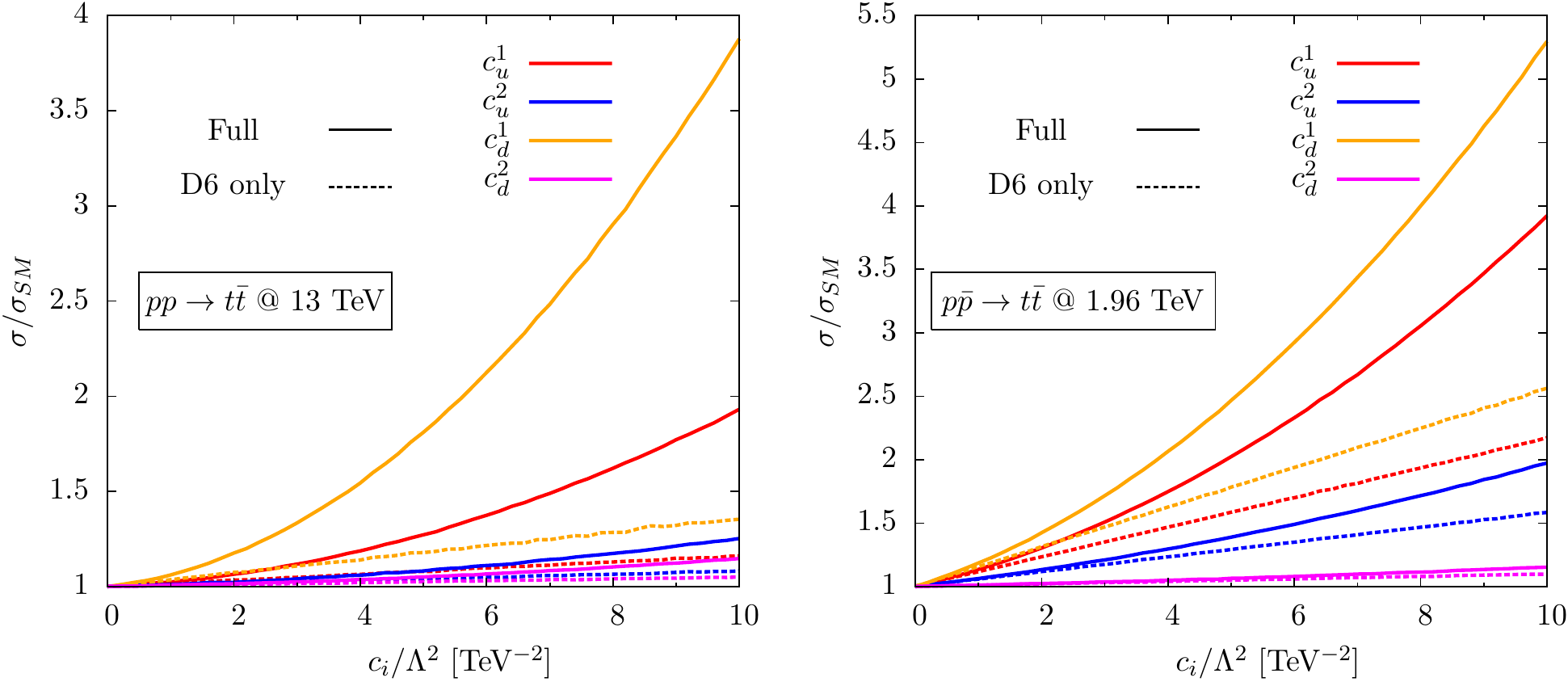}
\caption[\ttbar production cross-section with the four-fermion operators \op{4q}.]{Ratio of the full cross-section for $t\bar t$ production at the 13 \tev LHC (left) and Tevatron (right) for the four-fermion operators contributing to top pair production to the Standard model prediction, as a function of the operator Wilson coefficient. The dashed lines show the effects of the interference term only, the solid lines show the interference and quadratic terms.}
\label{fig:xsec_vs_4567}
\end{center}
\end{figure}

For the $q\bar q$ channel, the situation is slightly more complicated. In addition to the chromomagnetic operator \op[33]{uG} already mentioned, there is also a contribution from various four-fermion operators listed in Tab. \ref{tab:4fops}. Though there are many individual operators that contribute, at the level of observables their effects factorise into only four unique linear combinations of 4-quark operators~\cite{Zhang:2010dr}.  
\begin{equation}
\begin{split}
\co[1]{u} & = \co[1,1331]{qq}+ \co[1331]{uu}+ \co[3,1331]{qq} \\
\co[2]{u} & = \co[8,1133]{qu} + \co[8,3311]{qu} \\
\co[1]{d} & = 4\co[3,1331]{qq}+\co[8,3311]{ud} \\
\co[2]{d} & = \co[8,1133]{qu} + \co[8,3311]{qd} . \\
\label{eqn:4fs}
\end{split}
\end{equation}
The interference matrix element for these operators is given by
\begin{equation}
2\Re\m{4q}^*\m{\text{SM}} = \frac{g_s^2}{9\pi^2}\frac{s}{\Lambda^2}\left[\frac{1}{4}\left(\co[1]{u,d}-\co[2]{u,d}\right)\left(\tau_1-\tau_2\right)+\frac{1}{4}\left(\co[1]{u,d}+\co[2]{u,d}\right)\left(\tau_1^2+\tau_2^2+\frac{\rho}{2}\right)\right] .
\end{equation}
The ratio of the full $t\bar t$ cross-section (summed over the $u$ and $d$ production channels), including these operators to the SM estimate, at the LHC and Tevatron, are shown in Fig. \ref{fig:xsec_vs_4567}. We see that for larger values of the Wilson coefficient, the squared term dominates contributions to the cross-section, and that the \co[1]{u,d} type operators typically have a much larger effect on the cross-section, at both the interference and squared level.

\subsubsection{Single top production}
For single top production, we can split up the processes into $s$ and $t$ channel production, as discussed in the previous chapter. However, since the diagrams are related by crossing symmetry, the same operator set contributes to both. The interfering operators which have numerical significance are \op[33]{uW} ($\equiv$ \op{tW}), \op[(3)]{\varphi q}, and the linear combination of four-fermion operators
\begin{equation}
\op{t} \equiv \op[3,1133]{qq}+\frac{1}{6}\left(\op[1,1331]{qq}+\op[3,1331]{qq}\right) .
\end{equation}
For $t$-channel production, the interference term generated by these three operators can be more conveniently expressed directly in terms of the Mandelstam invariants and, for the $ub\to dt$ subprocess, takes the form
\begin{equation}
\begin{split}
2\Re\m{D6}^*\m{\text{SM}} = & \frac{\co[(3)]{\varphi q}}{\Lambda^2} \frac{V_{tb}|V_{ud}|^2g^2v^2 s(s-m_t^2)}{4(t-m_W^2)^2} - \frac{\co{tW}}{\Lambda^2}\frac{\sqrt{2}V_{tb}|V_{ud}|^2 m_t m_W st }{(t-m_W^2)^2} \\
					 &+ \frac{\co{t}}{\Lambda^2}\frac{9 V_{tb}V_{ud}  g^2 s(s-m_t^2)}{8(t-m_W^2)} .
\end{split}					
\end{equation}
The corresponding expression for the $\bar d b\to \bar u t$ subprocess is obtained by substituting $s\to u$ in each term. For $s$-channel production, the expression can be obtained by substituting $s\to u, t\to s$, making use of crossing symmetry. 

The interference of \op[(3)]{\varphi q} has the same kinematic dependence as the SM contribution, in fact it amounts to rescaling the CKM element $V_{tb}^2 \to V_{tb}^2+ 2\co[(3)]{\varphi q} V_{tb}/\Lambda^2$. Therefore it does not affect the shapes of distributions and merely rescales the overall cross-section. The other two operators, in addition to modifying the overall cross-section, modify the shapes of distributions, and so stronger bounds should be expected on their coefficients.

The relative contributions of each of these operators will serve as a rough guide for how strongly they can be constrained when their coefficients are fit to relevant measurements. The next chapter will discuss in detail such a fit. The same arguments can be made to calculate the effects of \D6 operators on the other top-related processes discussed in Chapter 1, such as associated production, charge asymmetries and decay observables. We will postpone such a discussion until the next chapter.

\subsection{Summary}
\label{sec:conc_ch2}
Despite the many successes of the Standard Model over the last forty years, there are several reasons to believe that it is only a stepping stone to a more fundamental theory of Nature, due to considerations such as the hierarchy problem, vacuum stability, and gauge coupling unification. Given its large Yukawa coupling, and as the only SM fermion with an electroweak scale mass, the top quark typically plays a special role in most of these scenarios, such as supersymmetry, extra dimensions and broken global symmetries such as the little Higgs family of models. These extensions generically predict direct and/or indirect modifications to top quark collider observables, such as new resonances decaying to top pairs, enhanced top quark production cross-sections, new top decay modes and modified decay distributions.

A generic way to parametrise the potential effects of heavy new physics on low energy observables is to regard the Standard Model as the leading order piece of an effective theory, where the non-SM interactions are encoded in higher-dimensional ($D>4$) operators. We formulated the \D6 extension of the Standard Model, and discussed the sector of this EFT that can potentially impact top quark observables at hadron colliders. In the next chapter we will perform a global fit of the Standard Model EFT to Tevatron and LHC Run I data, and will discuss the implications of the results of this fit for constraints on new physics.

\newpage

\section{A global fit of top quark effective theory to data}

\subsection{Introduction}
In chapter 1, we discussed the role of the top quark in the Standard Model of particle physics, and the unique properties of its production and decay mechanisms that can be measured with precision at hadron colliders. In chapter 2, we showed that the top quark also plays a special role in scenarios of physics Beyond the Standard Model, and outlined the formulation of the Standard Model as the leading part of an effective theory. We then briefly touched upon the sector of this effective theory that could be probed with top quark measurements at hadron colliders.

The motivations for this formulation are manifold: Firstly, with the LHC Run II well underway, the main take-home message is that, apart from a few scattered anomalies, all measurements are in agreement with Standard Model predictions. This implies that, if new heavy degrees of freedom exist at all, they are decoupled~\cite{Appelquist:1974tg,Wilson:1983dy} from the electroweak scale (either there is a large mass gap or very weak coupling between the SM and new physics sectors), in which case they will necessarily integrate out into higher-dimensional operators~\cite{Buchmuller:1985jz,Hagiwara:1986vm,Burges:1983zg,Leung:1984ni} in the low energy limit\footnote{Current collider measurements, however, cannot rule out the existence of light degrees of freedom, see e.g. Ref.~\cite{King:2012tr}.}. Secondly, faced with the large number of hypothesised new physics scenarios and the frequent degeneracy in their experimental signatures, it is prudent to describe deviations from the Standard Model in as \emph{model-independent} a way as possible. The differences in early inclusive Higgs production cross-section measurements from their SM values, for instance, are often described by `signal strength' ratios. Likewise, electroweak observables are also phrased in the language of anomalous couplings.

From a phenomenological perspective, the EFT description is nothing more than another model-independent way of asking `which self-consistent Lagrangian best describes the data?', but it has the advantage over other approaches such as signal strengths in that it can also accommodate differential distributions and angular observables, because the higher-dimensional operators lead to new vertex structures which can impact event kinematics. They also have the advantage over `form-factors' in that they manifestly preserve the full \smgg gauge symmetry, and so can be more straightforwardly linked to concrete ultraviolet completions.

This merits have not gone unnoticed, and EFT techniques have received much attention in interpreting available Higgs results~\cite{Azatov:2012bz,Espinosa:2012im,Plehn:2012iz,Carmi:2012in,Peskin:2012we,Dumont:2013wma,Djouadi:2013qya,Lopez-Val:2013yba,Englert:2014uua,Ellis:2014dva,Ellis:2014jta,Falkowski:2014tna,Corbett:2015ksa,Buchalla:2015qju,Aad:2015tna,Berthier:2015gja,Falkowski:2015jaa,Englert:2015hrx}, although this area is still in its infancy, and the corresponding bounds (and thus the conclusions that one can draw about extended Higgs sectors) are limited by low statistics on the experimental side. Top quark physics, by contrast, has entered a precision era, and data from the LHC and Tevatron are far more abundant. As already discussed, the top quark plays a special role in most scenarios of Beyond the Standard Model physics, motivating scrutiny of its phenomenology.  Furthermore, the top sector is strongly coupled to Higgs physics owing to the large top quark Yukawa coupling, and so represents a complementary window into physics at the electroweak scale. Thus, it is timely to review the constraints on new top interactions through a global fit of all dimension-six operators relevant to top production and decay at hadron colliders. This is the subject of this chapter.

There have been several studies of the potential for uncovering new physics effects in the top quark sector at the LHC and Tevatron, phrased in model-independent language, either through anomalous couplings~\cite{AguilarSaavedra:2008zc,Bernardo:2014vha,Grzadkowski:2003tf,Nomura:2009tw,Hioki:2009hm,Hioki:2010zu,Hioki:2013hva,Aguilar-Saavedra:2014iga,Chen:2005vr,AguilarSaavedra:2008gt,AguilarSaavedra:2010nx,AguilarSaavedra:2011ct,Fabbrichesi:2014wva,Fabbrichesi:2013bca,Cao:2015doa,GonzalezSprinberg:2011kx} or higher-dimensional operators~\cite{Davidson:2015zza,Jung:2014kxa,Cao:2007ea,Degrande:2010kt,Zhang:2010dr,Greiner:2011tt,Degrande:2012wf}. Though there is a one-to-one correspondence between these two approaches (for the reasons discussed below) the latter is the approach taken in this analysis. Other studies have also set limits on top dimension-six operators, but by considering different physics, such as precision electroweak data~\cite{deBlas:2015aea}, or flavour-changing neutral currents~\cite{Aguilar:2015vsa,Durieux:2014xla}.

The chapter is structured as follows. In Section~\ref{sec:ops} we review the
higher-dimensional operators relevant for top quark physics and in
Section~\ref{sec:fit} we review the experimental measurements entering our fit,
as well as the limit-setting procedure we adopt. In Section~\ref{sec:fitresults} we
present our constraints, and discuss the complementarity of LHC and Tevatron
analyses, and the improvements obtained from adding differential distributions
as well as inclusive rates. In Section \ref{sec:validity} we discuss issues relating to the validity of the EFT framework and how our constraints look in the context of specific new physics models. Finally, in
Section~\ref{sec:conc_ch3} we discuss our results and conclude.

\subsection{Higher-dimensional operators}
\label{sec:ops}
In order to keep this chapter self-contained, in this section we briefly revisit the operators relevant for top observables at hadron colliders. As discussed in chapter 2, the leading contributions to
\lag{eff} at collider energies enter at dimension \D6
\begin{equation}
 \lag{eff} = \lag{SM} + \frac{1}{\Lambda^2}\sum_{i}\co{i}\op{i}(G^A_\mu,W^I_\mu,B_\mu,\varphi,Q_L,u_R,d_R,L_L,e_R) +\mathcal{O}(\Lambda^{-4})\, ,
\end{equation}
where \op{i} are \D6 operators made up of SM fields, and \co{i} are
dimensionless Wilson coefficients. At dimension-six, assuming minimal flavour
violation and Baryon number conservation, there are 59 independent operators. Clearly, allowing 59 free
parameters to float in a global fit is intractable. Fortunately, for any given
class of observables, only a smaller subset is relevant. In top physics, for the observables we consider, we
have the following effective operators, expressed in the so-called `Warsaw
basis' of Ref.~\cite{Grzadkowski:2010es}\footnote{Given the simplicity of how it
  captures modifications to SM fermion couplings, this basis is well-suited to
  top EFT. For basis choices of interest in Higgs physics, see
  e.g. Refs.~\cite{Gupta:2014rxa,Giudice:2007fh,Contino:2013kra,Masso:2014xra,Pomarol:2014dya},
  and Ref.~\cite{Falkowski:2015wza} for a tool for translating between them.}
\begin{align}
\op[(1)]{qq} &= (\bar{Q}\gamma_{\mu}Q)( \bar{Q}\gamma^{\mu}Q)  &  \op{uW} &= (\bar{Q}\sigma^{\mu \nu} \tau^I u)\tilde \varphi W_{\mu\nu}^{I}  & \op[(3)]{\varphi q} &= i(\varphi^\dagger \overleftrightarrow{D}^I_\mu \varphi )(\bar{Q}\gamma^\mu \tau^I Q) \nonumber \\
\op[(3)]{qq} &= (\bar{Q}\gamma_{\mu}\tau^IQ)( \bar{Q}\gamma^{\mu}\tau^I Q) &  \op{uG} &= (\bar{Q}\sigma^{\mu \nu} T^A u)\tilde \varphi G_{\mu\nu}^{A}  & \op[(1)]{\varphi q} &= i(\varphi^\dagger \overleftrightarrow{D}_\mu \varphi )(\bar{Q}\gamma^\mu Q)    \nonumber \\
\op{uu} &= (\bar{u}\gamma_{\mu}u)( \bar{u}\gamma^{\mu} u) &  \op{G} &= f_{ABC} G_{\mu}^{A \nu}G_{\nu}^{B \lambda} G_{\lambda}^{C \mu}  &  \op{uB} &= (\bar{Q}\sigma^{\mu \nu}u)\tilde \varphi B_{\mu\nu}       \nonumber \\
\op[(8)]{qu} &=  (\bar{Q}\gamma_{\mu}T^AQ)( \bar{u}\gamma^{\mu} T^Au) &  \op{\tilde G} &= f_{ABC} \tilde G_{\mu}^{A \nu}G_{\nu}^{B \lambda} G_{\lambda}^{C \mu}  & \op{\varphi u} &= (\varphi^\dagger i \overleftrightarrow{D}_\mu\varphi)(\bar{u}\gamma^\mu u)    \nonumber \\
\op[(8)]{qd} &= (\bar{Q}\gamma_{\mu}T^AQ)( \bar{d}\gamma^{\mu} T^Ad)  &  \op{\varphi G} &= (\varphi^\dagger \varphi)G_{\mu\nu}^{A}G^{A \mu\nu} \nonumber &  \op{\varphi \tilde G} &= (\varphi^\dagger \varphi) \tilde G_{\mu\nu}^{A}G^{A \mu\nu} \\
\op[(8)]{ud} &= (\bar{u}\gamma_{\mu}T^Au)( \bar{d}\gamma^{\mu} T^Ad) \label{eqn:allops} \,.
\end{align}
%
We adopt the same notation as Ref.~\cite{Grzadkowski:2010es}, where $T^A=\tfrac{1}{2}\lambda^A$ are the $SU(3)$ generators, and $\tau^{I}$ are the Pauli matrices, related to the generators of $SU(2)$ by $S^I=\tfrac{1}{2}\tau^I$. For the four-quark operators on the left column of Eq.~\eqref{eqn:allops}, we denote a specific flavour combination $(\bar{Q}_i...Q_j)(\bar{Q}_k...Q_l)$ by e.g. \op[4q]{ijkl}. It should be noted that the operators \op{uW}, \op{uG} and \op{uB} are not hermitian and so may
have complex coefficients which, along with \op{\tilde G} and
\op{\varphi \tilde G}, lead to $\mathcal{CP}$-violating effects. These do not
contribute to the observables built out of spin-averaged matrix elements that we consider, but they are in
principle sensitive to polarimetric information such as spin correlations, and
should therefore be treated as independent operators. However, currently available
measurements that would be sensitive to these degrees of freedom have been extracted by making model-specific assumptions that preclude their usage in the fit, e.g. by assuming that the tops are produced with either SM-like spin correlation or no spin correlation at all, as in Refs.~\cite{Chatrchyan:2013wua,Aad:2014pwa,Aad:2014mfk}. We will discuss this issue in more detail in the next section. With these caveats, a
total of 14 constrainable \cp-even dimension-six operators contribute to top quark
production and decay at leading order in the SMEFT.

\subsection{Methodology}
\label{sec:fit}

\subsubsection{Experimental inputs}
The experimental measurements used in the
fit~\cite{Aad:2014kva,ATLAS:2012aa,Aad:2012mza,Aad:2012qf,Aad:2012vip,Aad:2014jra,Aad:2015pga,Chatrchyan:2013ual,Chatrchyan:2012bra,Chatrchyan:2012ria,Chatrchyan:2012vs,Chatrchyan:2013kff,Chatrchyan:2013faa,Khachatryan:2015uqb,Aaltonen:2013wca,Aad:2014fwa,Aaltonen:2014qja,Khachatryan:2014iya,Abazov:2009pa,Abazov:2011rz,Aad:2014zka,Aaltonen:2009iz,Chatrchyan:2012saa,Khachatryan:2015oqa,Abazov:2014vga,Aad:2013cea,Chatrchyan:2014yta,Aaltonen:2012it,Abazov:2014cca,Aaltonen:2013kna,Abazov:2012vd,Aad:2012ky,Aaltonen:2012lua,Chatrchyan:2013jna,Abazov:2010jn,Aad:2015uwa,Aad:2015eua,Khachatryan:2014ewa}
are included in Table~\ref{table:measurements}. All these measurements are
quoted in terms of `parton-level' quantities; that is, top quarks and their
direct decay products. Whilst it is possible to include particle-level
observables, these are far less abundant and they are beyond the scope of the
present study.

The importance of including kinematic distributions is manifest here. For top
pair production, for instance, we have a total of 195 measurements, 174 of which
come from differential observables. This size of fit is unprecedented in top
physics, which underlines the need for a systematic fitting approach, as
provided by \textsc{Professor}~\cite{Buckley:2009bj}. Indeed top pair production cross-sections make up
the bulk of measurements that are used in the fit. Single top production
cross-sections comprise the next dominant contribution. We also make
use of data from charge asymmetries in top pair production, as well as inclusive measurements of top pair production in association with a photon or a $Z$ ($t\bar{t}\gamma$ and $t\bar{t}Z$) and
observables relating to top quark decay. We take each of these categories of
measurement in turn, discussing which operators are relevant and the constraints
obtained on them from data.

\begin{table*}[!t]
\begin{center}
\begin{adjustbox}{max width=\textwidth}
\begin{tabular}{l r  l  l || l  r  l  l  }
\toprule
Dataset & $\sqrt{s}$ (\tev) & {Measurements} & {arXiv ref.} & Dataset & $\sqrt{s}$ (\tev) & {Measurements} & {arXiv ref.} \\
\midrule
\multicolumn{8}{l}{\textit{Top pair production}} \\
\multicolumn{4}{l}{Total cross-sections:} & \multicolumn{4}{l}{Differential cross-sections:} \\
ATLAS & 7  & lepton+jets   & 1406.5375 & ATLAS & 7  &  $p_T (t),m_{t\bar{t}},|y_{t\bar{t}}|$  & 1407.0371   \\
ATLAS & 7  & dilepton   &  1202.4892 & CDF    & 1.96   & $m_{t\bar{t}}$  & 0903.2850  \\
ATLAS & 7  & lepton+tau   & 1205.3067 & CMS    & 7      &  $p_T (t),m_{t\bar{t}},y_t,y_{t\bar{t}} $  &  1211.2220  \\
ATLAS & 7  & lepton w/o $b$ jets   & 1201.1889 & CMS    & 8      &  $p_T (t),m_{t\bar{t}},y_t,y_{t\bar{t}}$  & 1505.04480  \\
ATLAS & 7  & lepton w/ $b$ jets   & 1406.5375 & \dzero      & 1.96  & $m_{t\bar{t}},p_T(t),|y_t|$ &  1401.5785  \\
ATLAS & 7  & tau+jets   & 1211.7205 & & & \\
ATLAS & 7  & $t\bar{t},Z\gamma,WW$   & 1407.0573 & \multicolumn{4}{l}{Charge asymmetries:}  \\
ATLAS & 8  & dilepton  & 1202.4892 & ATLAS & 7  &  \Ac (inclusive+$m_{t\bar{t}},y_{t\bar{t}}$)  & 1311.6742  \\
CMS & 7  & all hadronic   & 1302.0508 & CMS & 7  &  \Ac (inclusive+$m_{t\bar{t}},y_{t\bar{t}}$)  & 1402.3803 \\
CMS & 7  & dilepton  & 1208.2761 & CDF & 1.96 & \Afb   (inclusive+$m_{t\bar{t}},y_{t\bar{t}}$) & 1211.1003  \\
CMS & 7  & lepton+jets   & 1212.6682 &  \dzero & 1.96 & \Afb  (inclusive+$m_{t\bar{t}},y_{t\bar{t}}$) & 1405.0421 \\
CMS & 7  & lepton+tau   & 1203.6810 &  & & \\
CMS & 7  & tau+jets   & 1301.5755 & Top widths: &  & \\
CMS & 8  & dilepton  &  1312.7582 & \dzero & 1.96 & $\Gamma_{\!\mathrm{top}}$& 1308.4050 \\
CDF + \dzero & 1.96  & Combined world average   & 1309.7570  & CDF & 1.96 & $\Gamma_{\!\mathrm{top}}$ & 1201.4156  \\
\addlinespace
\multicolumn{4}{l}{\textit{Single top production}} & \multicolumn{4}{l}{ $W$\!-boson helicity fractions:} \\
ATLAS & 7  & $t$-channel (differential)   & 1406.7844 & ATLAS & 7  & & 1205.2484  \\
CDF & 1.96  & $s$-channel (total)   & 1402.0484 & CDF & 1.96 & & 1211.4523  \\
CMS & 7  & $t$-channel (total)   & 1406.7844 & CMS  & 7  & & 1308.3879  \\
CMS & 8  & $t$-channel (total)   & 1406.7844 & \dzero & 1.96  & & 1011.6549   \\
\dzero & 1.96  & $s$-channel (total)   & 0907.4259 & & & \\
\dzero & 1.96  & $t$-channel (total)   & 1105.2788 & & & \\
\addlinespace
\multicolumn{4}{l}{\textit{Associated production}} & \multicolumn{4}{l}{\textit{Run~II data}}\\
ATLAS & 7  & $t\bar{t}\gamma$   & 1502.00586 & CMS & 13 & $t\bar{t}$ (dilepton) & 1510.05302 \\
ATLAS & 8  & $t\bar{t}Z$   & 1509.05276 &&&& \\
CMS & 8  & $t\bar{t}Z$   & 1406.7830 &&&& \\
\bottomrule
\end{tabular}
\hspace{0.2cm}
\end{adjustbox}
\end{center}
\caption[Measurements used in the {\sc{TopFiiter}} fit.]{\label{table:measurements}
 The measurements entering the fit. Details of each are described in the text.}
\end{table*}

\subsubsection{Treatment of uncertainties}
The uncertainties entering the fit can be classed into three categories:

\begin{mydescription}
\item[Experimental uncertainties:]
  We generally have no control over these. In cases where statistical and
  systematic (and luminosity) errors are recorded separately, we add them in
  quadrature. Correlations between measurements are also an issue: the unfolding
  of measured distributions to parton-level introduces some correlation
  between neighbouring bins. If estimates of these effects have been provided in
  the experimental analysis, we use this information in the fit, if they are
  not, we assume zero correlation. However, we have checked that bin
  correlations have little effect on our numerical results.

  There will also be correlations between apparently separate measurements. The
  multitude of different top pair production cross-section measurements will
  clearly be correlated due to overlapping event selection criteria and detector
  effects, etc. Without a full study of the correlations between different decay
  channels measured by the same experiment, these effects cannot be completely
  taken into account, but based on the negligible effects of the bin-by-bin correlations on
  our numerical results we can expect these effects to be small as well.

\item[Standard Model theoretical uncertainties:]
 These stem from the choice of parton distribution functions (PDFs), as well as
  neglected higher-order perturbative corrections. As discussed in chapter 1, we model
  the latter by varying the renormalisation and factorisation scales
  independently in the range $\mu_0/2\leq\mu_\mathrm{R,F}\leq 2\mu_0$, where
  we use $\mu_0=m_t$ as the default scale, and take the envelope as our
  uncertainty. For the PDF uncertainty, we follow the PDF4LHC
  recommendation~\cite{Butterworth:2015oua} of using
  CT10~\cite{Nadolsky:2008zw}, MSTW~\cite{Martin:2009iq} \&
  NNPDF~\cite{Ball:2010de} NLO fits, each with associated scale uncertainties,
  then taking the full width of the scale+PDF envelope as our uncertainty
  estimate -- i.e. we conservatively assume that scales and parton densities are
  100\% correlated. Unless otherwise stated, we take the top quark mass to be
  $m_t = 173.2 \pm 1.0~\gev$. We do not consider electroweak corrections.

  Only recently has a lot of progress been made in extending the dimension
  six-extended SM to higher order in $\alpha_s$, see
  Refs.~\cite{Passarino:2012cb,Mebane:2013zga,Jenkins:2013zja,Jenkins:2013sda,Jenkins:2013wua,Alonso:2013hga,Hartmann:2015oia,Ghezzi:2015vva,Zhang:2013xya,Englert:2014cva,Hartmann:2015aia,Cheung:2015aba,Drozd:2015rsp,Gauld:2015lmb}. Including
  these effects is beyond the scope of this work, also because we work to
  leading order accuracy in the electroweak expansion of the SM. QCD corrections to four fermion operators included via renormalisation group equations are typically of the order of 15\%, depending on the resolved phase space~\cite{Englert:2014cva}.  As pointed out in Ref.~\cite{Berthier:2015oma}, these effects can be important in
  electroweak precision data fits.

\item[Interpolation error:]
  A small error relating to the Monte Carlo interpolation (described in more
  detail in the next section) is included. This is estimated to be 5\% at a
  conservative estimate, as discussed in the following section, and thus subleading compared to the previous two categories.

\end{mydescription}

\subsubsection{Fitting procedure}

Our fitting procedure, briefly outlined in Ref.~\cite{Buckley:2015nca}, uses the
\textsc{Professor} framework. The first step is to construct an $N$-dimensional
hypercube in the space of dimension six couplings, compute the
observables at each point in the space, and then to fit an {\it interpolating
  function} $f(\mathbf{c})$ that parametrises the theory prediction as a
function of the Wilson coefficients $\mathbf{c}=\{\co{i}\}$. This can then be used
to rapidly generate theory observables for arbitrary values of the
coefficients. Motivated by the dependence of the total cross-section with a
Wilson coefficient (also shown in Eq.~\ref{eqn:d6xsec}):
\begin{equation}
\sigma \sim \sigma_\mathrm{SM} + \co{i}\sigma_{D6}+ \co[2]{i}\sigma_{D6^2}\, ,
\label{eqn:sigmad6}
\end{equation}
\noindent the fitting function is chosen to be a second-order or higher polynomial:
\begin{equation}
f_b(\{\co{i}\}) = \alpha_0^b + \sum_{\substack{i}} \beta_i^b\co{i}  + \sum_{\substack{i\leq j}}\gamma^b_{i,j}\co{i}\co{j}+\ldots\,.
\label{eqn:ipol}
\end{equation}

\begin{figure}[!t]
\begin{center}
\includegraphics[width=\textwidth]{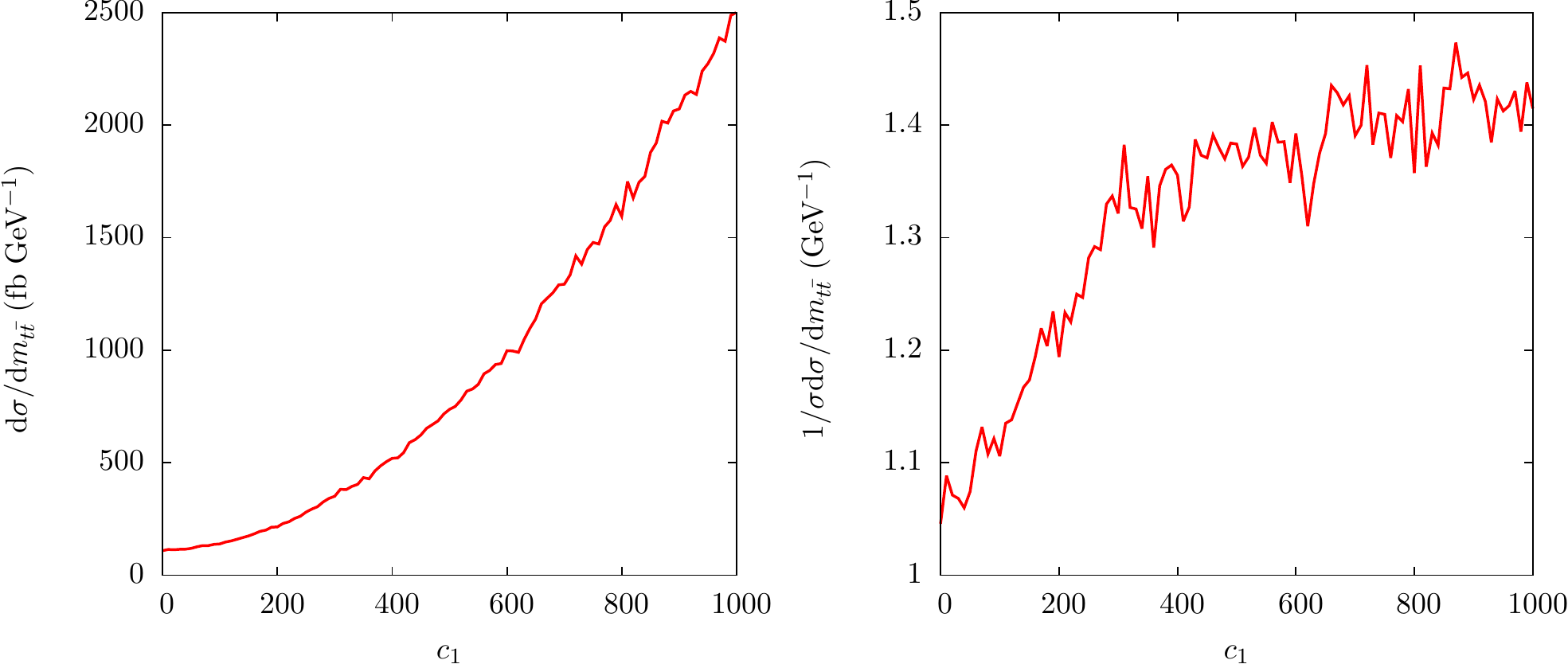}
\caption[Dependence on a Wilson coefficient for normalised and unnormalised distributions.]{Dependence of a bin in the unnormalised (left) and normalised (right) $t\bar{t}$ invariant mass distribution on one of the dimension-six coefficients considered in this fit.}
\label{fig:rawvsnorm}
\end{center}
\end{figure}

In the absence of systematic uncertainties, each unnormalised observable would exactly follow
a second-order polynomial in the coefficients, and higher-order terms capture
bin uncertainties which modify this. The polynomial also serves as a useful
check that the dimension-six approximation is valid. By comparing
Eq.~\eqref{eqn:sigmad6} with Eq.~\eqref{eqn:ipol}, we see that the terms
quadratic in \co{i} are small provided that the coefficients in the interpolating
function $\gamma_{i,j}$ are small. This is a more robust way to ensure validity
of the dimension-six approximation than to assume a linear fit from the
start.

The simple quadratic dependence on the coefficients is not guaranteed to propagate into every observable, however.
Many of the differential distributions recorded by the LHC experiments (and used in this fit) are normalised to unity. This has the advantage of dividing out many of the systematic uncertainties entering the extraction of the cross-section, but has the practical  disadvantage from a fitting perspective that the observables follow a much more complicated polynomial dependence on the coefficients. Consider, for example, a bin in the normalised differential cross-section in some observable $X$, which will schematically take the form

\begin{equation}
\frac{1}{\sigma(\co{i})}\frac{d\sigma(\co{i})}{dX} \sim \frac{1}{f+g\co{i}+h\co[2]{i}} \times (f'+g'\co{i}+h'\co[2]{i}),
\end{equation}

where $\{f,g,h,f',g',h'\}$ are dimensionful functions of kinematic and Standard Model parameters. It is clear to see that this function will not be quadratic in \co{i}. This is exemplified in Fig.~\ref{fig:rawvsnorm}, where we show the dependence of a given bin in the top pair invariant mass distribution, both unnormalised and normalised. A clear quadratic dependence is seen in the former, whereas the latter is much more irregular and should be modelled by a higher-order function.

\begin{figure}[!t]
\begin{center}
\includegraphics[width=0.45\textwidth]{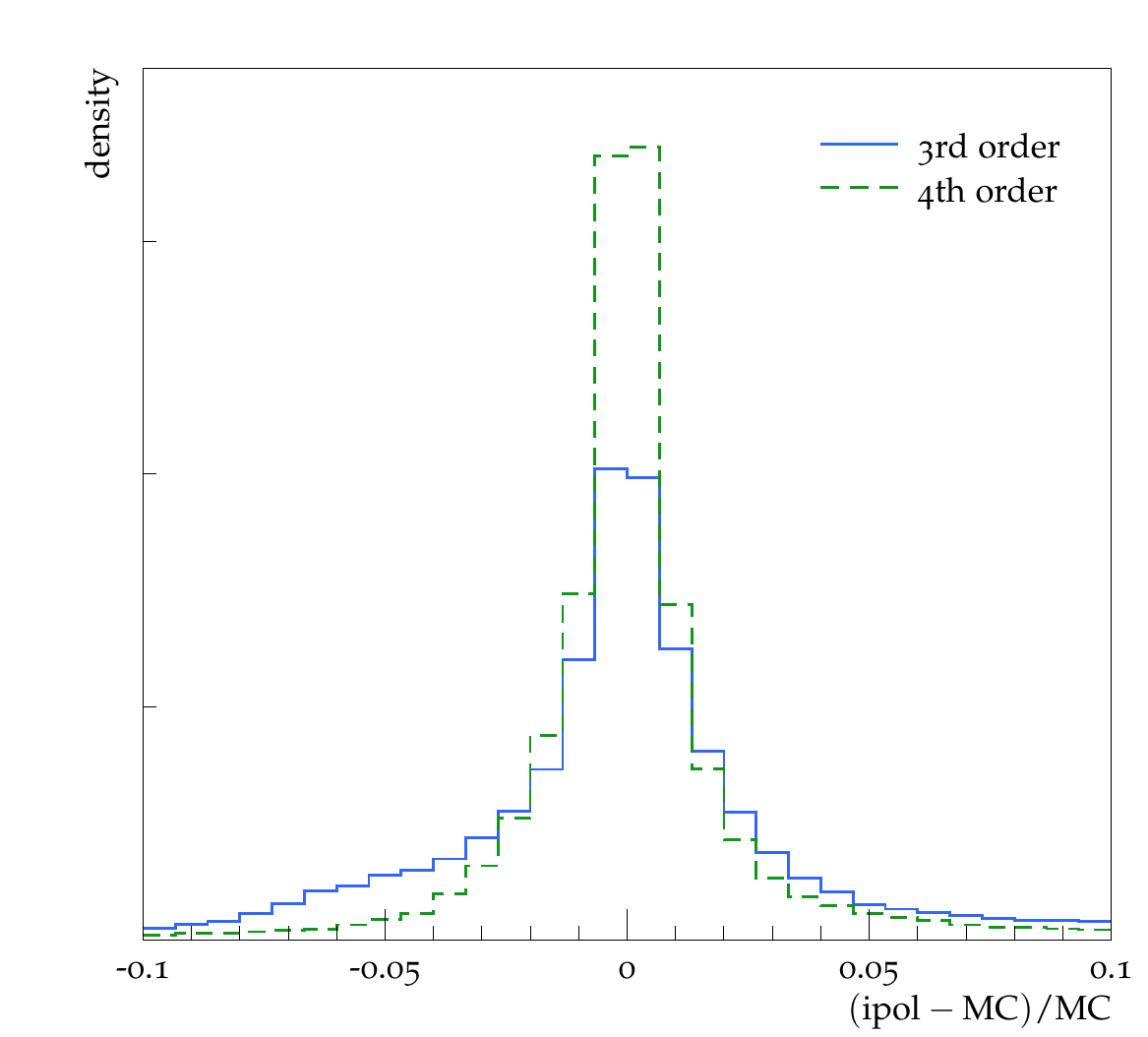}
\quad
\includegraphics[width=0.45\textwidth]{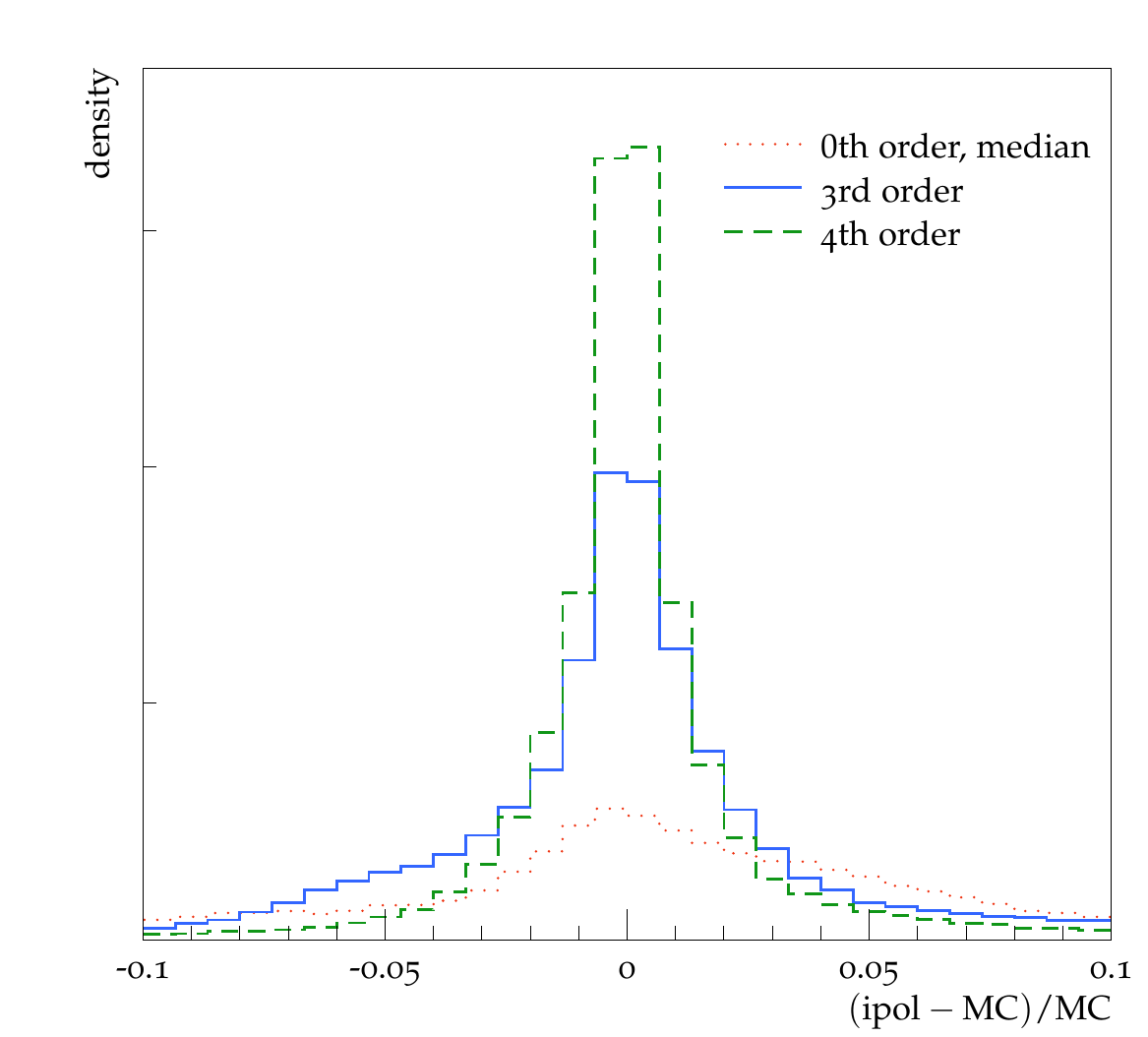}
\caption[Residuals distributions from {\sc{Professor}}.]{Residuals distributions for interpolated observable values~(left) and
  uncertainties~(right), evaluated over all input MC runs and all observables. The 4th
  order polynomial parameterisation gives the best performance and the vast
  majority of entries are within 5\% of the explicit MC value. The poor
  performance of a constant uncertainty assumption based on the median input
  uncertainty is evident -- since all three lines have the same normalisation, the
  majority of residual mismodellings for the median approach are (far) outside
  the displayed 10\% interval.}
\label{fig:residuals}
\end{center}
\end{figure}

In principle there is no limit on how high a polynomial order we may use. The limiting factor is the increased computation time at each successive order, and the inefficiency of overfitting to statistical noise, which very high order polynomials almost certainly do. In practice, to minimise the interpolation uncertainty, we use up to a 4th order
polynomial in Eq.~\eqref{eqn:ipol}, depending on the observable of interest. The
performance of the interpolation method is shown in Fig.~\ref{fig:residuals},
which depicts the fractional deviation of the polynomial fit from the explicit
MC points used to constrain it. The central values and the sizes of the
modelling uncertainties may both be parameterised with extremely similar
performance, with 4th order performing best for both. The width of this residual
mismodeling distribution being $\sim \text{3\%}$ for each of the value and
error components is the motivation for a total 5\% interpolation uncertainty to
be included in the goodness of fit of the interpolated MC polynomial $f(\mathbf{c})$ to the experimentally measured value $E$:
\begin{equation}
  \chi^2(\mathbf{c}) = \sum_{\substack{\mathcal{O}} } \sum_{\substack{i,j} }\frac{(f_i(\mathbf{c}) - E_i)\rho_{i,j} (f_j(\mathbf{c}) - E_j)}{\Delta_i\Delta_j} \,,
\label{eqn:chi2}  
\end{equation}
where we sum over all observables $\mathcal{O}$ and all bins in that observable
$i$. We include the correlation matrix $\rho_{i,j}$ where this is provided by
the experiments, otherwise $\rho_{i,j}=\delta_{ij}$. The uncertainty on each bin
is given by $\Delta_i = \sqrt{\Delta_{\mathrm{th},i}^2+\Delta_{\mathrm{exp},i}^2}$, i.e. we treat
theory and experimental errors as uncorrelated. The
parameterisation of the theory uncertainties is restricted to not become larger
than in the training set, to ensure that polynomial blow-up of the uncertainty
at the edges of the sampling range cannot produce a spuriously low $\chi^2$ and
disrupt the fit.

\begin{figure}[!t]
\begin{center}
\includegraphics[width=0.7\textwidth]{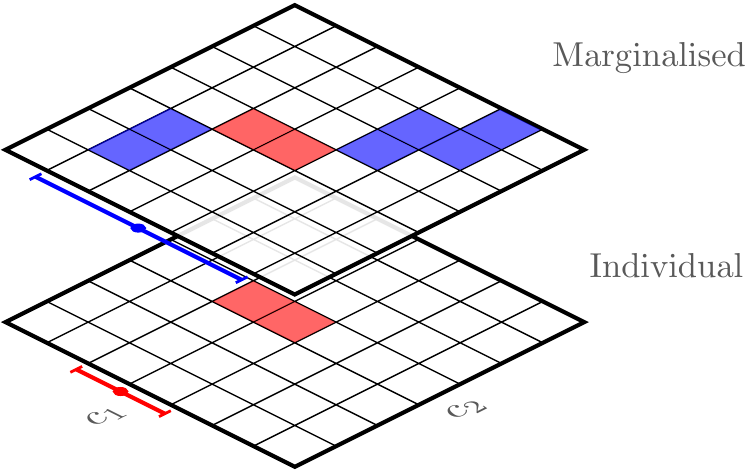}
\caption[Visualisation of individual and marginalised constraints.]{Illustration of the difference between individual and marginalised constraints for a two-parameter fit. The allowed region for $c_1$ with all other coefficients set to zero (red squares) can be made larger by varying $c_1$ and $c_2$ simultaneously and tuning them so that the theory prediction is close to the data (blue squares).}
\label{fig:marginalise}
\end{center}
\end{figure}

We hence have constructed a fast parameterisation of model goodness-of-fit as a
function of the EFT operator coefficients. This may be used to produce $\chi^2$
maps in slices (where all but one parameter is fixed, typically to zero) or marginalised (where all parameters are varied simultaneously) projections of the operator space. Since in the marginalised case, a pull of the theory prediction in one direction by one parameter can be compensated by tuning another parameter to pull it back into agreement with the data, the net result is that marginalised confidence intervals are wider than individual `slices'. A visualisation of this is shown in Fig.~\ref{fig:marginalise} for a two parameter fit.  These projections are transformed to confidence intervals on the coefficients \co{i}, defined by the
regions for which
\begin{equation}
 1 - \mathrm{CL} \geq \int^{\infty}_{\chi^2(\co{i})} f_k(x) dx \,,
\end{equation}
where typically $\mathrm{CL} \in \{0.68,0.95,0.99\}$ and $f_k(x)$ is the
$\chi^2$ distribution for $k$ degrees of freedom, which we define as
$k = N_\mathrm{measurements} - N_\mathrm{coefficients}$.
 
\begin{figure}[!t]
\begin{center}
\includegraphics[width=\textwidth]{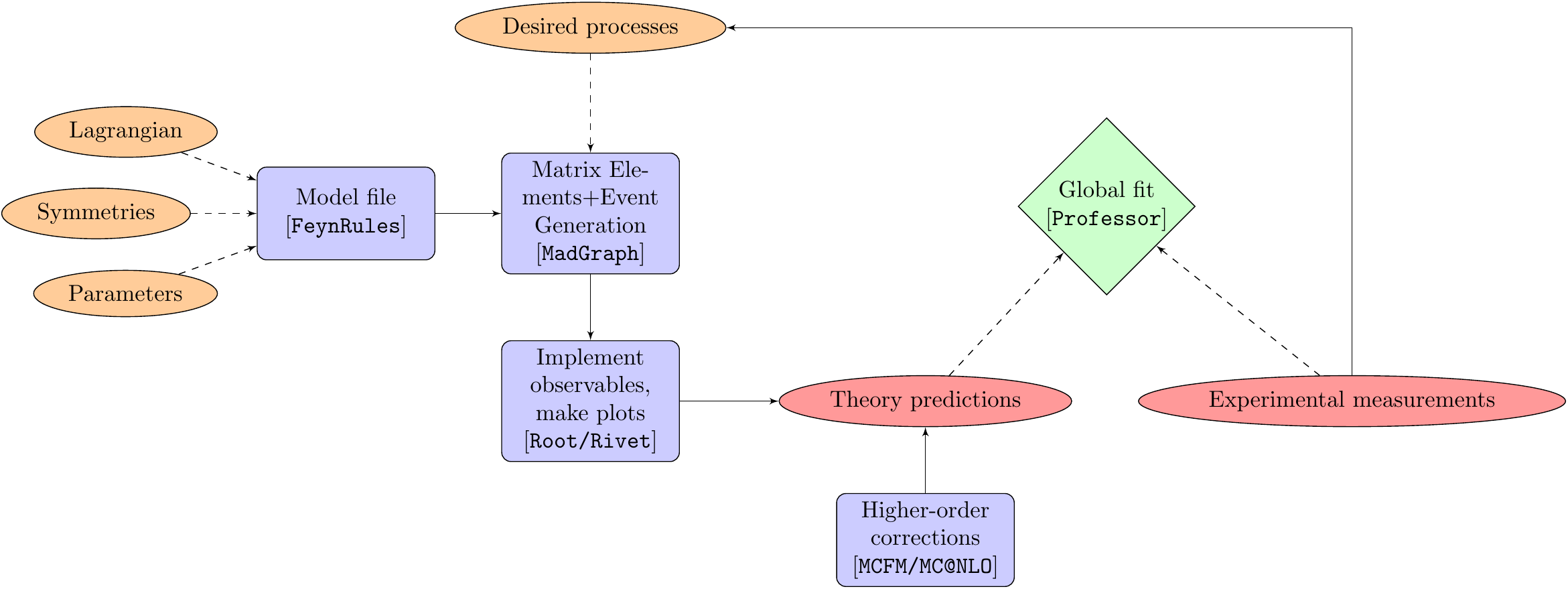}
\caption[{\sc{TopFitter}} flowchart.]{Flowchart illustrating the fitting procedure and relevant software used at each step of the analysis.}
\label{fig:flowchart}
\end{center}
\end{figure}

 A flowchart of the fitting procedure is shown in Fig.~\ref{fig:flowchart}.

\subsection{Results}
\label{sec:fitresults}

The entire 59 dimensional operator set of
Ref.~\cite{Grzadkowski:2010es} was implemented in a
FeynRules~\cite{Christensen:2008py} model file, with care taken to ensure consistent redefinitions of the relation between SM input parameters and observables (see appendix A for details). The contributions to
parton level cross-sections and decay observables from the above
operators were computed using
\textsc{MadGraph/Madevent}~\cite{Alwall:2014hca}, making use of the
Universal FeynRules Output (UFO)~\cite{Degrande:2011ua} format. We
model NLO~QCD corrections by including Standard Model $K$-factors
(bin-by-bin for differential observables), where the NLO observables
are calculated using MCFM~\cite{Campbell:2010ff}, cross-checked with
MC@NLO~\cite{Frixione:2002ik,Frixione:2010wd}. These $K$-factors are used for arbitrary
values of the Wilson coefficients, thus modelling NLO effects in the
pure-SM contribution only.  More specifically, this amounts to
performing a simultaneous expansion of each observable in the strong
coupling $\alpha_s$ and the (inverse) new physics scale
$\Lambda^{-1}$, and neglecting terms $\sim{\cal
  O}(\alpha_S\Lambda^{-2})$. For $t$-channel single top production, all our results are presented in a five-flavour scheme for the incoming pdfs, however we have cross-checked our results against a four-flavour scheme and found good agreement. Our final 95\% confidence limits for
each coefficient are presented in Fig.~~\ref{fig:constraints}; we
discuss them in more detail below.

\subsubsection{Top pair production}
\label{sec:toppair}
By far the most abundant source of data in top physics is from the production of
top pairs. The \cp-even dimension-six operators that interfere with the
Standard Model amplitude are 

\begin{figure}[!t]
\begin{center}
\includegraphics[width=0.95\textwidth]{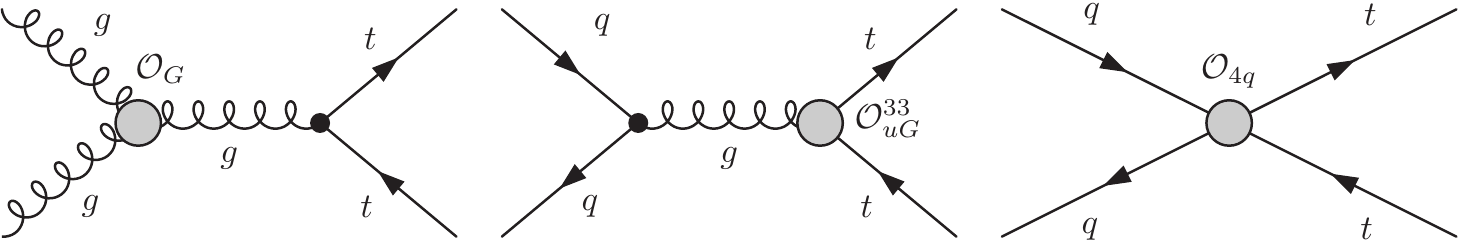}
\caption[Feynman diagrams for \ttbar production with \D6 operators.]{Sample Feynman diagrams for the interference of the leading-order SM amplitudes for top pair production with the operators of Eq.~\eqref{eqn:ttbarops}. \op{4q} denotes the insertion of any of the four-quark operators. }
\label{fig:ttbarfeyn}
\end{center}
\end{figure}

%
\begin{equation}
\begin{split}
\lag{D6} & \supset  \frac{\co{uG}}{\Lambda^2} (\bar{Q}\sigma^{\mu \nu} T^A u)\tilde \varphi G_{\mu\nu}^{A} +  \frac{\co{G}}{\Lambda^2}  f_{ABC} G_{\mu}^{A \nu}G_{\nu}^{B \lambda} G_{\lambda}^{C \mu} + \frac{\co{\varphi G}}{\Lambda^2} (\varphi^\dagger \varphi)G_{\mu\nu}^{A}G^{A \mu\nu} \\
& + \frac{\co[1]{qq}}{\Lambda^2}(\bar{Q}\gamma_{\mu}Q)( \bar{Q}\gamma^{\mu}Q) +  \frac{\co[3]{qq}}{\Lambda^2}(\bar{Q}\gamma_{\mu}\tau^IQ)( \bar{Q}\gamma^{\mu}\tau^I Q) +  \frac{\co{uu}}{\Lambda^2}(\bar{u}\gamma_{\mu}u)( \bar{u}\gamma^{\mu} u) \\
& + \frac{\co[8]{qu}}{\Lambda^2}(\bar{Q}\gamma_{\mu}T^AQ)( \bar{u}\gamma^{\mu} T^Au) +  \frac{\co[8]{qd}}{\Lambda^2}(\bar{Q}\gamma_{\mu}T^AQ)( \bar{d}\gamma^{\mu} T^Ad) +  \frac{\co[8]{ud}}{\Lambda^2}(\bar{u}\gamma_{\mu}T^Au)( \bar{d}\gamma^{\mu} T^Ad) \,.
\end{split}
\label{eqn:ttbarops}
\end{equation}

As pointed out in Ref.~\cite{Buckley:2015nca}, the operator \op{\varphi G} cannot be
bounded by top pair production alone, since the branching ratio to virtual top
pairs for a 125~\gev Higgs is practically zero, therefore we do not consider it
here. For a recent constraint from Higgs physics see
e.g. Ref.~\cite{Corbett:2015ksa,Ellis:2014jta,Falkowski:2015jaa,Englert:2015hrx}. We further ignore the contribution of the operator \op[11]{uG}\,, as this operator is a direct mixing of the left- and right- chiral $u$ quark fields, and so contributes terms proportional to $m_u$. We
also note that the six four-quark operators of Eq.~
\eqref{eqn:ttbarops} interfere with the Standard Model QCD processes $\bar{u}u,\,\bar{d}d\,\rightarrow\,\bar{t}t$ to produce terms dependent only on the four linear
combinations of Wilson Coefficients: \co[1,2]{u,d}, displayed in Eq.~\eqref{eqn:4fs}.

It is these four that are constrainable in a dimension-six analysis. Finally, we
note that the operator \op{G}, whilst not directly coupling to the top at tree-level, should not be neglected. Since it modifies the triple gluon vertex, and the $gg$ channel contributes $\sim 75\%$ $(90\%)$ of the total top pair production cross-section at the 8 (13) \tev LHC, moderate values of its Wilson
coefficient can substantially impact total rates, as we already saw in chapter 2. We note, however, that in this special case, the cross section modifications are driven by the squared dimension six terms instead of the linearised interference with the SM. Nonetheless, in the interests of generality, we choose to include this operator in the fit at this stage, noting that bounds on its Wilson coefficient should be interpreted with caution.\footnote{We have observed that excluding this operator actually tightens the bounds on the remaining ones, so choosing to keep it is the more conservative option.} Representative Feynman diagrams for the interference of these operators are
shown in Fig.~\ref{fig:ttbarfeyn}.

\begin{figure}[!t]
\begin{center}
\includegraphics[width=\textwidth]{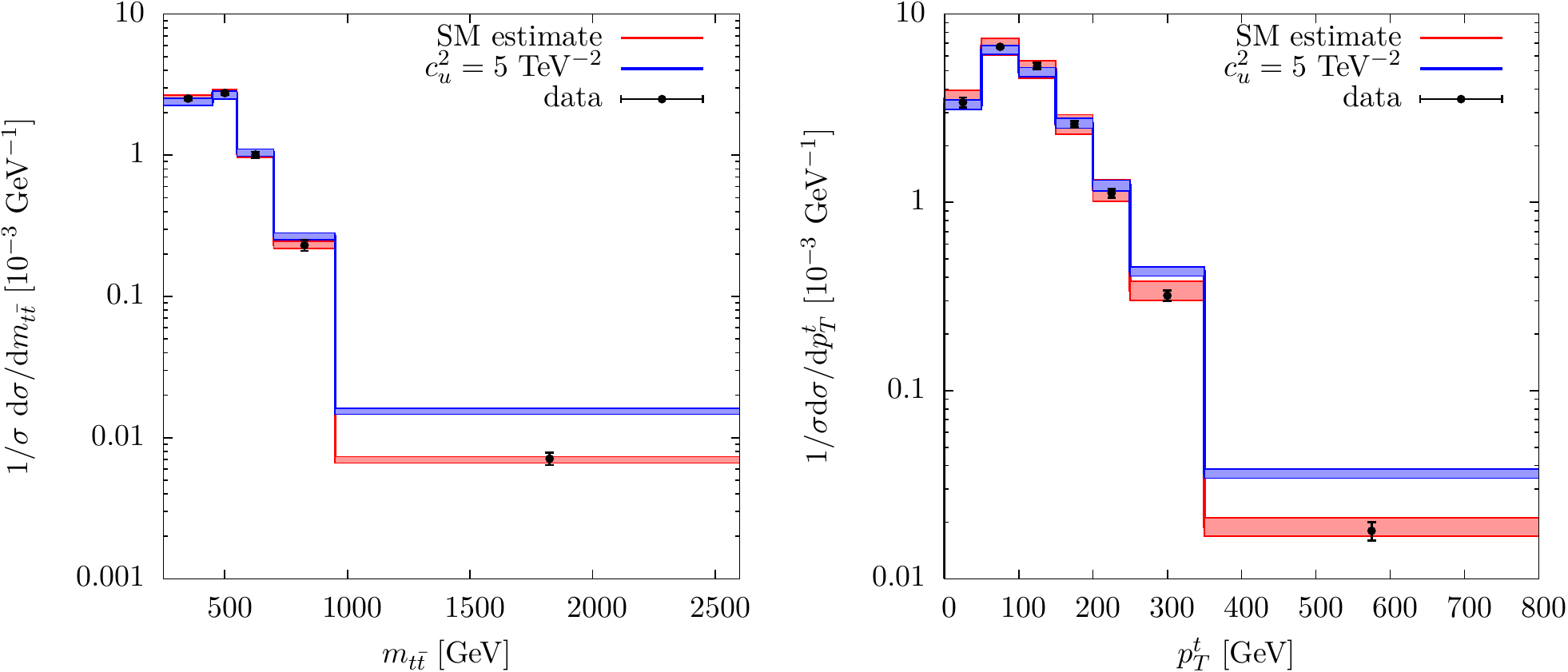}
\caption[\ttbar differential distributions and ATLAS data.]{Parton level differential distributions in top pair production, considering SM only (red) and the effects of the four-quark operator \op[2]{u}, showing the enhancement in the tails of the distributions. Data taken from Ref.~\cite{Aad:2014zka}.}
\label{fig:distributions}
\end{center}
\end{figure}

The most obvious place to look for the effects of higher-dimensional terms is
through the enhancement (or reduction, in the case of destructive interference)
of total cross-sections. Important differences between SM and dimension-six
terms are lost in this approach, however, since operators can cause
deviations in the shape of distributions without substantially impacting event
yields. This is highlighted in Fig.~\ref{fig:distributions}, where we plot our
NLO SM estimate for two top pair kinematic distributions, vs.  one with a large new physics
interference term. Both are consistent with the data in the threshold region,
which dominates the cross-section, but clear discrimination between SM and
dimension-six effects is visible in the high-mass region, which simply originates from the scaling of dimension-six operator effects as $s/\Lambda^2$.

\begin{figure}[!t]
\begin{center}
\includegraphics[width=0.325\textwidth]{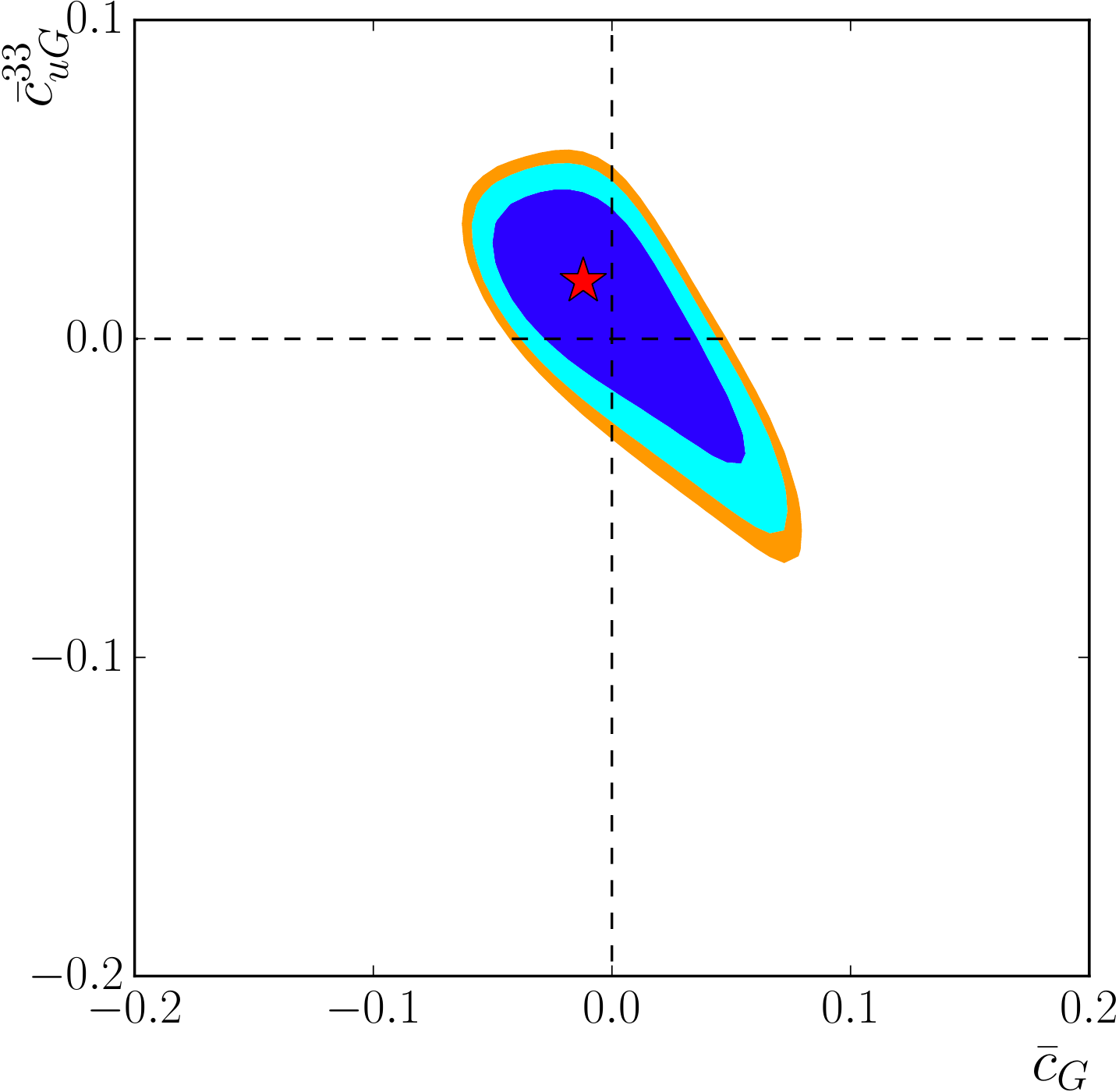}
\includegraphics[width=0.325\textwidth]{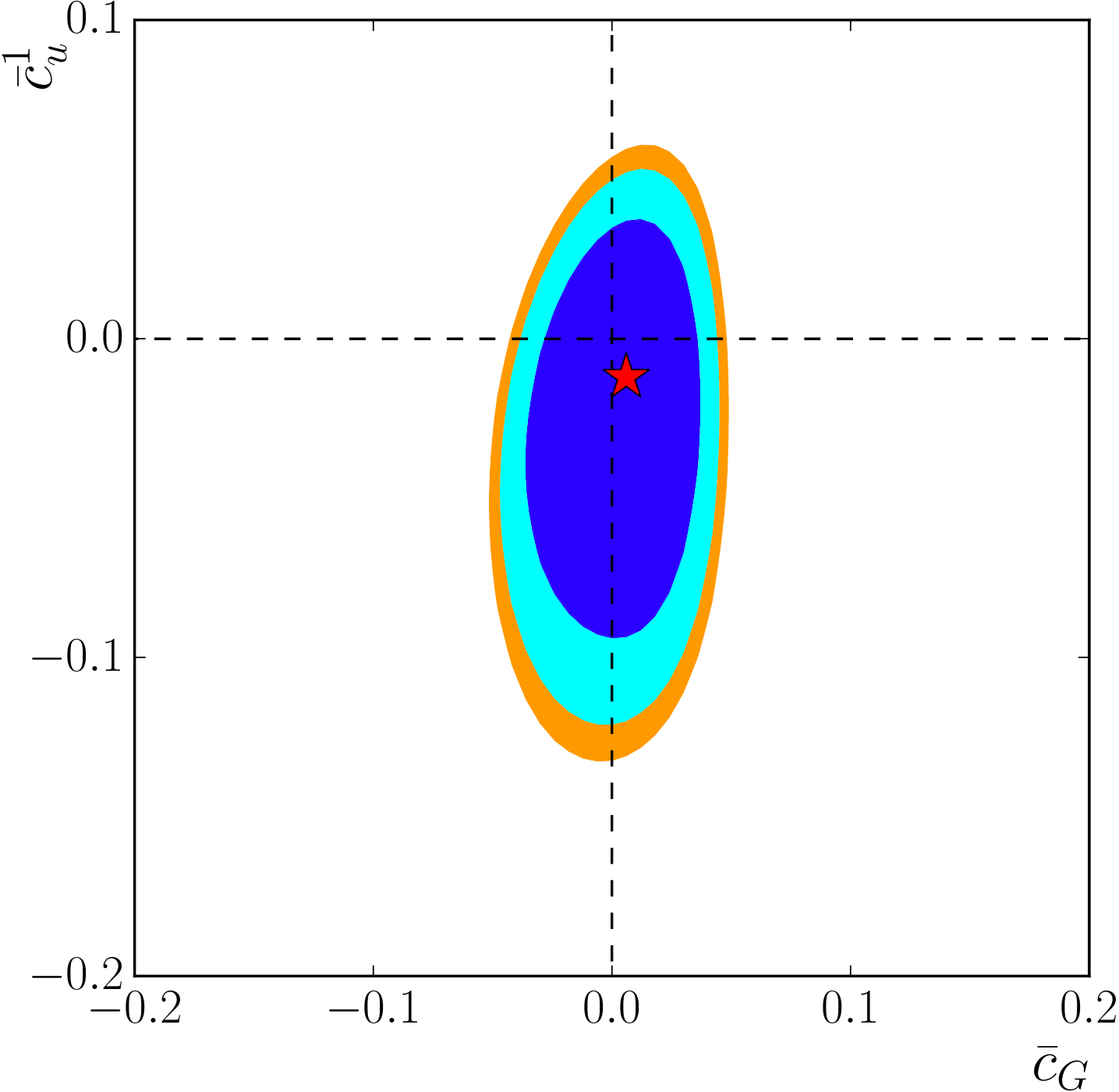}
\includegraphics[width=0.325\textwidth]{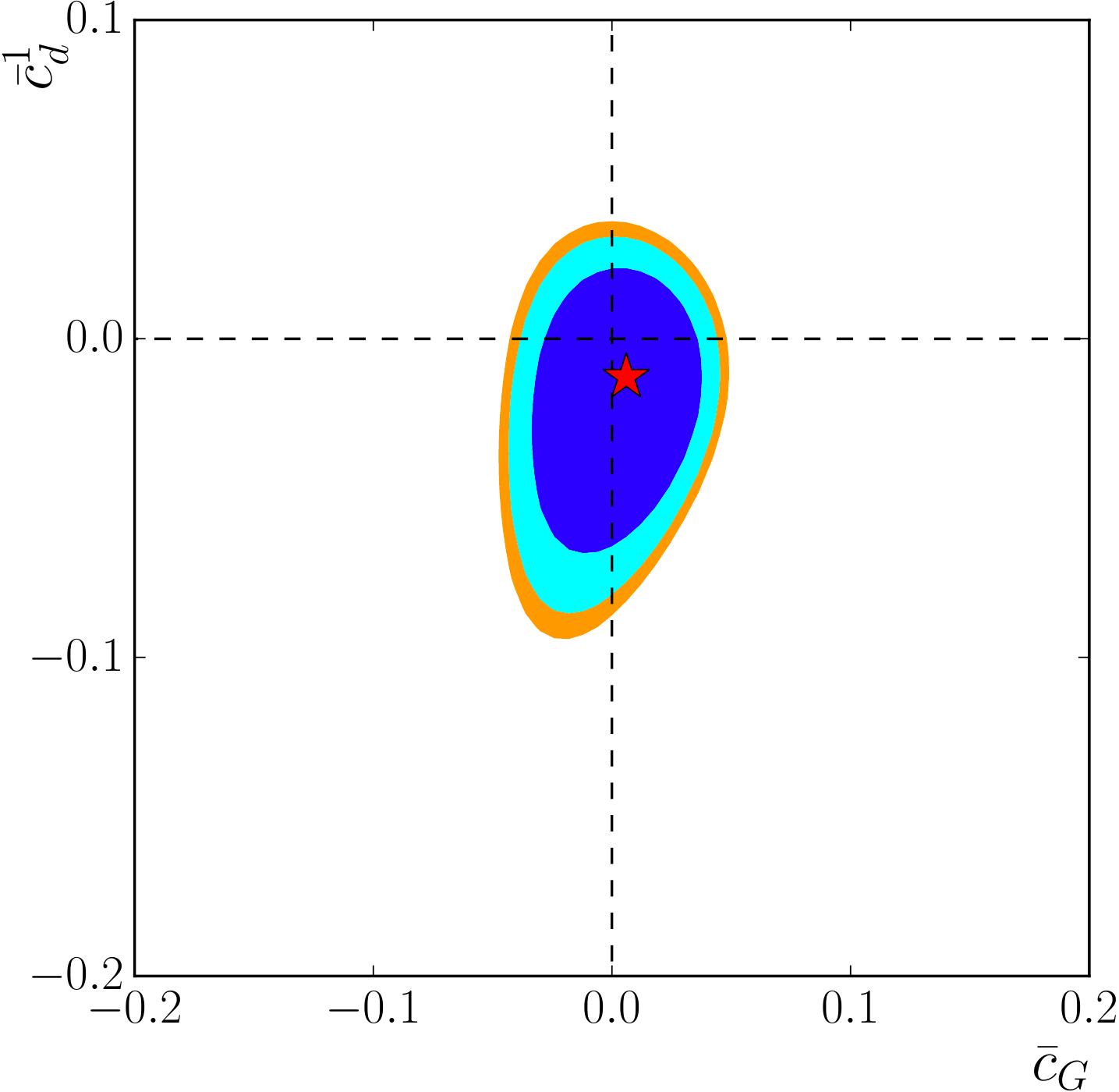}
\includegraphics[width=0.325\textwidth]{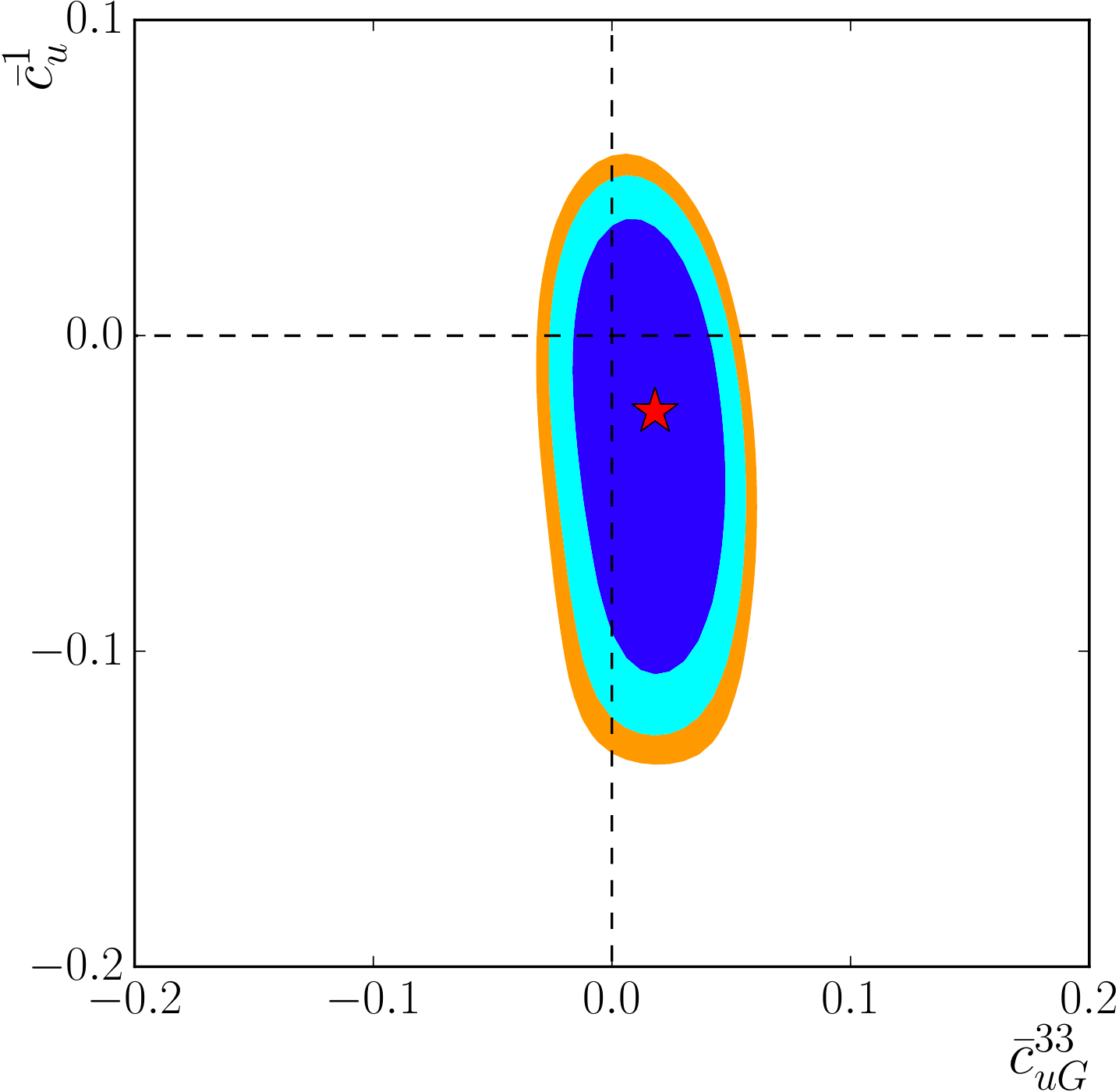}
\includegraphics[width=0.325\textwidth]{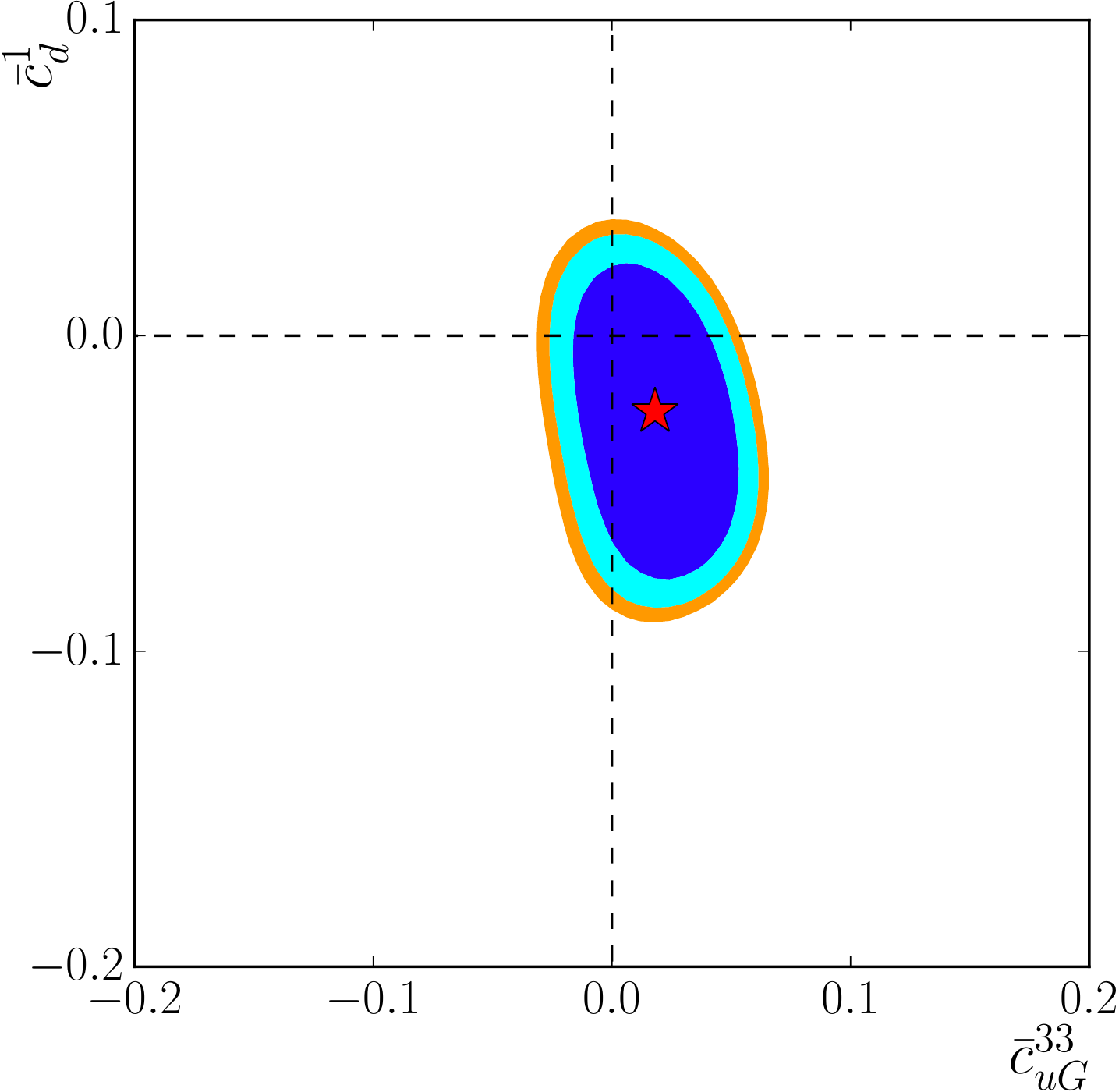}
\includegraphics[width=0.325\textwidth]{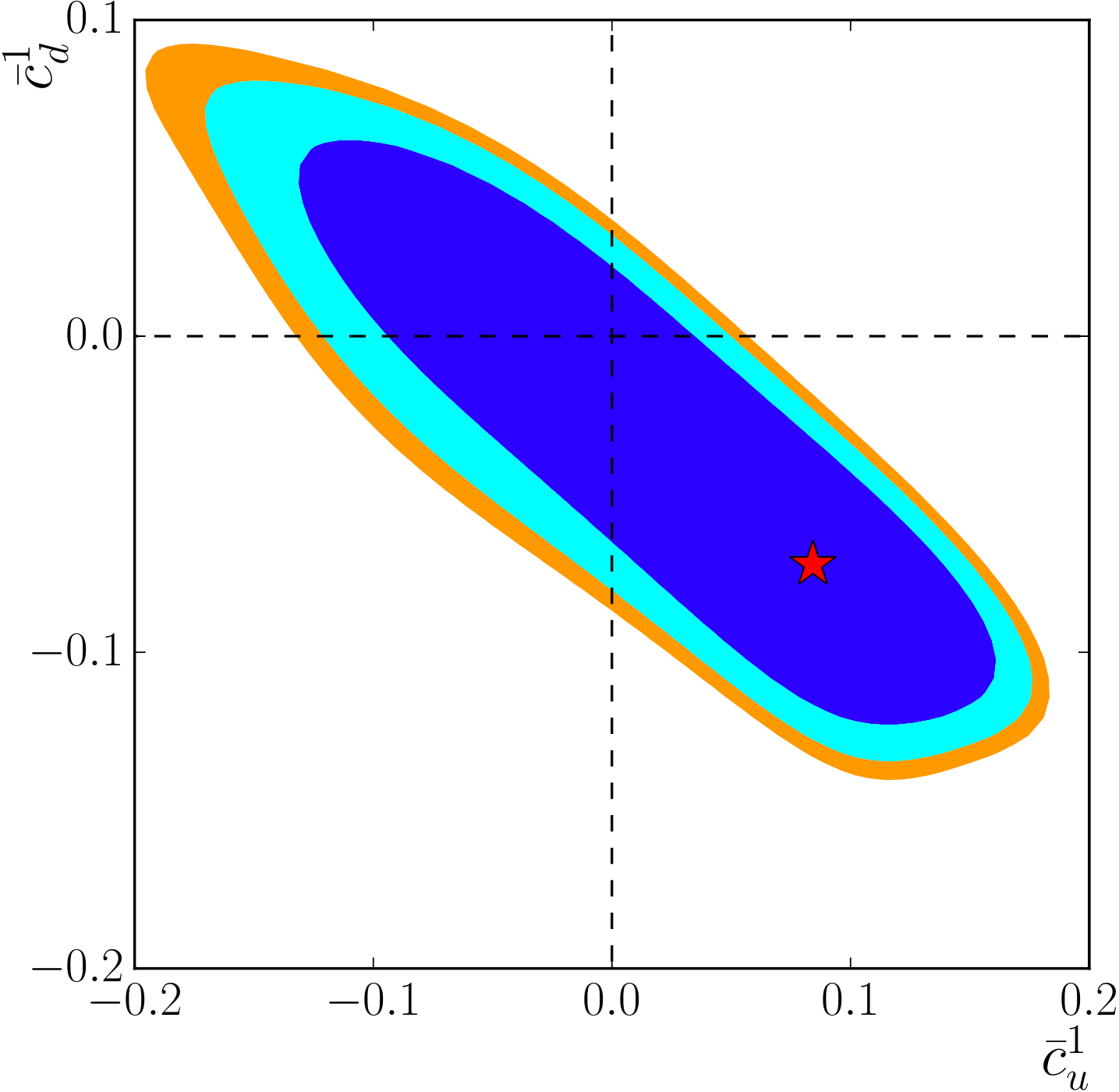}
\caption[2D exclusion contours from \ttbar measurements.]{68\%, 95\% and 99\% confidence intervals for selected combinations of operators contributing to top pair production, with all remaining operators set to zero. The star marks the best fit point, indicating good agreement with the Standard Model. Here $\cb{i} = \co{i}v^2/\Lambda^2$.}
\label{fig:toppairops}
\end{center}
\end{figure}

\begin{figure}[!t]
\begin{center}
\includegraphics[width=\textwidth]{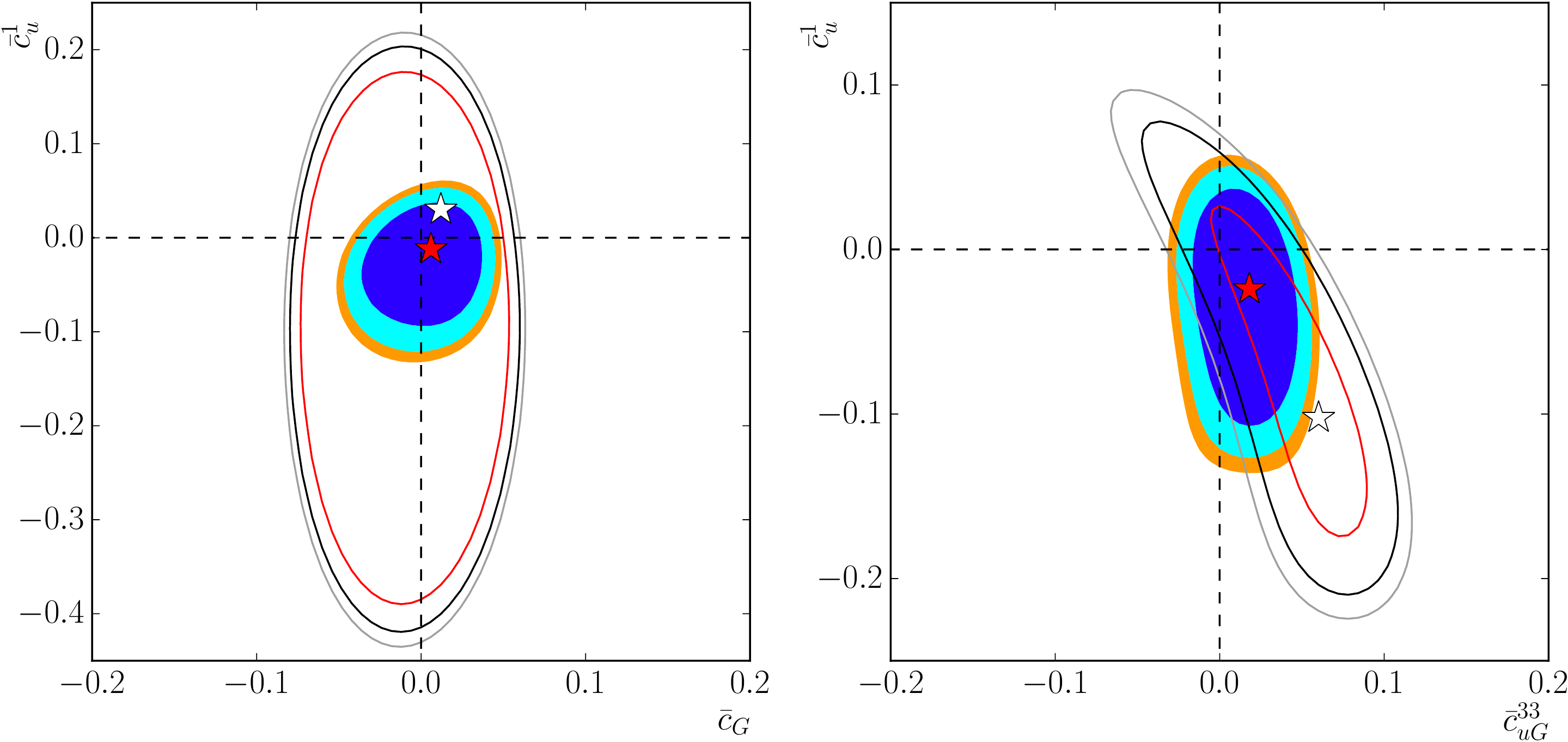}
\caption[2D exclusion contours from \ttbar measurements for total rates and distributions.]{Left: 68\%, 95\% and 99\% confidence intervals on the operators \co{G} vs. \co[1]{u} , considering differential and total cross-sections (contours, red star), and total cross-sections only (lines, white star). Right: Limits on \co[33]{uG} vs. \co[1]{u}, considering both Tevatron and LHC data (contours) and Tevatron data only (lines).}
\label{fig:c3c4diff}
\end{center}
\end{figure}

Limits on these operators can be obtained in two ways; by setting all other
operators to zero, and by marginalising over the other parameters in a global
fit. In Fig.~\ref{fig:toppairops} we plot the
allowed 68\%, 95\% and 99\% confidence intervals for various pairs of operators, with all others set to zero,
showing correlations between some coefficients. Most of these operators appear
uncorrelated, though there is a strong correlation between \co[1]{u} and \co[1]{d},
due to a relative sign between their interference terms. Given the
lack of reported deviations in top quark measurements, it is perhaps unsurprising to
see that all Wilson coefficients are consistent with zero within the 95\% confidence intervals, and that
the SM hypothesis is an excellent description of the data. In Fig.~\ref{fig:c3c4diff}, the stronger joint constraints on \co{G} vs \co[1]{u} obtained from including differential measurements make manifest the importance of utilizing all available cross-section information.

It is also interesting to note the complementarity between measurements from the LHC
and Tevatron, as illustrated in Fig.~\ref{fig:c3c4diff}. It is interesting to see
that although Tevatron data are naively more sensitive to
four-quark operators, after the LHC Run~I and early into Run~II, the LHC data size and probed energy transfers lead to comparably stronger constraints. In the fit this is highlighted by the simple fact that LHC data comprise
more than 80\% of the bins, so have a much larger pull. This
stresses the importance of collecting large statistics as well as using
sensitive discriminating observables.

\subsubsection{Single top production}
The next most abundant source of top quark data is from single top
production. In the fit we consider production in the $t$ and $s$
channels, and omit $Wt$-associated production. Though measurements of
the latter process have been published, they are not suitable for
inclusion in a fit involving parton level theory predictions. As discussed in chapter 1, $Wt$ production interferes with top pair production at NLO
and beyond in a five-flavour
scheme~\cite{Zhu:2001hw,Campbell:2005bb,Cao:2008af}, or at LO in a
four-flavour one. Its separation from top pair production is then a
delicate issue, discussed in detail in
Refs.~\cite{Frixione:2008yi,White:2009yt,Kauer:2001sp,Kersevan:2006fq}. We
thus choose to postpone the inclusion of $Wt$ production to a future
study, going beyond parton level. The operators that could lead to
deviations from SM predictions are shown in Eq.~\eqref{eqn:singletopops}.
\begin{equation}
\begin{split}
\lag{D6}	& \supset  \frac{\co{uW}}{\Lambda^2} (\bar{Q}\sigma^{\mu \nu} \tau^I u)\,\tilde \varphi\, W_{\mu\nu}^{I} + \frac{\co[(3)]{\varphi q}}{\Lambda^2} i(\varphi^\dagger \overleftrightarrow{D}^I_\mu \varphi )(\bar{Q}\gamma^\mu \tau^I Q) \\
			& + \frac{\co{\varphi ud}}{\Lambda^2} (\varphi^\dagger  \overleftrightarrow{D}_\mu \varphi )(\bar{u}\gamma^\mu d) +  \frac{\co{dW}}{\Lambda^2} (\bar{Q}\sigma^{\mu \nu} \tau^I d)\,\tilde \varphi \,W_{\mu\nu}^{I}  \\
			& + \frac{\co[3]{qq}}{\Lambda^2}(\bar{Q}\gamma_{\mu}\tau^IQ)( \bar{Q}\gamma^{\mu}\tau^I Q) + \frac{\co[1]{qq}}{\Lambda^2}(\bar{Q}\gamma_{\mu}Q)( \bar{Q}\gamma^{\mu}Q)  + \frac{\co[1]{qu}}{\Lambda^2}(\bar{Q}\gamma_{\mu}Q)( \bar{u}\gamma^{\mu} u)\,.
\end{split}
\label{eqn:singletopops}
\end{equation}

As in top~pair production there are several simplifications which reduce
this operator set. The right-chiral down quark fields appearing in \op{uW} and \op{\varphi ud} cause these operators' interference with the left-chiral SM weak interaction to be proportional to the relevant down-type quark mass. For example, an operator insertion of \op[33]{\varphi ud} will always contract with the SM $Wtb$\,-vertex to form a term of order $m_b\, m_t\, \co[33]{\varphi ud}/\Lambda^2$. Since $m_b$ is much less than both $\hat{s}$ and the other dimensionful parameters that appear, $v$ and $m_t$, we may choose to neglect these operators. By the same rationale we neglect \op[(1)]{qu} as its contribution to observables is proportional to $m_u$. We have further checked numerically that the contribution of these operators is practically negligible.  Finally, all contributing four-fermion partonic subprocesses depend only on the linear combination of Wilson Coefficients:

\begin{equation}
\begin{split}
 \co{t} = &~\co[3,1133]{qq} + \tfrac{1}{6}(\co[1,1331]{qq}- \co[3,1331]{qq}).
\end{split}
\label{eqn:4fs2}
\end{equation}

\begin{figure}[!t]
\begin{center}
\includegraphics[width=\textwidth]{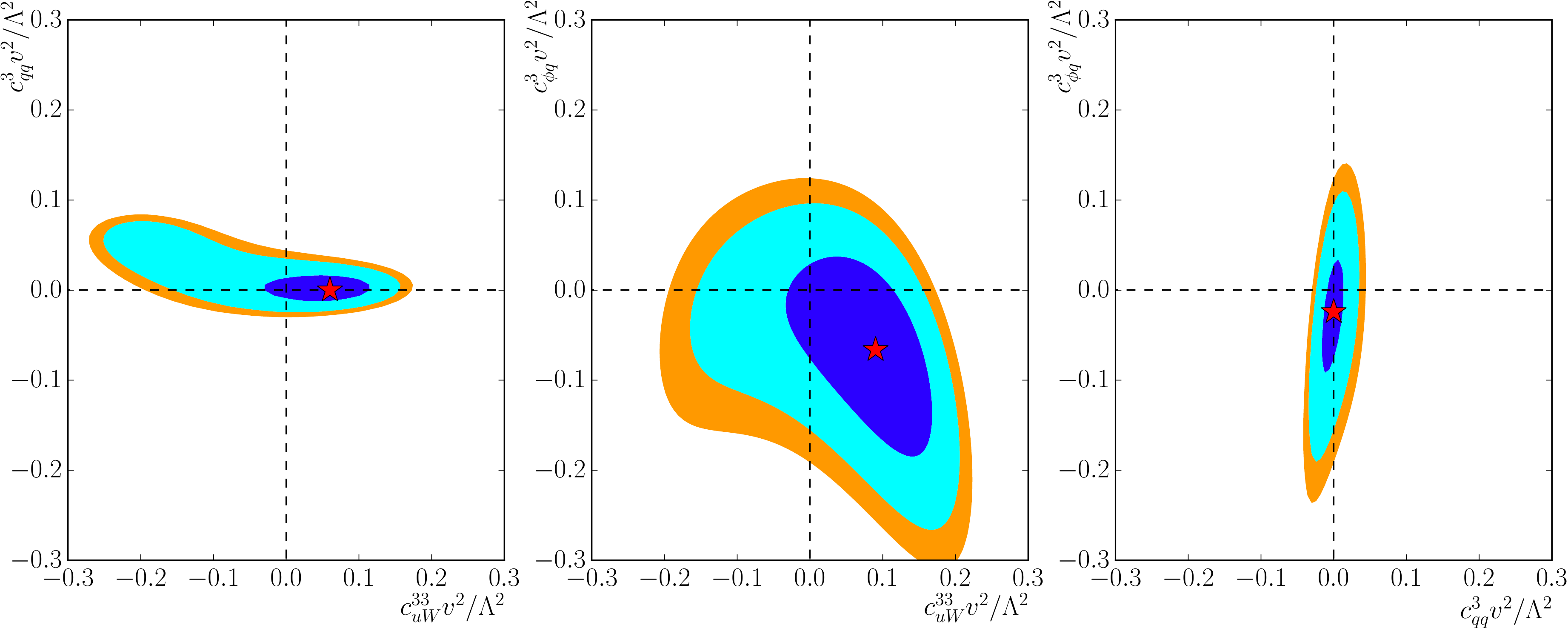}
\caption[2D exclusion contours from single top measurements.]{Marginalised 68\%, 95\% and 95\% confidence intervals on dimension-six operators in single top production.}
\label{fig:singletopcorrs}
\end{center}
\end{figure}

Single top production can thus be characterised by the three dimension-six operators \op{uW}, \op[(3)]{\varphi q} and \op{t}. The correlations among the constraints of these operators are displayed in  Fig.~\ref{fig:singletopcorrs}.

As noted in the introduction to this chapter, several model-independent studies have noted the
potential for uncovering new physics in single top production, though these have
typically been expressed in terms of anomalous couplings, via the Lagrangian
\begin{equation}
\mathcal{L}_{Wtb} = \frac{g}{\sqrt{2}} \bar{b} \gamma^\mu (V_L P_L + V_R P_R) t W^{-}_{\mu} +  \frac{g}{\sqrt{2}}\bar{b}\frac{i\sigma^{\mu\nu}q_\nu}{M_W}(g_LP_L + g_RP_R)tW^-_\mu + h.c. \,
\end{equation}
where $q = p_t - p_b$. There is a one-to-one mapping between this Lagrangian and those dimension-six
operators that modify the $Wtb$ vertex:
%
\begin{align}
  V_L & \to V_{tb} + \co[(3)]{\varphi q}v^2/\Lambda^2 & V_R & \to \frac{1}{2}\co{\varphi ud}v^2/\Lambda^2 \nonumber \\
  g_L &  \to \sqrt{2}\co{uW}v^2/\Lambda^2 & g_R &  \to \sqrt{2}\co{dW}v^2/\Lambda^2
\end{align}

\begin{figure}[!t]
\begin{center}
\includegraphics[width=\textwidth]{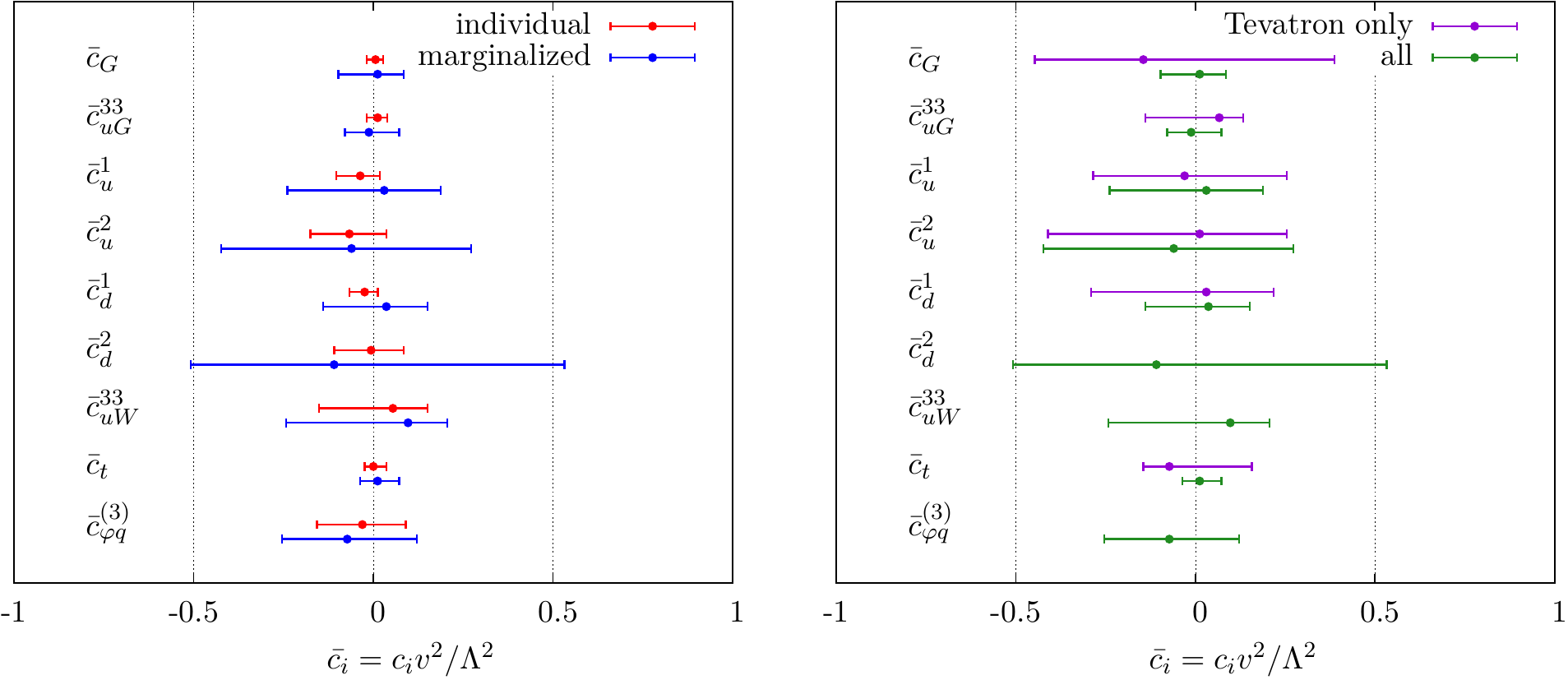}
\caption[LHC and Tevatron \ttbar and single top constraints.]{Left: Individual (red) and marginalised (blue) 95\% confidence intervals on dimension-six operators from top pair production and single top production (bottom three). Right: Marginalised 95 \% bounds considering all data from LHC and Tevatron (green) vs Tevatron only (purple).}
\label{fig:allvstev}
\end{center}
\end{figure}

Although anomalous couplings capture most of the same physics, the advantages of using higher-dimensional operators are
manifold. Firstly, the power-counting arguments of the previous paragraph that
allowed us to reject the operators \op{dW}, \op{\varphi ud} at order
$\Lambda^{-2}$ would not be clear in an anomalous coupling framework. In
addition, the four-quark operator $\op[(3)]{qq}$ in Eq.~\eqref{eqn:singletopops} can
have a substantial effect on single-top production, but this can only be
captured by an EFT approach. For a detailed comparison of these approaches, see e.g.~Ref.~\cite{Zhang:2010px}. The 95\% confidence limits on these operators from
single top production are shown in Fig. (\ref{fig:allvstev}), along with those operators previously
discussed in top pair production.

Let us compare these results to our findings of Section~\ref{sec:toppair}. The bounds on operators from top pair production are typically stronger. The so-called chromomagnetic moment operator \op{uG} is also tightly constrained, owing to its appearance in both the $q\bar{q}$ and $gg$ channels,
i.e. it is sensitive to both Tevatron and LHC measurements.  For the four-quark
operators, the stronger bounds are typically on the \co[1]{i}-type. This
originates from the more pronounced effect on kinematic distributions that they
have. The phenomenology of the $\co[2]{i}$-type operators is SM-like, and their effect becomes only visible in the tails of distributions.

The much wider marginalised bounds on these two operators come
from the relative sign between their interference term and those of the other
operators, which results in cancellations in the total cross-section that
significantly widen the allowed ranges of \co{i}. With the exception of \co{t}, which strongly modifies the single top production cross-section, the individual bounds on the operator coefficients from single top
production are typically weaker. This originates from the larger experimental
uncertainties on single top production, that stem from the multitude of
different backgrounds that contaminate this process, particularly top pair
production. For the Tevatron datasets this is particularly telling: the few
measurements that have been made, with no differential distributions, combined
with the large error bars on the available data, mean that two of the three operators are
not constrained at dimension-six\footnote{Our bounds on these two operators are
  of the same order, but wider, than a pre-LHC phenomenological study~\cite{Cao:2007ea}, owing to
  larger experimental errors than estimated there.}. Still, as before, excellent
agreement with the SM is observed.

In addition to single-top production, the operator $\op{uW}$ may be constrained
by distributions relating to the kinematics of the top quark decay. The matrix
element for hadronic top quark decay $t\to Wb \to b q q'$, for instance, is
equivalent to that for $t$-channel single top production via crossing symmetry,
so decay observables provide complementary information on this operator. We will
discuss the bounds obtainable from decay observables in Section~\ref{sec:decays}.

\subsubsection{Associated production}
\label{sec:assoc}
\begin{figure}[!t]
\begin{center}
\includegraphics[width=0.5\textwidth]{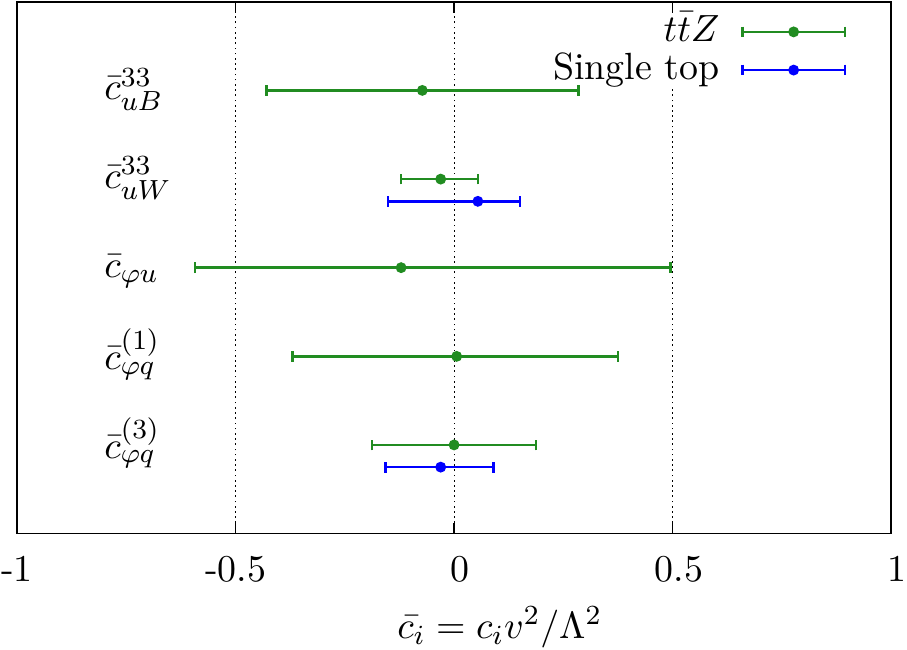}
\caption[\ttz 1D constraints.]{Individual 95\% confidence intervals for the operators of Eq.~\eqref{eqn:ttzops} from $t\bar{t}\gamma$ and $t\bar{t}Z$ production (green) and in the two cases where there is overlap, from single top measurements (blue). }
\label{fig:ttzconstraints}
\end{center}
\end{figure}

In addition to top pair and single top production, first measurements have been
reported~\cite{Aad:2015uwa,Aad:2015eua,Khachatryan:2014ewa} of top pair
production in association with a photon and with a $Z$ boson ($t\bar{t}\gamma$
and $t\bar{t}Z$)\footnote{Early measurements of top pair production in
  association with a $W$ has also been reported by ATLAS and CMS, but the
  experimental errors are too large to say anything meaningful about new physics
  therein; the measured cross-sections are still consistent with zero.}. The
cross-section for these processes are considerably smaller, and statistical
uncertainties currently dominate the quoted measurements. Still, they are of
interest because they are sensitive to a new set of operators not previously
accessible, corresponding to enhanced top-gauge couplings which are ubiquitous
in simple $W'$ and $Z$ models, and which allow contact to be made with
electroweak observables. The operator set for $t\bar{t}Z$, for instance,
contains the 6 top pair operators in Eq.~\eqref{eqn:ttbarops}, plus the
following
\begin{equation}
\begin{split}
\lag{D6}	& \supset  \frac{\co{uW}}{\Lambda^2} (\bar{q}\sigma^{\mu \nu} \tau^I u)\,\tilde  \varphi \, W_{\mu\nu}^{I} + \frac{\co{uB}}{\Lambda^2} (\bar{Q}\sigma^{\mu \nu}  u)\,\tilde \varphi \,B_{\mu\nu}+ \frac{\co[(3)]{\varphi q}}{\Lambda^2} i(\varphi^\dagger  \overleftrightarrow{D}^I_\mu \varphi )(\bar{Q}\gamma^\mu \tau^I Q) \\
			&+ \frac{\co[(1)]{\varphi q}}{\Lambda^2} i(\varphi^\dagger \overleftrightarrow{D}_\mu \varphi )(\bar{Q}\gamma^\mu  Q) + \frac{\co{\varphi u}}{\Lambda^2}(\varphi^\dagger i \overleftrightarrow{D}_\mu\varphi)(\bar{u}\gamma^\mu u)  \,.
\end{split}
\label{eqn:ttzops}
\end{equation}
There is therefore overlap between the operators contributing to associated production, and those contributing to both top pair and single top. In principle, one should include all observables in a global fit, fitting all coefficients simultaneously. However, the low number of individual $t\bar{t}V$ measurements, coupled with their relatively large uncertainties, means that they do not have much effect on such a fit. Instead, we choose to present individual constraints on the operators from associated production alone, comparing these with top pair and single top in what follows. For the former, we find that the constraints on the operators of Eq.~\eqref{eqn:ttzops} obtained from $t\bar{t}\gamma$ and $t\bar{t}Z$ measurements are much weaker than those obtained from top pair production, therefore we do not show them here. The constraints on the new operators of Eq.~\eqref{eqn:ttzops} are displayed in Fig.~\ref{fig:ttzconstraints}. It is interesting to note that the constraints from associated production measurements are comparable with those from single top production, despite the relative paucity of the former.

\subsubsection{Decay observables}
\label{sec:decays}
This completes the list of independent dimension-six operators that affect top
quark production cross-sections. However, dimension-six operators may also
contribute (at interference level) to observables relating to top quark decay.
Top quarks decay almost 100\% of the time to a $W$ and $b$ quark. The fraction
of these events which decay to $W$-bosons with a given helicity: left-handed,
right-handed or zero-helicity, can be expressed in terms of helicity fractions,
which for leading order with a finite $b$-quark mass were shown in Eq.~\eqref{eqn:helfrac}:
\begin{equation}
\begin{split}
F_0 &=\frac{ (1-y^2)^2- x^2(1+y^2)}{(1-y^2)^2+x^2(1-2x^2+y^2)} \\
F_L &= \frac{x^2(1-x^2+y^2)+\sqrt{\lambda}}{(1-y^2)^2+ x^2(1-2x^2 + y^2)} \\
F_R &=  \frac{x^2(1-x^2+y^2) -\sqrt{\lambda}}{(1-y^2)^2+ x^2(1-2x^2 + y^2)}
\end{split}
\end{equation}
%
\begin{figure}[t]
\begin{center}
\includegraphics[width=0.5\textwidth]{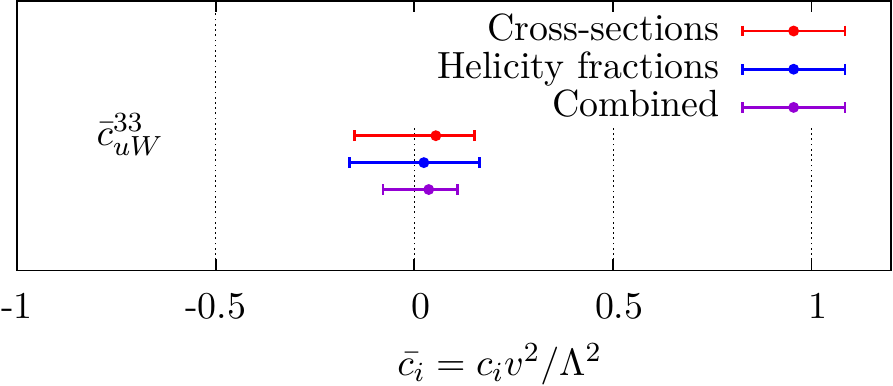}
\caption[Helicity fraction constraints.]{95\% bounds on the operator $\op{uW}$ obtained from data on top quark helicity fractions (blue) vs. single top production cross-sections (red), and both sets of measurements combined (purple).  }
\label{fig:helfrac}
\end{center}
\end{figure}
%
where $x=M_W/m_t$, $y=m_b/m_t$ and
$\lambda = 1 + x^4 + y^4 - 2x^2y^2 - 2x^2 - 2y^2 $. As noted
in Ref.~\cite{Zhang:2010dr}, measurements of these fractions can be translated into
bounds on the operator $\op{uW}$. The operator \op[(3)]{\varphi q} cannot be accessed in this way, since its only effect is to rescale the
$Wtb$ vertex $V^2_{tb} \to V_{tb} \left(V_{tb}+v^2\co[(3)]{\varphi q}/\Lambda^2\right)$, therefore it has no effect on
event kinematics. The desirable feature of
these quantities is that they are relatively stable against higher order
corrections, so the associated scale uncertainties are small. The Standard Model
NNLO estimates for these are: $\{F_0,F_L,F_R \} = \{0.687 \pm 0.005, 0.311
\pm 0.005, 0.0017 \pm 0.0001 \}$~\cite{Czarnecki:2010gb}, i.e. the
uncertainties are at the per mille level. It is interesting to ask whether the
bound obtained on $\op{uW}$ in this way is stronger than that obtained from
cross-section measurements. In Fig.~\ref{fig:helfrac} we show the constraints
obtained in each way. Although they are in excellent agreement with each other,
cross-section information gives a slightly stronger bound, mainly due to the
larger amount of data available, but also due to the large experimental
uncertainties on $F_i$. Still, these measurements provide complementary
information on the operator $\op{uW}$, and combining both results in a stronger constraint than either alone, as expected. 

\subsubsection{Charge asymmetries}
Asymmetries in the production of top quark pairs have received a lot of
attention in recent years, particularly due to an apparent discrepancy between
the Standard Model prediction for the so-called `forward-backward' asymmetry
$\Afb$ in top pair production of Eq.~\eqref{eqn:afb}
\begin{equation}
\Afb = \frac{N(\Delta y > 0)-N(\Delta y < 0)}{N(\Delta y> 0) + N(\Delta y < 0)}
\end{equation}
where $\Delta y = y_t - y_{\bar{t}}$, and a measurement by CDF~\cite{Aaltonen:2011kc}. This discrepancy
was most pronounced in the high invariant mass region, pointing to potential
$\tev$-scale physics at play. However, recent work has cast doubts on its
significance for two reasons: Firstly, an updated analysis with higher
statistics~\cite{Aaltonen:2012it} has slightly lowered the excess. Secondly, a full NNLO QCD
calculation~\cite{Czakon:2014xsa} of $\Afb$ showed that, along with NLO QCD + electroweak
calculations~\cite{Hollik:2011ps,Kuhn:2011ri,Bernreuther:2012sx} the radiative corrections to $\Afb$ are large. The current
measurements for the inclusive asymmetry are now consistent with the Standard Model within 2$\sigma$. Moreover, the
\dzero experiment reports~\cite{Abazov:2014cca} a high-invariant mass measurement
\textit{lower} than the SM prediction. From a new physics perspective, it is
difficult to accommodate all of this information in a simple,
uncontrived model without tension.

\begin{figure}[!t]
\begin{center}
\includegraphics[width=0.5\textwidth]{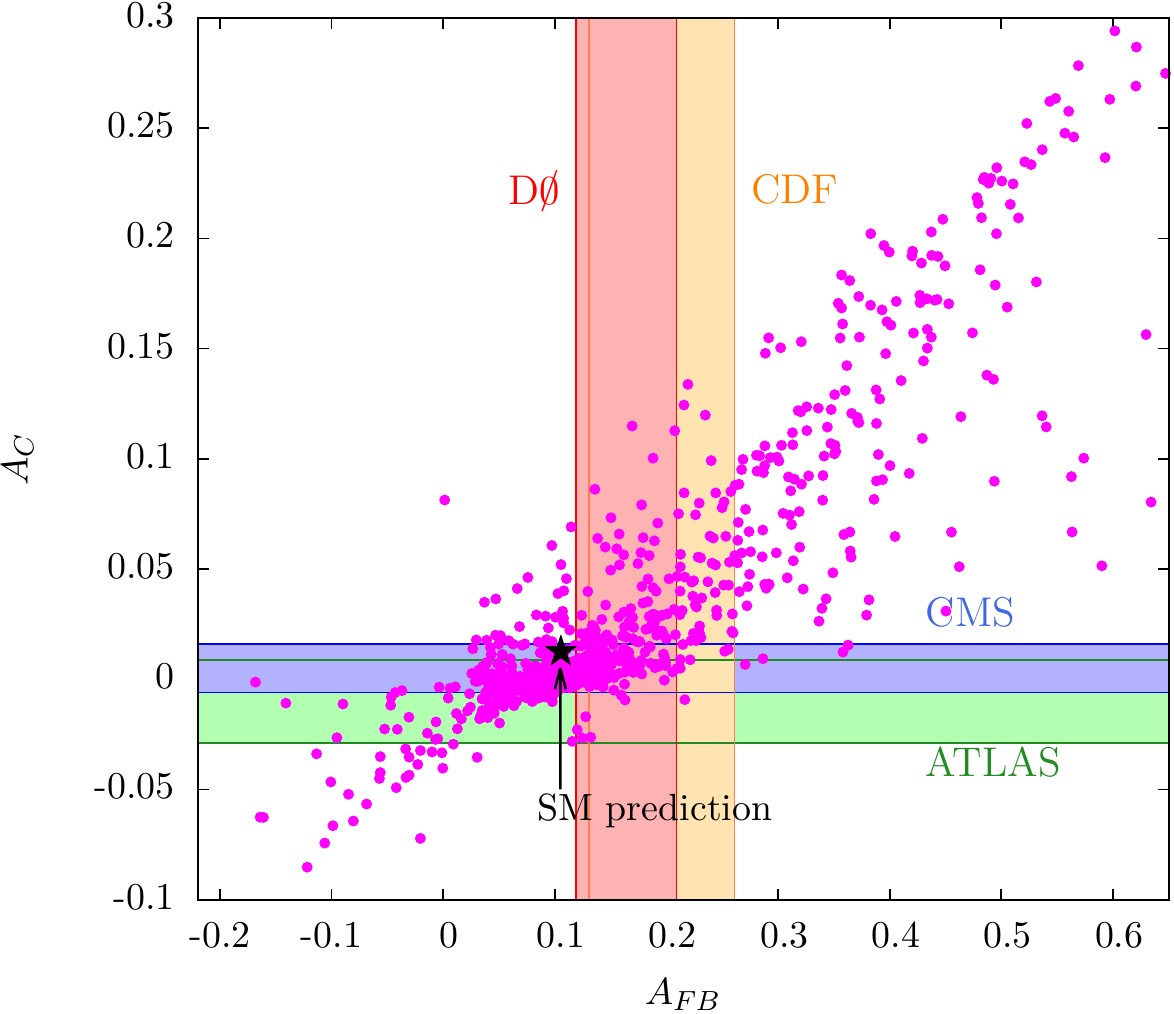}
\caption[\Afb vs \Ac plane.]{Results of a 1000 point parameter space scan over   -10 TeV $^{-2}  < \co[1,2]{u,d}/\Lambda^2 < $ 10 $TeV ^{-2}$  overlaid with the most up to date measurements of $\Afb$ and \Ac, showing clearly the correlation between them.}
\label{fig:afbvsac}
\end{center}
\end{figure}

Still, in an effective field theory approach, deviations from the Standard Model
prediction of $\Afb$ take a very simple form. A non-zero asymmetry arises from
the difference of four-quark operators:
\begin{equation}
\Afb = (\co[1]{u}-\co[2]{u}+\co[1]{d}-\co[2]{d})\frac{3s\beta}{4g_s^2\Lambda^2(3-\beta^2)},
\end{equation}
where $\beta = \sqrt{1-s/4m_t^2}$ is the velocity of the $t\bar{t}$ system\footnote{Contributions to $A_{FB}$ also arise from the normalisation of $A_{FB}$ and the dimension-six squared term\cite{Bauer:2010iq,AguilarSaavedra:2011vw,Delaunay:2011gv}, which we keep, as discussed in Sections 3.3 and 4.}. Combining this inclusive measurement with differential measurements such as $d\Afb/dm_{t\bar{t}}$ allows simultaneous bounds to be extracted on all four of these operators. Therefore it is instructive to compare the bounds obtained on \co[1,2]{u,d} from charge asymmetries to those obtained from $t\bar{t}$ cross-sections.  Again it is possible to (indirectly) investigate the complementarity between Tevatron
and LHC constraints. Though the charge symmetric initial state of the LHC does
not define a `forward-backward' direction, a related charge asymmetry can be (Eq.~\ref{eqn:ac})
defined as:
\begin{equation}
A_{C} = \frac{N(\Delta |y| > 0)-N(\Delta |y| < 0)}{N(\Delta |y|> 0) + N(\Delta |y| < 0)},
\end{equation}
making use of the fact that tops tend to be produced at larger rapidities than
antitops. This asymmetry is
diluted with respect to $\Afb$, however. The most up-to-date SM
prediction is $A_C = 0.0123 \pm 0.005$~\cite{Bernreuther:2012sx} for $\sqrt{s} =$ 7 \tev. The experimental status of these measurements is illustrated in Fig.~\ref{fig:afbvsac}. The inclusive measurements of
$\Afb$ are consistent with the SM expectation, as are those of
\Ac. The latter, owing to large statistical errors, are also consistent with zero, however, so this result is not particularly conclusive. Since these are different measurements, it is also
possible to modify one without significantly impacting the other. Clearly they
are correlated, however, as evidenced in Fig.~\ref{fig:afbvsac}, where the most up to
date measurements of $\Afb$ and \Ac are shown along with the results
of a 1000 point parameter space scan over the four-quark operators. This
highlights the correlation between the two observables: non-resonant new physics
which causes a large $\Afb$ will also cause a large \Ac, provided it generates a dimension-six operator at low energies.

We have used both inclusive measurements of the charge asymmetries \Ac and $\Afb$, and measurements as a function of the top pair invariant mass $m_{t\bar{t}}$ and rapidity difference $|y_{t\bar{t}}|$. In addition, ATLAS has published measurements of \Ac with a longitudinal `boost' of the $t\bar{t}$ system: $\beta = (|p^z_t+p^z_{\bar{t}})|/(E_t+E_{\bar{t}}) > 0.6 $, which may enhance sensitivity to new physics contributions to \Ac, depending on the model~\cite{AguilarSaavedra:2011cp}. Since $\Afb = 0$ at leading-order in the SM, it is not possible to define a
$K$-factor in the usual multiplicative sense. Instead we take higher-order QCD effects into
account by adding the NNLO QCD prediction to the dimension-six terms. In the case of \Ac, we normalise the small (but non-zero) LO QCD piece, to the NLO prediction, which has been calculated with a Monte Carlo and cross-checked with a dedicated NLO calculation~\cite{Bernreuther:2012sx}.

\begin{figure}[t!]
\begin{center}
\includegraphics[width=0.5\textwidth]{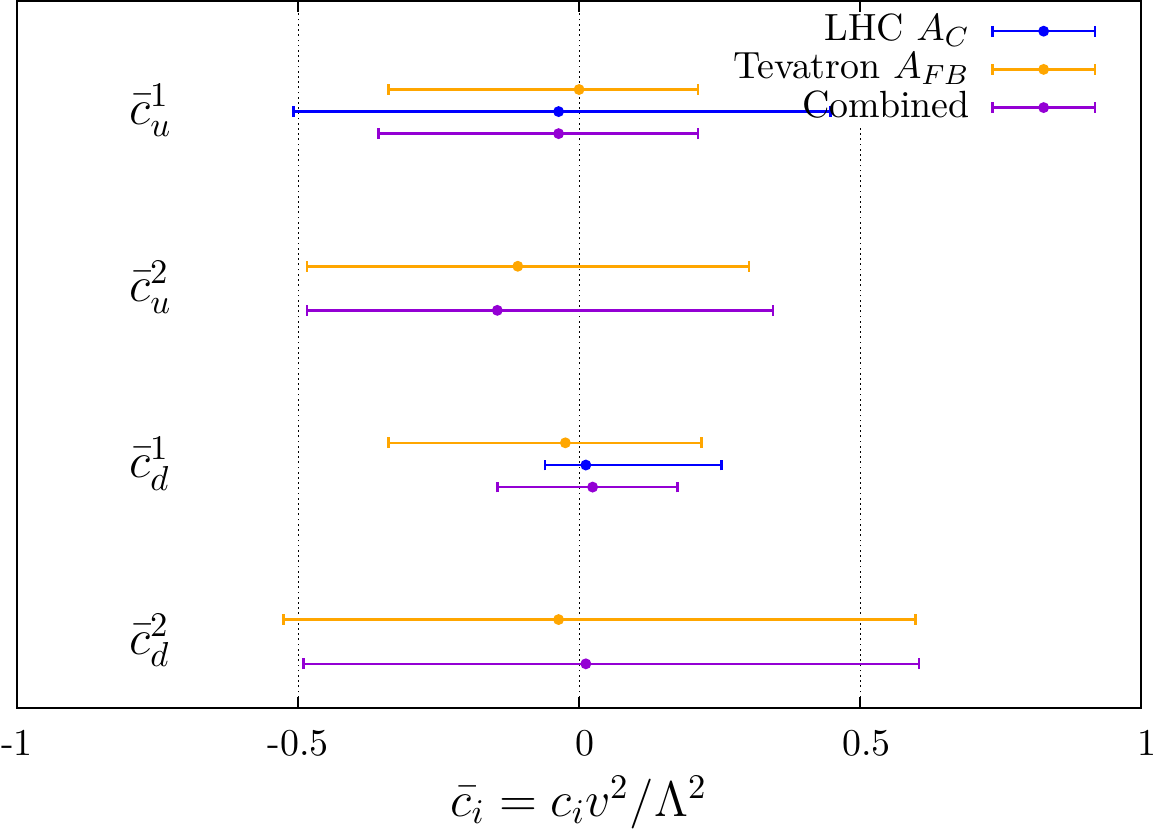}
\caption[Charge asymmetry constraints.]{Marginalised 95\% confidence intervals on top pair four quark operators from charge asymmetries at the LHC and Tevatron.}
\label{fig:asymms}
\end{center}
\end{figure}

The above asymmetries have been included in the global fit results presented in Fig.~\ref{fig:constraints}. However, it is also interesting to see what constraints are obtained on the operators from asymmetry data alone. To this end, the 95\% confidence intervals on the coefficients of the operators \op[1,2]{u,d} from purely charge
asymmetry data are shown in Fig.~\ref{fig:asymms}. Unsurprisingly, the bounds are much weaker than
for cross-section measurements in Fig.~\ref{fig:allvstev}, with the \op[2]{i}-type operators unconstrained by LHC data alone.  Despite the small discrepancy between the measured $\Afb$ and its SM value, this does not translate into a non-zero Wilson coefficient; as before, all operators are zero within the 95\% confidence intervals.  At 13~\tev, the asymmetry \Ac will be diluted
even further, due to the increased dominance of the $gg\to t\bar{t}$ channel,
for which $A_C = 0 $. It is therefore possible that charge asymmetry measurements (unlike cross-sections) will not further tighten  the bounds on these operators during LHC Run~II.

\subsubsection{Contribution of individual datasets}

\begin{figure}[p!]
\begin{center}
\includegraphics[width=0.9\textwidth]{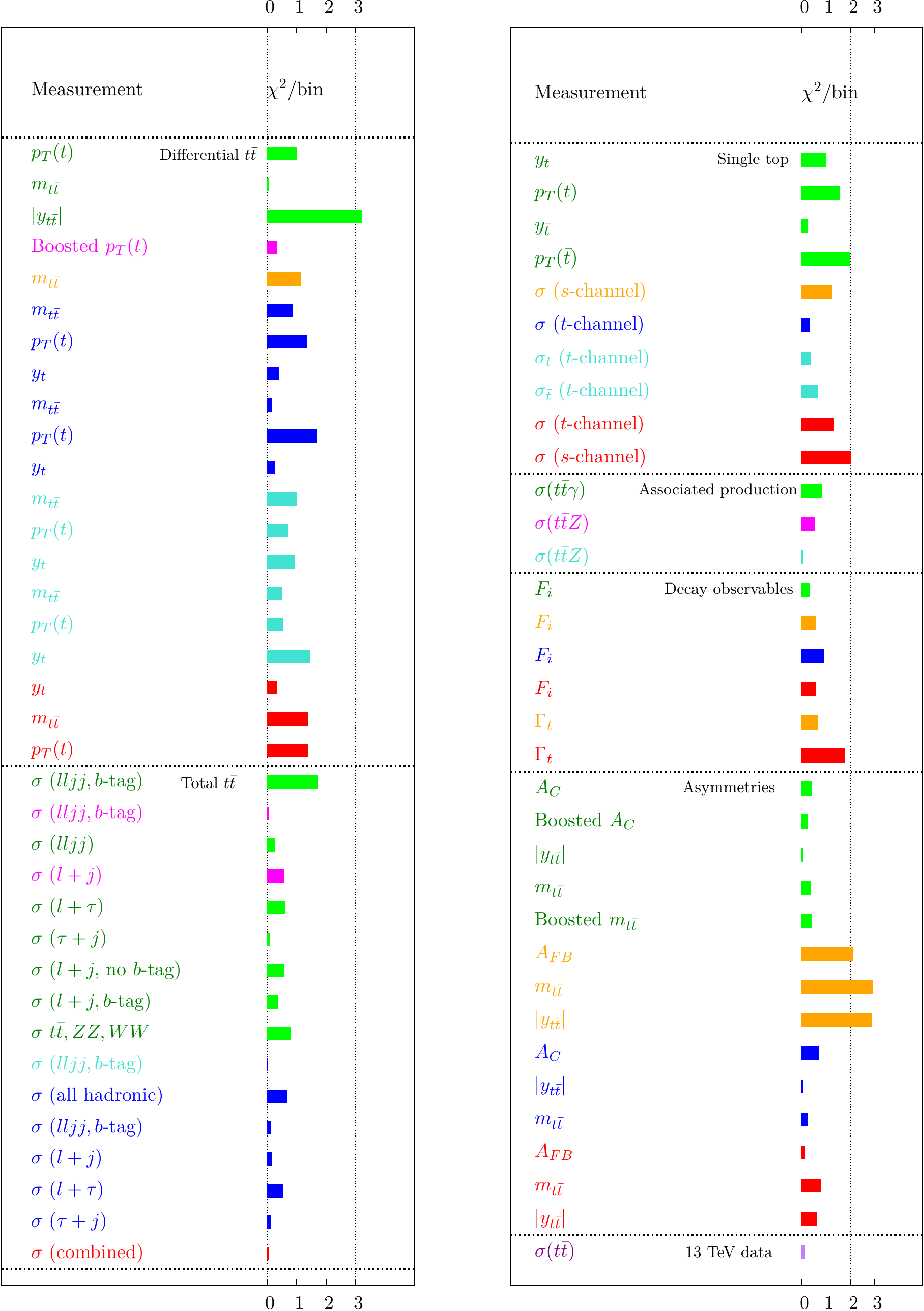}
\caption[$\chi^2$ per bin for measurements used in the fit.]{$\chi^2$ per bin between measurement and the interpolated best fit point, for measurements considered in this fit. Colours: Green: ATLAS 7 \tev, Magenta: ATLAS 8 \tev, Blue: CMS 7 \tev, Turquoise: CMS 8 \tev, Red: \dzero, Orange: CDF, Purple: CMS 13 \tev.}
\label{fig:pullfactors}
\end{center}
\end{figure}

Counting each bin independently, there are a total of 234 measurements entering our fit, giving a total $\chi^2$ of 206.1 at the best fit point. It is instructive to examine how this is distributed across the individual datasets. We quantify this by
calculating the $\chi^2$ per bin between the data and the global best fit point,
as shown in Fig.~\ref{fig:pullfactors}. Overall, excellent agreement is seen across the board, with no measurement in obvious tension with any other. The largest single contributors to the $\chi^2$ come from the rapidity distributions in top pair production. It has been known for some time that there is some tension between data and Monte Carlo generators for this observable, especially in the forward region (see e.g. the rapidity distributions in Ref.~\cite{Khachatryan:2015oqa}). It is quite likely that this discrepancy stems from the QCD modelling of the event kinematics, rather than potential new physics. Moreover, in a fit with this many measurements, discrepancies of this magnitude are to be expected on purely statistical grounds.

At the level of total $\ttbar$ cross-sections, the vanishingly small contributions to the $\chi^2$ stem from the simple fact that the total rate is well-described by the SM. Single top production measurements are also in good agreement with the SM.  The associated production processes $tt\gamma$ and $ttZ$, along with the charge asymmetry measurements from the LHC, have a very small impact on the fit, owing to the large statistical uncertainties on the current measurements. For the former, this situation will improve in Run~II, for the latter the problem will be worse. The forward-backward asymmetry measurements from CDF remain the most discrepant dataset used in the fit, even though the inclusive asymmetry is in good agreement with the NNLO SM asymmetry, because the NNLO corrections shrink the scale uncertainty band whilst not enlarging the central value, thus enhancing the CDF excess (see Ref.~\cite{Czakon:2014xsa}).

\subsection{Validity of the EFT approach}
\label{sec:validity}
As we have just shown, collider measurements can be used to extract bounds on Wilson coefficients in a completely model-independent way. However, if the corresponding dimension-six operators are to be interpreted as the leading terms in a consistent effective field theory, then care must be taken to ensure the constraints are valid. The first question one might ask is if the dimension-six truncation is valid, and more broadly one can ask if the measurements used in the fit are probing kinematic regimes that make the entire effective description invalid.

\subsubsection{Impact of quadratic terms}
As discussed in section \ref{sec:topeft}, at the level of observables adding \D6 operators to the Standard Model Lagrangian amounts to replacing the Standard Model matrix element with

\begin{equation}
|\mathcal{M}_{\text{full}}|^2 = |\mathcal{M}_{\text{SM}}|^2+ 2\Re\mathcal{M}_{\text{D6}}^*\mathcal{M}_{\text{SM}}+|\mathcal{M}_{\text{D6}}|^2.
\end{equation}

The last term on the right-hand side is of order $\mathcal{O}(1/\Lambda^4)$. However, to have a fully consistent description of an observable at this order, one should include \textit{all} terms proportional to $\mathcal{O}(1/\Lambda^4)$, i.e. also include the interference terms between the Standard Model and dimension-eight operators. In the strictest interpretation of the effective Lagrangian approach, neglecting these terms renders the EFT description of such an observable meaningless. This is an overly restrictive viewpoint, however, as there exist several prescriptions for ensuring that the dimension-six approximation is under control.

One might, for example, restrict the parameter space of Wilson coefficients to regions for which the linear interference terms dominate over the quadratic terms. This ensures that all constraints are valid in a dimension-six effective field theory interpretation. However, it may reduce the sensitivity to the operators, and weaken (or wipe out altogether) the obtained constraints. Moreover, there is an ambiguity in how tolerant of the squared terms one should be: should one, for example, only cut off the parameter space when they become \textit{larger} than the interference terms, in which case their effects will still be considerable, or when they reach 10, 20 or 50\% of the linear piece? There is no first principles answer to this.

\begin{figure}[t!]
\begin{center}
\includegraphics[width=\textwidth]{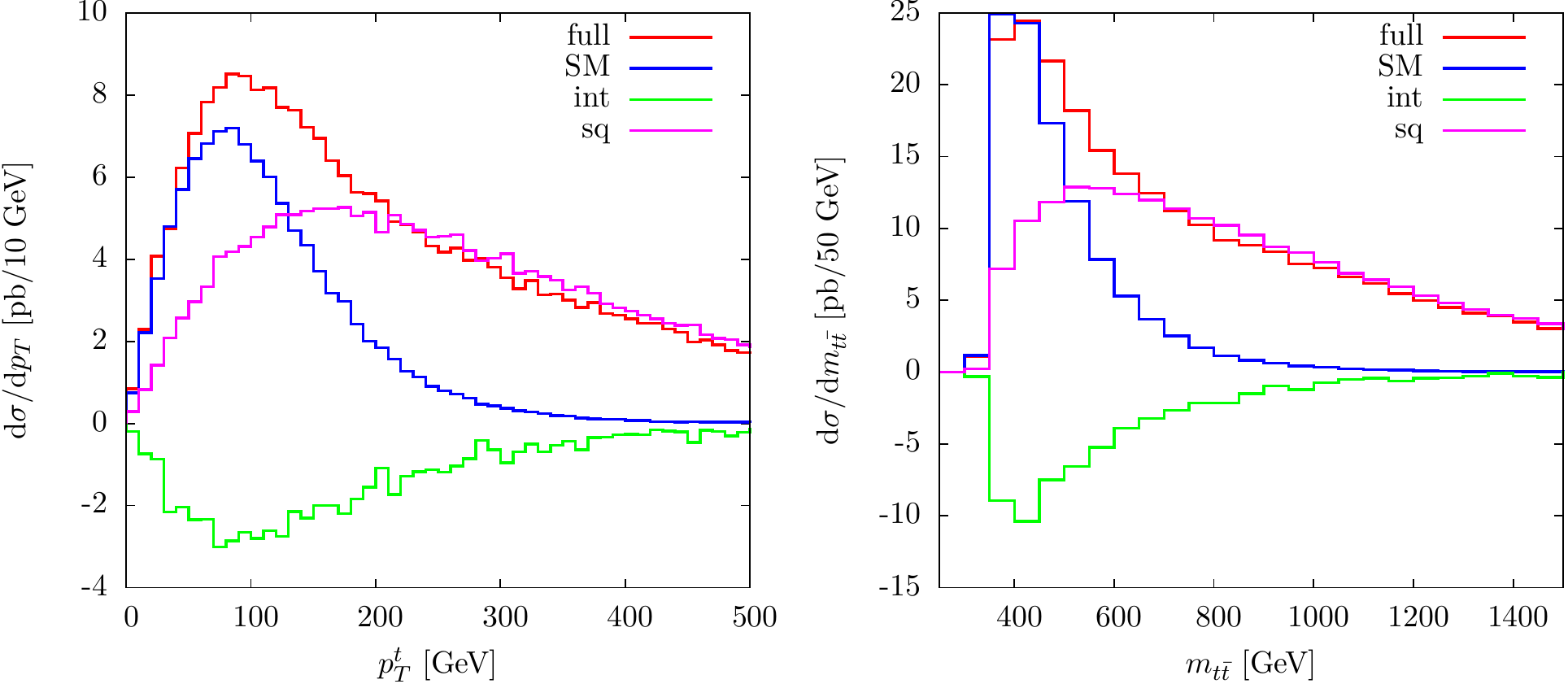}
\caption[\ttbar distributions with interference and quadratic terms.]{Distributions in top quark pair production for a given point in the parameter space of $t\bar{t}$ Wilson coefficients considered in this fit, namely $\{c_i\} = \{\co{G},\co[33]{uG},\co[1]{u},\co[2]{u},\co[1]{d},\co[2]{d}\} = \{0.01,0.01,0,0.9,0,-9.2,-9.2\}$ TeV$^{-2}$. In blue is the SM prediction, in green the pure dimension-six contribution, in pink the pure squared term, and in red the sum of the three. Subtracting off the square terms leads to negative cross-sections in the intermediate mass range. All distributions are leading-order in $\alpha_s$.}
\label{fig:subtracted}
\end{center}
\end{figure}

Alternatively, one could subtract off the squared terms altogether, either at the level of matrix elements or of observables. By definition this ensures that the constraints can be interpreted in terms of a dimension-six EFT. This treatment, however, lacks a clear physical motivation. If a non-zero Wilson coefficient were explicitly measured, it would have to correspond to the low energy limit of some ultraviolet completion. Matching the EFT constraints onto a specific UV model can be done in a general way, and will determine if the data favours a `natural' region of that model's parameter space. However, subtracting off quadratic effects will in general disrupt the accuracy of the matching procedure, as the most dominant effects of a particular model on a certain observable may be from the non-interfering piece\footnote{See also Refs.~\cite{Biekotter:2016ecg,Freitas:2016iwx} for a discussion of this.}. Furthermore, subtracting off some contributions to the cross-section might lead to unphysical effects in distributions, as shown in Fig. \ref{fig:subtracted}.

By looking at the range $p^t_T \gtrsim $ 200 GeV, and $m_{t\bar{t}} \gtrsim $ 600 GeV, we see that considering only the interference term leads to negative predictions for the differential cross-section: a clearly unacceptable result. Here the terms proportional to $\mathcal{O}(1/\Lambda^4)$ keep the cross-section physically meaningful. One could take this as evidence that the dimension-six approximation is then not valid at all, because the squared term is dominating. However, if one would like to compare the EFT constraints to those obtained in a specific new physics  model, it is necessary to include these to ensure an accurate matching condition. The importance of keeping the squared term or not is then ultimately a model-dependent question.

Finally, one could simply adopt the pragmatic approach of viewing the dimension-six framework not as the leading part of a consistent effective theory, but as a model-independent way of parameterising how well the Standard Model describes the data. If a non-zero Wilson coefficient was to be measured, it would still be evidence for new physics, even if it could not be simply linked to a particular new physics model. This is the approach we adopt in this fit, and we leave interpretational issues aside. 

Even if one considers a fit with quadratic terms completely removed from all observables, $\mathcal{O}(\Lambda^{-4})$ terms manifest in another way when likelihood contours are drawn. Consider the likelihood function of Eq.~\eqref{eqn:chi2}, considering just one bin and one operator for simplicity. For a linear fitting function $f$, it will have a polynomial dependence on \co{i} proportional to

\begin{equation}
\chi^2(\co{i}) = \frac{(f(\co{i})-E)^2}{\Delta^2} \sim  \co{i} + \co[2]{i} - E.
\end{equation}

To ensure the $\chi^2$ has a local minimum, the squared term must be kept. This provides another argument in favour of keeping the quadratic terms throughout the fit.

\subsubsection{Overflow bins}
\begin{figure}[t!]
\begin{center}
\includegraphics[width=\textwidth]{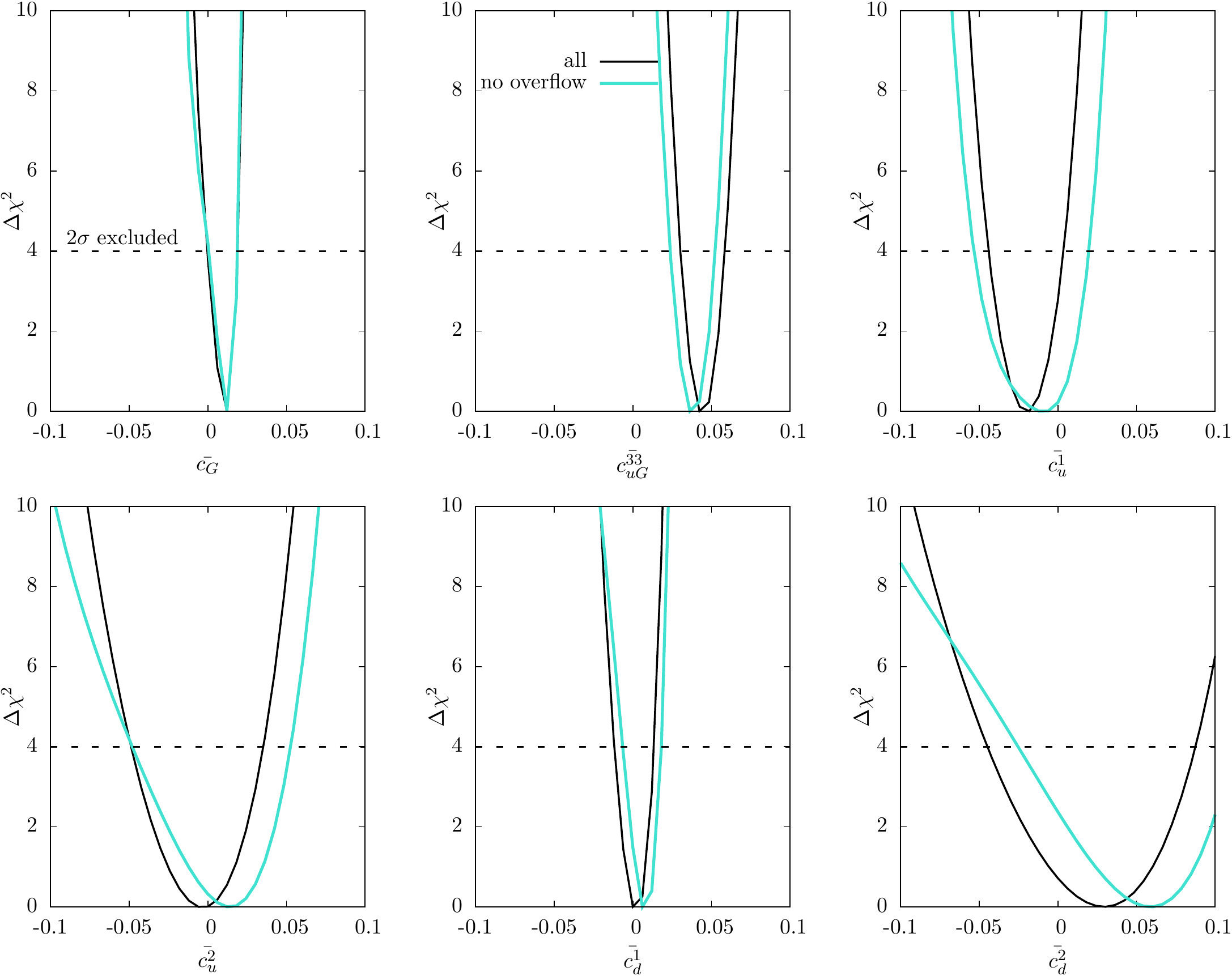}
\caption[One-dimensional likelihood plots with and without overflow bins.]{Individual one-dimensional likelihood plots for the top pair Wilson coefficients considered in this fit, containing all differential top pair measurements (black), and omitting the final bin in the $m_{t\bar{t}}$ and $p^t_T$ distributions (turquoise).}
\label{fig:chi2s}
\end{center}
\end{figure}

Related, but not identical, to the question of whether one should omit or keep the quadratic terms, is whether one should worry about events in the tails of distributions that might invalidate the EFT treatment. For inclusive observables this is less of a problem, because they tend to be dominated by electroweak scale thresholds, well within the valid region of the phase space. However, by na\"ive power counting, the convergence of the EFT expansion rests on the two conditions

\begin{equation}
\begin{split}
\frac{g_*^2v^2}{\Lambda^2} < 1 \hspace{10pt} \text{and}  \hspace{10pt} \frac{E^2}{\Lambda^2} < 1,
\end{split}
\label{eqn:powcount}
\end{equation}

where $g_*$ is a generic new physics coupling, $v$ is the Higgs vev, and $E$ is the maximum energy scale probed by the process. The first condition can be ensured for any weakly coupled UV completion. The second condition is troublesome when differential distributions are included in the fit, however. The final bin in experimentally measured distributions, such as those in Fig. \ref{fig:distributions}, typically contains not only events in that phase space region, but also so-called overflow events to the right of the plot. If the experiment has not included information on the maximum momentum transfer probed in the published dataset, i.e. the maximum value of the `overflow' entry to the right of the plot, it is difficult to consistently interpret the resulting bounds on the Wilson coefficients in an underlying UV model, because those overflow events may violate the power counting conditions of Eq.~\eqref{eqn:powcount}. As a test of how much pull they have on the fit, in Fig. \ref{fig:chi2s} we plot the 1-dimensional likelihood distributions (equivalent to $\Delta\chi^2 = \chi^2-\chi^2_{min}$) for the Wilson coefficients relevant for top pair production, considering the full $t\bar{t}$ dataset, and omitting the overflow bins in the kinematic distributions.

The differences in the constraints are small, typically at the order of a few percent, showing that the fit is not unduly biased by phase space points that undermine the validity of the effective field theory. One may take the approach of omitting these overflow bins altogether, to ensure control over the scales involved in the fitted measurements. The subsequent limits are then slightly weaker, due to reduced statistical power and sensitivity to the operators.
However, the exclusion of certain data points undermines the `global' nature of a global fit, so they are included for full generality. Indeed, this is again a model-dependent question. To illustrate this, in the next section we will convert our EFT constraints onto specific UV models.

Before doing this, a final comment is in order about Fig. \ref{fig:chi2s}. The limit setting shown there, is obtained from a likelihood ratio test, rather than the raw $\chi^2$ we employ elsewhere in the fit. This is so that the 2$\sigma$ constraints for both datasets can be easily shown on the same plot, as they both correspond to $\Delta\chi^2(\co{i}) < 4 $. For the $\chi^2$ test both datasets have different numbers of degrees of freedom, corresponding to the number of input measurements they contain. However, the same results apply in this case as well. The limits obtained in the latter approach (i.e. the one we adopt in the rest of the fit) are actually weaker. Without a compelling reason to adopt either approach, it thus seems sensible to take the more conservative option.

\subsubsection{Constraining UV models}
\label{sec:uvmodels}
As an illustration of the wide-ranging applicability of EFT techniques, we
conclude by matching our effective operator constraints to the low-energy regime
of some specific UV models. These models do not necessarily represent concrete UV scenarios, but serve as illustrative examples of how EFT constraints could map onto the parameter space of more fundamental theories.

\subsubsection*{$s$-channel axigluon:}

Considering top pair production, one can imagine the four operators of Eq.~\eqref{eqn:4fs} as being generated by integrating out a heavy $s$-channel resonance which interferes with the QCD $q\bar{q}\to t\bar{t}$ amplitude. One particle that could generate such an interference is the so-called axigluon. These originate from models with an extended strong sector with gauge group $SU(3)_{c1}\times SU(3)_{c2}$ which is spontaneously broken to the diagonal subgroup $SU(3)_{c}$ of QCD. In the most minimal scenario, this breaking can be described by a non-linear sigma model
\begin{equation}
\mathcal{L} = -\frac{1}{4}G_{1\mu\nu}G_{1}^{\mu\nu} -\frac{1}{4}G_{2\mu\nu}G_{2}^{\mu\nu} + \frac{f^2}{4}\text{Tr}D_\mu \Sigma D^\mu \Sigma^\dagger \hspace{10pt},\hspace{10pt} \Sigma=\exp\left(\frac{2i\pi^a t^a}{f}\right) \hspace{10pt},\hspace{10pt} a=1,...,8.
\end{equation}
Here $\pi^a$ represent the Goldstone bosons which form the longitudinal degrees of freedom of the colorons, giving them mass, $t^a$ are the Gell-Mann matrices, and $f$ is the symmetry breaking scale. The nonlinear sigma fields transform in the bifundamental representation of $SU(3)_{c1}\times SU(3)_{c2}$:
\begin{equation}
\Sigma \to U_L \Sigma U^\dagger_R \hspace{10pt},\hspace{10pt} U_L = \exp\left(\frac{i\pi^a\alpha_L^a }{f}\right) \hspace{10pt},\hspace{10pt} U_R = \exp\left(\frac{i\pi^a\alpha_R^a }{f}\right).
\end{equation}
The physical fields are obtained by rotating the gauge fields $G_1$ and $G_2$ to the mass eigenstate basis
\begin{equation}
\left(\begin{array}{c} G^a_{1\mu} \\ G^a_{2\mu} \end{array}\right) = \left(\begin{array}{c c} \cos\theta_c &  -\sin\theta_c \\  \sin\theta_c &  \cos\theta_c  \end{array}\right)\left(\begin{array}{c} G^a_\mu \\ C^a_\mu \end{array}\right),
\end{equation}
where the mixing angle $\theta_c$ is defined by
\begin{equation}
\sin\theta_c = \frac{g_{s1}}{\sqrt{g^2_{s1}+g^2_{s2}}}.
\end{equation}
The case of an axigluon corresponds to maximal mixing $\theta = \pi/4$, i.e. $g^2_{s1}=g^2_{s2}=g^2_s/2$. Taking the leading-order interference with the SM amplitude for $q\bar{q}\to t\bar{t}$, in the limit $s \ll M_A^2$, we find that the axigluon induces the dimension-six operators
\begin{equation}
\frac{\co[1]{u}}{\Lambda^2} = \frac{g^2_s}{M_A^2},\hspace{10pt}\hspace{10pt} \frac{\co[1]{d}}{\Lambda^2} = \frac{5g^2_s}{4M_A^2},\hspace{10pt}\hspace{10pt} \frac{\co[2]{u}}{\Lambda^2} =  \frac{\co[2]{d}}{\Lambda^2} = \frac{2 g^2_s}{M_A^2}.
\end{equation}
Substituting the marginalised constraints on the 4-quark operators, we find this translates into a lower bound on an axigluon mass. $M_A \gtrsim 1.4 $ \tev at the 95\% confidence level. Since this mass range coincides with the overflow bin of Fig.~\ref{fig:distributions}, this bound creates some tension with the validity of the EFT approach in the presence of resonances in the $t\bar t$ spectrum (for a general discussion see Ref.~\cite{Englert:2014cva,Brehmer:2015rna,Isidori:2013cga}); at this stage in the LHC programme indirect searches are not sensitive enough to compete with dedicated searches.

\subsubsection*{$s$-channel $W'$:}
Turning our attention to single top production, we consider the example of the operator $\op[(3)]{qq}$ being generated by a heavy charged vector resonance ($W'$) which interferes with the SM amplitude for $s$-channel single top production: $u\bar{d}\to W\to t\bar{b}$. The most general Lagrangian for such a particle (allowing for left and right chiral couplings) is (see e.g. Ref.~\cite{Boos:2006xe}):
\begin{equation}
\mathcal{L} = \frac{1}{2\sqrt{2}}V_{ij}g_{W'}\bar{q}_i\gamma_\mu(f^R_{ij}(1+\gamma^5)+f^L_{ij}(1-\gamma^5))W^\mu q_j + h.c.
\end{equation}
We take the generic coupling $g_{W'} = g_{SM}$. Since we are considering the interference term only, which must have the same $(V-A)$ structure as the SM, we can set $f^R=0$. Considering the tree-level interference term for between the diagrams for $u\bar{d}\to W', W'\to t\bar{b}$, and taking the limit  $s \ll M_W'^2$ (we also work in the narrow-width approximation $\Gamma_{W}, \Gamma_{W'} \ll M_W, M_{W'} $), we find
\begin{equation}
\label{eq:wprime}
\frac{\co[3,1133]{qq}}{\Lambda^2} = \frac{g^2}{4M^2_{W'}},
\end{equation}
which, using our global constraint on \op{t}, translates into a bound $M_{W'} \gtrsim 1.2 $ \tev.

These bounds are consistent with, but much weaker than, constraints from direct searches for dijet resonances from ATLAS~\cite{Aad:2011fq,Aad:2014aqa} and CMS~\cite{Khachatryan:2015sja}, which report lower bounds of $\{M_A,M_{W'}\} > \{2.72,3.32\} $ \tev and $\{M_A,M_{W'}\} > \{2.2,3.6\} $ \tev respectively. It is unsurprising that these dedicated analyses obtain stronger limits, given the generality of this fit. Again this energy range is resolved by the fit thus in principle invalidating the EFT approach to obtain Eq.~\eqref{eq:wprime}. Nonetheless, these bounds provide an interesting comparison of our numerical results, whilst emphasising that for model-specific examples, direct searches for high-mass resonances provide stronger limits than general global fits.

\subsection{Summary}
\label{sec:conc_ch3}

In this chapter, we performed a global fit of top quark effective
field theory to experimental data, including all constrainable operators at
dimension six. For the operators, we use the `Warsaw basis' of
Ref.~\cite{Grzadkowski:2010es}, which has also been widely used in the context
of Higgs and precision electroweak physics. We use data from the Tevatron and
LHC experiments, including LHC Run~II data, up to a centre of mass energy of
13~\tev. Furthermore, we include fully inclusive cross-section measurements, as
well as kinematic distributions involving both the production and decay of the
top quark. Counting each bin independently, the total number of observables
entering the fit is 234, with a total of 12 contributing operators. Constraining
the coefficients of these operators is then a formidable computational task. To
this end we use the parametrisation methods in the \textsc{Professor} framework,
first developed in the context of Monte Carlo generator
tuning~\cite{Buckley:2009bj}, and discussed here in Section~\ref{sec:fit}.

\begin{figure}[!t]
\begin{center}
\includegraphics[width=0.7\textwidth]{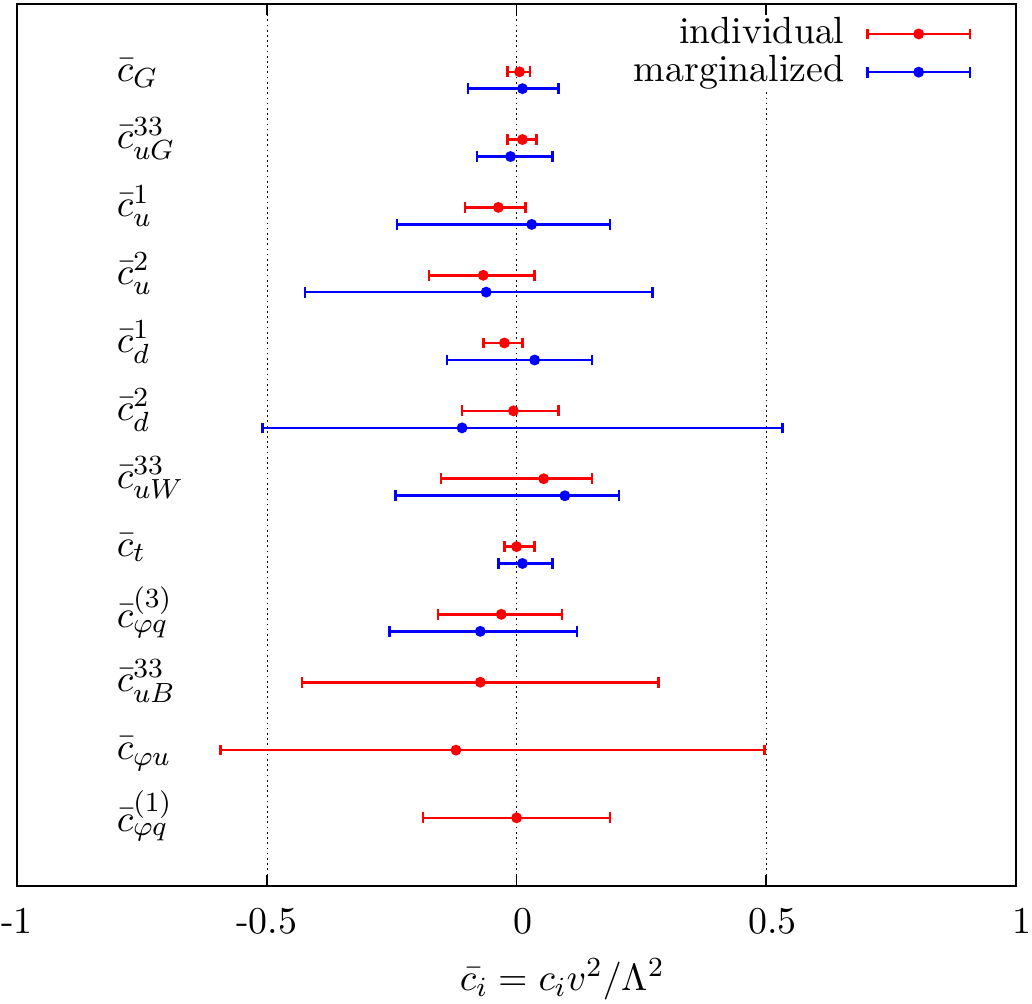}
\caption[Bottom-line 95\% confidence intervals on the operators considered.]{95\% confidence intervals for the dimension-six operators that we consider here, with all remaining operators set to zero (red) and marginalised over (blue). In cases where there are constraints on the same operator from different classes of measurement, the strongest limits are shown here. The lack of marginalised constraints for the final three operators is discussed in Section~\ref{sec:assoc}.}
\label{fig:constraints}
\end{center}
\end{figure}

\begin{table}[t!]
\begin{center}
\setlength\extrarowheight{2.5pt}
\begin{tabular}{| c | c | c | } \hline 

 \textbf{Coefficient} & 		 \textbf{Individual constraint} & 		 \textbf{Marginalised constraint} \\ \hline  
$\co{G}v^2/\Lambda^2$ & 		 (---0.018, 0.027) & 		 (---0.097, 0.085) \\ \hline 
$\co[33]{uG}v^2/\Lambda^2$ & 		 (---0.018, 0.039) & 		 (---0.079, 0.073) \\ \hline 
$\co[1]{u}v^2/\Lambda^2$ & 		 (---0.103, 0.018) & 		 (---0.236, 0.188) \\ \hline 
$\co[2]{u}v^2/\Lambda^2$ & 		 (---0.175, 0.036) & 		 (---0.424, 0.272) \\ \hline 
$\co[1]{d}v^2/\Lambda^2$ & 		 (---0.067, 0.121) & 		 (---0.139, 0.151) \\ \hline 
$\co[2]{d}v^2/\Lambda^2$ & 		 (---0.109, 0.085) & 		 (---0.508, 0.533) \\ \hline 
$\co[33]{uW}v^2/\Lambda^2$ & 		 (---0.151, 0.151) & 		 (---0.242, 0.206) \\ \hline 
$\co{t}v^2/\Lambda^2$ & 		 (---0.024, 0.036) & 		 (---0.036, 0.073) \\ \hline 
$\co[(3)]{\varphi q}v^2/\Lambda^2$ & 		 (---0.157, 0.091) & 		 (---0.254, 0.121) \\ \hline 
$\co[33]{uB}v^2/\Lambda^2$ & 		 (---0.430, 0.284) & 		 (---, ---) \\ \hline 
$\co{\varphi u}v^2/\Lambda^2$ & 		 (---0.593, 0.496) & 		 (---, ---) \\ \hline 
$\co[(1)]{\varphi q}v^2/\Lambda^2$ & 		 (---0.188, 0.188) & 		 (---, ---) \\ \hline 
\end{tabular}
\caption[95\% confidence intervals on the Wilson coefficients considered in the fit.]{Numerical values of the individual and marginalised 95\% confidence intervals on the operators presented here.} 
\label{table:numbers}
\end{center}
\end{table}

We perform a $\chi^2$ fit of theory to data, including appropriate correlation
matrices where these have been provided by the experiments. We obtain bounds on
the Wilson coefficients of various operators contributing to top quark
production and decay, summarised in Fig.~\ref{fig:constraints}, in two cases: (i) when all other coefficients are set to zero; (ii) when all other operators coefficients are marginalised over. The numerical values of these constraints are also shown in table \ref{table:numbers}. 

The strongest constraints are on operators involving the gluon, as expected given the
dominance of gluon fusion in top pair production at the LHC (for which there is
more precise data). Four fermion operators are constrained well in general, with
weaker constraints coming from processes whose experimental uncertainties remain
statistically dominated (e.g. $t\bar{t}V$ production). We have quantified the
interplay between the Tevatron and LHC datasets, as well as that between
different measurement types (e.g. top pair, single top).

The results are all in agreement with the SM only hypothesis, with no tensions beyond the 95\% confidence level, which is perhaps to be expected given the lack of reported deviations in previous studies. However, the
fact that this agreement is obtained, in a wide global fit, is itself testament
to the consistency of different top quark measurements, with no obvious tension
between overlapping datasets. New data from LHC Run~II is continuously
appearing, and can be implemented in our fit framework in a systematic way. 
Still, there are several potential refinements of our analysis that can be made
in order to improve the numerical constraints presented here, that go beyond simply replacing 8 \tev measurements with 13 \tev ones. A discussion of these issues is the subject of the next chapter.

\newpage
\null
\newpage

\newpage
\section{Future prospects for top quark EFT}

\subsection{Introduction}
\label{sec:intro}
In chapter 3 we performed a comprehensive global fit of the \D6 operators that can influence top quark observables at hadron colliders to all the published top measurements from the Tevatron and Run I of the LHC. These constituted (predominantly) top pair production in various decay channels, as well as single top, associated vector boson production and observables from top quark decay. Despite the impressive statistical sample that entered the fit, the subsequent bounds on the studied Wilson coefficients are rather weak, pointing to values of $\Lambda$ of order \ord{\lesssim 1 \tev}, depending on the assumed size of the UV couplings. While one can take care to ensure that all bounds are consistent within an EFT formulation, for example by cutting out the high-mass `overflow' bins in the differential distributions used, for which there is no control over scales, this is still a disappointingly low scale compared to the design mass reach of the LHC.

The wide allowed ranges for these operators stems not from a lack of sensitivity to the operators, but from the large experimental systematics and theory uncertainty bands from varying the scales and PDFs, and the more general problem of searching for precision deviations at a hadron collider, rather than `bump hunts'.  Still, given that we are at a very early stage of the full LHC programme, it is well-motivated to ask what improvements can be made over its lifetime as theory descriptions are improved and experimental error bars are shrank.  We saw that, for the case of \ttbar production, vast improvement could be achieved by adding differential distributions as well as total rates. Typically, however, the measurements used in the fit were based on standard top reconstruction techniques, which while providing good coverage of the low $p_T$ threshold region, suffer from large statistical and systematic uncertainties in the high $p_T$ tails, precisely the region where we want to be most sensitive to the effects of the operators.

Moreover, the distributions used were typically unfolded to parton level; that is, the final-state objects were corrected for detector effects and the actual measured cross-section in a fiducial volume of the detector extrapolated to the full phase-space, without cuts. This substantially eases the workflow of our fit, since the data can be compared directly to parton-level predictions without the need for the full parton shower, hadronisation and detector simulation chain to be implemented at each point in the parameter space. However, the extrapolation, which makes use of comparing to Monte Carlo simulations, necessarily biases the unfolded distributions towards SM-like shapes. It also introduces additional correlations between neighbouring bins which can broaden the $\chi^2$\footnote{Experimental resolved measurements are now provided at hadron level as well.}.

Both of these problems can be attacked by employing `boosted' top reconstruction techniques. Rather than the standard `resolved' reconstruction techniques used in the analyses of chapter 3, which require the decay products from the top to be relatively well-separated in the detector, these are optimised for events in which the top is produced at very large $p_T$ so that its decay products are collimated, and can be captured in a single large radius \emph{fat} jet, in contrast to the typical one-to-one parton-jet matching of a resolved event reconstruction. This has the potential to dramatically increase our sensitivity to the high-$p_T$ region. In addition, boosted reconstruction necessitates a hadron level description, so the model-dependence of the constraints induced by the unfolding procedure can also be mitigated. It is instructive to quantify how much they can improve the limits from the Run I fit. This is the subject of the first part of this chapter. 

In the second part of this chapter, we move away from hadron colliders and study the role that future lepton colliders can play in this endeavour, focusing on the two most mature proposed colliders: the International Linear Collider (ILC) and Compact Linear Collider (CLIC). While in general lepton collider measurements will be sensitive to a different set of operators, there is overlap with LHC measurements, so that the sensitivities can be directly compared.

This chapter is structured as follows: In section \ref{sec:boost}, we discuss the improvements on the \ttbar constraints that can be made by employing boosted jet substructure techniques. We analyse the importance of improving experimental systematics as well as collecting larger statistics, and the gain that can be made when theory uncertainties are improved beyond their present values. We also study the implications of our constraints for the reach of the LHC for generic (perturbative) UV completions. In section \ref{sec:ttz} we discuss the potential for improving the bounds on the electroweak operators in the top quark sector of the SMEFT, which can be accessed at hadron colliders through the process \pp $\to$ \ttz. In section \ref{sec:lepcoll} we switch to lepton colliders, and compare the bounds on the operators of section \ref{sec:ttz} with the bounds that can be achieved with the forecasted capabilities of the ILC and CLIC colliders, before summarising in section \ref{sec:conc_ch4}.

\subsection{Improving the fit with boosted reconstruction}
\label{sec:boost}
Top pair production is (at leading order in $\alpha_s$) a 2 $\to$ 2 process, so the relevant observables which span the partonic phase space are the momentum transfer $\hat t$ and the partonic centre-of-mass energy $\hat s$. All other observables are functions of these parameters, of which the top quark transverse momentum is the most crucial in determining the quality and efficiency of the boosted tagging approach which we will employ here~\cite{Plehn:2009rk,Plehn:2010st,Plehn:2011tg,Altheimer:2013yza,Schaetzel:2013vka,Plehn:2011sj,Backovic:2013bga}. As discussed in chapters 2 and 3, at leading order in the Standard Model EFT, the operators that contribute to top pair production are: the three-gluon vertex operator \op{G}, the top chromomagnetic dipole moment operator \op{uG}, as well as six four-quark operators \op{4q}, which contribute at interference level through four linear combinations \op[1,2]{u,d}. To keep this chapter self-contained, these operators are again displayed in Eq.~\eqref{eqn:ttbarops2}.

\begin{equation}
\begin{split}
\lag{\ttbar} & \supset  \frac{\co{uG}}{\Lambda^2} (\bar{Q}\sigma^{\mu \nu} T^A u)\tilde \varphi G_{\mu\nu}^{A} +  \frac{\co{G}}{\Lambda^2}  f_{ABC} G_{\mu}^{A \nu}G_{\nu}^{B \lambda} G_{\lambda}^{C \mu} + \frac{\co{\varphi G}}{\Lambda^2} (\varphi^\dagger \varphi)G_{\mu\nu}^{A}G^{A \mu\nu} \\
& + \frac{\co[1]{qq}}{\Lambda^2}(\bar{Q}\gamma_{\mu}Q)( \bar{Q}\gamma^{\mu}Q) +  \frac{\co[3]{qq}}{\Lambda^2}(\bar{Q}\gamma_{\mu}\tau^IQ)( \bar{Q}\gamma^{\mu}\tau^I Q) +  \frac{\co{uu}}{\Lambda^2}(\bar{u}\gamma_{\mu}u)( \bar{u}\gamma^{\mu} u) \\
& + \frac{\co[8]{qu}}{\Lambda^2}(\bar{Q}\gamma_{\mu}T^AQ)( \bar{u}\gamma^{\mu} T^Au) +  \frac{\co[8]{qd}}{\Lambda^2}(\bar{Q}\gamma_{\mu}T^AQ)( \bar{d}\gamma^{\mu} T^Ad) +  \frac{\co[8]{ud}}{\Lambda^2}(\bar{u}\gamma_{\mu}T^Au)( \bar{d}\gamma^{\mu} T^Ad) \, ,
\end{split}
\label{eqn:ttbarops2}
\end{equation}
and the four linear combinations of operators are
\begin{equation}
\begin{split}
 \op[1]{u} = &~\op[1,1331]{qq}+ \op[1331]{uu}+ \op[3,1331]{qq} \\
 \op[2]{u} = &~\op[8,1133]{qu} +  \op[8,3311]{qu} \\
 \op[1]{d} = &~4\op[3,1331]{qq}+\op[8,3311]{ud} \\
 \op[2]{d} = &~\op[8,1133]{qu} +  \op[8,3311]{qd}\,.
\end{split}
\label{eqn:linearcomb}
\end{equation}
\begin{figure}[t!]
\begin{center}
    \includegraphics[width=0.7\textwidth]{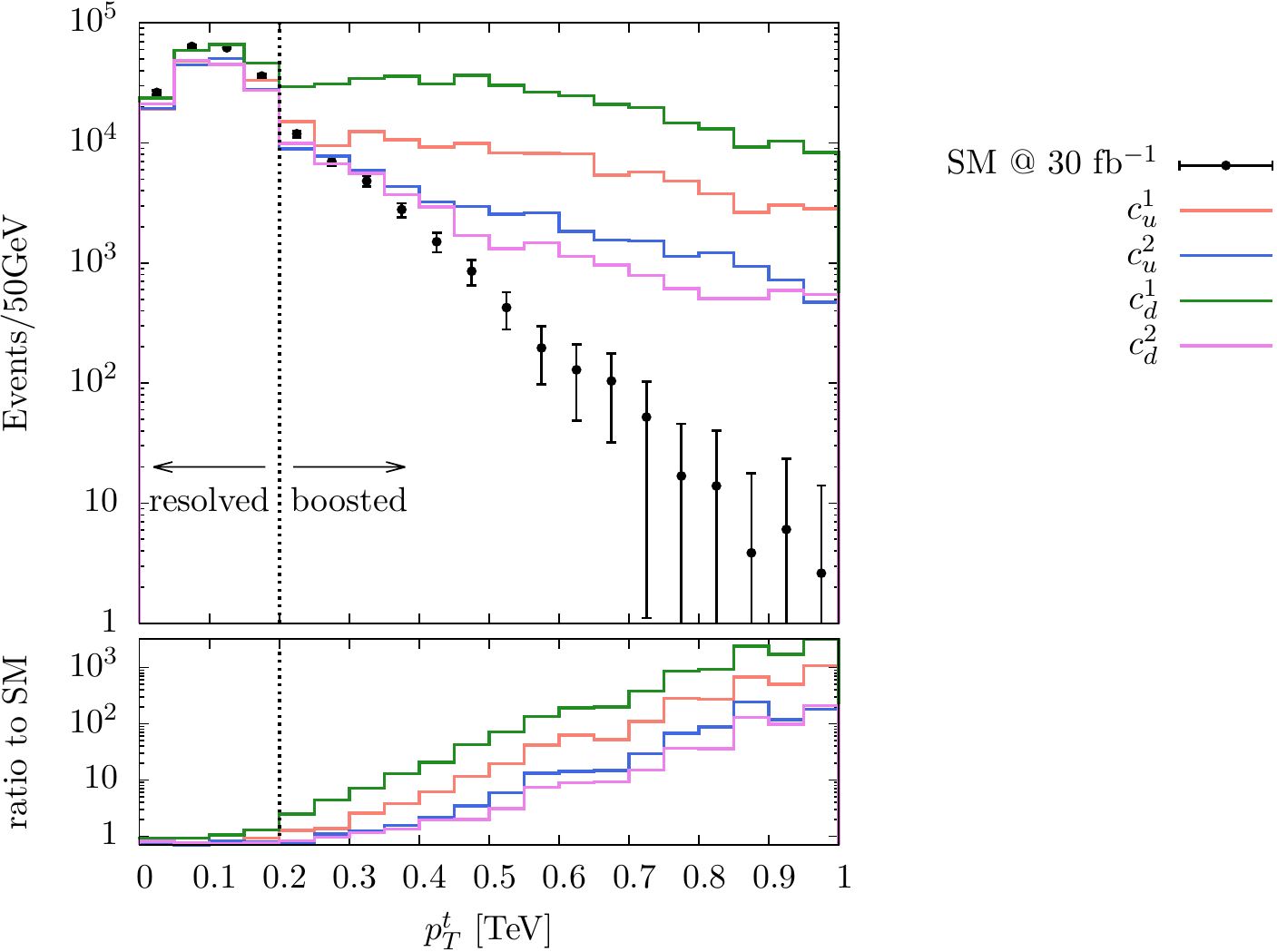}
    \caption[Hadronic top candidate transverse momentum in semileptonic \ttbar events.]{Transverse momentum distributions for the reconstructed hadronic top quark candidate. The bars represent 30~\ifb of pseudodata with \sqrts = 13 \tev constructed with the SM-only hypothesis, while the shaded curves include the effects of four-quark operators with Wilson coefficients \co{i} = 10 $TeV^{-2}$ for illustration. Details of the top quark reconstruction are described in the text. 
    \label{fig:boostedpt}}
\end{center}
\end{figure}
To emphasise that the effects of these operators are most pronounced at high $p_T$, in Fig.~\ref{fig:boostedpt} we plot the $p_T$ distribution of the hadronic top quark candidate (reconstructed as detailed below) in the Standard Model and for the four quark operator coefficients switched on to a (huge) value of 10 TeV$^{-2}$, showing the enhancement in the tail.

\subsubsection{Analysis details}

The events generated from {\textsc{MadEvent}} which sample the Wilson coefficient space are subsequently showered by {\textsc{Herwig++}}~\cite{Bahr:2008pv,Bellm:2015jjp}, which takes into account initial and final state radiation showering, as well as hadronisation and the underlying event. At this stage, all our predictions are at leading order in the Standard Model EFT. While considerable progress has recently been made in extending the effective Standard Model description of top quark physics to next-to-leading order~\cite{zhang,Degrande:2014tta}, the full description of top quark pair production is incomplete at this order. As in chapter 3, we take into account higher-order QCD corrections by re-weighting the Standard Model piece of our distributions to the NLO QCD prediction with $K$-factors, as obtained from {\textsc{Mcfm}}~\cite{Campbell:2010ff} and cross-checked with {\textsc{Mc@Nlo}}~\cite{Alwall:2014hca}. Recently, full NNLO results for top quark pair production have become available in~\cite{Czakon:2016ckf,Czakon:2013goa,Moch:2012mk}, we will comment on their potential for improving our results in Sec. \ref{sec:theory}.

We estimate scale uncertainties in the usual way: For the central value of the distributions we choose renormalisation and factorisation scales equal to the top quark mass $\mu_R = \mu_F = m_t$. Then we vary the scales independently over the range $m_t/2 < \mu_{R,F} < 2m_t$. PDF uncertainties are estimated by generating theory observables with the {\textsc{Ct14}}~\cite{Dulat:2015mca}, {\textsc{Mmht14}}~\cite{Harland-Lang:2014zoa} and {\textsc{Nnpdf3.0}}~\cite{Ball:2014uwa} as per the recommendations of the {\textsc{Pdf4Lhc}} working group for LHC run 2~\cite{Butterworth:2015oua}, and we take the full scale+PDF envelope as our theory band. This defines an uncertainty on the differential $K$-factor which we propagate into each observable. We treat theory uncertainties as uncorrelated with experimental systematics and take them to be fixed as a function of luminosity unless stated otherwise.

Our study focuses on \ttbar production at the LHC with \sqrts = 13 \tev. For simplicity we focus on the semileptonic decay channel $pp \to \ttbar \to q \bar q^\prime b l \nu_l b $, where $l \in \{e,\mu\}$ and $q \in \{u,d,s\}$, which strikes the best balance between the large rate but involved jet combinatorics of the fully hadronic channel and the clean signal but low rate and two-neutrino ambiguity of the dilepton case. The sting in the tail for analyses selecting high $p_T$ objects is, of course, low rates. In \ttbar production, for instance, only 15\% of the cross-section comes from the region $p_T \gtrsim 200$ \gev. We thus aim to quantify at what stage of the LHC programme, if any, the increased sensitivity in the boosted selection can overcome the poorer statistics relative to the resolved selection. We thus construct an analysis which targets both regions simultaneously. Our analysis setup, as implemented in {\sc{Rivet}}~\cite{Buckley:2010ar}, is as follows (also shown in Tab.~\ref{tab:cuts}). 

\begin{table}[!t]
\def\arraystretch{1.05}
\begin{center}
\begin{tabular}{ll@{\qquad}ll} \hline
\textit{Leptons} & $p_T > 30$ GeV & \\
			& $|\eta|  < 4.2$ \\ 
\textit{Missing energy} & $E_T^{\text{miss}} > 30$ GeV & \\
\textit{Small jets} & anti-$k_T$ $R = 0.4$ \\
			& $p_T > 30$ GeV , $|\eta| < 2 $ \\	
\textit{Fat jets} & anti-$k_T$ $R = 1.2$ \\
			& $p_T > 200$ GeV , $|\eta| < 2 $ \\	\hline
\textbf{Resolved} & $\geq$ 4 small jets w/$\geq$ 2 b-tags \\
\textbf{Boosted} & $\geq$ 1 fat jet, $\geq$ 1 small jet w/ b-tag \\	 \hline		
\end{tabular}
\caption[Event selection criteria in the boosted analysis.]{\label{tab:cuts} Summary of the physics object definitions and event selection criteria in our hadron-level analysis.}
\end{center}
\end{table}

Firstly, we require a single charged lepton with $p_T > 30$ GeV\footnote{We do not consider $\tau$ decays here to avoid the more involved reconstruction.}, and find the $E_T^{\text{miss}}$ vector as the negative vector sum of the reconstructed momenta, which we require to have a magnitude $ > 30$ GeV. The leptonic $W$-boson is reconstructed from these by assuming it was produced on-shell. Final state hadrons\footnote{We do not consider a detector simulation, and B-hadrons are kept stable.} are then clustered into jets using the anti-$k_T$ algorithm~\cite{Cacciari:2008gp} implemented in {\textsc{FastJet}}~\cite{Cacciari:2011ma} in two separate groups with $ R=(0.4,1.2) $ requiring $p_T >(30,200) $ \gev respectively, and jets which overlap with the charged lepton within $\Delta R = 0.3$ are removed. The constituents of the $R=1.2$ fat jets are reclustered with the Cambridge-Aachen algorithm~\cite{Dokshitzer:1997in,Wobisch:1998wt}, with all fat jets required to be within $|\eta| < 2$, and the $R=0.4$ small jets are b-tagged within the same $\eta$ range with an efficiency of $70\%$ and fake rate of $1\%$~\cite{ATLAS:2012ima}. 

If at least one fat jet and one b-tagged small jet which does not overlap with the leading fat jet exists, we perform a boosted top-tag of the leading fat jet using the \textsc{HEPTopTagger}~\cite{Plehn:2009rk,Plehn:2010st,Kasieczka:2015jma} algorithm.

The {\sc{HEPTopTagger}} procedure is a multistep algorithm optimised to isolate the characteristic three-prong pattern of a hadronically decaying top quark ($t \to Wb \to q\bar q^\prime b$). It can be used efficiently for top quark $p_T$ as low as 200 \gev, provided the radius of the large-$R$ jet is large enough to capture all the decay products. Beginning with a Cambridge-Aachen jet $J$ of radius $R_{\text{fat}}$ we undo the last step of the clustering, giving two subjets $j_1$ and $j_2$  (defined by $m_{j_1} > m_{j_2}$). A \emph{mass-drop} criterion is applied on the heavier subjet:
\begin{equation}
m_{j_1} /m_J < \mu_{\text{frac}},
\end{equation}
where $\mu_{\text{frac}}$ is a tuneable parameter. If this criterion is not met the jet is discarded. If it is met, the criterion is applied iteratively on both subjets until all subjets either have masses less than some input parameter $m_{\text{cut}}$ or the jet constituents (tracks, calorimeter deposits) are reached, in which case further unclustering is impossible. If at the end of this unclustering stage, there are less than three subjets, the jet $J$ is discarded. Among the subjets, all possible combinations of 3 subjets are formed (these are called \emph{triplets}). The constituents of the subjets in each triplet are reclustered using the C/A algorithm with a size parameter $R_{\text{filt}} = \min[0.3,\Delta R_{j_1,j_2}/2]$, where $\Delta R_{j_1,j_2}$ is the smallest separation between any two subjets in the triplet. Any constituents of the original jet $J$ that are left outside the reclustered triplets are discarded. This procedure is generically referred to as \emph{filtering}.

All triplets with mass outside the range 140 \gev $\leq m_j \leq$ 200 \gev are rejected. If more than one is inside the range, the one closest to the top quark mass is selected, this triplet (with 3 or more subjets) is referred to as the top quark candidate. The $N_{\text{subjet}}$ highest-$p_T$ subjets of this triplet are chosen. From these subjets, exactly 3 jets are constructed using the C/A algorithm with distance parameter $R_{\text{jet}}$ on their constituents. Finally, invariant mass and geometrical requirements on these 3 subjets are applied, to isolate the presence of a $W$ boson decay, namely:
\begin{equation}
\begin{split}
&R_- < \frac{m_{23}}{m_{123}} < R_+ \quad \text{and} \quad 0.2 < \arctan \frac{m_{13}}{m_{12}} < 1.3  \\
&R^2_-\left(1+\left(\frac{m_{13}}{m_{12}}\right)^2\right) < 1 - \left(\frac{m_{23}}{m_{123}}\right)^2 < R^2_+\left(1+\left(\frac{m_{13}}{m_{12}}\right)^2\right) \quad \text{and} \quad \frac{m_{23}}{m_{123}}  > 0.35 \\
&R^2_-\left(1+\left(\frac{m_{12}}{m_{13}}\right)^2\right) < 1 - \left(\frac{m_{23}}{m_{123}}\right)^2 < R^2_+\left(1+\left(\frac{m_{12}}{m_{13}}\right)^2\right) \quad \text{and} \quad \frac{m_{23}}{m_{123}}  > 0.35 ,
\label{eqn:htt}
\end{split}
\end{equation}
where $R_\pm = (1\pm f_W)(m_W/m_t)$ and $f_W$ is a tuneable parameter of the algorithm between 0 and 1. If at least one of the criteria in Eq.~\eqref{eqn:htt} are met, the top quark candidate is considered `tagged'. The HEPTopTagger algorithm can achieve stable efficiencies of around 30\% with background contamination of 1\% for top candidate $p_T$ ranges from 200 \gev to over 1 \tev (this is the $p_T$ range we consider). To summarise then, the tuneable parameters of the {\sc{HepTopTagger}} algorithm and their optimal values for our analysis are:
\begin{equation} 
\{\mu_{\text{frac}}, m_{\text{cut}}, N_{\text{subjet}}, R_{\text{jet}}, f_W \} = \{0.8,30 \text{ GeV}, 5, 0.3, 0.15 \} .
\end{equation}
The leptonic top candidate is reconstructed using the leading, non-overlapping (we require $\Delta R(l,j) > 0.4$ for all jets) b-tagged small jet and the reconstructed leptonic $W$. If no fat jet fulfilling all the criteria exists, we instead require at least 2 b-tagged small jets and 2 light small jets. If these exist we perform a resolved analysis by reconstructing the hadronic $W$-boson by finding the light small jet pair that best reconstructs the $W$ mass, and reconstruct the top candidates by similarly finding the pairs of reconstructed $W$-bosons and b-tagged small jets that best reconstruct the top mass.

Finally, regardless of the approach used, we require both top candidates to have $|m_\text{cand} - m_\text{top}| < 40$ GeV. If this requirement is fulfilled the event passes the analysis.

\subsubsection{Results}

\begin{figure}[t!]
\begin{center}
     \includegraphics[width=0.6\textwidth]{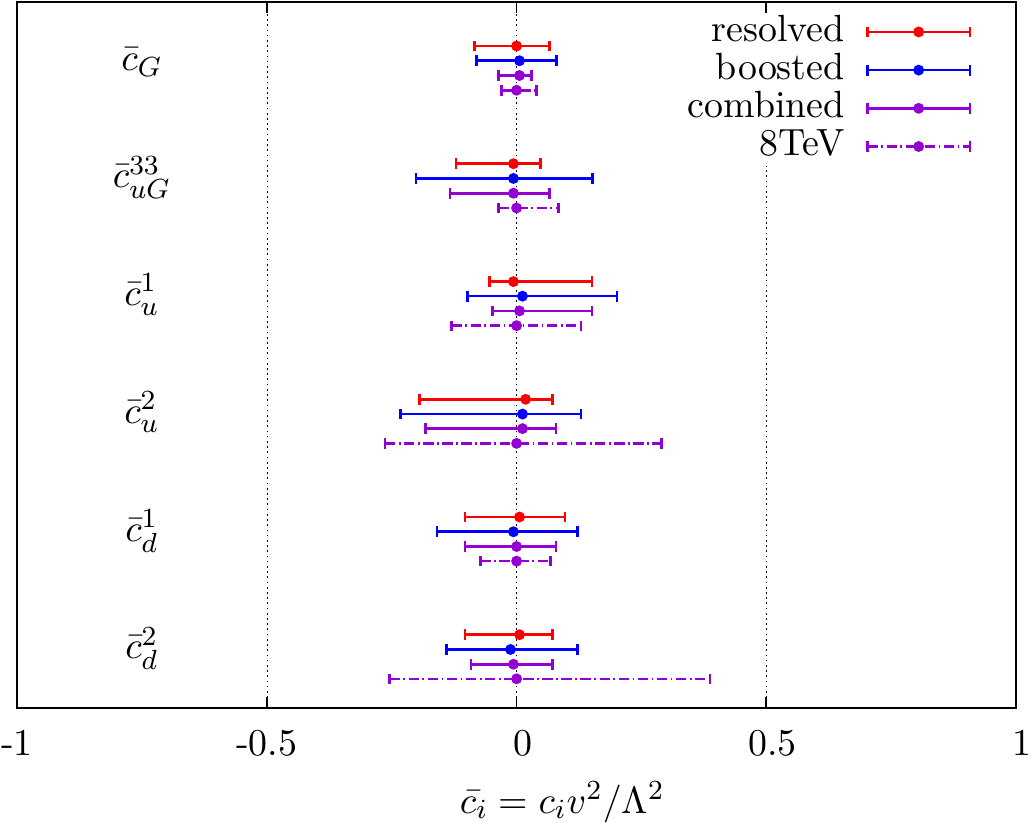}
         \caption[95\% confidence bounds from the boosted and resolved analysis.]{Individual 95\%~bounds on the operators considered here, from the boosted analysis and the resolved fat jet analysis, and the combined constraint from both, assuming 20\%~systematics and 30~\ifb of data. We also show existing constraints from unfolded 8 TeV $p_T$ distributions published in~\cite{Khachatryan:2015oqa} and~\cite{Aad:2015mbv}, showing the sizeable improvement even for a modest luminosity gain.} \label{fig:constraintsnew}
  \end{center}
\end{figure}

\begin{figure*}[t!]
\begin{center}
     \includegraphics[width=0.95\textwidth]{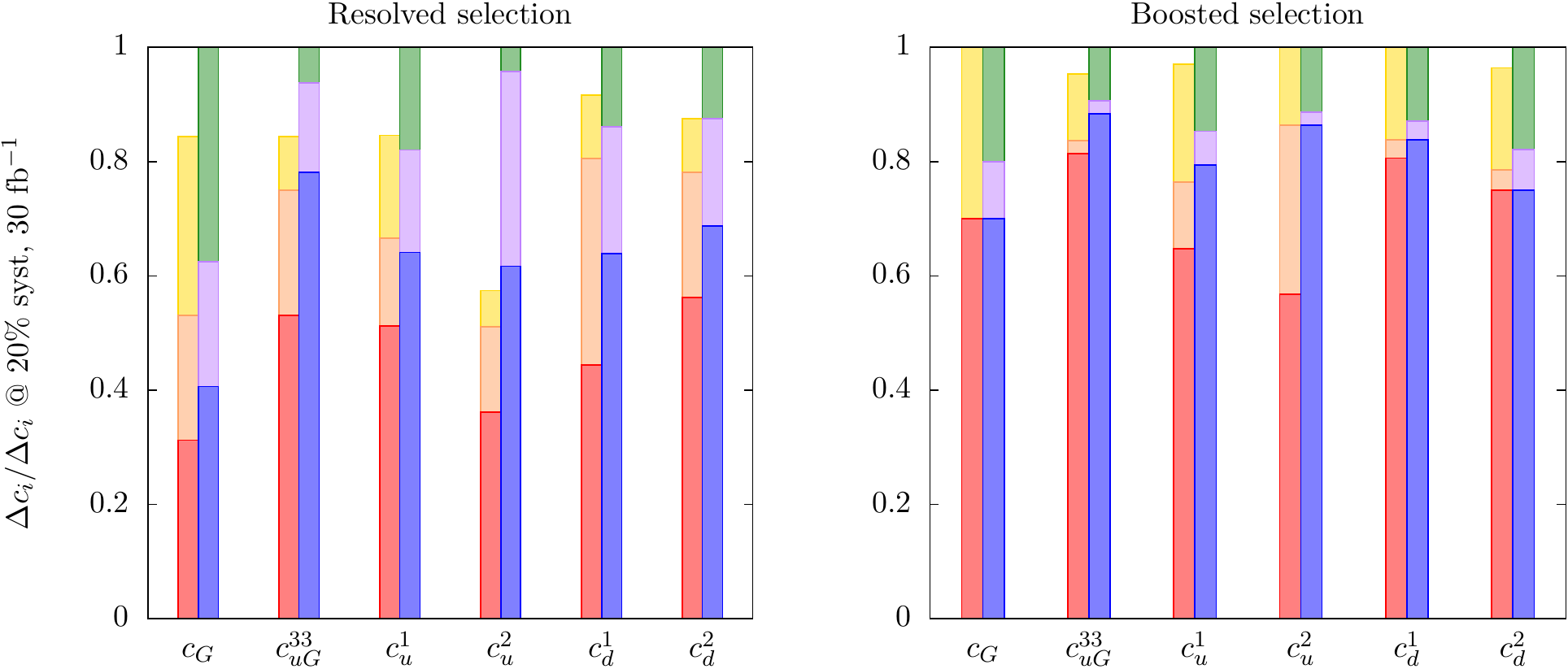}
         \caption[Fractional improvement on the constraints for various systematics and luminosity benchmarks.]{Fractional improvement on the 95\%~confidence intervals for the operators considered here, with various combinations of luminosity and experimental systematics considered. We take the width of the 95\% confidence limit obtained from 20~\% systematic uncertainty and 30~\ifb of data as a baseline (green bar), and normalise to this, i.e. we express constraints as a fractional improvement on this benchmark. The purple and blue bars represent respectively, 300~\ifb and 3~\iab of data, also at  20\% systematics, while the yellow, orange and red are the analogous data sample sizes for 10\% systematics. \label{fig:norm_constraints}}
  \end{center}
\end{figure*}

\subsubsection*{Impact of experimental precision}
Using a sample size of 30~\ifb with a flat 20\% systematic uncertainty (motivated by typical estimates from existing experimental analyses by ATLAS~\cite{Aad:2015hna} and CMS~\cite{Khachatryan:2016gxp}) on both selections as a first benchmark, and the $p_T$ distribution of Fig.~\ref{fig:boostedpt}, the 1-dimensional 95\% confidence intervals on the operators considered here are presented in Fig.~\ref{fig:constraintsnew}. All the bounds presented here are `one-at-a-time', i.e. we do not marginalise over the full operator set. Our purpose here is to highlight the relative contributions to the allowed confidence intervals here, rather than to present a global operator analysis.

As a general rule, the increased sensitivity to the Wilson coefficients offered by the boosted selection is overpowered by the large experimental systematic uncertainties in this region, and the combined limits are dominated by the resolved top quarks. The exception to this rule is the coefficient \co{G} from the operator \op{G} = $f_{ABC} G^{\mu,A}_\nu G^{\nu,B}_\lambda G^{\lambda,C}_\mu.$ Expanding out the field strength tensors leads to vertices with up to six powers of momentum in the numerator, more than enough to overcome the na\"{i}ve $1/\hat{s}^2$ unitarity suppression. Large momentum transfer final states thus give stronger bounds on this coefficient, even with comparatively fewer events.

With these constraints as a baseline, it is then natural to ask by how much they can be improved upon when refinements to experimental precision are made. The constraints are presented in Fig.~\ref{fig:norm_constraints} for different combinations of systematic and statistical uncertainties. We take the width of the 95\% confidence interval in Fig.~\ref{fig:constraintsnew} as our normalisation (the green bars), and express the fractional improvements on the limits that can be achieved relative to this baseline, for each operator. The right bars (green, purple, blue) represent 20\% systematic uncertainties with, respectively 30, 300 and 3~\iab of data. The left bars (yellow, orange, red) represent the same respective data sample sizes, but with 10\% systematic uncertainties. 

Beginning with the resolved selection, we find that the limits on the
coefficient \co{G} can be improved by 40\% by going from 30~\ifb to
300~\ifb, and by a further 20\% when the full LHC projected data
sample is collected. Systematic uncertainties have a more modest
effect on this operator: at 3~\iab the limit on \co{G} is only
marginally improved by a 10\% reduction in systematic
uncertainty. This merely reflects that \co{G} mostly impacts the high
$p_T$ tail, so it can only be improved upon in the threshold region by
collecting enough data to overcome the lack of sensitivity. 8
TeV measurements are already constraining the relevant phase space
region efficiently and the expected improvement at 13 TeV is only mild
(see below).

For the chromomagnetic dipole operator \op[33]{uG}, improving the experimental systematics plays much more of a role. A 10\% improvement in systematics, coupled with an increase in statistics from 30~\ifb to 300~\ifb leads to stronger limits that maintaining current systematics and collecting a full 3~\iab of data. Similar conclusions apply for the four-quark operators, to varying degrees, i.e. reducing systematic uncertainties can provide comparable improvements to collecting much larger data samples.

For the boosted selection, the situation is quite different. For all the operators we consider, improving systematic uncertainties by 10\% has virtually no effect on the improvement in the limits. This simply indicates that statistical uncertainties dominate the boosted region at 30~\ifb. For \co{G}, at 300~\ifb some improvement can be made if systematics are reduced, however we then see that systematic uncertainties saturate the sensitivity to \co{G}, i.e. there is no improvement to be made by collecting more data. For \co[33]{uG}, a modest improvement can also be made both by reducing systematics by 10\% and by increasing the dataset to 300~\ifb. However, going beyond this, the improvement is minute. The four-quark operators again follow this trend, although \co[2]{u} shows much more of an improvement when going from 300~\ifb to 3~\iab. 

\begin{figure*}[!t]
\begin{center}
 \includegraphics[width=0.42\textwidth]{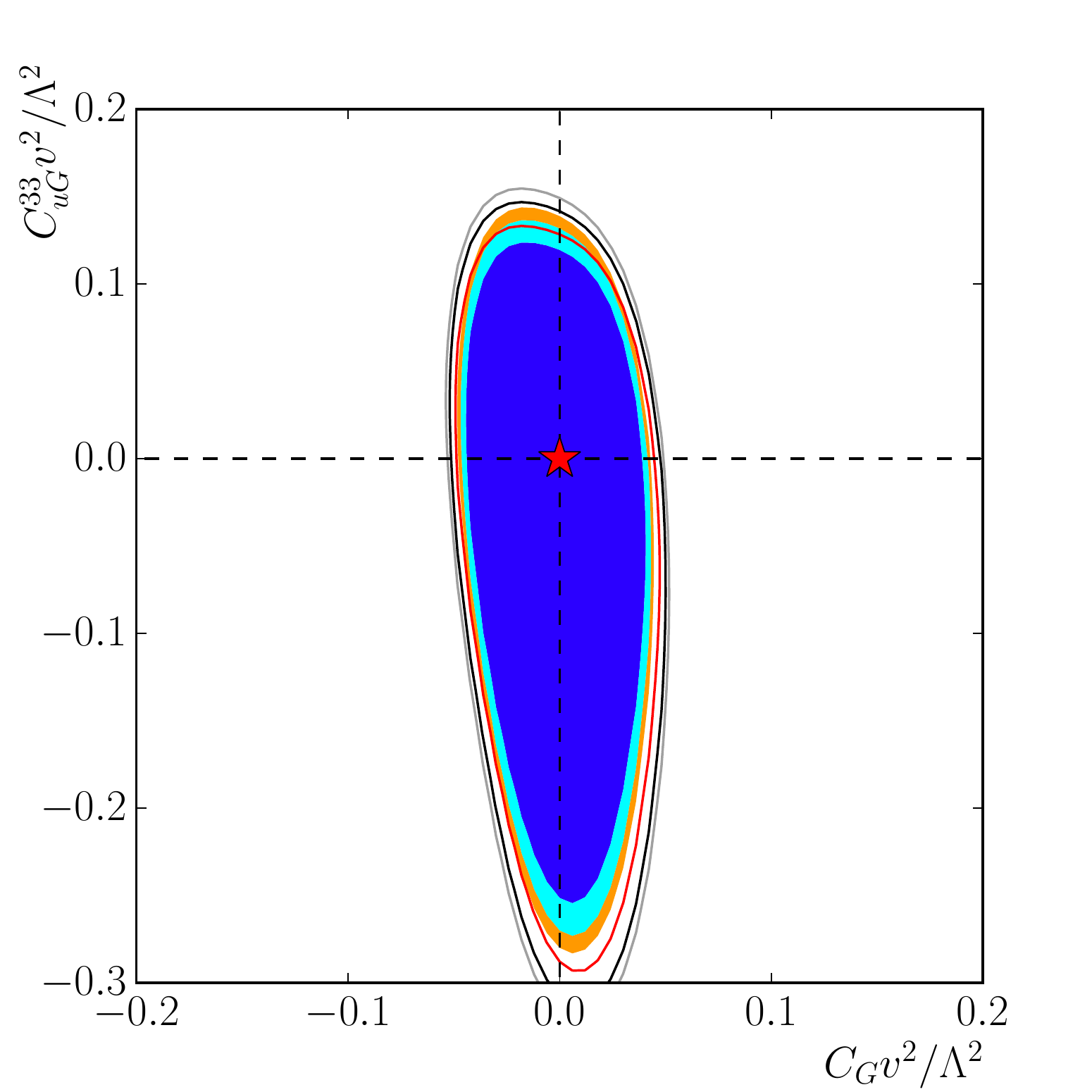}
  \hspace{1cm}
  \includegraphics[width=0.42\textwidth]{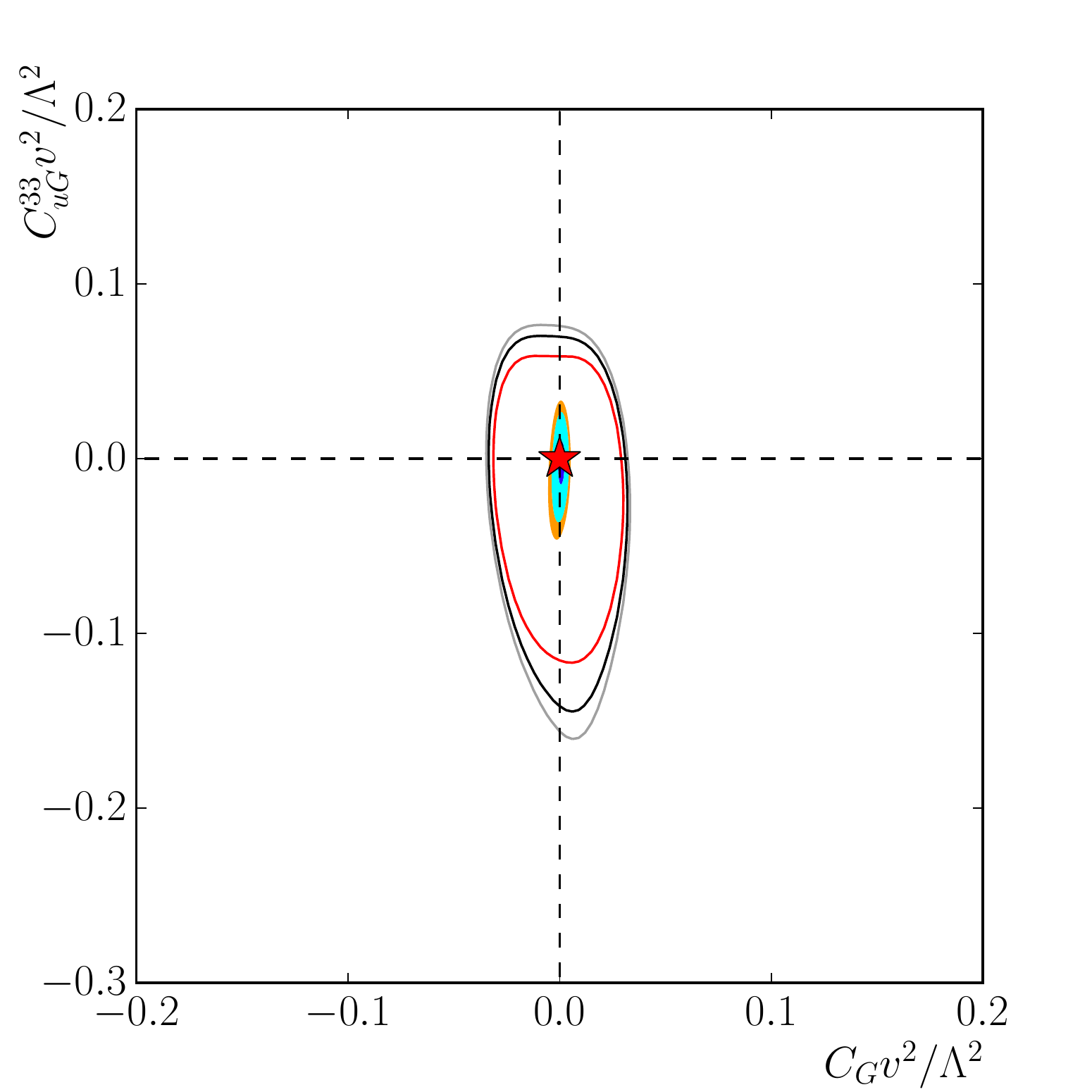}
  \caption[2D confidence intervals with and without theory uncertainties.]{Left: 68\%, 95\% and 99\% confidence intervals for \co{G} and \co[33]{uG}, the lines are obtained using experimental (20\% systematics and 30~\ifb of data) uncertainties along with theoretical uncertainties, the filled contours using only experimental uncertainties. Right: the same plot, but using 10\% systematics and 3~\iab of data, showing the much stronger impact of theory uncertainties in this region.}
\label{fig:contours}
\end{center}
\end{figure*}

\subsubsection*{The role of theory uncertainties}
\label{sec:theory}
The other key factor in the strength of our constraints is the uncertainties that arise from theoretical modelling. The scale and PDF variation procedure typically leads to uncertainties in the 10-15\% range. Fully differential $K$-factors for top pair production at NNLO QCD (i.e. to order $\mathcal{O}(\alpha_s^4)$) have become available, which have substantially reduced the scale uncertainties. The numbers quoted in Refs. \cite{Czakon:2015owf,Czakon:2016ckf} are for the Tevatron and 8 TeV LHC, and available only for the low to intermediate $p^t_T$ range ($p^t_T < 400$ \gev). Updated results for 13 TeV have become available only recently~\cite{Czakon:2016dgf}. It is worthwhile to ask what impact such an improvement could have on the constraints.

We put this question on a firm footing by showing in Fig. \ref{fig:contours} the 2D exclusion contours for the coefficients \co{G} and \co[33]{uG}, as obtained from combining the boosted and resolved limits, at fixed luminosity and experimental systematics, first using our NLO theory uncertainty, and also using \textit{no theory uncertainty at all}. For 30~\ifb the improvement is limited, indicating that at this stage in the LHC programme the main goal should be to first improve experimental reconstruction of the top quark pair final state. However, at 3~\iab the improvement is substantial, indicating that it will also become necessary to improve the theoretical modelling of this process, if the LHC is to augment its kinematic reach for non-resonant new physics. 

In addition to SM theoretical uncertainties, there are uncertainties relating to missing higher-order terms in the EFT expansion. Uncertainties due to to loop corrections and renormalisation-group flow of the operators \op[(6)]{i} are important for measurements at LEP-level precision~\cite{Berthier:2015gja,Berthier:2016tkq} where electroweak effects are also resolved. However, at the LHC we find them to be numerically insignificant compared to the sources of uncertainty that we study in detail here.  In addition, there is also the possibility of large effects due to \D8 operators, particularly owing to additional derivatives in the EFT expansion. Since the interference effects of omitted \D8 operators are formally of the same order as the retained quadratic terms in the \D6 operators, we emphasise that the numerical constraints presented here should be treated with caution. The only way to be certain that the omission of these terms is justified is to compute the effects of the interference of the relevant \D8 operators to a given process and demonstrate them to be small. This has been shown to be true for the $gg\to t\bar{t}$ subprocess\cite{Cho:1993eu,Cho:1994yu}. However, due to the large number of operators present there, this has not been studied for the $q\bar{q}\to t\bar{t}$ process. We leave a full computation of these effects as a future direction of study.

\subsubsection{Interpreting the results}

\begin{figure}[t!]
\begin{center}
     \includegraphics[width=0.55\textwidth]{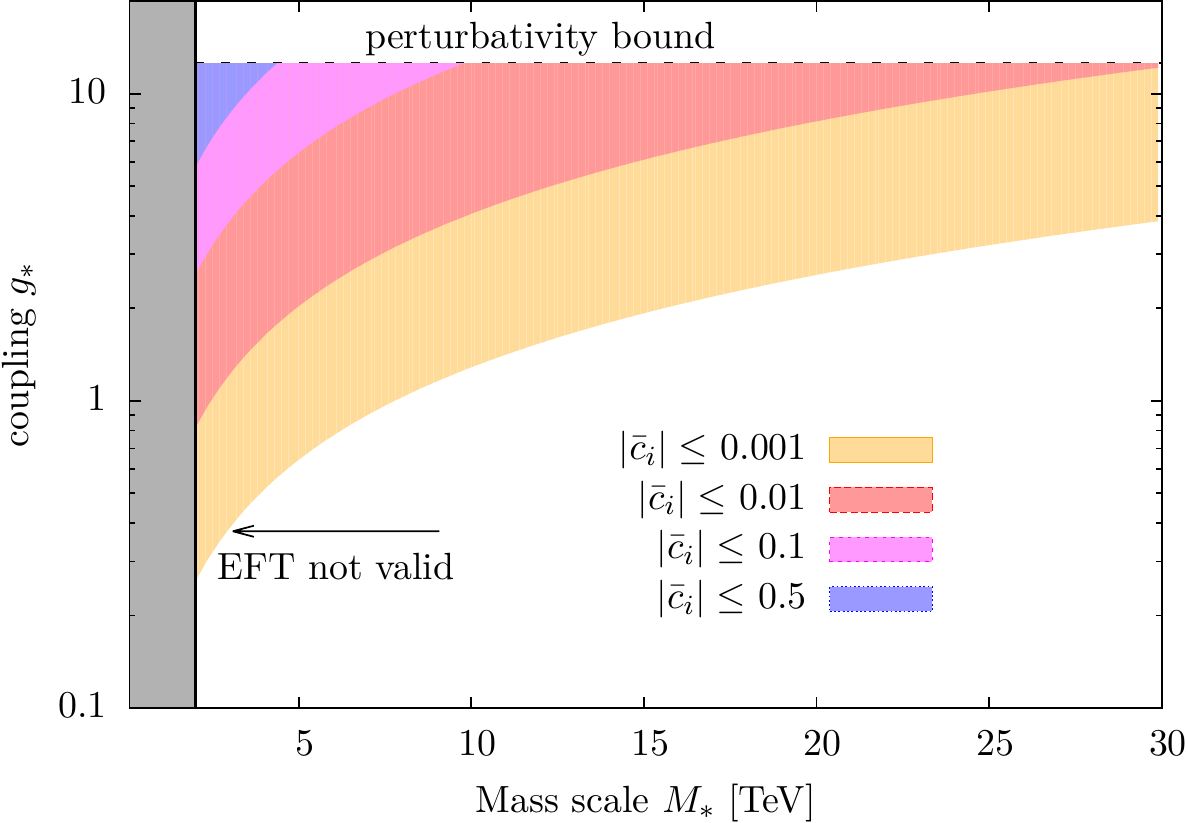}
         \caption[Impact of constraints on a generic new physics $g_*$-$M_*$ plane.]{Areas in the new coupling-BSM mass scale plane (see also \cite{Englert:2014cva}), resulting from our fit coverage. Shaded areas are constrained in perturbative UV completions at a scale $M_\ast$, subject to the boundary condition Eq.~\eqref{eqn:matching}. The shaded grey area is probed by the pseudodata of our fit. We do not consider unitarity bounds in this work.} 
         \label{fig:eft_validity}
  \end{center}
\end{figure}

The whole purpose of the EFT approach is to serve as a bridge between the Standard Model and heavy degrees of freedom residing at some unknown mass scale $M_*$. Connecting the EFT to this scale, however, necessarily involves making assumptions about the couplings of this new physics. We can make statements about the relation between the constraints presented here and such a scale, however, by making general assumptions, such as perturbativity of the underlying new physics. 

Consider, for example, the simple case where the perturbative UV physics is characterised entirely by a single coupling $g_*$ and a unique mass scale $M_*$. Such a scenario could arise from integrating out a heavy, narrow resonance. In this case we have the simple tree-level matching condition
\begin{equation}
\frac{\co{i}}{\Lambda^2} = \frac{g_*^2}{M_*^2}.
\label{eqn:matching}
\end{equation}
Constraints on \co{i} then map onto allowed regions in the $g_*$-$M_*$ plane. In Fig. \ref{fig:eft_validity} we sketch these regions for illustrative values of \co{i}. In order for the EFT description of a given mass region to be valid, we must not resolve it our measurement. Therefore we impose a hard cut at $\sqrt{s} = 2$ TeV, obtained from the maximum $t\bar{t}$ invariant mass probed in our SM pseudodata. We also impose a generic perturbativity restriction $g_* \lesssim 4\pi$ to ensure that our EFT expansion is well-behaved and higher-dimensional operators do not affect the power counting.

We see that for large Wilson coefficients $\cb{i} \gtrsim 0.5 $ only a very small window of parameter space may be constrained, but the weak limits push the underlying coupling to such large values that loop corrections are likely to invalidate the simple relation of Eq.~\eqref{eqn:matching}, making it hard to trust these limits. However, at 3 \iab, the projected constraints are typically $\cb{i} \lesssim 0.01 $, therefore, even for moderate values of the coupling $g_*$, our constraints are able to indirectly probe mass scales much higher than the kinematic reach of the LHC.

\subsubsection{Discussion}
The special role of the top quark in BSM scenarios highlights the importance of searches for new interactions in the top sector. Taking the lack of evidence of resonant new physics in the top sector at face value \cite{Aad:2012em,Aad:2012ans,Chatrchyan:2012yca}, we can assume that new interactions are suppressed by either weak couplings or large new physics scales. In both cases we can analyse the presence of new physics using effective field theory techniques. A crucial question that remains after the results from the LHC run 1 is in how far a global fit from direct search results will improve with higher statistics and larger kinematic coverage. We address this question focusing on the most abundant top physics-related channel $pp\to t\bar t$, which probes a relevant subset of top quark effective interactions. 
In particular, we focus on complementary techniques of fully-resolved vs. boosted techniques using jet-substructure technology, which are affected by different experimental systematic uncertainties. Sensitivity to new physics is a trade off between small statistical uncertainty and systematic control for low $p_T$ final states at small new physics-induced deviations from the SM expectation (tackled in fully-resolved analyses) and the qualitatively opposite situation at large $p_T$. For the typical parameter choices where top-tagging becomes relevant and including the corresponding efficiencies, we can draw the following conclusions:

\begin{itemize}
\item Boosted top kinematics provide a sensitive probe of new interactions in $t\bar t$ production mediated by modified trilinear gluon couplings. In particular, this observation shows how differential distributions help in breaking degenerate directions in a global fit by capturing sensitivity in phenomenologically complementary phase space regions.

\item The sensitivity to all other operators detailed in Eq.~\eqref{eqn:ttbarops2} is quantitatively identical for boosted and fully-resolved analyses for our choice of $p_T^{\text{boost}}\geq 200~\text{GeV}$. Increasing the boosted selection to higher $p_T$ (where the top tagging will become more efficient) will quickly move sensitivity to new physics effects to the fully resolved part of the selection. The boosted selection is saturated by large statistical uncertainties for the for the typical run 2 luminosity expectation. These render systematic improvements of the boosted selection less important in comparison to the fully resolved selection, which provides an avenue to set most stringent constraint from improved experimental systematics. Similar observations have been made for boosted Higgs final states~\cite{Butterworth:2015bya} and are supported by the fact that the overflow bins in run 1 analyses provide little statistical pull~\cite{Buckley:2015lku}.

\item Theoretical uncertainties that are inherent to our approach are
  not the limiting factors of the described analysis in the forseeable
  future, but will become relevant when statistical uncertainties
  become negligible at very large integrated luminosity.
  
\end{itemize}

Boosted analyses are highly efficient tools in searches for resonant
new
physics~\cite{Joshi:2012pu,Aad:2012em,Aad:2012ans,Chatrchyan:2012yca}. Our
results show that similar conclusions do not hold for non-resonant new
physics effects when the degrees of freedom in question do not fall inside the
kinematic coverage of the boosted selection anymore. Under these
circumstances, medium $p_T$ range configurations which maximise new
physics deviation relative to statistical and experimental as well as
theoretical uncertainty are the driving force in setting limits on
operators whose effects are dominated by interference with the SM
amplitude in the top sector. This also implies that giving up
  the boosted analysis in favor of a fully resolved analysis extending
  beyond $p_T^t\geq 200~\gev$ will not improve our results
  significantly. The relevant phase space region can be accessed with fully resolved techniques, with a large potential for improvement from the
experimental systematics point of view.

\subsection{Associated $Z$ production projections}
\label{sec:ttz}
We have seen that there are good improvement prospects for the \ttbar \D6 operators. Of the constraints listed in chapter 3, however, by far the weakest are those extracted from top quark neutral couplings in \ttz and $t\bar t\gamma$ production. This is because these processes have a much smaller rate, so at this stage of the LHC programme their measurements are currently statistics dominated. It is also natural to then ask how they may be improved over the lifetime of the LHC. This is the subject of this section.

\subsubsection{Top electroweak couplings}
\label{sec:topew}

In the SM, the electroweak \ttz coupling is given by the vector-axial-vector coupling 
\begin{equation}
\lag{ttZ} = e \bar t\left[ \gamma^\mu(v_t-\gamma_5 a_t)\right] t Z_\mu
\label{eqn:ttz}
\end{equation}
where
\begin{equation}
\begin{split}
v_t &= \frac{T_t^3 - 2Q_t \sin^2 \theta_W}{2\sin\theta_W\cos\theta_W} \simeq 0.24, \\
a_t &= \frac{T_t^3}{2\sin\theta_W\cos\theta_W} \simeq 0.60.
\end{split}
\end{equation}
To capture effects beyond the SM in this Lagrangian there are two approaches: one can write down anomalous couplings for the \ttz vertex, such that \lag{ttZ} receives a term
\begin{equation}
\Delta \lag{ttZ} =  e \bar t  \left[   \gamma^\mu(C_{1V} + \gamma_5C_{1A}) 
			  +   \frac{i\sigma^{\mu\nu}q_\nu}{2M_Z}(C_{2V} +\gamma_5C_{2A}) \right]t Z_\mu ,
\label{eqn:anomcoups}
\end{equation}
where $q = p_t -p_{\bar t}$. While this has the advantage of elucidating the various spin structures that can impact the \ttz vertex, it has the drawback that it does not allow for a simple power counting of which anomalous couplings would have the strongest effect. For example, the coefficient $C_{2A}$ is zero in the Standard Model, so that any corrections to it come solely from new physics contributions, which should be smaller than couplings that have SM interference. 

To augment this description, one can instead supplement Eq.~\eqref{eqn:ttz} with higher-dimensional operators. At leading order in the SMEFT, the list of operators that generate modifications to the \ttz vertex is, expressed in the basis and notation of Ref.~\cite{Grzadkowski:2010es}:
\begin{equation}
\begin{split}
\op{uW} &= (\bar Q\sigma^{\mu\nu}u)\tau^I\tilde \varphi W^I_{\mu\nu} \\
\op{uB}  &= (\bar Q\sigma^{\mu\nu}u)\tilde \varphi B_{\mu\nu} \\
\op[(3)]{\varphi q} &= (\varphi^\dagger i\overleftrightarrow{D^I_\mu} \varphi)(\bar Q \tau^I \gamma^\mu Q) \\
\op[(1)]{\varphi q} &= (\varphi^\dagger i\overleftrightarrow{D_\mu} \varphi)(\bar Q\gamma^\mu Q) \\
\op{\varphi u} &= (\varphi^\dagger i\overleftrightarrow{D_\mu} \varphi)(\bar u \gamma^\mu u) . \\
\end{split}
\label{eqn:ttzd6ops}
\end{equation}
The dictionary between the \D6 operators of Eq.~\eqref{eqn:ttzd6ops} and the anomalous couplings of Eq.~\eqref{eqn:anomcoups} is
\begin{equation}
\begin{split}
C_{1V} &= \frac{v^2}{\Lambda^2}\Re\left[\co[(3)]{\varphi q}-\co[(1)]{\varphi q}-\co{\varphi u}\right]^{33} \\
C_{1A} &= \frac{v^2}{\Lambda^2}\Re\left[\co[(3)]{\varphi q}-\co[(1)]{\varphi q}+\co{\varphi u}\right]^{33} \\
C_{2V} &= \sqrt{2}\frac{v^2}{\Lambda^2}\Re\left[\cos\theta_W\co{uW}-\sin\theta_W\co{uB}\right]^{33} \\
C_{2A} &= \sqrt{2}\frac{v^2}{\Lambda^2}\Im\left[\cos\theta_W\co{uW}+\sin\theta_W\co{uB}\right]^{33},
\end{split}
\end{equation}
where the superscript 33 denotes that we are considering the 3rd generation only in the fermion bilinears of Eq.~\eqref{eqn:ttzd6ops}. Since \co[(3)]{\varphi q} and \co[(1)]{\varphi q} only appear with an overall opposite sign, we can only constrain the operator $\op[(3)]{\varphi q} - \op[(1)]{\varphi q} \equiv \op {\varphi q}$ from \ttz couplings. We will discuss a method for bounding the two operators independently later in the chapter.

\begin{figure}[t!]
  \begin{center}
    \includegraphics[width=0.6\textwidth]{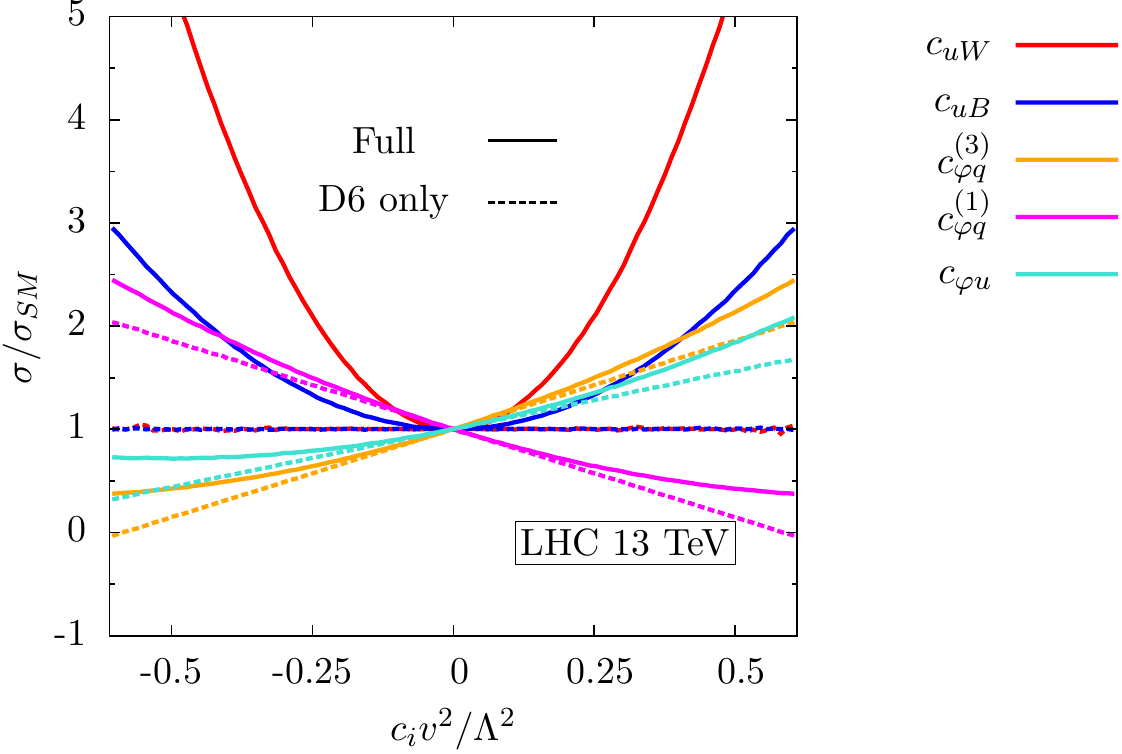}
    \end{center}
    \caption[\ttz cross-section in the presence of \D6 operators.]{Ratio of the full SM $pp \to \ttz$ cross-section with the operators of Eq.~\eqref{eqn:ttzd6ops} switched on individually to the NLO Standard Model estimate. The dashed lines show the contribution from the interference term, and the solid lines show the full dependence.\label{fig:ttzxsec}}
\end{figure}

$C_{2A}$ is generated by a \cp-odd combination of operators, therefore it does not interfere with SM amplitudes and so its effects are expected to be smaller. Since in this study we are more interested in the absolute mass scales of these operators, we set all Wilson coefficients to be real, however we note that \cp-sensitive observables such as angular distributions can also distinguish the \cp character of the Wilson coefficients. We also assume that the new physics solely impacts the \ttz vertex, so we do not consider operators which modify the $Ze\bar e$ vertex, nor four-fermion operators which can contribute to the $q\bar q \to \ttbar$ or $\ep \to \ttbar$ processes (see e.g. Refs.~\cite{Buckley:2015lku,Rosello:2015sck} for constraints on the former).

\subsubsection{Total rates}

To appreciate the impact of the operators of Eq.~\eqref{eqn:ttzd6ops}, in Fig.~\ref{fig:ttzxsec} we plot the ratio of the full \ttz cross-section with each operator switched on individually, to the NLO SM prediction, taken from Ref.~\cite{Maltoni:2015ena}. For ease of interpretation, we split up the cross-section into the contribution from the interference term and the quadratic term. We see firstly that the operators \op{uW} and \op{uB} have the strongest impact on the total cross-section, but this comes purely from the squared term (this was also noted in Ref.~\cite{Bylund:2016phk}). The remaining operators have a milder effect on the cross-section, but their interference term dominates. We also see that the operators \op[(3)]{\varphi q} and \op[(1)]{\varphi q} contribute the same dependence but with an opposite sign, as discussed in Sec.~\ref{sec:topew}, therefore we can only bound the linear combination \op{\varphi q}. 

\begin{figure}[t!]
\begin{center}
 \includegraphics[width=0.55\textwidth]{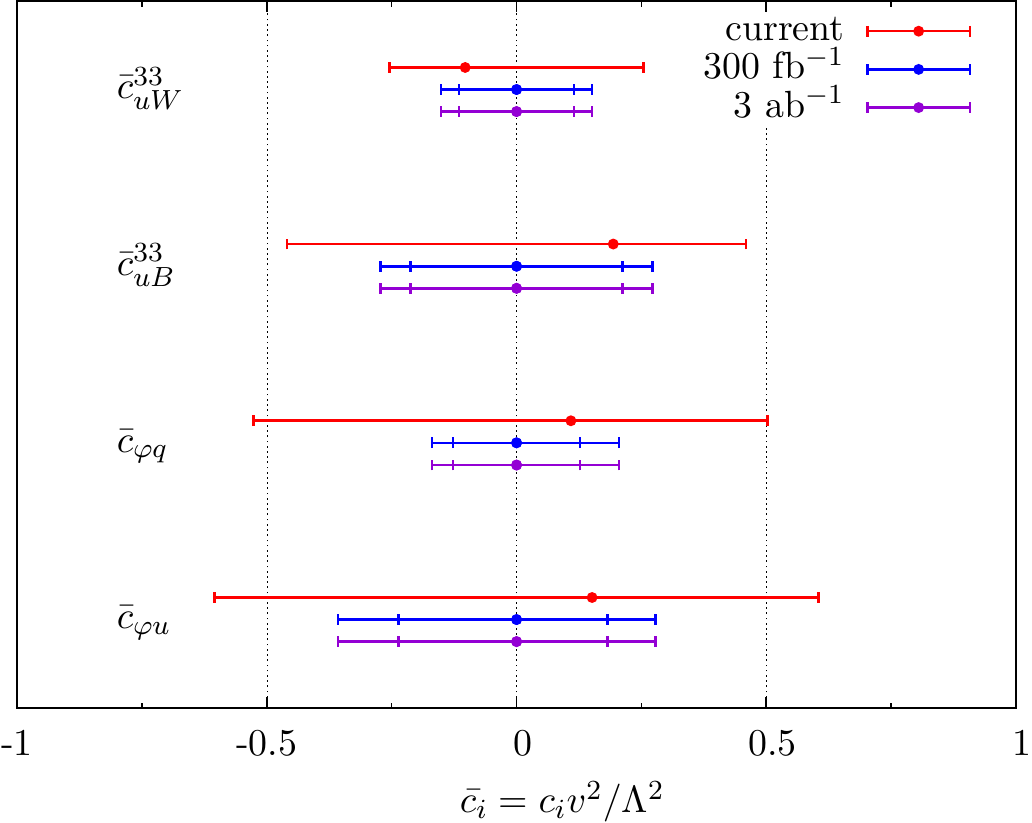}
 \caption[Current and projected constraints from \ttz production.]{Individual 95\% confidence intervals on the coefficients of the operators of Eq.~\eqref{eqn:ttzd6ops} using the current 13~\tev measurements (red bars). Also shown are the projected constraints using 300~\ifb (blue) and 3~\iab (purple) of SM pseudodata. For the latter two cases, the inner bars show the improvement when theory uncertainties are reduced to 1\%.} \label{fig:ttzprojections}
 \end{center}
 \end{figure}

The LHC bounds on the coefficients of these operators from 8 \tev \ttz production cross-sections were presented in chapter 3. The current constraints are weak. Since then, ATLAS and CMS have presented measurements using 13 \tev  collision data, with measured values 0.9 $\pm$ 0.3 \pb~\cite{Aaboud:2016xve} and 0.7 $\pm$ 0.21 \pb~\cite{CMS-PAS-TOP-16-017}, respectively. The constraints on the operators using these two measurements are shown in Fig.~\ref{fig:ttzprojections}, where the coefficients are normalised to the `bar' notation \cb{i} = $\co{i}v^2/\Lambda^2$, and the operators are switched on individually. 

We see that the current constraints are still quite weak, mainly due to the large ($\sim$ 30\%) experimental uncertainties. These measurements are currently statistics dominated, so it is instructive to ask what the expected improvement is over the lifetime of the LHC. Using a constant systematic uncertainty of 10\% based on the current estimate, we also plot in Fig.~\ref{fig:ttzprojections} the constraints using 300 \ifb and 3 \iab of SM pseudodata. We see that there will be an improvement by factors of $1.5$ to $2$ by the end of Run III, but after this the measurement is saturated by systematics.

To highlight the benefits of improving the theory description in tandem, we also show in Fig.~\ref{fig:ttzprojections} the projected constraints if theory uncertainties are improved to 1\% from the current \ord{10\%} precision, which does not seem unreasonable over the timescales we are considering. We see again that there will be no subsequent improvement after 300~\ifb unless experimental systematics are reduced.

\subsubsection{Impact of differential distributions}

\begin{figure}[t!]
\begin{center}
 \includegraphics[width=0.95\textwidth]{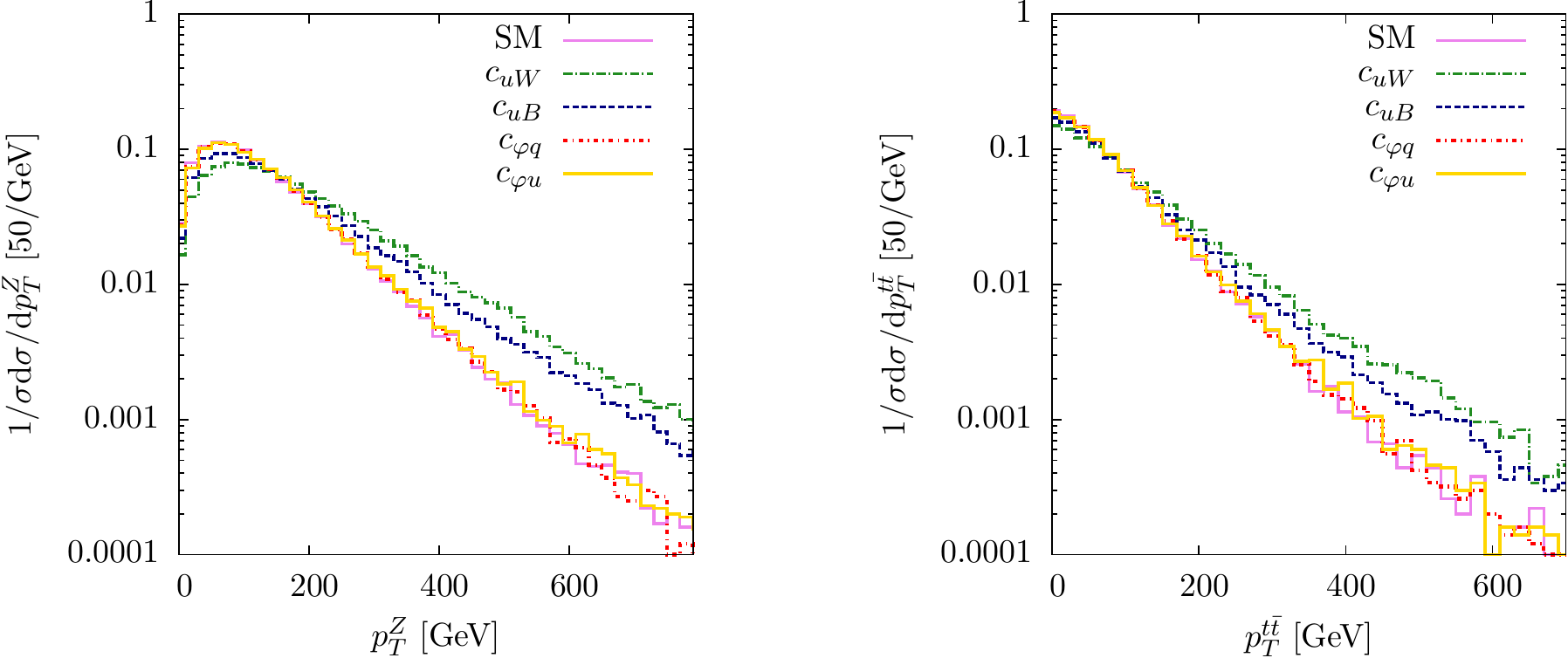}
 \caption[Differential distributions in \ttz production.]{Kinematic distributions in $pp \to$ \ttz production at 13 \tev for the SM prediction and for the operators of Eq.~\eqref{eqn:ttzd6ops} switched on to their maximum value allowed by current data. Left: the $Z$ boson transverse momentum spectrum. Right: $p_T^{t\bar t} = p_T^t -p_T^{\bar t}$ spectrum. All distributions are normalised to the total cross-section. Shape differences can be seen in the tails for the operators \op{uW} and \op{uB}, showing that differential distributions provide complementary information to overall rates.}
\label{fig:ttzdistributions}
 \end{center}
 \end{figure}

Finally, it should be noted that as more data becomes available, it may be possible to measure \ttz cross-sections differentially in final state quantities. Since cuts on the final state phase space can enhance sensitivity to the region where na\"ive power counting says \D6 operators become more important, differential distributions could substantially improve the fit prospects, as has already been demonstrated for \ttbar production.~\cite{Buckley:2015lku,Englert:2016aei}. 

To illustrate this, in Fig.~\ref{fig:ttzdistributions} we plot the distributions for the $Z$ boson transverse momentum and top pair transverse momentum, both for the SM only case and with each operator switched on to a value of $\cb{i} \simeq 0.3$; approximately the maximum allowed by current constraints in Fig.~\ref{fig:ttzprojections}. We see that extra enhancement in the tail is visible for the field strength tensor operators \op{uW} and \op{uB}, due to the extra momentum dependence in the numerator from the field strength tensor. For the $\varphi$-type operators, since the interference is solely proportional to $\varphi^\dagger\varphi \to v^2/\Lambda^2$, there is no extra enhancement at high $p_T$.

We do not estimate the improvement of the fit by taking these distributions into account, since this would require proper estimates of experimental systematics and tracking the nontrivial correlations between the kinematic quantities in the massive 3-body final state. Here, we merely comment that it may be an avenue worth pursuing as more data becomes available.

\subsection{Future collider prospects}
\label{sec:lepcoll}

We see that despite the impressive statistical sample of top quark data that enters these fits, the subsequent direct bounds on the operator coefficients, and, by extension, the scale of new physics that would generate those operators, are rather weak~\cite{Buckley:2015lku}. There are few top quark measurements at the LHC that can be considered ``precision observables" (helicity fractions in top decays are an exception~\cite{AguilarSaavedra:2006fy}). By dimensional analysis, the strength of the interference of these operators with energy \sqrts typically scales as $s/\Lambda^2$, and the areas of phase space that are most sensitive are plagued by correlated experimental and theoretical systematics. Moreover, the associated weak limits translate into values of $\Lambda$ that are probed by the high energy bins of the measurement, bringing into question the validity of the truncated EFT description~\cite{Contino:2016jqw} and care needs to be taken when combining measurements of different exclusive energy ranges of a binned distribution~\cite{Englert:2014cva}. Inclusive cross-sections, being typically dominated by the threshold region $\sqrts \sim (2)m_t $, are under more theoretical control, but bring far less sensitivity.

Lepton colliders are not vulnerable to either of these problems. Firstly, there is excellent control over the hard scale of the interaction $\sqrt{s}$, so one can always ensure that the limits on the \D6 operators are consistent with a well-behaved EFT expansion.\footnote{Consistently improving the perturbative precision within the dimension 6 framework, however, makes the truncation of the perturbative series necessary as corrections to $(D=6)^2$ operators will typically require unaccounted for $D=8$ counterterms.} Secondly, the theoretical uncertainties from Standard Model calculations are much smaller: there are no PDFs, and the current state of the art precision for \ttbar production is N$^3$LO QCD at fixed-order~\cite{Kiyo:2009gb}, and NNLO+NNLL including threshold resummation, which bring SM scale uncertainty variation bands to the percent level~\cite{Chen:2016zbz}. 

The physics case for a \ep collider is by now well-established. The principal motivation is to perform a detailed precision study of the couplings of the Higgs boson in the much cleaner environment that a lepton collider affords, which will bring Higgs coupling measurements to an accuracy that will not be challenged by the LHC, even after it collects 3~\iab of data~\cite{Klute:2013cx}.  The electroweak couplings of the top quark are also clearly within the remit of such a collider. Currently, the only handle on top quark electroweak couplings from the LHC is through the associated production $pp \to \ttbar V$ where $V \in \{Z,W,\gamma\}$. Whilst measurements of these processes are now approaching the 5$\sigma$ level, the pull that they have on a global fit is small~\cite{Buckley:2015lku}. Measurements of electroweak single top production bring stronger bounds, but are sensitive to a smaller subset of operators. 

At a lepton collider, on the other hand, the process $\ep \to Z^*/\gamma \to \ttbar$ is extremely sensitive to top electroweak couplings. While the overall rate is more modest than at the LHC due to the parametric $\alpha_{EW}/\alpha_s$ and $s$-channel suppression, the process is essentially background-free, and would constitute the first true precision probe of the electroweak sector of the top quark, and open up a new avenue for top quark couplings, complementary to the well-studied top QCD interactions. Several studies of the prospects for improvement of top measurements at future colliders have already been undertaken (see for example Refs.~\cite{Vos:2017ses,Vos:2016til,Amjad:2015mma,Rontsch:2015una,Coleppa:2017rgb,Cao:2015qta}), in particular for the proposed International Linear Collider (ILC), but none have explicitly quantified the gain in the constraints on the top electroweak sector of the SMEFT, nor provided a comparative study of different collider options. The remainder of this chapter provides such a study. 

Going beyond the LHC, currently, the most mature proposal is for a linear \ep collider with a centre of mass energy ranging from 250 \gev to up to 1 \tev. There are several scenarios for integrated luminosity and CM energy combinations. The most-studied is the so-called H-20 option, which involves running at 500 \gev for 500 \ifb of data, followed by 200 \ifb of data at the \ttbar threshold to perform detailed measurements of the top quark mass, and 300 \ifb of data at \sqrts = 250 \gev to maximise the machine's Higgs potential with high precision. After a luminosity upgrade, a further 3.5 \iab is gathered at \sqrts = 500 \gev, followed by another \sqrts = 250 \gev run at 1.5 \iab. Since we are most interested in the ILC mass reach for new physics, in this study we focus on the 500 \gev  ILC running.

An important parameter for lepton colliders is the energy spread of electron and positron beams (see e.g.~\cite{Boogert:2002jr}). In order to estimate the effect on our results, we use the results of \cite{Boogert:2002jr} to calculate the expected change in the cross-section by including the effects of initial state radiation, beam spread and beamstrahlung. We find that for the typical beam profile, the associated uncertainty is not a limiting factor and we neglect these effects in the following.

\begin{figure*}[t!]
\begin{center}
 \begin{subfigure}{0.42\textwidth}
 \includegraphics[width=\textwidth]{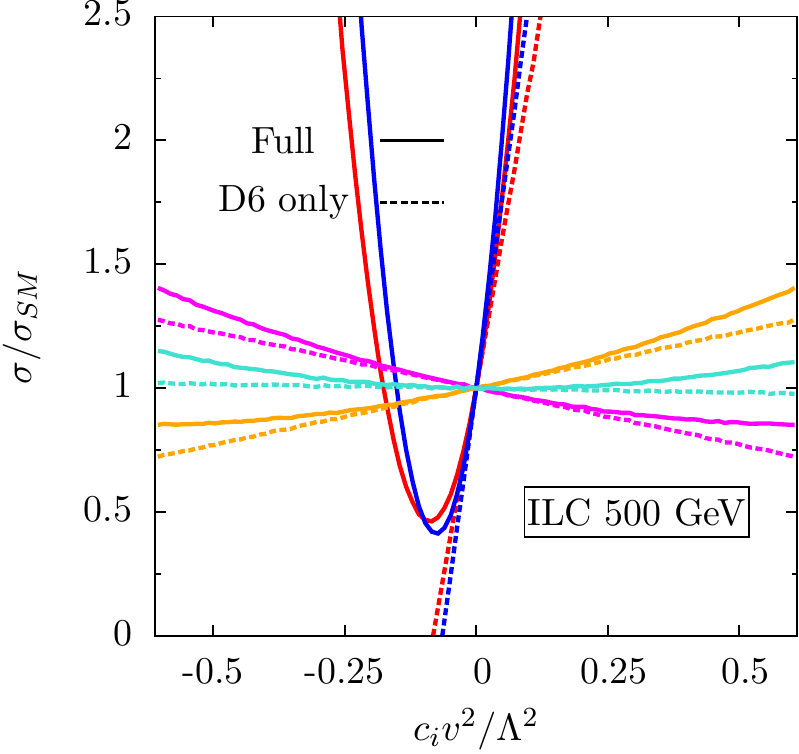}
 \end{subfigure}
\qquad
  \begin{subfigure}{0.42\textwidth}
 \includegraphics[width=\textwidth]{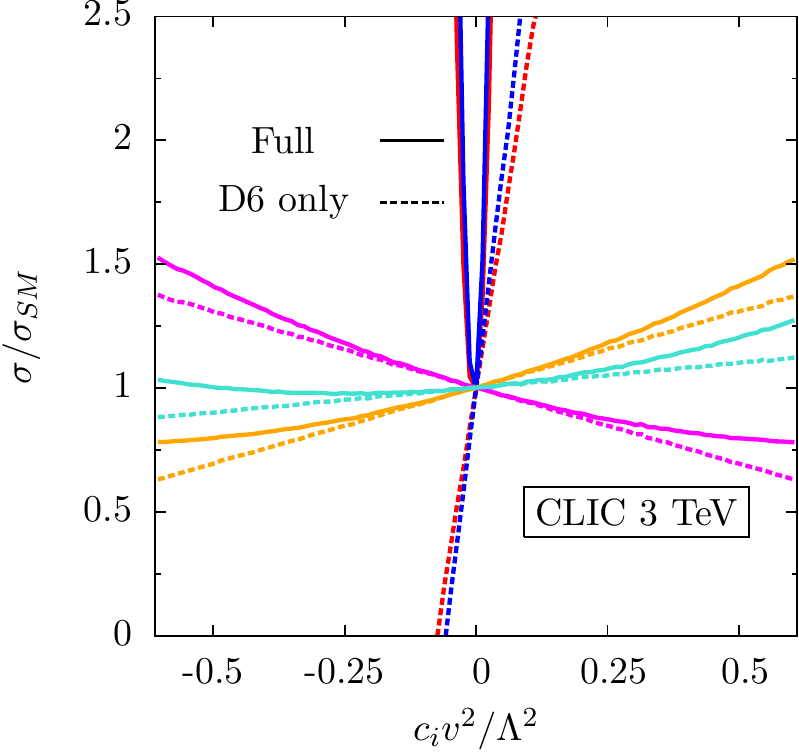}
 \end{subfigure}
   \caption[ILC and CLIC \ttbar cross-sections.]{Left: Ratio of the full SM \ep to \ttbar cross-section at \sqrts = 500 \gev with the operators of Eq.~\eqref{eqn:ttzd6ops} switched on individually to the NLO Standard Model estimate. The dashed lines show the contribution from the interference term, and the solid lines show the full dependence The operator colour-coding is the same as Fig.~\ref{fig:ttzxsec}. Right: Likewise for CLIC running at \sqrts = 3 \tev. \label{fig:ilcxsec}}
  \end{center}
\end{figure*}

\subsubsection{The \ttbar total cross-section}
Top pair production has a more modest rate here than at a hadron collider. The state-of-the-art Standard Model calculations for (unpolarised) \ep $\to$ \ttbar production are at N$^3$LO QCD~\cite{Kiyo:2009gb,Chen:2016zbz}, and at NLO EW~\cite{Fleischer:2003kk} (with partial NNLO results in Ref.~\cite{Gao:2014nva}) and predict a cross-section $\sigma \simeq $ 0.57 pb. The conventional scale variation gives a QCD uncertainty at the per-mille level. While this rate is more than a factor of a thousand smaller than at the 13 \tev LHC, the process is essentially background free. Thus, after even 500 \ifb of data the statistical uncertainty will be approximately  0.2\%, and so completely subdominant to the systematics. 

We can thus repeat the exercise of extracting the bounds on the coefficients of the operators of Eq.~\eqref{eqn:ttzd6ops} using SM pseudodata. As a guide for the expected numerical constraints, we also plot the ratio of the total cross-section in the presence of the operators to the SM prediction, this time using the total (unpolarised) cross-section at the 500 \gev ILC. This is shown on the left of Fig.~\ref{fig:ilcxsec}.

We see again that the operators \op{uW} and \op{uB} are the strongest, however, unlike the case of \ttz production the interference term dominates at small $\co{i}/\Lambda^2$. The result of this is that there is a cancellation between the interference and quadratic terms at approximately $\co{i}/\Lambda^2 \simeq -3~\tev^{-2}$, leading to a SM-like cross-section and a second, degenerate minimum in the $\chi^2$. The constraints obtained from a one-at-a-time fit of these operators to the 
SM pseudodata is shown in the red bars on the right of Fig.~\ref{fig:ilcglobal}. 

The operators \op{uW} and \op{uB} are very tightly constrained, due to their much stronger impact on the cross-section stemming from the extra momentum dependence flowing through the vertex. The $\varphi$-type operators are more weakly constrained, but on the whole the constraints are typically 100 times stronger than for the LHC \ttz production projections in Sec.~\ref{sec:ttz}, which is unsurprising giving the difference in precision. 

\begin{figure*}[t!]
\begin{center}
 \includegraphics[width=0.9\textwidth]{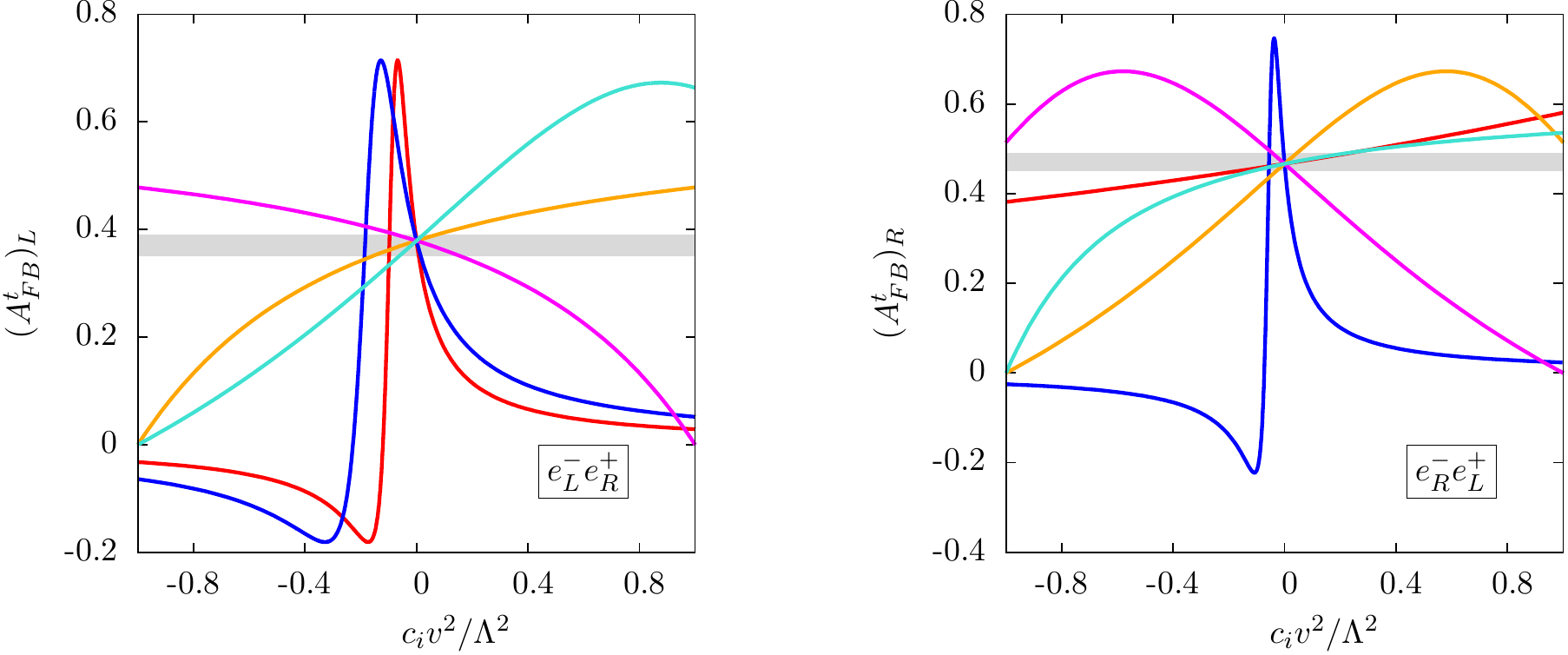}
 \caption[\ttbar forward-backward asymmetry at an \ep collider.]{Full dependence of the \ttbar forward-backward asymmetry of Eq.~\eqref{eqn:asymm} on the operators  of Eq.~\eqref{eqn:ttzd6ops} for left-handed polarised electrons (left) and right-handed polarised electrons (right) at \sqrts = 500 \gev, the operator colour coding is the same as Fig.~\ref{fig:ttzxsec}. We also show a 5\% uncertainty band around the SM prediction, to estimate the expected constraints. }
\label{fig:epasymmetries}
 \end{center}
 \end{figure*}

Individual constraints are less useful in practice, however. Firstly, in a plausible UV scenario that would generate these operators one would typically expect more than one to be generated at once, so that one-at-a-time constraints cannot be straightforwardly linked to a specific `top-down' model. Secondly, there can in general be cancellations between different operators for a given observable that can yield spurious local minima and disrupt the fit. This would not be visible in the individual constraints, and so would obscure degeneracies in the operator set that could be broken by considering different observables. Therefore, we also consider constraints where we marginalise over the remaining three coefficients in the fit, as also discussed in chapter 3. These are shown in the blue bars on the right of Fig.~\ref{fig:ilcglobal}.\footnote{Note, however, that a full marginalisation will be overly conservative when confronting a concrete UV model.}

We see that, with the exception of \op{uW} and \op{uB}, marginalising over the full operator set wipes out the constraints. This is because even for large values of coefficients, the pull that a particular operator has on the cross-section can easily be cancelled by another operator. We can conclude that, despite the impressive precision that can be achieved in extracting the cross-section, it has limited use in constraining new physics in a simultaneous global fit of several operators. It is worthwhile to  make use of other measurements.

\begin{figure}[t!]
\begin{center}
\begin{subfigure}{0.48\textwidth}
 \includegraphics[width=\textwidth]{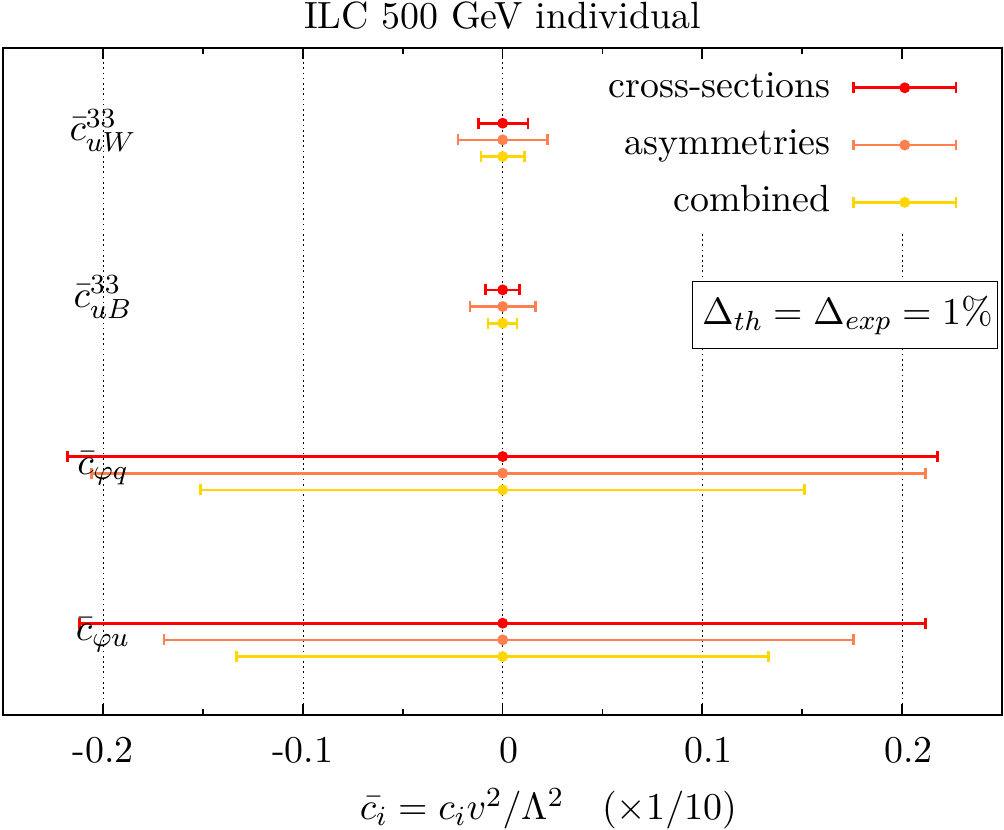}
 \end{subfigure}
\quad
  \begin{subfigure}{0.48\textwidth}
 \includegraphics[width=\textwidth]{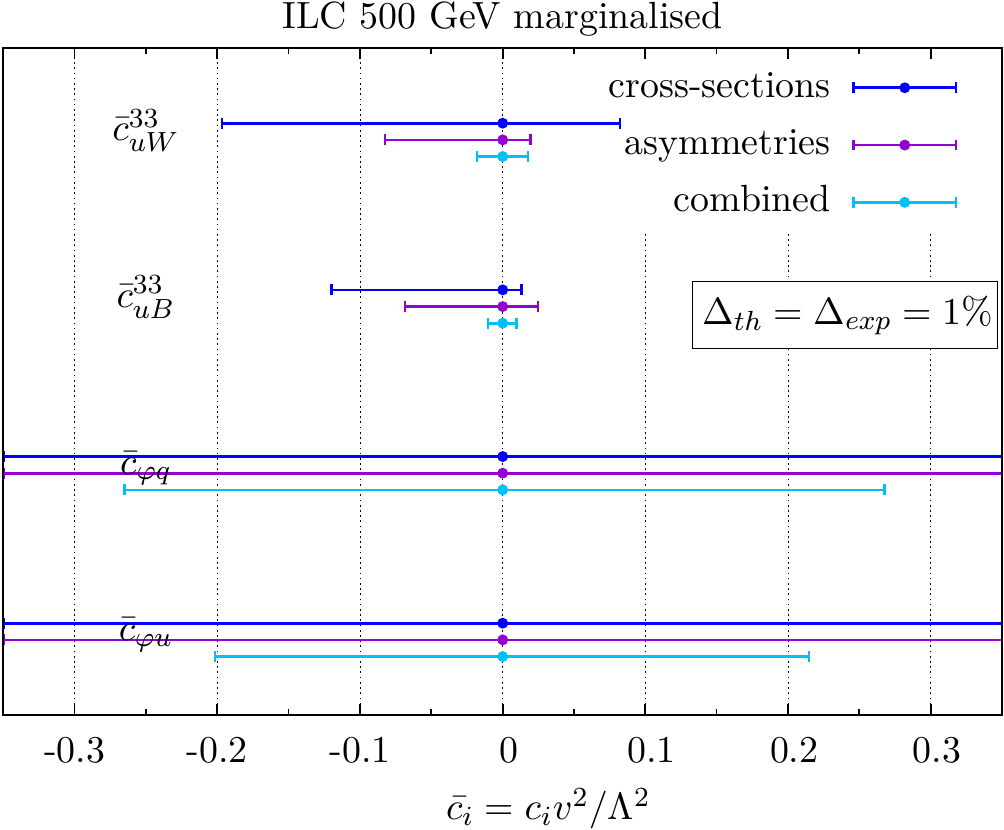}
 \end{subfigure}
   \caption[ILC constraints on top electroweak operators.]{95\% confidence ranges for the operators we consider here, from the 500 \gev ILC, assuming 1\% theoretical and experimental uncertainties, by fitting to cross-sections, asymmetries, and the combination, with each operator considered individually (left) or in a 5D fit (right). To display both on the same plot, we scale the individual constraints up by a factor of 10, so that the bottom axis is actually \cb{i}/10.\label{fig:ilcglobal}}
  \end{center}
\end{figure}

\subsubsection{Polarised beams}
One of the principal strengths of lepton colliders is that the polarisation of the incoming beams can be finely controlled, so that the relative contributions between different subprocesses to a given final state can be tuned. Moreover, because the dependence of top observables on the operators of Eq.~\eqref{eqn:ttzd6ops} depends strongly on the initial state polarisation, varying the settings increases the number of independent measurements that can be used to place bounds in a global fit.

To emphasise this point, we study the forward-backward asymmetry, defined as
\begin{equation}
A^t_{FB} = \frac{N(\cos\theta_t > 0)-N(\cos\theta_t < 0)}{N(\cos\theta_t > 0)+N(\cos\theta_t < 0)},
\label{eqn:asymm}
\end{equation}
where $\theta_t$ is the polar angle between the top quark and the incoming electron, for three incoming beam polarisation settings: unpolarised beams, denoted $(A^t_{FB})_U$; a fully left-handed initial polarised electron beam and fully right-handed polarised positron beam, denoted $(A^t_{FB})_L$; and vice versa, denoted $(A^t_{FB})_R$. The SM predictions for these settings at tree level are $\{(A^t_{FB})_U,(A^t_{FB})_L,(A^t_{FB})_R\} \simeq \{0.40, 0.37, 0.47\}$, which agree well with the full NNLO QCD estimates~\cite{Bernreuther:2006vp,Gao:2014eea}. The dependence of these asymmetries on the operators of Eq.~\eqref{eqn:ttzd6ops} is shown in Fig.~\ref{fig:epasymmetries}.

We see that the dependence on the operators distinctively depends on the initial state polarisations. For the $(A^t_{FB})_L$ case, we again see the large interference-square cancellation in the gauge-type operators \op{uW} and \op{uB}. For the right-handed case the impact of \op{uW} is much milder. For both cases we see that the operators \op[(3)]{\varphi q} and \op[(1)]{\varphi q} pull the prediction in opposite directions. Most encouragingly, we see that the departure from the SM prediction is now much stronger for the $\varphi$-type operators than for the total cross-section, which should lead to a sizeable improvement in the final constraints.

To generate these constraints, we consider a global fit of the four operators to six observables: 
\begin{equation}
\{(A^t_{FB})_U,(A^t_{FB})_L,(A^t_{FB})_R, (\sigma^{\ttbar}_{\text{tot}})_U, (\sigma^{\ttbar}_{\text{tot}})_L, (\sigma^{\ttbar}_{\text{tot}})_R \}. 
\end{equation}

In extracting the constraints, we consider the more realistic ILC polarisation capabilities $\mathcal{P}_{e^-} = \pm$ 0.8, $\mathcal{P}_{e^+} = \mp$ 0.3, noting that the cross-section for arbitrary \ep polarisations is related to the fully polarised one by~\cite{Hikasa:1985qi,MoortgatPick:2005cw}
\begin{equation}
\sigma_{\mathcal{P}_{e^-}\mathcal{P}_{e^+}} = \frac{1}{4} \{ (1+\mathcal{P}_{e^-})(1-\mathcal{P}_{e^+})\sigma_{\text{RL}}
+ (1-\mathcal{P}_{e^-})(1+\mathcal{P}_{e^+})\sigma_{\text{LR}} \} ,
\end{equation}
where $\sigma_{\text{RL}}$ is the cross-section for fully right-handed polarised electrons and fully left-handed polarised positrons and $\sigma_{\text{RL}}$ is vice versa (the $\sigma_{\text{RR}}$ and $\sigma_{\text{LL}}$ components vanish for $p$-wave annihilation into spin-1 bosons). Performing a $\chi^2$ fit of the full analytic expression for each observable, using SM pseudodata with 1\% experimental error bars (based on studies in Refs.~\cite{Amjad:2013tlv,Amjad:2015mma}) and SM theory uncertainties of 1\% (based on the calculations of Refs.~\cite{Kiyo:2009gb,Chen:2016zbz,Bernreuther:2006vp,Fleischer:2003kk,Gao:2014eea}) the individual and marginalised constraints on these operators are shown in Fig.~\ref{fig:ilcglobal}.

At the level of individual operators, the constraints are not improved drastically by adding in asymmetry information. For the global fit, however, the constraints lead to much stronger bounds than for fitting to cross-sections (although the marginalisation typically weakens the overall constraints by a factor \ord{100}).

We see that the constraints are again much stronger for the field strength operators \op{uW} and \op{uB}, where the constraints are at the $|\cb{i}|  \lesssim 10^{-4}$ level for the individual constraints and $|\cb{i}|  \lesssim 10^{-2}$ for the marginalised case, corresponding to a mass reach of $\Lambda \gtrsim 10$ \tev and $\Lambda \gtrsim 2.16$ \tev, respectively, assuming $\co{i} \simeq 1 $. The weakest constraints are on the operators \op[(3)]{\varphi q} (\op[(1)]{\varphi q}), which translate into bounds on $\Lambda$ of roughly 700 \gev. 

While it is encouraging that the bounds are consistent with an EFT formulation, in the sense that $\Lambda \gg \sqrt{s}$, the ILC mass reach for the scale of new physics that would generate these operators is still low. We note, however, that these bounds are on the conservative side, since other observables such as oblique parameters and LEP asymmetries contribute complementary information that will in general tighten them. To keep this fit self-contained, we postpone this discussion until Sec.~\ref{sec:lep}.

\subsubsection{CLIC constraints}
\label{sec:clic}

\begin{figure}[t!]
\begin{center}
\begin{subfigure}{0.48\textwidth}
 \includegraphics[width=\textwidth]{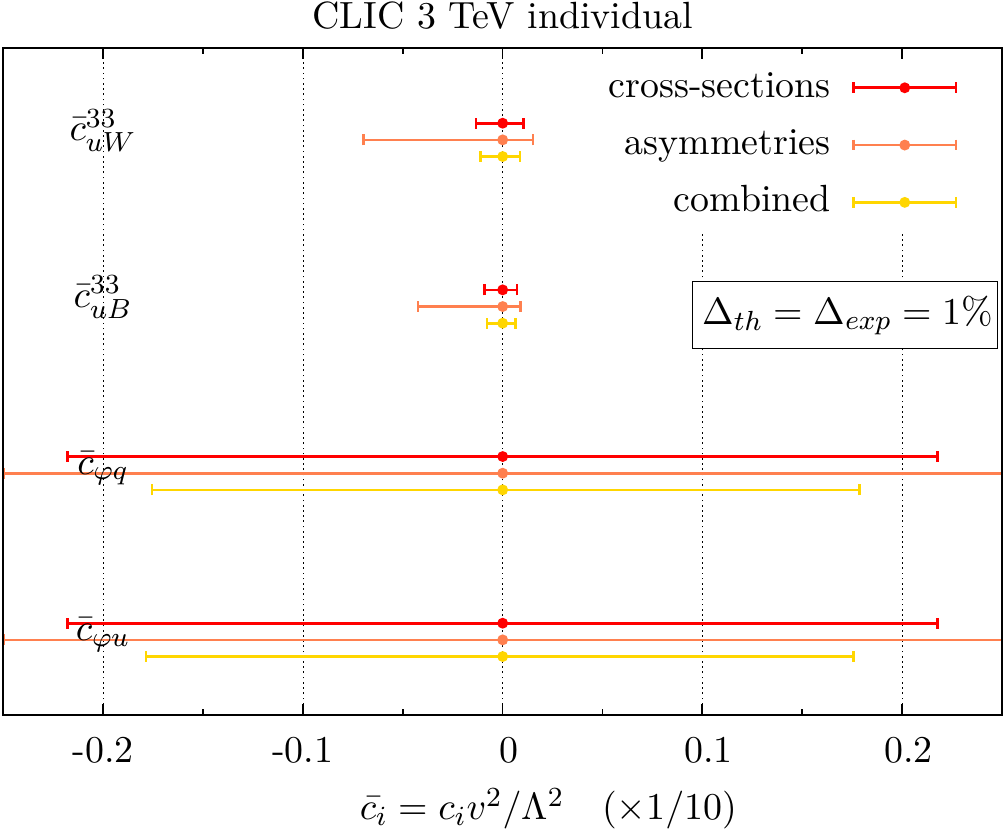}
 \end{subfigure}
\quad
  \begin{subfigure}{0.48\textwidth}
 \includegraphics[width=\textwidth]{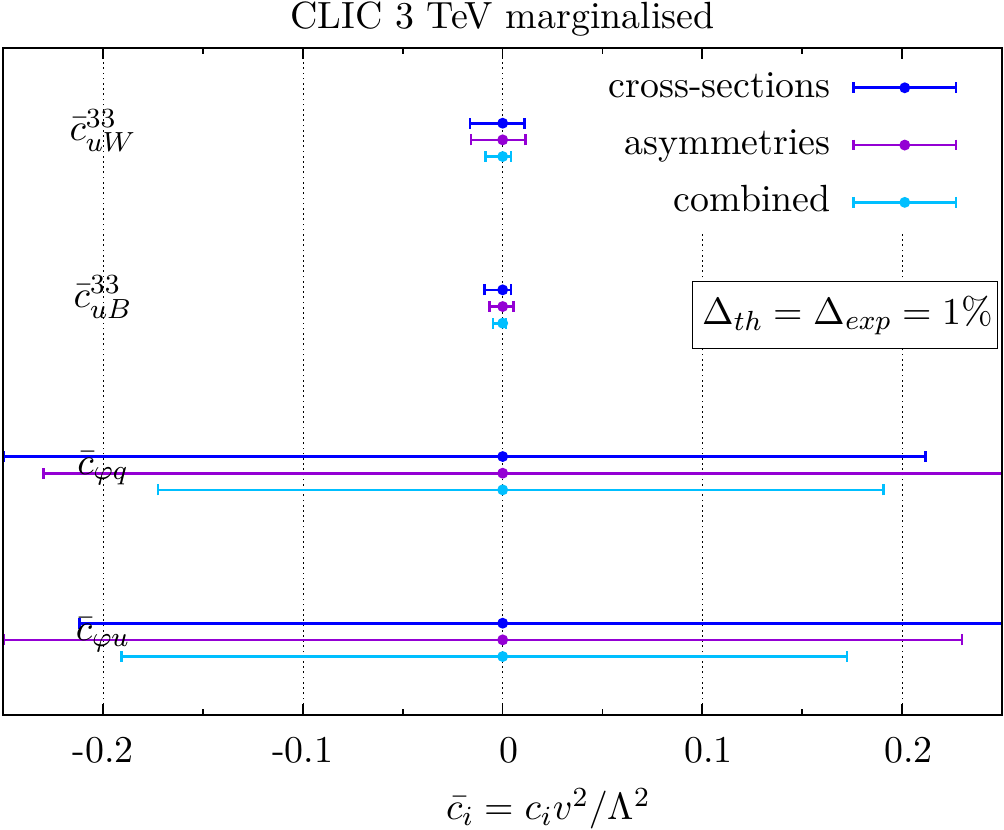}
 \end{subfigure}
   \caption[CLIC constraints from top electroweak operators.]{95\% confidence ranges for the operators we consider here, from CLIC running at \sqrts = 3 \tev, assuming 1\% theoretical and experimental uncertainties, by fitting to cross-sections, asymmetries, and the combination, with each operator considered individually (a) or in a 5D fit (b). To display both on the same plot, we scale the individual constraints up by a factor of 10, so that the bottom axis is actually \cb{i}/10.\label{fig:clicglobal}}
  \end{center}
\end{figure}

The Compact Linear Collider (CLIC) project~\cite{CLIC:2016zwp,Abramowicz:2016zbo}, with its larger maximum centre-of-mass energy \sqrts = 3 \tev, will be in a stronger position to discover the effects of some higher-dimensional operators, whose effects na\"ively scale with the CM energy as $s/\Lambda^2$. There are two main running scenarios, but both envisage total integrated luminosities of 500 \ifb at \sqrts = 500 \gev, 1.5 \iab at 1.4 or 1.5 \tev, and 2 \iab at 3 \tev. Again, we focus on the highest energy setting \sqrts = 3 \tev, to maximise discovery potential for non-resonant new physics through \D6 operators.

Moving further away from the \ttbar threshold, the total \ep $\to Z^*/\gamma \to$ \ttbar rate is smaller than at the ILC; at \sqrts = 3 \tev it is around 20 \fb, which means for the total forecast integrated luminosity at this energy  there will be a statistical uncertainty of $\simeq$ 0.5\%. A total experimental uncertainty of 1\% may therefore be too optimistic an estimate once systematics are fully itemised. Nonetheless, for ease of comparison with the ILC figures, we take this as a baseline, and the corresponding constraints, using the same observables and beam settings, are shown in Fig.~\ref{fig:clicglobal}.

We see that for the individual fit, CLIC constraints are of the same order of magnitude worse than ILC ones.\footnote{This is in contrast to Higgs sector constraints from \ep $\to hZ$, where the projected sensitivity is extremely dependent on the momentum flow through the vertex, leading to better overall CLIC constraints~\cite{Ellis:2017kfi}.} Although the direct sensitivity to the operators is enhanced, we see that as we move away from the \ttbar threshold, the interference effect of the $\varphi$-type operators is much smaller. This is not the case for the operators \op{uW} and \op{uB}, whose contributions stem mainly from the $(\D6)^2$ term, as seen on the right of Fig.~\ref{fig:ilcxsec}, which receives no suppression. Their individual constraints are close to the ILC values, indicating that energy scale is not the dominant factor driving these limits, but rather the theory and experimental uncertainties which saturate the sensitivity, which we do not vary.

For the more general marginalised fit, we see again that combining cross-section and asymmetry measurements will break blind directions in the fit, leading to much more powerful overall constraints. Unlike for the case of the ILC, however, care must be taken in interpreting these limits in terms of the mass scale of a particular UV model. The marginalised constraint $|\cb{\varphi u}| \lesssim 0.05$, for example, corresponds to a mass scale $\Lambda/\sqrt{c} \gtrsim $ 1.1 \tev, which is less than the energy scale probed in the interaction, so that the constraint can only be linked to a particular model if it is very weakly coupled: $g_* \ll 1$. 

\subsection{Beyond $e^+e^-\to t\bar t$: Precision Observables}
\label{sec:lep}
Obviously the direct constraints that we have focused on in this work do not exist in a vacuum and the interplay of direct and indirect sensitivity plays an important part in ultimately obtaining the best constraints for a given model (see \cite{Berthier:2015oma,Ghezzi:2015vva}). To put the expected direct constraints detailed above into perspective we analyse the impact of the considered operators on LEP precision observables. Note, that these $Z$ resonance observables are sensitive to a plethora of other new interactions and a direct comparison is not immediately straightforward~\cite{Berthier:2015oma}. Nonetheless, there is significant discriminative power that is worthwhile pointing out, which we will discuss in the following.

\begin{figure}[t!]
\begin{center}
 \includegraphics[width=0.325\textwidth]{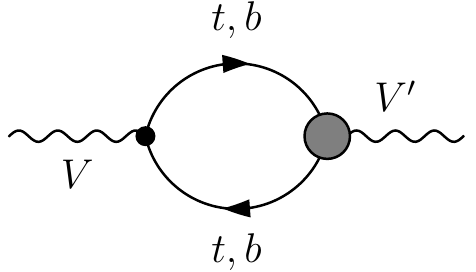}
  \caption[Insertions of \D6 operators into Peskin-Takeuchi $S,T,U$ parameters.]{\label{fig:pols} Representative Feynman diagram contributing to the $S,T,U$ parameters at one-loop. The grey-shaded area marks a possible dimension six insertion while the black dot represents a SM vertex of the $V-V'$ polarisation function, $V,V'=W^\pm, Z,\gamma$.}
\end{center}
\end{figure}

\subsubsection{Oblique corrections}
The $S,T,U$ parameters~\cite{Peskin:1990zt,Peskin:1991sw} (see also \cite{Ross:1975fq}) are standard observables that capture oblique deviations in the SM electroweak gauge sector from the SM paradigm~\cite{Grojean:2013kd,Barbieri:2004qk} through modifications of the gauge boson two-point functions. The operators considered in this work modify these at the one-loop level through diagrams of the type shown in Fig.~\ref{fig:pols}. Throughout we perform our calculation in dimensional regularisation.

\begin{figure}[t!]
\begin{center}
 \includegraphics[width=0.6\textwidth]{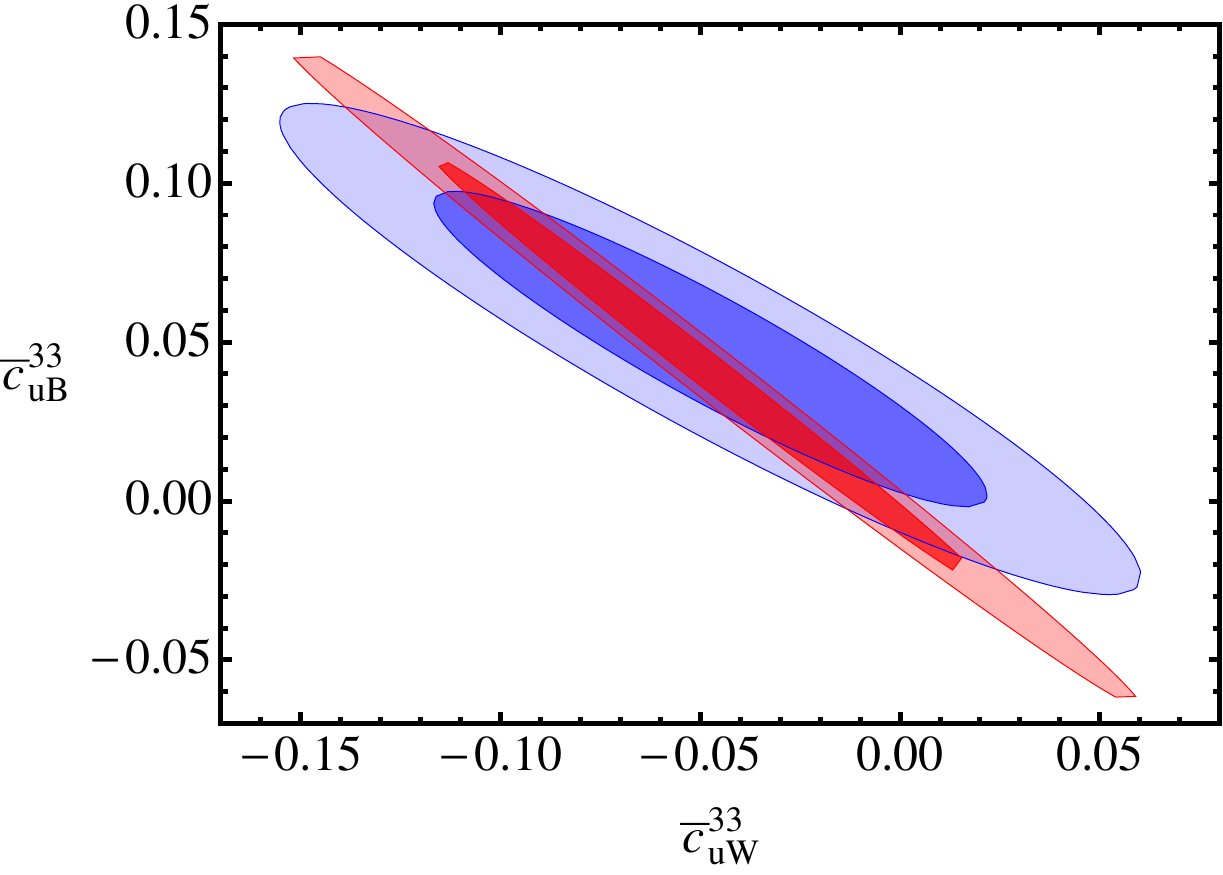}
  \caption[Allowed regions in the \op{uW}-\op{uB} plane from the $S,T,U$ fit.]{\label{fig:stu} Contour of the $S,T,U$ fit reported in \cite{Baak:2014ora} for specifically the operators $\op{uW},\op{uB}$, which are unconstrained by down-sector measurements. All other Wilson coefficients are chosen to be zero. The dark and light shaded areas represent 68\% and 95\% confidence levels for this projections, while the blue contour uses $\mu_R=m_Z$ and the red contour $\mu_R=1~\text{TeV}$.}
\end{center}
\end{figure}

The definitions of $S,T,U$, see~\cite{Peskin:1990zt,Peskin:1991sw,Barbieri:2004qk}, are such that in the SM all divergencies cancel when replacing the renormalised polarisation functions by their bare counterparts. The modifications of Fig.~\ref{fig:pols}, however, induce additional divergencies due to the dimension 6 parts and the introduction of two-point function counterterms is essential to obtain a UV-finite result, see also~\cite{Ghezzi:2015vva}. This leads to a regularisation scale $\mu_R$ dependence of the \D6 amplitude parts after renormalisation, as shown in Fig.~\ref{fig:stu}. It is this part which we focus on as we choose the SM with a 125 GeV Higgs as reference point~\cite{Baak:2014ora}.

\subsubsection{Non-oblique corrections}
A well-measured quantity at LEP is the $Zb \bar b$ vertex, which enters the prediction of the bottom forward-backward asymmetry $A_{FB}^{b\bar b}$, see e.g.~\cite{Abdallah:2003gp}. Similar to the operators in Eq.~\eqref{eqn:ttzd6ops}, in the generic dimension six approach we can expect similar operators for the down-sector of the 3rd fermion family. These will modify the interactions along the same lines as we focused on above for the top sector. However, due to the different isospin properties, the bottom forward backward asymmetry is now sensitive to the sum $\co[(3)]{\varphi q}+\co[(1)]{\varphi q}$. This leads to a complementary constraint by the LEP forward backward asymmetry compared to the direct measurements in $t\bar t$, as shown on the left of Fig.~\ref{fig:lepilc}.

Moreover, the constraints on $\co[(3)]{\varphi q}+\co[(1)]{\varphi q}$ from $A_{FB}^{b\bar b}$ can be combined with the constraints on  $\co[(3)]{\varphi q} - \co[(1)]{\varphi q}$ to extract independent bounds on \co[(3)]{\varphi q} and \co[(1)]{\varphi q}. This is shown on the right of Fig.~\ref{fig:lepilc}. Care should be taken when interpreting these constraints individually, however. We are considering marginalised bounds for the ILC constraint but only one operator combination for the LEP bound. In general, other operators that we do not consider here will impact the $Zb\bar b$ vertex at tree level and in general weaken the bound. This serves as a useful visualisation, however, of the complementarity between past and future colliders in constraining these operators.

\subsection{Summary}
\label{sec:conc_ch4}
Given the unsatisfactory precision of current probes of top quark electroweak couplings from hadron collider measurements, they must be a key priority in the physics agenda of any future linear \ep collider. By parameterising non-standard top couplings through \D6 operators, we have analysed the potential for the ILC and CLIC to improve the current precision of the top electroweak sector. Unsurprisingly, if experimental precision would match current estimates, and theory uncertainties can be brought to the same level, the current constraints can be drastically improved by both colliders, with associated bounds on the scale of new physics typically in the 1 \tev to few \tev range, depending on the assumed coupling structures of the underlying model. Using asymmetry measurements as well as cross-sections will be crucial to this endeavour, as will collecting large datasets with several incoming beam polarisations. 

\begin{figure}[t!]
\begin{center}
\begin{subfigure}{0.475\textwidth}
 \includegraphics[width=\textwidth]{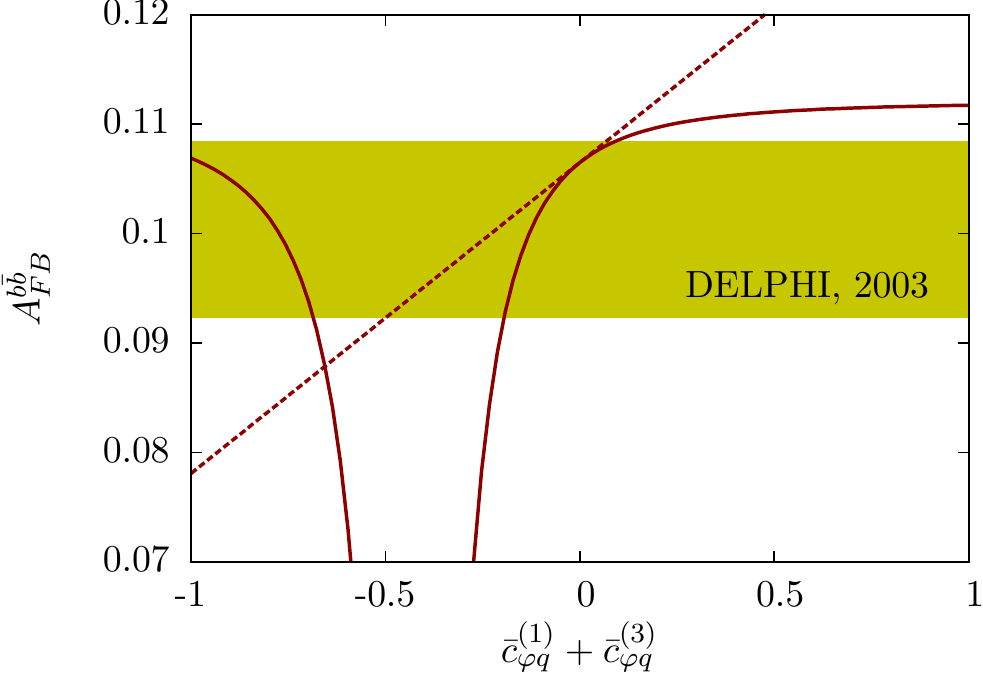}
  \end{subfigure}%
  \quad
  \begin{subfigure}{0.47\textwidth}
  \vspace{5pt}
 \includegraphics[width=\textwidth]{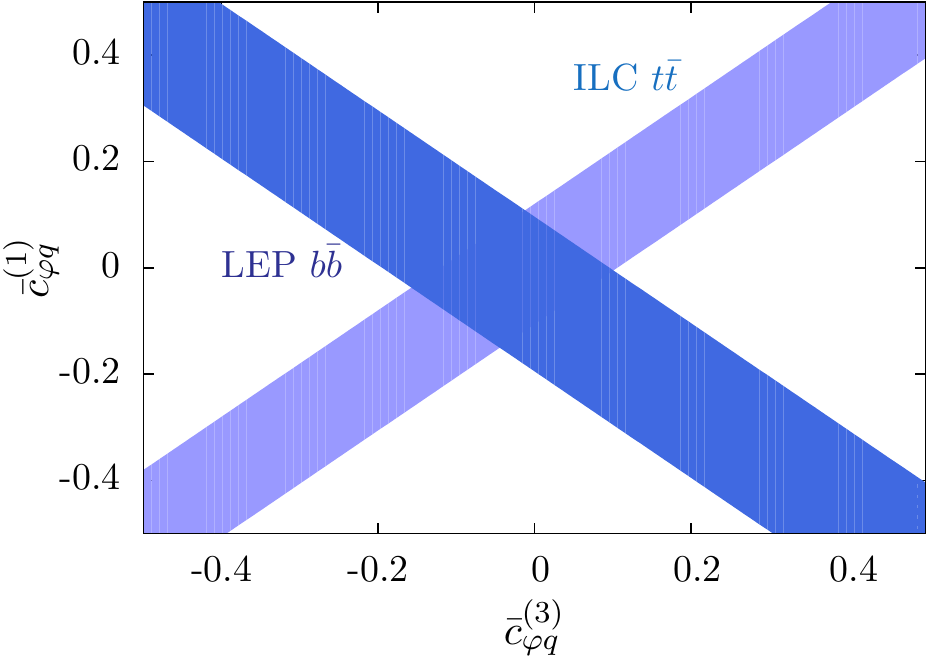}
  \end{subfigure}
   \caption[LEP $b\bar b$ forward-backward asymmetry and complementarity with ILC \ttbar asymmetry.]{\label{fig:lepilc} Left: Forward-backward asymmetry (linearised, dashed, and full results, solid) as a function of $\co[(3)]{\varphi q}+ \co[(1)]{\varphi q}$. The exclusion contour is taken from DELPHI collaboration's Ref.~\cite{Abdallah:2003gp} for the most constraining measurement at \sqrts=91.26~\gev. Right: Allowed 95\% confidence regions for the Wilson coefficients \cb[(3)]{\varphi u} and \cb[(1)]{\varphi u} obtained from combining the information from ILC \ttbar asymmetries and cross-sections (dark blue) and LEP1 $b\bar b$ measurements (lighter blue).}
\end{center}
\end{figure}

We have found that, unlike for the Higgs sector, the large increase in centre-of-mass energy at CLIC does not necessarily offer a competitive advantage over the ILC for bounding new top interactions by the operators we consider, and bounds on the operators that we consider are typically stronger at the latter, though in simultaneous 4D fits the difference is not striking. For some of the operators we consider, the bounds derived from CLIC fits correspond to mass scales smaller than the CM energy that we consider, which can call into question the validity of the EFT description, unless the CLIC sensitivity can exceed the expectations we forecast here.

By combining $Z$-pole measurements from LEP1 with \ttbar measurements (and future improved electroweak precision measurements), one can in principle break degeneracies in the operator set and disentangle individual operators that could previously be only bounded in combinations. We showed this for the LEP forward-backward asymmetry, this could be improved by fitting other precision electroweak observables too. Care must be taken in interpreting the associated constraints, however, as both sets of measurements will in general talk to other operators for which there is no complementarity, and a more systematic approach taking into account EFT loop corrections would have be undertaken before these numerical bounds can be taken at face value.

\newpage

\section{Top tagging with Deep Neural Networks}

\subsection{Introduction}
In the last chapter, we discussed how the sensitivity to non-resonant new physics in the top quark sector could be improved by making use of reconstruction techniques for `boosted'~\cite{Seymour:1993mx,Butterworth:2008iy,Kaplan:2008ie,Thaler:2008ju,Almeida:2008tp,Cui:2010km} top final states; that is, top quarks whose decay products are highly collimated in the detector. Since we generally expect on dimensional grounds that the effects of heavy new physics coupling to the top quark would be most prominent in the high $p_T$ tail, top tagging is well suited to this task. There are many other applications of boosted tagging, however, and as higher momentum transfer final states begin to be probed more regularly at the LHC, jet substructure methods are becoming an increasingly crucial component of its analysis program. 

In top pair production for instance, 15\% of the (Standard Model) total cross-section comes from the region where $p_T^t \gtrsim 200 $ \gev, and in new physics scenarios where, for example, a heavy resonance decays into a top pair, the boosted regime is the region of interest~\cite{Anders:2013oga}. Because they rely on the final state objects being well-separated in the detector, standard `resolved' reconstruction techniques begin to falter here. Boosted reconstruction techniques, on the other hand, which allow for more than one of the hard partons in the final state to be captured by the same large-radius jet, and then analyse the substructure of those jet(s), are more efficient in this regime. Improving the efficiency of boosted top tagging in the increasingly challenging hadronic environment of the LHC detectors will thus be of vital importance over the LHC lifetime. This is the subject of this chapter.

There exist several well-studied examples of these boosted techniques, for example mass-drop filtering~\cite{Butterworth:2008iy}, trimming~\cite{Krohn:2009th}, pruning~\cite{Ellis:2009su,Ellis:2009me}, shower deconstruction~\cite{Soper:2011cr} and the HEPTopTagger~\cite{Plehn:2009rk,Plehn:2010st,Kasieczka:2015jma} algorithm, which was used in chapter 4. These taggers have typically been designed to focus on the hard substructures of the jet, and veto the softer activity in the detector. This approach, whilst well-motivated from a QCD perspective, in principle throws out valuable information about the jet's properties, and it is interesting to ask whether the softer constituents of the jets can also offer powerful discriminating features between signal and background, as has already been demonstrated for the hard substructures.

An intriguing new avenue in this direction has recently been opened, that makes use of machine learning algorithms known as Convolutional Neural Networks (ConvNets). These \emph{deep learning} techniques are routinely used in computer vision problems such as image/facial recognition as well as natural language processing. As applied to boosted jet finding, the basic idea is to view the calorimeter $(\eta,\phi)$ plane as a sparsely filled image, where the filled pixels correspond to the calorimeter cells with non-zero energy deposits and the pixel intensities to the energy or $E_T$ deposited. After some image preprocessing, a training sample of these \emph{jet images}, with signal and background events, is fed through a ConvNet designed to learn the signal-like and background-like features of the images; shapes, clusters, edges, voids,  etc. The final layer of the network converts the learned features of the image into a probability of it being either signal or background. One then feeds a test sample through the network to quantify its performance, usually expressed in terms of its receiver operator characteristic (ROC) curve: signal efficiency vs. background rejection. This method will be described in detail later in the chapter.

Using Monte Carlo simulated data, these networks have been shown to be comparable in performance to various already well-established QCD-inspired taggers, even in the presence of pileup~\cite{Cogan:2014oua,deOliveira:2015xxd,Komiske:2016rsd,Barnard:2016qma,Almeida:2015jua,Baldi:2016fql,Guest:2016iqz,Baldi:2016fzo,Louppe:2017ipp}. However, these analyses have typically focused on hadronically decaying $W$-boson tagging, and top tagging has not yet been studied. Whilst a conceptually similar problem, the extra prong in the top decay corresponding to the $b$-quark, and the extra mass splitting in the $tWb$ vertex, lead to a different structure of the final large-$R$ jet. It would be desirable to show that jet image techniques can be applied here with similar efficiency. A broader goal would be to demonstrate that these techniques have applicability across the realm of jet substructure, not just in the narrow example of $W$-tagging. This analysis is a step in that direction.

This chapter is structured as follows: In section \ref{sec:qcd} we outline the main jet substructure and top tagging concepts visited throughout the analysis. In section \ref{sec:machines} we discuss the two machine learning algorithms that we utilise: boosted decision trees (BDTs) and convolutional neural networks. In section \ref{sec:analysis} we discuss the details of how we build a jet image and the specific network architectures we test against. In section \ref{sec:results} we discuss the performance of the neural network compared to standard QCD based taggers, and investigate the physics that the network learns, before summarising in section \ref{sec:conc_ch6}.

\subsection{QCD-inspired top tagging}
\label{sec:qcd}

The goal of any QCD-based tagger is to use an understanding of perturbative QCD to construct observables and algorithms that offer the best discrimination between jets from signal processes; typically involving tops, $W$ and $Z$ bosons or Higgses, and background; typically originating from QCD dijet or multijet processes. The simplest such observable is the \emph{jet mass}, defined simply as the mass of the Lorentz vector sum of all of the constituent momenta in a jet.
\begin{equation}
m_J^2 = \summ{i}^{\text{constits}} p^\mu_i p_{\mu i} .
\end{equation}

Clearly, the exact value of the reconstructed jet mass will be dependent on the algorithm used to cluster the jet, but it should be close to the mass of the parton(s) it originated from. In the case of signal jets, this should be fairly close to the mass of the decaying object, as shown on the left of Fig.~\ref{fig:jetmass}, where the reconstructed fat jet mass for jets within the range $ 350 \gev \leq p_T^J \leq 450 \gev $ is shown (to ensure the jets are tagged as tops we require that they are matched to a parton-level top within $\Delta R = 1.2$). In the case of QCD jets, hard perturbative emissions push the reconstructed jet mass to higher values than one would na\"ively expect from its massless constituents, as shown on the right hand side of Fig. \ref{fig:jetmass}, where the QCD jet mass is broadly peaked at $m_J \sim$ 100 \gev, rather than towards the origin. Some additional processing of the reconstructed jets therefore must be done before the jet mass can be considered a reliable observable, i.e. a detector-level observable that captures the kinematics of the hard process. These techniques are generally referred to as \emph{grooming}.

\subsubsection{Jet grooming}

The three most commonly used jet groomers are \emph{filtering},  \emph{pruning} and \emph{trimming}. Filtering a jet is aimed at keeping its hard constituents originating from the hard interaction whilst rejecting as much as possible the softer constituents originating from QCD radiation. Beginning with a large-$R$ jet, its constituents are reclustered with the Cambridge-Aachen algorithm (C/A)~\cite{Dokshitzer:1997in,Wobisch:1998wt}, with radius parameter $R_{\text{filt}}$. Then, all constituents outside of the $N$ hardest subjets are filtered out. That is, the $N$ subjets with the largest transverse momenta after filtering are kept, the rest of the jet is discarded. For a hadronically decaying top quark $t\to Wb\to q\bar q^\prime b$, there are three hard subjets associated with the decay, and one typically allows tolerance for some extra QCD emission, to avoid mis-filtering out the decay products. The 5 hardest subjets are usually kept. A typical radius  $R_{\text{filt}}$ is 0.3, for an original jet of size $R=1.2$ or 1.5. The effects of filtering on the jet mass for top and QCD jets are shown in Fig. \ref{fig:jetmass}. We see that the mass peak for the signal jets has become much sharper, and is symmetric around the top mass. For the QCD jets, we see a moderate shift towards zero.

\begin{figure}[t!]
  \begin{center}
  \includegraphics[width=\textwidth]{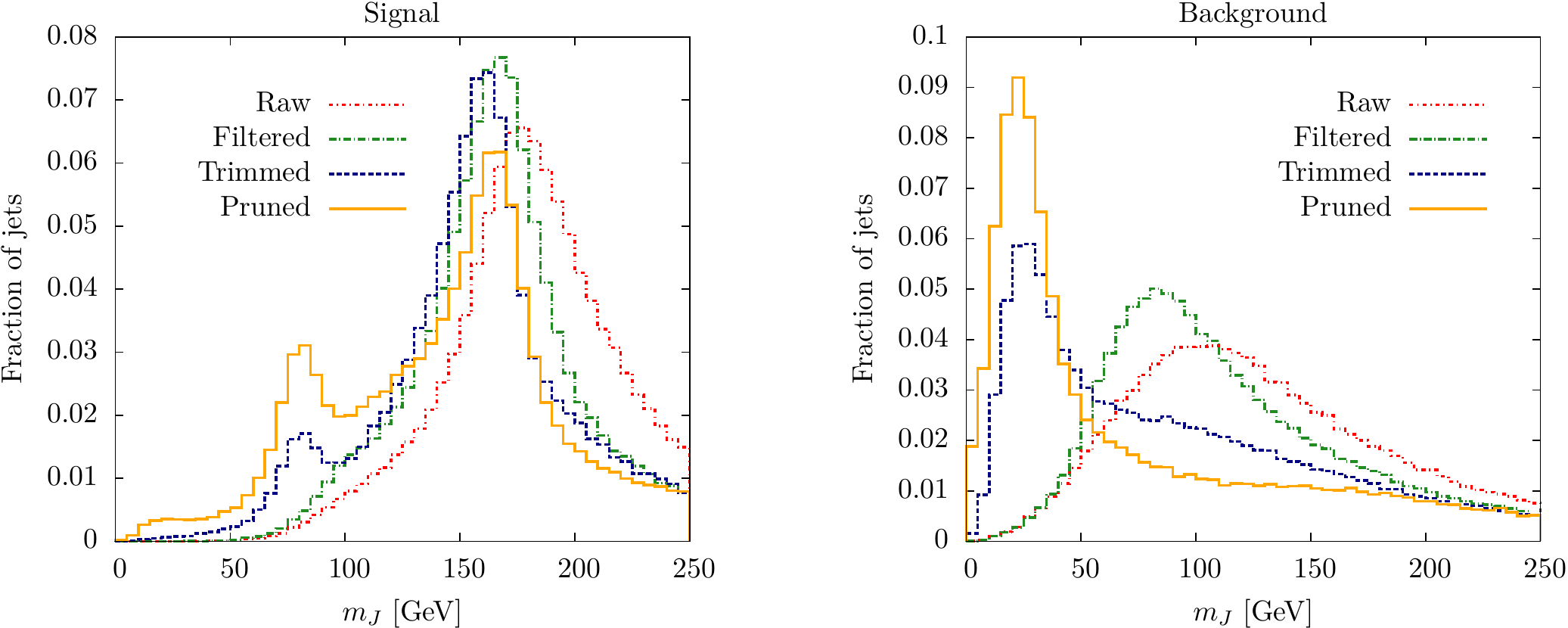}
  \end{center}
  \caption[Top and QCD jet mass distributions after the application of various grooming algorithms.]{Top (left) and QCD (right) jet mass distributions after the application of various grooming algorithms. The original jets are constructed with the Cambridge-Aachen algorithm with distance parameter 1.5. Filtering is applied with radius parameter $R_{\text{filt}}=0.3$ and the 5 hardest subjets are kept. The trimming procedure uses $R_{\text{sub}}=0.35$ and$f_{\text{cut}} = 0.03$, and the pruned jets are reclustered with the C/A algorithm with $z_{\text{cut}} = 0.1$, $R_{\text{cut}} = 0.5$ (see text for details).  }
\label{fig:jetmass}
\end{figure}

For trimming, rather than keeping the $N$ hardest subjets, one cuts on the fraction of the subjet $p_T$ to the original fat jet $p_T$. Namely, one reclusters the fat jet constituents with the $k_T$ algorithm using a radius $R_{\text{sub}}$, thus clustering softer energy deposits first and harder activity last. Any subjets $i$ with $p_{Ti}/p_T^J < f_{\text{cut}}$ are removed, and the final trimmed jet is composed of the remaining subjets. The jet mass distributions for tops and QCD after trimming, using $R_{\text{sub}}=0.35$, $f_{\text{cut}} = 0.03$, are also shown in Fig. \ref{fig:jetmass}. Comparing to the raw distributions, we see that QCD jets typically lose 30\%-50\% of their mass, whereas top jets keep most of their mass, therefore trimming is a robust procedure for isolating the hard components of a jet that one is most interested in. A drawback is that sometimes the $b$ quark will be vetoed from the jet if it is very soft, and so the jet mass will be spuriously reconstructed to the $W$ mass.

Pruning is related to trimming, but along with kinematic requirements on the $p_T$, it applies an additional geometrical cut, designed to remove wide-angle radiation. Starting with a large-$R$ jet, one reapplies either the C/A or $k_T$\cite{Catani:1993hr,Ellis:1993tq} jet clustering algorithm on its constituents, but at each clustering step, the criterion that a) the softer of the constituents $j_1$ and $j_2$ (taken to be $j_2$, so that $p_T^{j_1} > p_T^{j_2}$) has $p_T^{j_2}/p_T^{j_1+j_2} < z_{\text{cut}}$, and b) $\Delta R_{j_1,j_2} < R_{\text{cut}} \times (2m_J/p_T^J)$, are both tested. 

At least one of a) or b) must be true, otherwise $j_1$ and $j_2$ are both removed. Typical values used in real jet substructure analyses are $z_{\text{cut}} = \{0.05,0.1\}$ and $R_{\text{cut}}= \{0.1,0.2,0.3\}$. The effects of pruning on the signal and background jets we are using are also shown in Fig. \ref{fig:jetmass}. We see that pruning is particularly effective for reducing the QCD jet mass, but the price paid is that it is also more likely to erroneously reconstruct the signal jet mass to the $W$ mass instead of the top, i.e. it is more likely to veto the $b$ quark.

\subsubsection{$N$-subjettiness}
As well as the absolute mass scales entering the jet, it is also useful to construct observables that capture the likely number of hard subjets within a jet, i.e. that gives a measure of how `prongy' a fat jet is. One could imagine, for instance, a $W^\prime$ or $Z^\prime$ boson with a mass close to the top quark. It would have a characteristic two-prong decay, but from jet mass distributions alone it would appear very similar to a top jet. $N$-subjettiness~\cite{Thaler:2010tr,Thaler:2011gf} is an observable designed to measure this property. For a fat jet of radius $R_0$, $N$-subjettiness is defined as
\begin{equation}
\tau_N = \frac{\summ{i} p_{Ti} \min(\Delta R_{i1},\ldots, \Delta R_{iN})^\beta}{\summ{i} p_{Ti} R_0} .
\end{equation}

The sum is over the jet constituents $i$, and $N$ is the number of candidate subjets. To define the distance measures $\Delta R$, one must define candidate subjet axes. This can be done either by summing over all possible candidate subjet directions and taking the minimum, or, in a computationally simpler approach, by defining the subjets by applying the $k_T$ clustering algorithm on the jet constituents and truncating after $N$ subjets are generated. The exponent $\beta$ is a free parameter that can be optimised for a given analysis. It is also implicitly assumed that the fat jet radius $R_0 > \Delta R_{ij}$ for all jet constituents.

\begin{figure}[t!]
  \begin{center}
  \includegraphics[width=\textwidth]{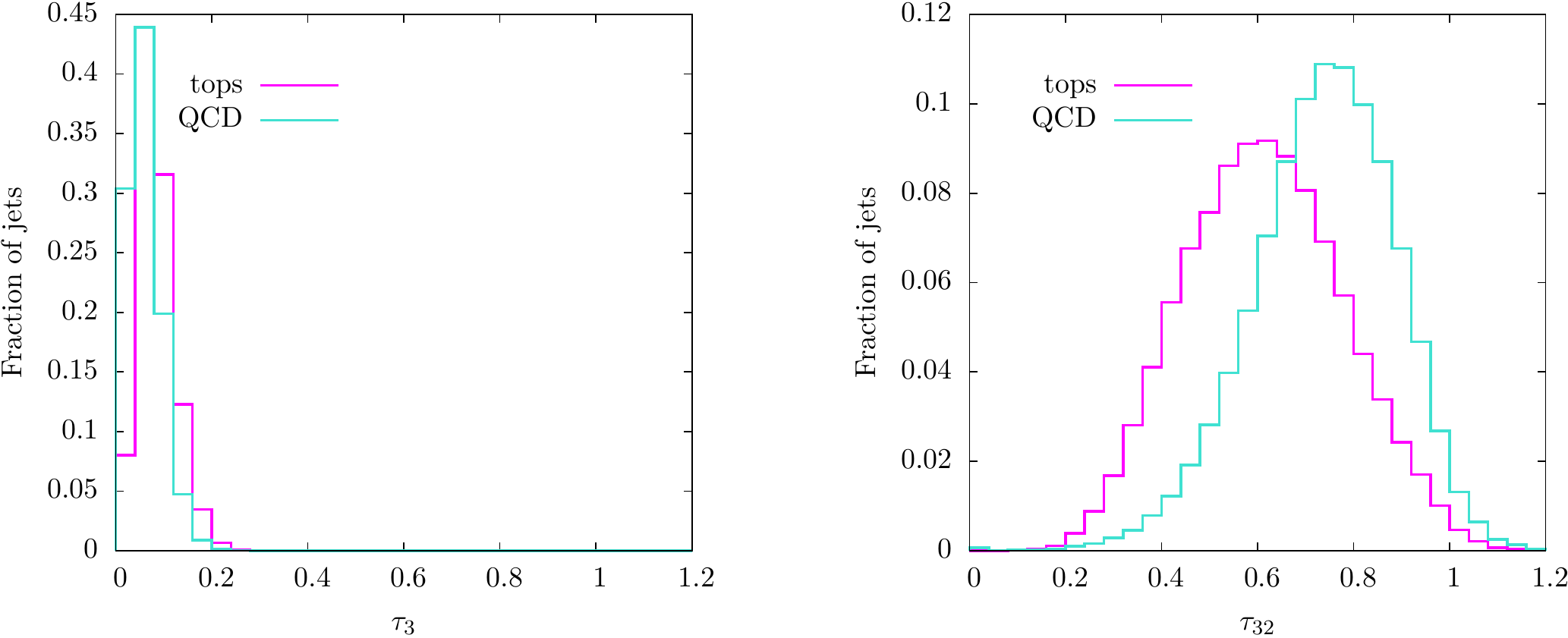}
  \end{center}
  \caption[$N$-subjettiness distributions for top and QCD jets.]{Distributions of the $N$-subjettiness parameter $\tau_3$ (left) and ratio $\tau_3/\tau_2$ (right) for top and QCD jets, clustered using the C/A algorithm with $R=1.5$ as before. The subjet axes are defined by $k_T$ clustering, and $\beta = 1$. Better discriminating power is offered by the ratio $\tau_3/\tau_2$, as seen on the right plot.}
\label{fig:tauhisto}
\end{figure}

In the limit $\tau_N \to 0$, the jet must have $\Delta R_{iN} = 0$ for all constituents, i.e. all constituents must be perfectly aligned with the candidate subjets, so the jet has exactly $N$ subjets. In the case of  $\tau_N \to 1$, the jet must have at least $N+1$ subjets, i.e. the minimisation missed some subjet axes. Therefore jets with small $\tau_N$ are said to be $N$-subjetty, whereas jets with larger $\tau_N$ have more than $N$ subjets. For top jets the ratio $\tau_3$ is clearly the one of interest. In fact, it has been shown that the ratio $\tau_3/\tau_2 \equiv \tau_{32}$ has the most discriminating power between tops and QCD, because several QCD uncertainties are present in the value of $\tau_3$ that drop out when taking the ratio $\tau_{32}$\footnote{See Ref.~\cite{Plehn:2011tg} for a discussion of this, and as a useful review of top tagging in general.}. The distributions of $\tau_{32}$ for tops and QCD dijets are shown in Fig. \ref{fig:tauhisto}, showing clear separation between signal and background. $N$-subjettiness is thus an extremely useful discriminating variable and will be utilised throughout this analysis.

\subsubsection{SoftDrop}
Another useful substructure algorithm for top tagging is {\sc{SoftDrop}}~\cite{Larkoski:2014wba}, which is designed to iteratively remove soft, wide-angle activity corresponding to contamination from the underlying event from the jet. The algorithm is as follows:

\begin{itemize}

\item For a fat jet $J$, undo the last step of the clustering so that there are two subjets $j_1$ and $j_2$.

\item If the softer of the two subjets has a fraction of the total jet $p_T$ greater than
\begin{equation}
\frac{\min(p_T^1,p_T^2)}{p_T^1+p_T^2} > z\left(\frac{\Delta R_{12}}{R_0} \right)^\beta,
\end{equation}
then $J$ is the final jet. 

\item If not, keep harder of the two jets and continue until the final jet is reached.

\end{itemize}

The parameters of {\sc{SoftDrop}} are thus the exponent $\beta$ and the $p_T$ fraction $z$. The limit $\beta \to 0$ removes the geometrical dependence of the cut, e.g. $\beta =0$, $z = 0.1$ removes subjets with less than 10\% of the total jet $p_T$. The additional $\Delta R$ cut shows that softer activity is much more likely to be cut out if it has a large angular separation from the rest of the jet. {\sc{SoftDrop}} is a useful procedure for defining a jet mass that is robust against contamination from soft radiation, and a `softdropped' jet mass, in conjunction with $\tau_{32}$ provides powerful discrimination between tops and QCD background.

There are many other observables used in top tagging, and we will not review them all here, we merely briefly discuss the ones that will be utilised in this analysis. The other key component of the analysis is the use of machine learning techniques. In the form of multivariate techniques such as Boosted Decision Trees, these techniques have been well-established as useful in jet substructure classification. Deep Neural Networks are a recent development. Since we will benchmark the performance of our neural network against various standard BDTs, in the next section we will outline the basics of a BDT architecture, before discussing the structure of a deep neural network in detail.

\subsection{Machine learning inspired top tagging}
\label{sec:machines}

\subsubsection{Boosted decision tree approaches}
Starting with a sample of signal and background jets that one wants to classify, a standard decision tree begins by ordering the input jets by the value of some discriminator variables, typically referred to as \emph{features} in BDT parlance. For our case, the features will be quantities such as the jet mass and $N$-subjettiness ratios. For each feature, the sample is split into two parts, based on the value of the feature that best separates the signal and background. An initial node, containing all the events, has thus been separated into two branches. The branches continue to be split until the final branches, called \emph{leaves} are either pure signal or pure background, or contain too few samples of either kind to continue splitting.

Each sample jet/event is given a weight $W_i$, where the weighting procedure is specific to the precise type of decision tree classifier. The splitting criterion is defined by the purity of the sample in each branch,
\begin{equation}
P = \frac{\summ{s} W_s}{\summ{s} W_s + \summ{b} W_b} ,
\end{equation}

For each branch, the optimal splitting of the parent branch into two daughter branches is defined by maximising the \emph{Gini} function
\begin{equation}
[(\sum_{\substack{i=1}}^n W_i)P(1-P)]_{\text{parent}} - [(\sum_{\substack{i=1}}^n W_i)P(1-P)]_{\text{daughter 1}} - [(\sum_{\substack{i=1}}^n W_i)P(1-P)]_{\text{daughter 2}} .
\end{equation}

If the events all have unit weight, then if a final leaf has a purity $P \geq 1/2$ it is called a signal leaf, and if $P \leq 1/2$ it is a background leaf\footnote{In fact the purity demanded can be varied between zero and one, which gives a smooth curve of background contamination for a given signal efficiency, i.e. an ROC curve.}. All events that finish on a signal leaf are classified as signal, likewise for background. The signal efficiency is defined as the number of signal events that landed on signal leaves divided by the initial number of signal samples, and the corresponding background contamination or `mistag' rate is the number of background events that landed on signal leaves divided by the initial number of background samples. A schematic diagram of a simple two-feature decision tree is shown in Fig. \ref{fig:classifier}.

\begin{figure}[t!]
  \begin{center}
  \includegraphics[width=0.76\textwidth]{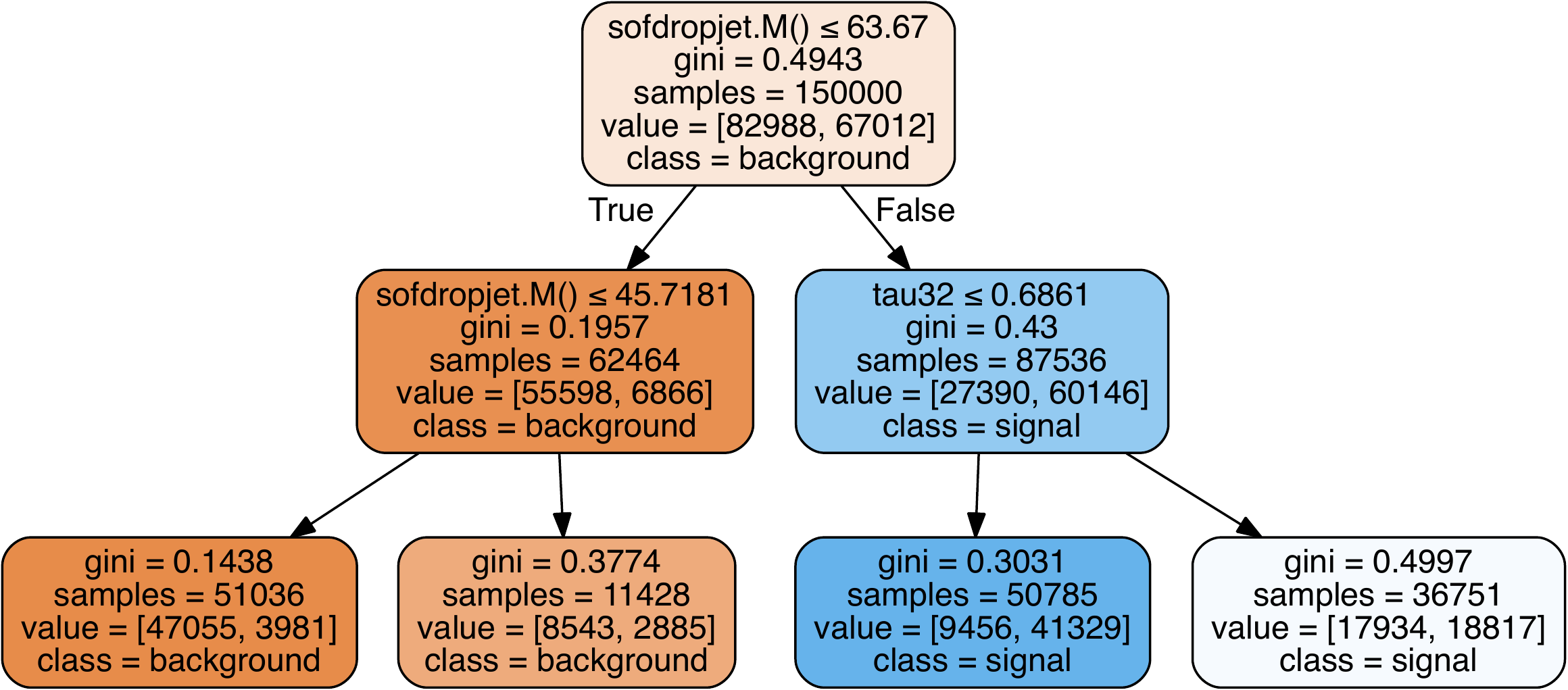}
  \end{center}
  \caption[A simple decision tree classifier.]{A simple decision tree classifier with a maximum tree depth of two, using just two features: the $N$-subjettiness ratio $\tau_{32}$ and the mass of the fat jet after applying the {\sc{SoftDrop}} algorithm. The tree first cuts on the jet mass, with jets above the cut going into the signal category, a further cut on $\tau_{32}$ is then applied to the signal jets, and a further cut on the jet mass to the background jets. Figure generated with \cite{sklearn}.}
\label{fig:classifier}
\end{figure}

The performance of a decision tree classifier can be significantly enhanced by applying `boosting' criteria at each branch. A boosted decision tree proceeds as above, except if a signal event is misclassified as background or vice versa, the weight of that event is \emph{boosted}. The entire sample of weighted events is fed through a second tree, and the procedure repeated for several trees. After each tree, the event is given a score of 1 if it lands on a signal leaf and -1 if it lands on a background leaf. The total BDT score for that event is the sum of its scores after each tree.

A commonly used boosting classifier is the adaptive boosting method or AdaBoost. Starting with a sample of $N$ events labeled $i$, each with a weight $1/N$, give signal events an initial score $y_i = 1$ and background events $y_i = -1$. For a set of input features $x_i$ for each event $i$, define $T_m (x_i) = 1$ if the event lands on a signal leaf at the end of tree $m$ and $T_m (x_i) = -1$ if it lands on a background leaf. A `misclassify' score $I$ is defined as $I(y_i \neq T_m(x_i)) = 1$ and $I(y_i = T_m(x_i)) = 0$, so that the total error on the $m$th tree is just the weighted sum of the misclassified event scores:
\begin{equation}
err_{m} = \frac{\sum_{\substack{i}}^N w_i I(y_i \neq T_m(x_i))}{\sum_{\substack{i}}^N w_i} .
\end{equation}
The misclassified events are reweighted by
\begin{equation}
w_i \to w_i \times \exp{(\alpha_m I(y_i \neq T_m(x_i)))} ,
\end{equation}
where
\begin{equation}
\alpha_m = \beta \times \log\left(\frac{1-err_m}{err_m}\right) .
\end{equation}
The parameter $\beta$ can be chosen freely; typical values used are $\beta=1$ or $\beta=1/2$. The goal of the decision tree is then to simply construct the weights at each tree so that the error function is minimised for each event. Boosted decision trees 
are typically more powerful because their performance is less sensitive to changing the model parameters, and they are less prone to overfitting. For this reason they are already widely used in signal/background discrimination in real particle physics analyses, and will thus provide an appropriate comparison for quantifying the performance of a deep neural network.

\subsubsection{Deep learning}

Deep neural network algorithms originate from computer vision, namely the problem of trying to correctly label an image from a fixed set of categories. The network consists of three steps: input, learning and evaluation.

An input image can be thought of as a vector of pixel densities $\{x_i\}$. The goal is then to construct a classifier which converts this into a label from a predefined set $\{y_i\}$. In a binary classification problem such as signal vs. background, $\{y_i\} \in \{0,1\} $. The simplest such classifier is a linear mapping, where each pixel is multiplied by a series of weights $W$ and added with some bias $b$: $y_i = W_{ij}x_j + b_i$. For a binary classifier over a (flattened) $N\times N$ pixel image, $W$ is then a $2\times N^2$ matrix and $b$ a 2D vector. This transformation would constitute a single \emph{layer} of a neural network. A typical NN has many such layers, with in general more complicated mappings between layers.

Each layer in a network is made of of \emph{units} or \emph{neurons} which receive as input the output from the previous layer. If every neuron in the layer is connected to every neuron in the previous layer, the layer is \emph{fully-connected} or \emph{dense}. The connections between neurons in neighbouring layers may be thought of as synapses, in analogy with neurons in the brain. In each neuron, the outputs from the previous layer are combined with the weights and biases of the current layer and compared to an \emph{activation function}, which models whether the signal entering the neuron is strong enough for the neuron to `fire', i.e. to activate the next layer. After passing the image through several such \emph{hidden layers}, the final layer of the network is a classifier which converts the inputs to a probability that the original image $\vec{x}$ belonged to class $y_i$. The goal of a neural network is to construct appropriate weights at each layer such that the distinctions between the different classes are maximally `learned' by the neurons.

In order to construct the weights and biases, one trains the network by feeding through a set of input images for which the classes are known. A \emph{loss function} quantifies how close the network class prediction for the image is to the truth label on the data. The entire network can thus be thought of as a single, differentiable, highly nonlinear function acting on the input pixel densities to produce an output classification. There are many choices of network available. `Deep learning' is distinguished from ordinary neural networks by large numbers of layers, but each layer is typically connected to a few neighbouring neurons in the previous layer, rather than shallow networks with small numbers of dense layers. As it is believed to simulate the process of image recognition in the human brain, it is well suited to the problem of image classification. Convolutional Neural Networks are among the most widely used deep learning architectures. Here we discuss  the main building blocks of a ConvNet and their effects and discuss the training of the network. 

The convolutional neural
network starts from a two-dimensional input image and identifies
characteristic patterns using a stack of convolutional layers. We use
a set of standard operations, starting from the $n\times n$ image
input $I$:

\begin{itemize}
\item[--] ZeroPadding: $(n\times n) \to (n+2 \times n+2)$\\ We
  artificially increase the image by adding zeros at all boundaries in order to
  remove dependence on non-trivial boundary conditions,
  \begin{align} 
    I \to  
    \begin{pmatrix} 
      0 & \cdots &0 \\
      \vdots & I & \vdots\\
      0 & \cdots & 0
    \end{pmatrix} \; .
  \end{align}

\item[--] Convolution: $n'_\text{c-kernel} \times (n\times n) \to
  n_\text{c-kernel} \times((n-n_\text{c-size}+1) \times
  (n-n_\text{c-size}+1))$\\ To identify features in an $n \times n$
  image or feature map we linearly convolute the input with
  $n_\text{c-kernel}$ kernels of size $n_\text{c-size} \times
  n_\text{c-size}$. If in the previous step there are
  $n'_\text{c-kernel}>1$ layers, the kernels are moved over all input
  layers.  For each kernel this defines a feature map $\widetilde
  F^{k}$ which mixes information from all input layers
  \begin{align}
    \widetilde F^{k}_{ij} = \sum_{l=0}^{n'_\text{c-kernel}-1} \quad  \sum_{r,s=0}^{n_\text{c-size}-1} 
    \widetilde{W}^{kl}_{rs}  \;
    I^{l}_{i+r,j+s} + b_k
    \qquad \text{for} \quad
    k = 0,...,n_\text{c-kernel}-1 \; .
  \label{eq:def_conv}
  \end{align}

\item[--] Activation: $(n\times n) \to (n \times n)$\\ This non-linear
  element allows us to create more complex features. A common choice is
  the rectified linear activation function (ReL) which sets pixel with
  negative values to zero, $f_\text{act}(x) = \max (0, x)$. In this
  case we define for example
  \begin{align} 
    F^{k}_{ij} = f_\text{act}(\widetilde F^{k}_{ij})
              = \max \left( 0, \widetilde F^{k}_{ij} \right) \; .
  \end{align}
  Instead of introducing an additional unit performing the activation,
  it can also be considered as part of the previous layer.

\item[--] Pooling: $(n\times n) \to (n/p \times n/p)$\\ We can reduce
  the size of the feature map by dividing the input into patches of
  fixed size $p \times p$ (sub-sampling) and assign a single value to
  each patch
  \begin{align} 
    F'_{ij} = f_\text{pool}(F_{(ip\dots (i+1)p-1,jp\dots (j+1)p-1})
    \; .
  \end{align}
  MaxPooling returns the maximum value of the subsample
  $f_\text{pool}(F)=\max_\text{patch}(F_{ij})$.

\end{itemize}
\medskip

A convolutional layer consists of a ZeroPadding, Convolution, and
Activation step each. We then combine $n_\text{c-layer}$ of these
layers, followed by a pooling step, into a block. Each of our
$n_\text{c-block}$ blocks therefore works with essentially the same
size of the feature maps, while the pooling step between the blocks
strongly reduces the size of the feature maps. This ConvNet setup
efficiently identifies structures in two-dimensional jet images,
encoded in a set of kernels $W$ transforming the original picture into
a feature map. In a second step of our analysis the ConvNet output
constitutes the input of a fully connected DNN, which translates the
feature map into an output label $y$:

\begin{itemize}
\item[--] Flattening: $(n\times n) \to (n^2 \times 1)$\\ While the
  ConvNet uses two-dimensional inputs and produces a set of
  corresponding feature maps, the actual classification is done by a
  DNN in one dimension. The transition between the formats reads
  \begin{align} 
    x = \left( F_{11},\dots,F_{1n},\dots,F_{n1},\dots,F_{nn} 
        \right) \; .
  \end{align}

\item[--] Fully connected (dense) layers: $n^2 \to
  n_\text{d-node}$\\ The output of a standard DNN is the weighted sum
  of all inputs, including a bias, passed through an activation
  function. Using rectified linear activation it reads
  \begin{align}
    y_i 
        = \max \left( 0, \sum_{j=0}^{n^2-1} W_{ij} x_j + b_i \right) \; .
  \label{eq:def_dnn}
  \end{align}
  For the last layer we apply a specific SoftMax activation function 
  \begin{align}
    y_i = \frac{\exp \left( W_{ij} x_j + b_i \right)}
               { \sum_i \exp \left( W_{ij} x_j + b_i \right)} \; .
  \end{align}
   It ensures $y_i \in [0,1]$, so the label can be interpreted as a
   signal or background probability.

\end{itemize}
\medskip

In a third step we define a cost or loss function, which we use to
train our network to a training data set. For a fixed architecture a
parameter point $\theta$ is given by the ConvNet weights
$\widetilde{W}_{rs}^{kl}$ defined in Eq.\eqref{eq:def_conv} combined
with the DNN weights $W_{ij}$ and biases $b_i$ defined in
Eq.\eqref{eq:def_dnn}.  The performance of the training is quantified by minimising the mean squared error
\begin{align}
  L(\theta) = \frac 1 N \sum_{i=0}^{N} \left( y(\theta; x_i) - y_i \right)^2
  \;, 
\end{align}
where $y(\theta;x_i)$ is the predicted binary label of the input $x_i$
and $y_i$ is its true value.
For a given parameter point $\theta$ we compute the gradient of the
loss function $L(\theta)$ and first shift the parameter point from
$\theta_n$ to $\theta_{n+1}$ in the direction of the gradient $\nabla
L(\theta_{n})$.  In addition, we can include the direction of the
previous change such that the combined shift in parameter space is
\begin{align}
  \theta_{n+1} = \theta_{n} - \eta_L \nabla L(\theta_{n}) + \alpha
  (\theta_{n} - \theta_{n-1}) \;.
\end{align}
The learning rate $\eta_L$ determines the step size and can be chosen to
decay with each step (decay rate).  The parameter $\alpha$, referred
to as momentum, dampens the effect of rapidly changing gradients and
improves convergence. The Nesterov algorithm changes the point of
evaluation of the gradient to
\begin{align}
  \theta_{n+1} = \theta_{n} - \eta_L \nabla L(\theta_{n} + \alpha
  (\theta_{n} - \theta_{n-1})) + \alpha
  (\theta_{n} - \theta_{n-1}) \;.
\end{align}
Each training step (epoch) uses the full set of training events.

\subsection{Analysis setup}
\label{sec:analysis}

\subsubsection{Building a jet image}
The analysis objects for the ConvNet are jet images, constructed from Monte Carlo simulations. For signal events we use 14 \tev LHC \ttbar samples in the all hadronic decay channel~\footnote{One could also consider the semileptonic \ttbar channel, which has a much cleaner signature, but since we are interested in the properties of the fat jets rather than the event as a whole, we make use of the much higher statistics of the all-hadronic channel.}. For background we consider a QCD dijet sample, which constitutes the dominant background to this signature. All samples are simulated with \textsc{Pythia8}~\cite{Sjostrand:2014zea}, without the effects of multiparton interactions. The events are then passed through a fast detector simulation with \textsc{Delphes3}~\cite{deFavereau:2013fsa}, using the ATLAS card, and calorimeter towers of size $\Delta\eta \times \Delta\phi = 0.1 \times 5^\circ$. We cluster these towers with the anti-$k_T$~\cite{Cacciari:2008gp} algorithm with $R=1.5$, as implemented in \textsc{FastJet3}~\cite{Cacciari:2011ma}, requiring that all jets have $|\eta |< 1.0$. These anti-$k_T$ jets give us a smooth outer shape of the fat jet and a well-defined jet area for our jet image.  

To ensure that the jet substructure in the jet image is consistent with QCD, and to prepare the jets for the \textsc{HEPTopTagger} algorithm, we re-cluster constituents of the anti-$k_T$ jet using the Cambridge-Aachen (C/A) algorithm with $R=1.5$. Its substructures define the actual jet image pixels. A final comment is in order here, when we identify these calorimeter towers with pixels, it is not clear whether the information used should be the energy $E$ or only
its transverse component $E_T$. We study the performance of the network in both cases.

The rather stringent geometrical cut $|\eta_\text{fat}| < 1.0$ guarantees that the fat jets are contained entirely in the central part of the detector and justifies our calorimeter tower size. For this study we focus on the range $p_{T,\text{fat}} = 350~...~450$~GeV, such that all top decay products can be easily captured in the fat jet.  For signal events, we additionally require that the fat jet can be associated with a Monte-Carlo truth top quark within $\Delta R < 1.2$.

\begin{figure}[t]
\begin{center}
  \includegraphics[width=0.48 \textwidth]{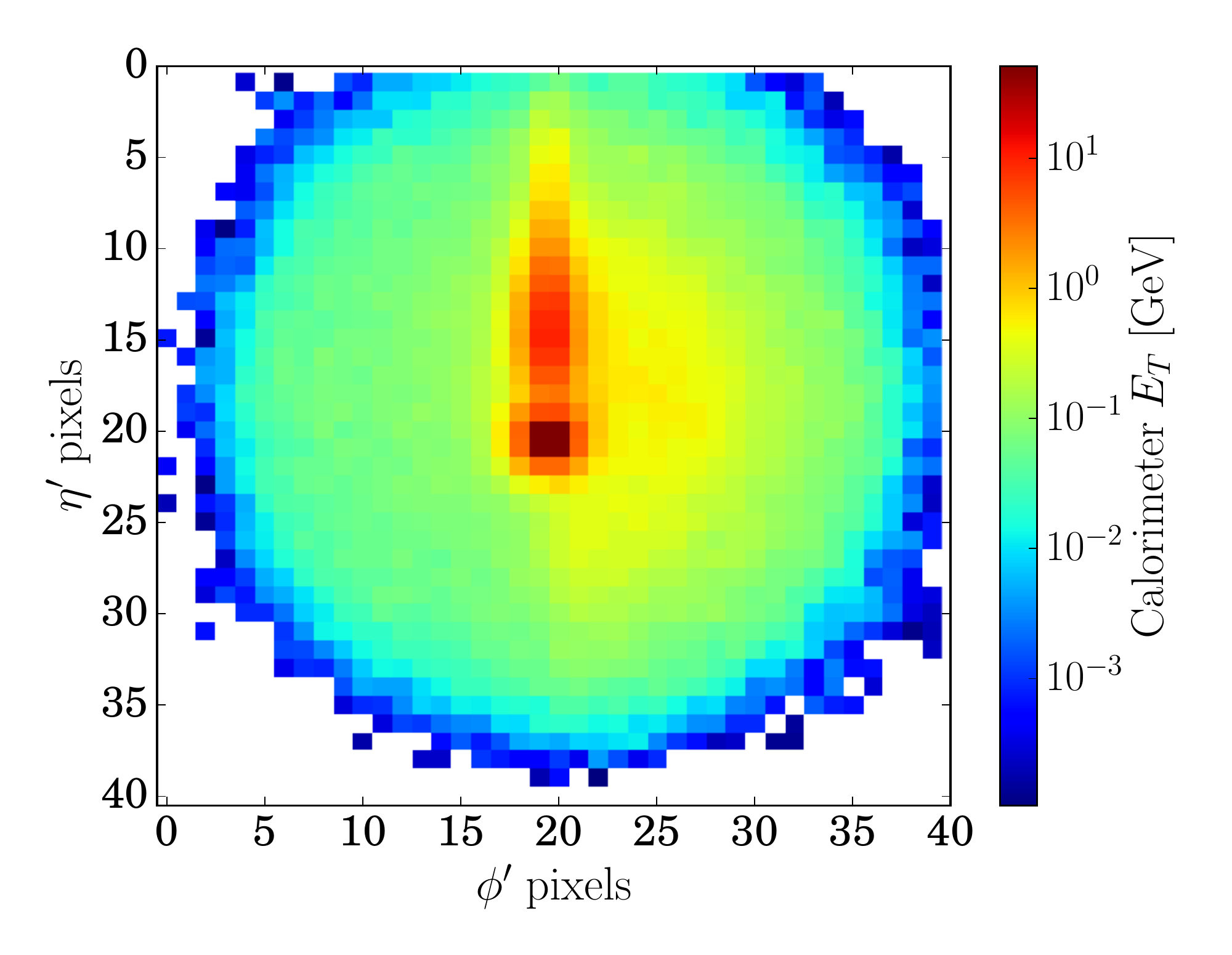}
   \hspace*{0.02\textwidth}
   \includegraphics[width=0.48 \textwidth]{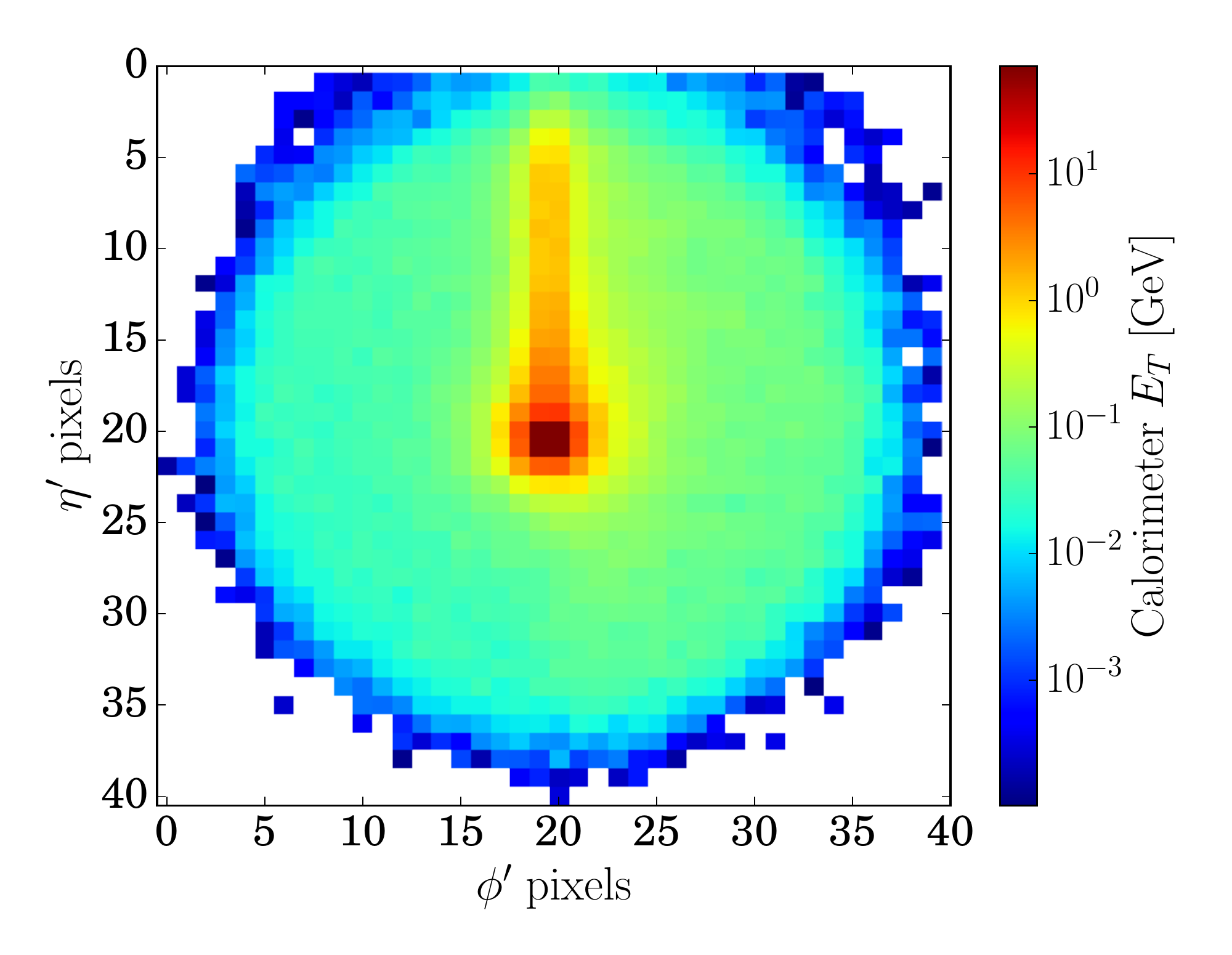}
  \caption[Jet images after preprocessing for top and QCD jets.]{Jet image after pre-processing for the signal (left) and
    background (right). Each picture is averaged over 10,000 actual
    images.}
  \label{fig:averaged_images}
\end{center}  
\end{figure} 

Before feeding in the images to the neural network, it is helpful to apply some preprocessing, such that the salient features of the image are contained in the same location\footnote{In facial recognition, this would be akin to shifting the image such that the eyes are always at the centre.}. The preprocessing steps are as follows:
\begin{enumerate}
\item Find maxima: before we can align any image we have to identify
  characteristic points. Using a filter of size $3\times3$ pixels, we
  localize the three leading maxima in the image.
\item Shift: we then shift the image to center the global maximum
  taking into account the periodicity in the azimuthal angle direction.
\item Rotation: next, we rotate the image such that the second maximum
  is in the 12 o'clock position. The interpolation is done linearly.
\item Flip: next we flip the image to ensure the third maximum
  is in the right half-plane.
\item Crop: finally, we crop the image to $40 \times 40$
  pixels.
\end{enumerate}

Throughout the analysis we will apply two pre-processing setups: for
minimal pre-processing we apply steps~1, 2 and~5 to define a centered
jet image of given size. Alternatively, for full pre-processing we
apply all five steps.  In Fig.~\ref{fig:averaged_images} we show
averaged signal and background images based on the transverse energy
from 10,000 individual images after full pre-processing.  The leading
subjet is in the center of the image, the second subjet is in the 12
o'clock position, and a third subjet from the top decay is smeared
over the right half of the signal images. These images indicate that
fully pre-processed images might lose a small amount of information
at the end of the 12 o'clock axis.

\begin{figure}[t]
\begin{center}
  \includegraphics[width=0.47\textwidth]{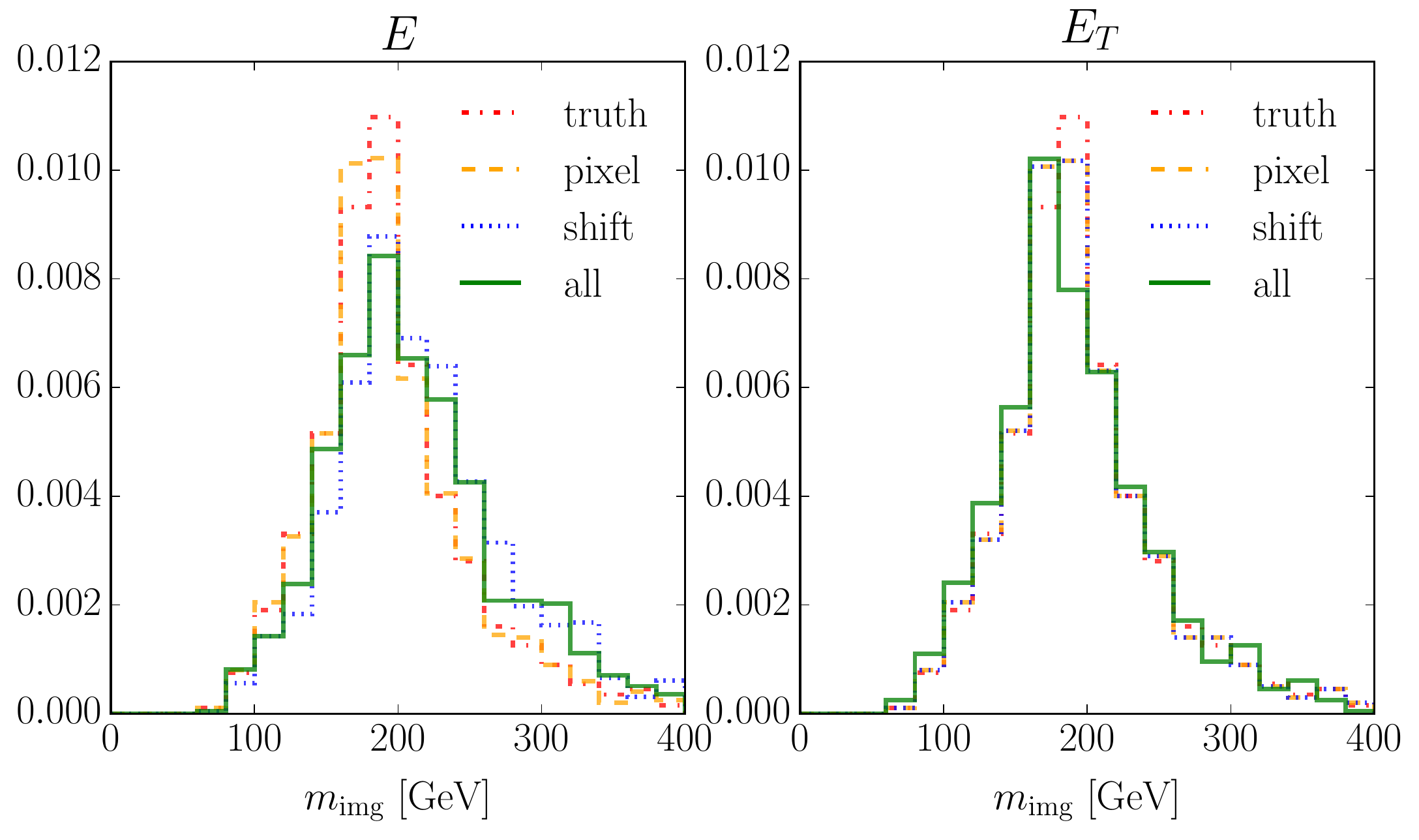}
  \quad
  \includegraphics[width=0.47\textwidth]{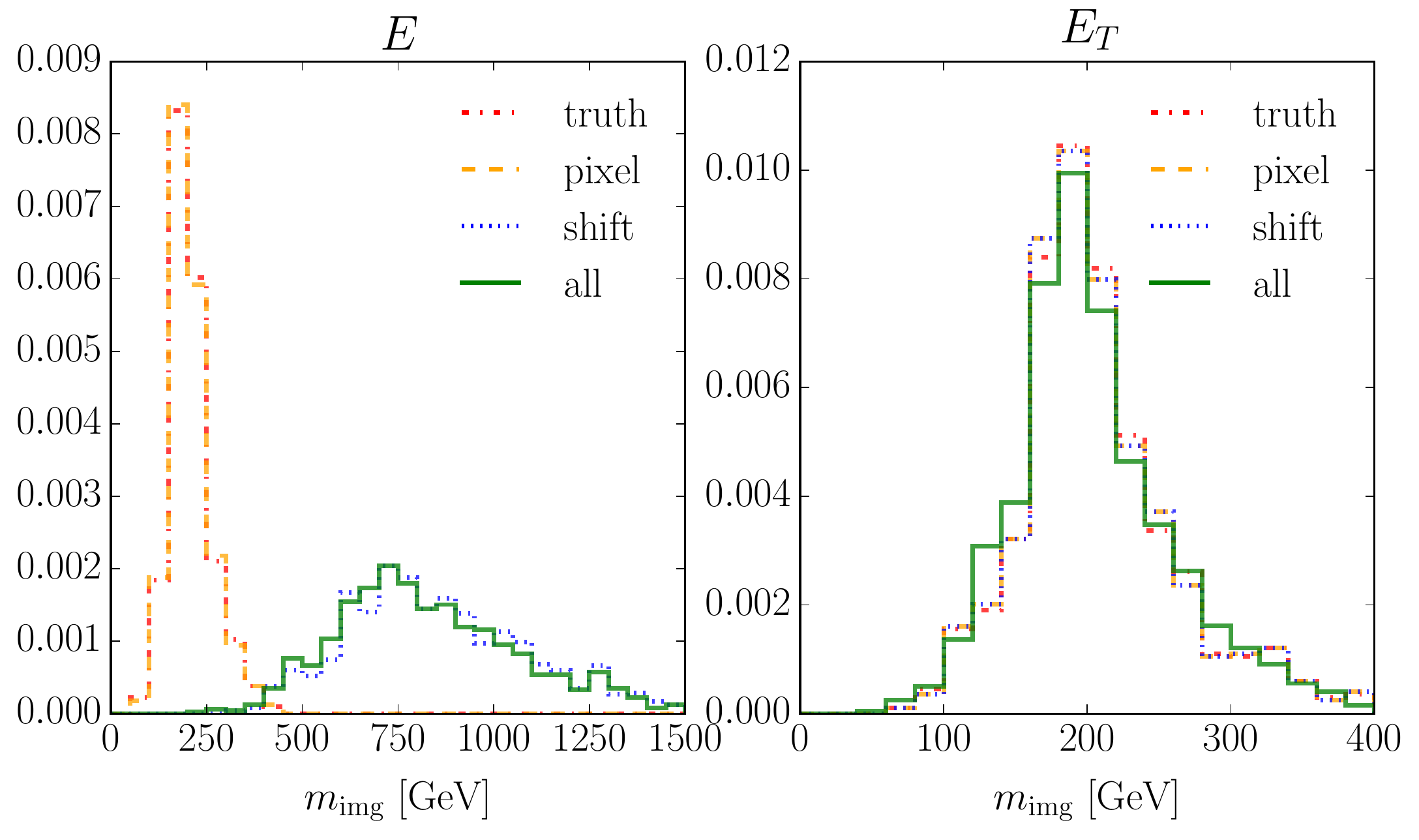}\\
  \includegraphics[width=0.47\textwidth]{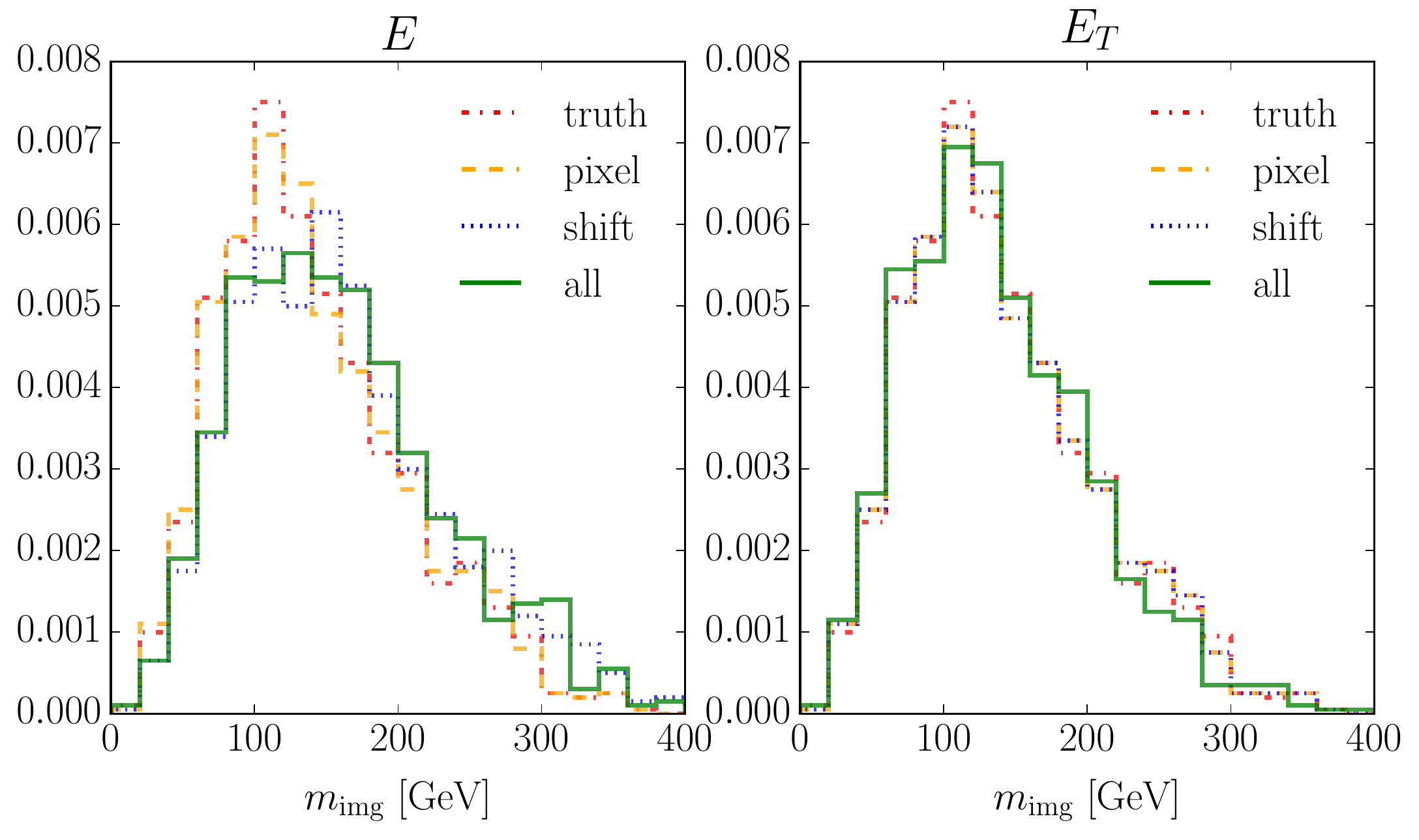}
  \quad
  \includegraphics[width=0.47\textwidth]{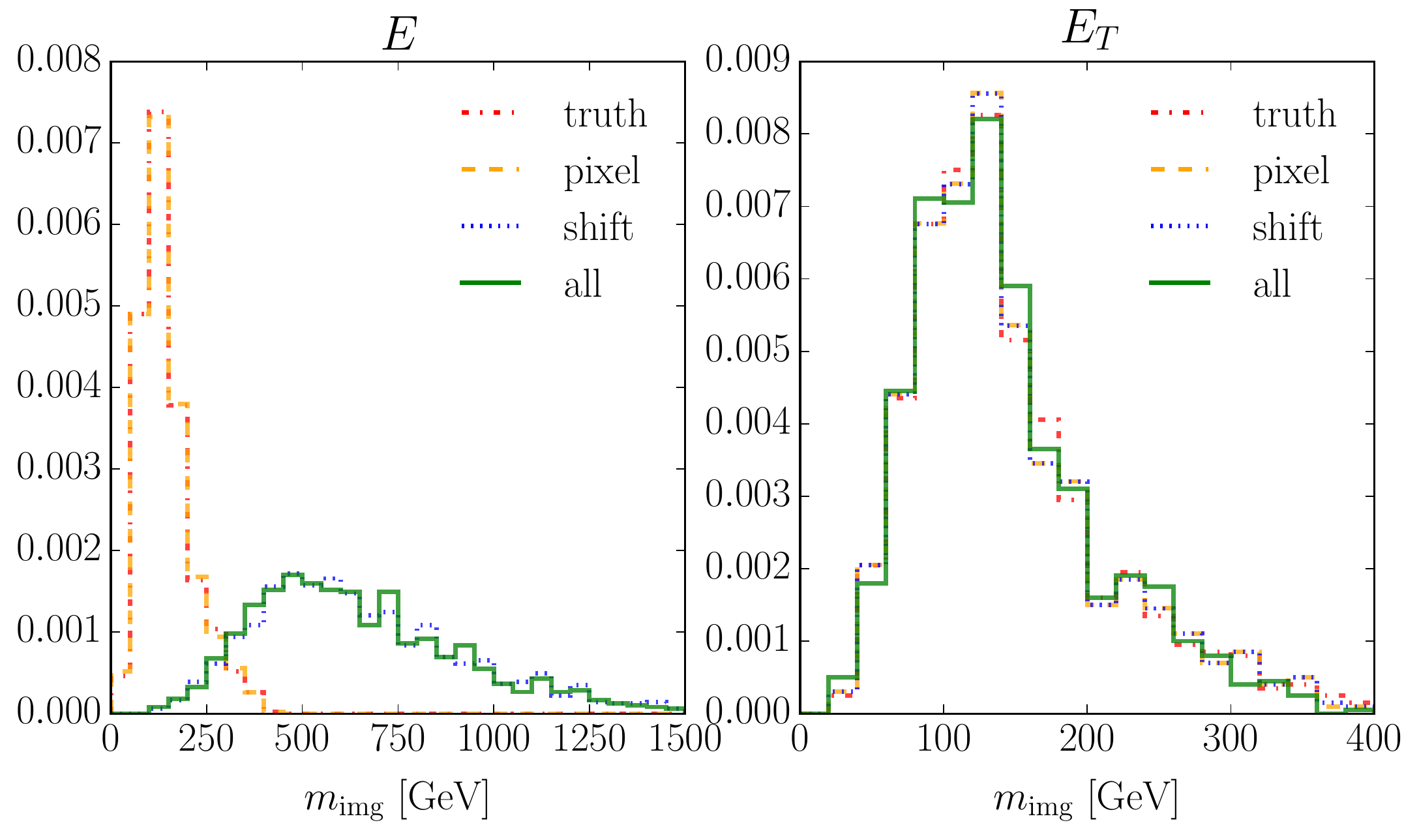}
  \caption[Distortion of jet mass distributions after preprocessing.]{Effect of the preprocessing on the image mass calculated
    from $E$-(left) and $E_T$-images (right) of signal (top) and
    background(bottom). The right set of plots illustrates the
    situation for forward jets with $|\eta| > 2$.}
  \label{fig:img_mass_step}
  \end{center}
\end{figure}

As mentioned above, a non-trivial question is whether one should use the calorimeter tower $E$ or $E_T$ as the pixel density, since the shift and rotation steps of pre-processing involve a longitudinal $\eta$ boost, under which $E$ is not invariant. Following Ref.~\cite{Cogan:2014oua} we investigate the effect on the mass information contained in the images,
\begin{align}
  m_\text{img}^2 = \left[ \sum_i E_i \left(1,\ \frac{\cos\phi'_i}{\cosh\eta'_i},\
      \frac{\sin\phi'_i}{\cosh\eta'_i},
      \frac{\sinh\eta'_i}{\cosh\eta'_i}\right)\right]^2 \qquad E_i =
  E_{T,i} \cosh\eta'_i \;,
\end{align}
where $\eta'_i$ and $\phi'_i$ are the centre of the $i$th pixel after
pre-processing.  The study of all pre-processing steps and their
effect on the image mass in Fig.~\ref{fig:img_mass_step} illustrates
that indeed the rapidity shift has the largest effect on the $E$
images, but this effect is not large.  For the $E_T$ images the jet
mass distribution is unaffected by the shift pre-processing step. The
reason why our effect on the $E$ images is much milder than the one
observed in Ref.~\cite{Cogan:2014oua} is our condition $|\eta_\text{fat}| <
1$. In the the right-hand panels of Fig.~\ref{fig:img_mass_step} we
illustrate the effect of pre-processing on fat jets with $|\eta| > 2$,
where the image masses changes dramatically. Independent of these
details we use pre-processed $E_T$ images as our machine learning
input~\cite{nature_rev}. Since, as a rule of thumb, neural networks perform better with small
numbers, we scale the images such that the pixel entries are between 0 and 1.

\subsubsection{Network architecture}

\begin{figure}[t]
\begin{center}
  \includegraphics[width=0.46\textwidth]{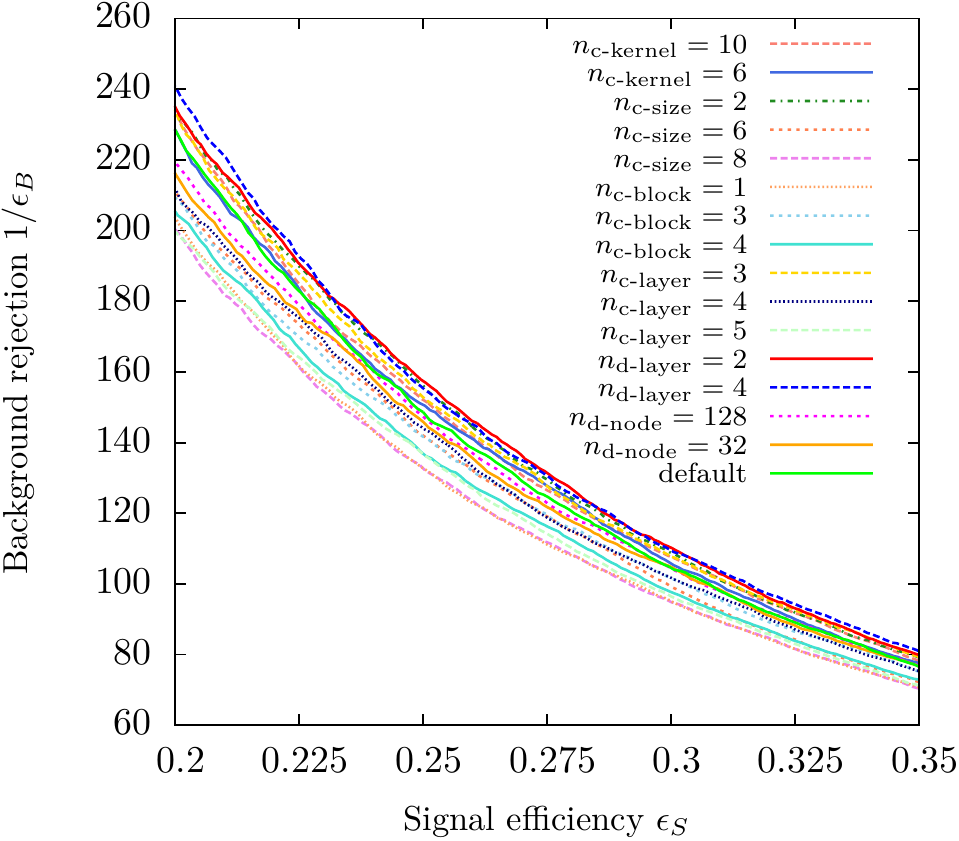}
  \hspace*{0.05\textwidth}
  \includegraphics[width=0.44\textwidth]{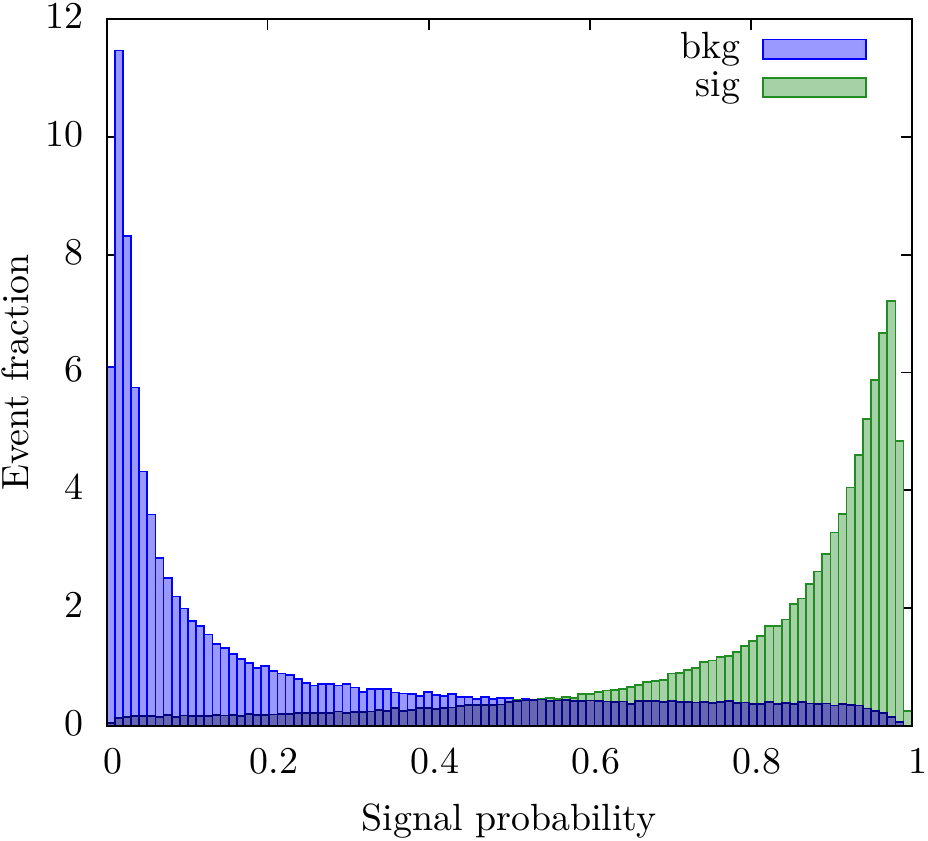}
  \caption[ROC curve for various \textsc{DeepTop} network architectures.]{Left: performance of some of our tested architectures for
    full pre-processing in terms of an ROC curve, including the
    default \textsc{DeepTop} network. Right: discrimination power or
    predicted signal probability for signal events and background probability for
    background events. We use the default network.}
  \label{fig:arc_scan}
  \end{center}
\end{figure}

The neural network architecture is constructed with {\sc{Keras}}~\cite{keras} using the {\sc{Theano}} backend~\cite{theano}.
To identify a suitable \textsc{DeepTop} network architecture, we scan
over several possible realizations or hyper-parameters.  As discussed
in the last section, we start with jet images of size $40 \times 40$.
For architecture testing we split our total signal and background
samples of 600,000 images each into three sub-samples.  After
independently optimizing the architecture we train the network with
150,000 events and after each training epoch test it on an independent
test sample of the same size. The relative performance on the training
and test samples allows us to avoid over-training. Finally, we
determine the performance of the default network on a third sample,
now with 300,000 events.

In a first step we need to optimize our network architecture.  The
ConvNet side is organized in $n_\text{c-block}$ blocks, each
containing $n_\text{c-layer}$ sequences of ZeroPadding, Convolution
and Activation steps.  For activation we choose the ReL step function.
Inside each block the size of the feature maps can be slightly reduced
due to boundary effects. For each convolution we globally set a filter
size or convolutional size $n_\text{c-size} \times n_\text{c-size}$.
The global number of kernels of corresponding feature maps is given by
$n_\text{c-kernel}$. Two blocks are separated by a pooling step, in
our case using MaxPooling, which significantly reduces the size of the
feature maps. For a quadratic pool size of $p \times p$ fitting into
the $n \times n$ size of each feature map, the initial size of the new
block's input feature maps is $n/p \times n/p$.  The final output
feature maps are used as input to a DNN with $n_\text{d-layer}$ fully
connected layers and $n_\text{d-node}$ nodes per layer.

\begin{table}[b!]
\begin{center}
\begin{tabular}{ l | c | c}
\hline
hyper-parameter & scan range & default \\
\hline
$n_\text{c-block}$  & 1,2,3,4  & 2 \\
$n_\text{c-layer}$  & 2,3,4,5  & 2 \\
$n_\text{c-kernel}$ & 6,8,10   & 8 \\
$n_\text{c-size}$   & 2,4,6,8   & 4 \\
$n_\text{d-layer}$  & 2,3,4     & 3 \\
$n_\text{d-nodes}$ & 32,64,128  & 64 \\
$p$                & 0,2,4     & 2  \\
\hline
\end{tabular}
\caption[Network hyperparameters varied in order to optimise performance.]{Range of parameters defining the combined ConvNet and DNN
  architecture, leading to the range of efficiencies shown in the left
  panel of Fig.~\ref{fig:arc_scan} for fully pre-processed images.}
\label{tab:arc_best}
\end{center}
\end{table}

In the left panel of Fig.~\ref{fig:arc_scan} we show the performance
of some test architectures. We give the complete list of tested
hyper-parameters in Tab.~\ref{tab:arc_best}. As our default we choose
one of the best-performing networks after explicitly ensuring its
stability with respect to changing its hyper-parameters. The
hyper-parameters of the default network we use for fully as well as
minimally pre-processed images are given in Tab.~\ref{tab:arc_best}.
In Fig.~\ref{fig:arc_best} we illustrate this default
architecture.

\begin{figure}[t]
  \includegraphics[width=\textwidth]{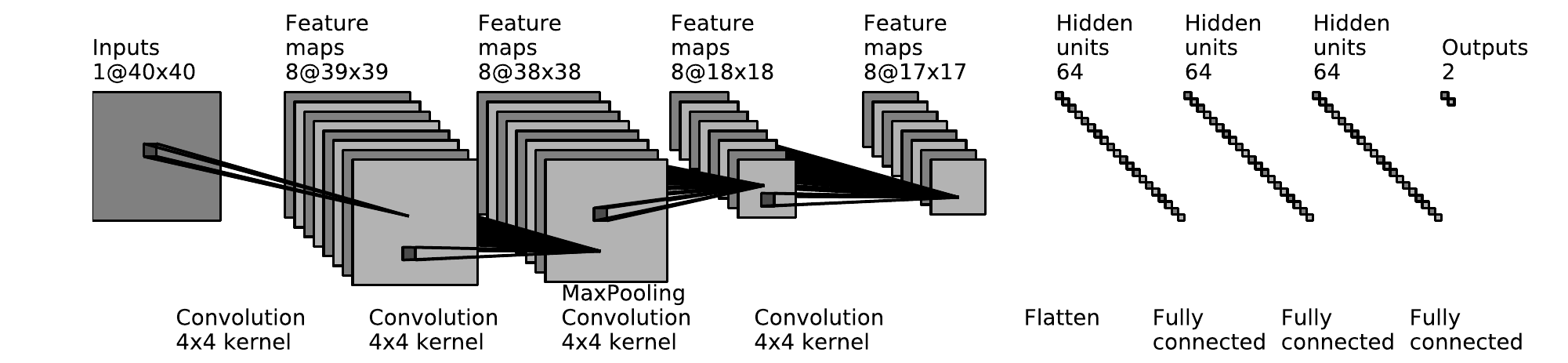} 
  \caption[Schematic of the default \textsc{DeepTop} network architecture.]{Architecture of our default networks for fully
    pre-processed images, defined in Tab.~\ref{tab:arc_best}. Figure drawn using Ref.~\cite{drawconvnet}.}
  \label{fig:arc_best}
\end{figure}

In the second step we train each network architecture using the mean
squared error as our loss function and the Nesterov algorithm with an
initial learning rate $\eta_L = 0.003$. We
train our default setup over up to 1000 epochs and use the network
configuration minimizing the loss function calculated on the test
sample. Different learning parameters were used to ensure convergence
when training on the minimally pre-processed and the scale-smeared
samples. Because the DNN output is a signal and background
probability, the minimum signal probability required for signal
classification is a parameter that allows to link the signal
efficiency $\epsilon_S$ with the mis-tagging rate of background events
$\epsilon_B$.

In Sec.~\ref{sec:results} we will use this trained network to test the
performance in terms of ROC curves, correlating the signal efficiency
and the mis-tagging rate.

\begin{figure}[t!]
  \includegraphics[width=0.119\textwidth]{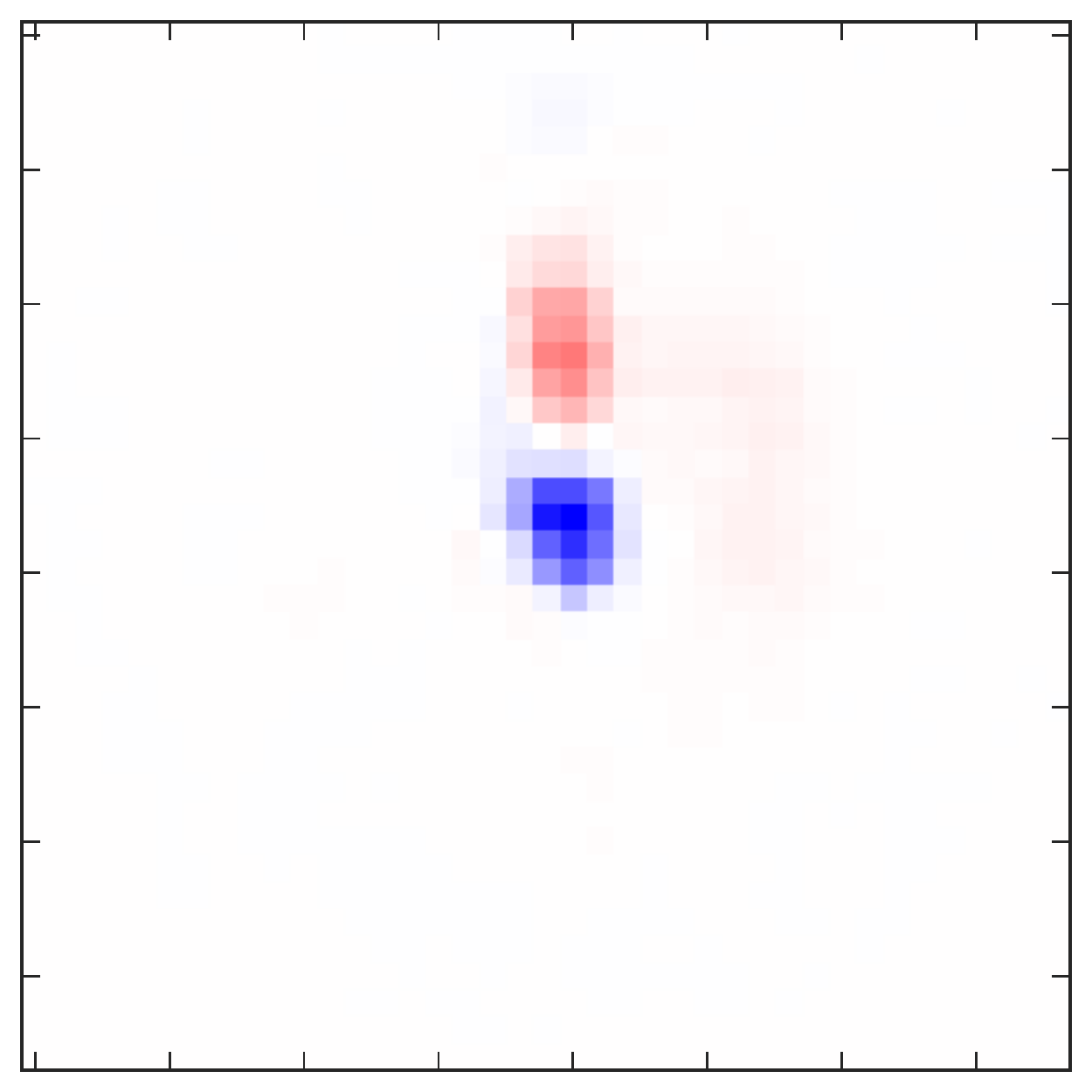}
  \includegraphics[width=0.119\textwidth]{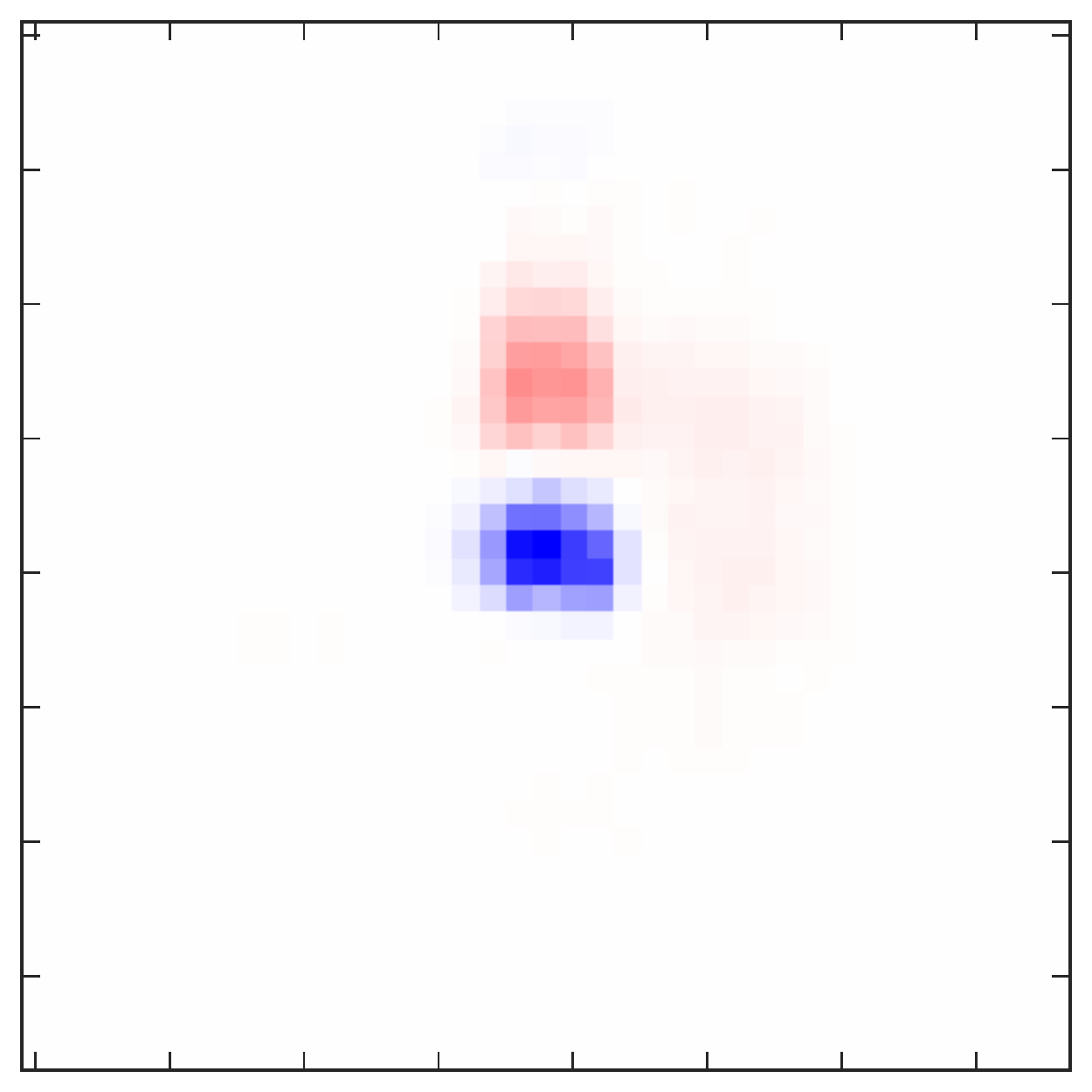}
  \includegraphics[width=0.119\textwidth]{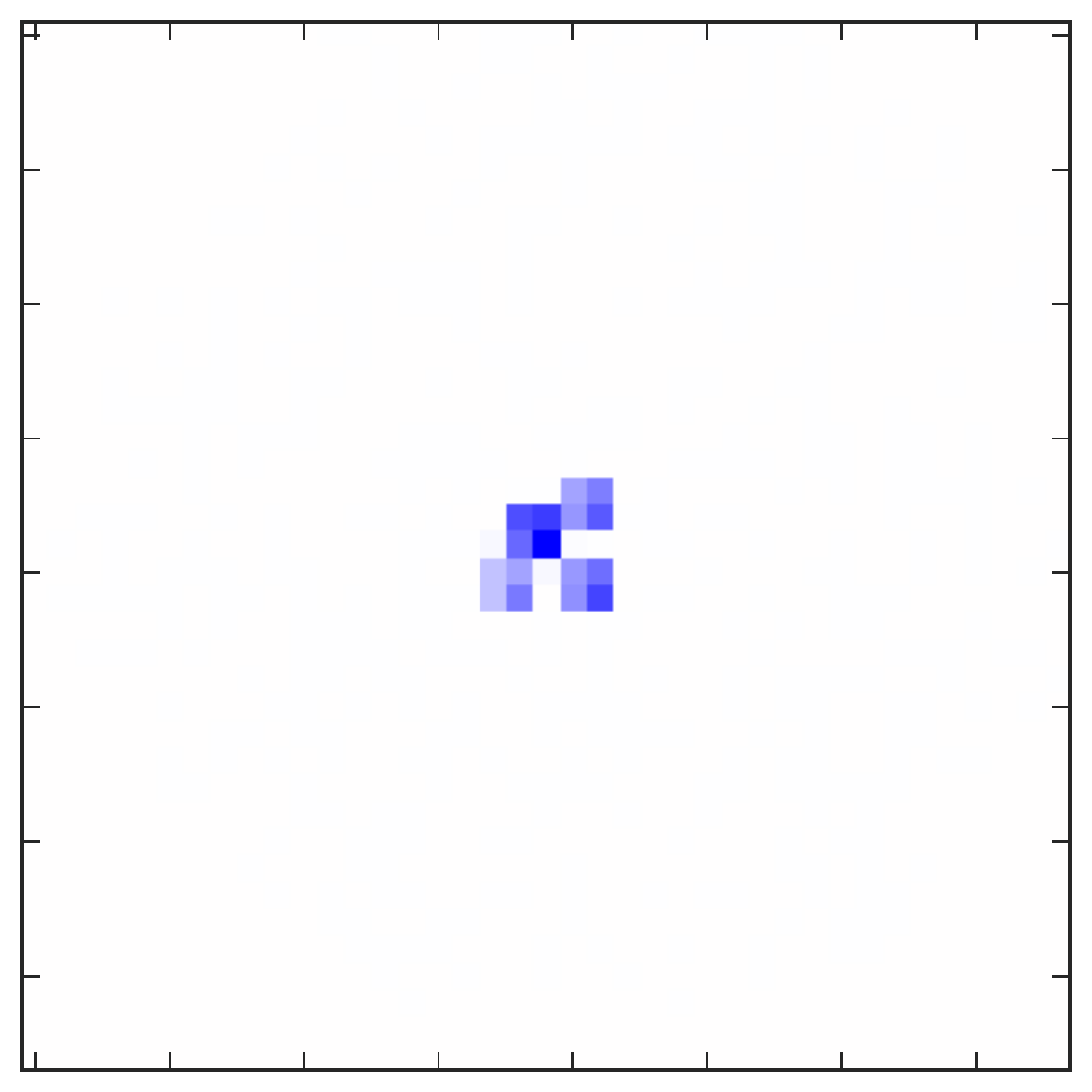}
  \includegraphics[width=0.119\textwidth]{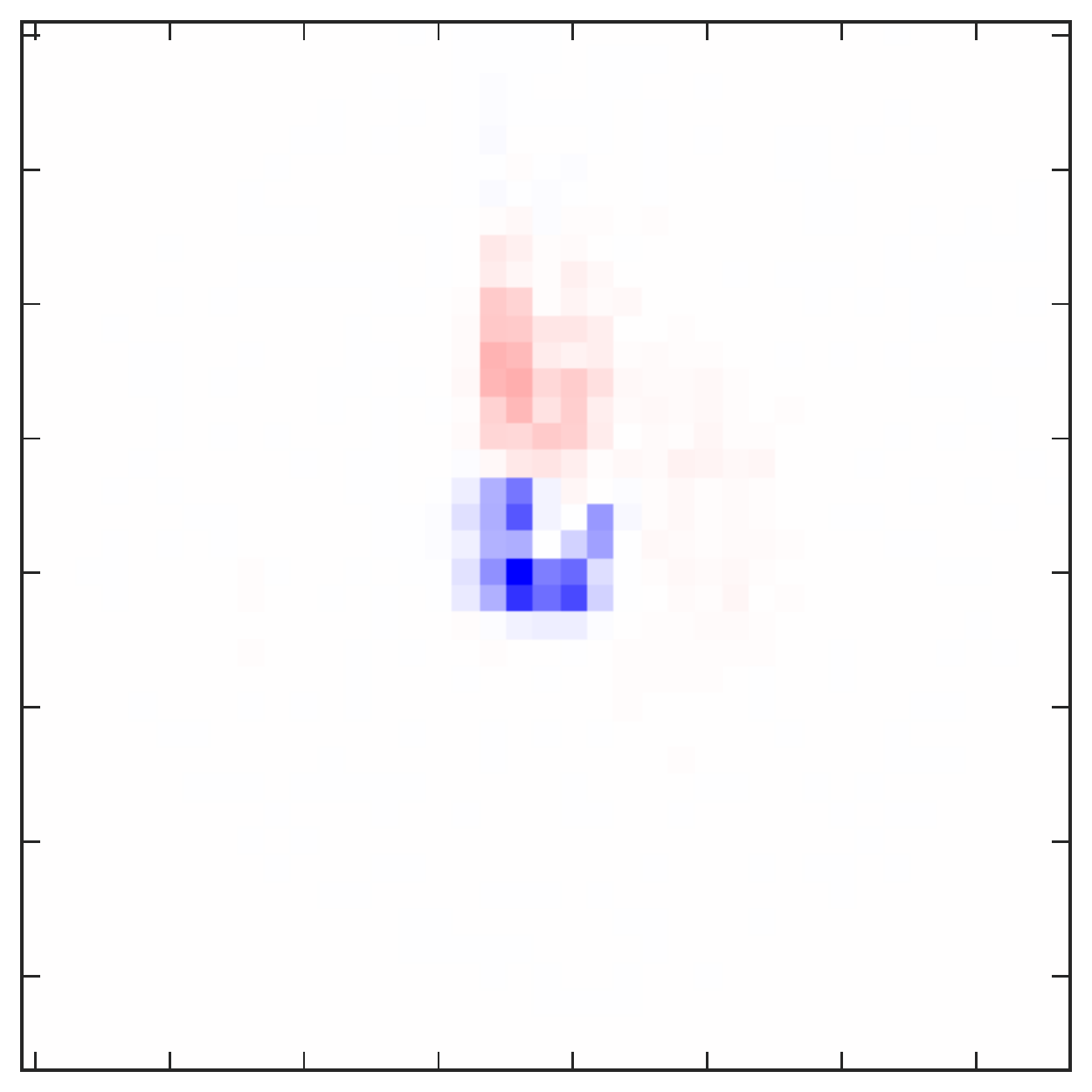}
  \includegraphics[width=0.119\textwidth]{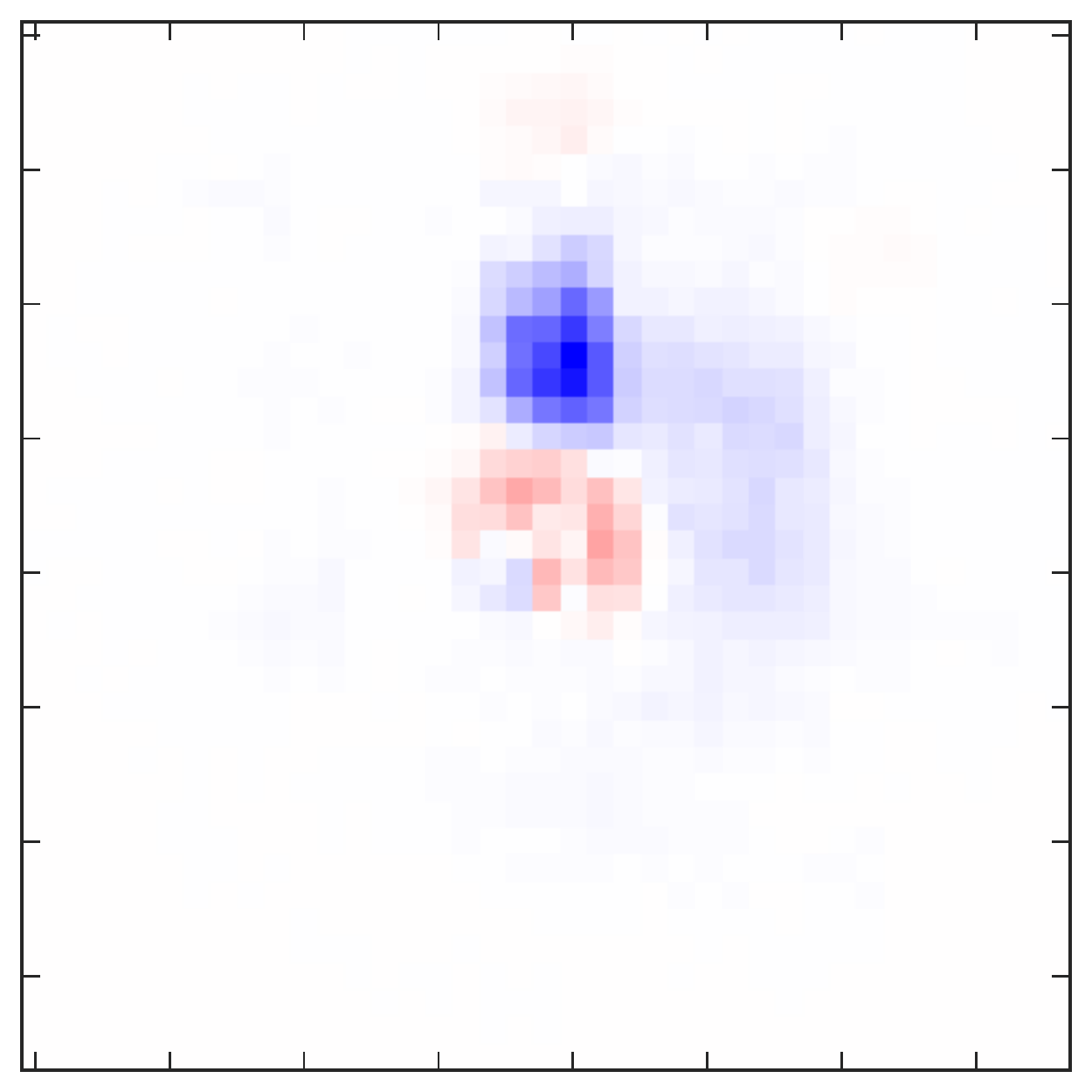}
  \includegraphics[width=0.119\textwidth]{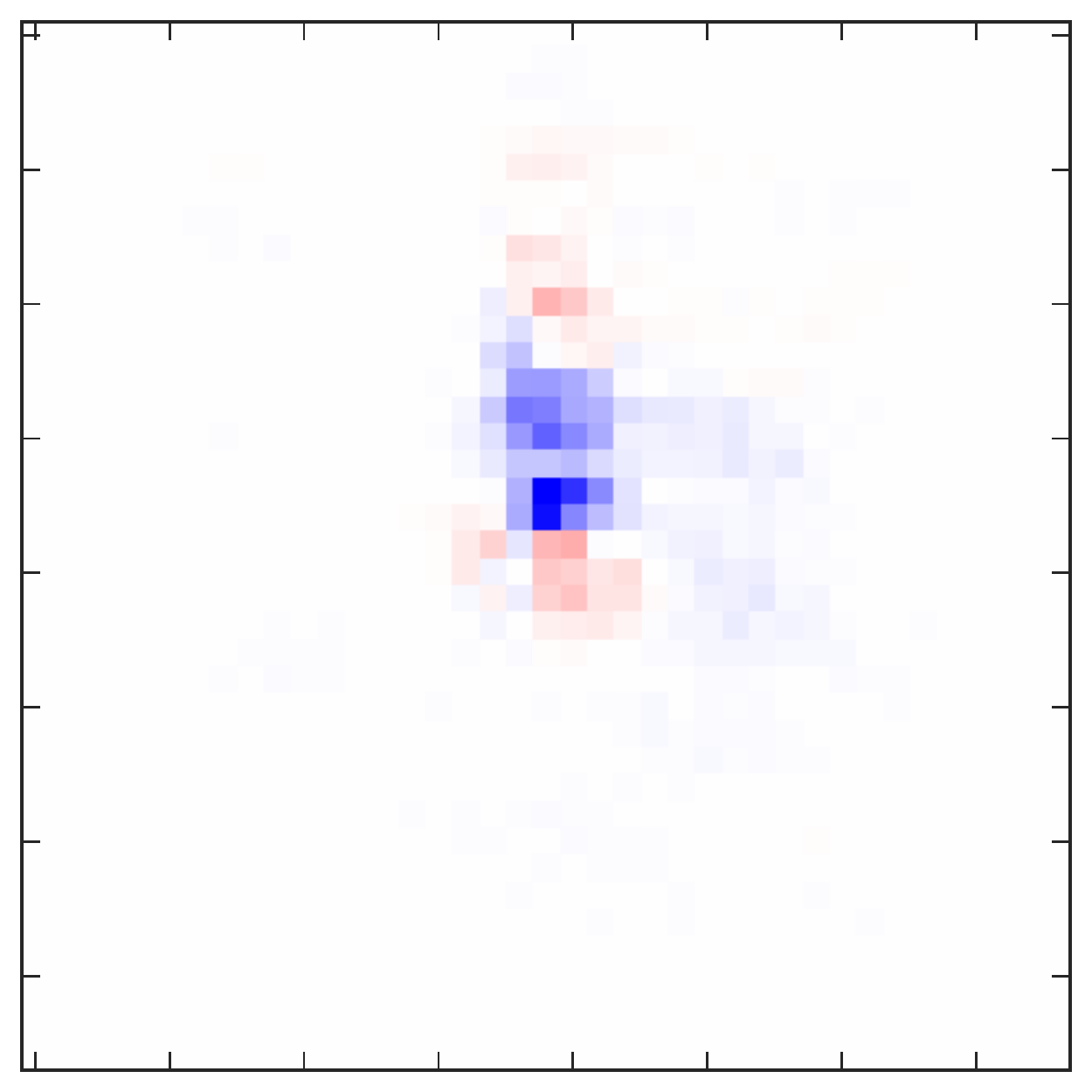} 
  \includegraphics[width=0.119\textwidth]{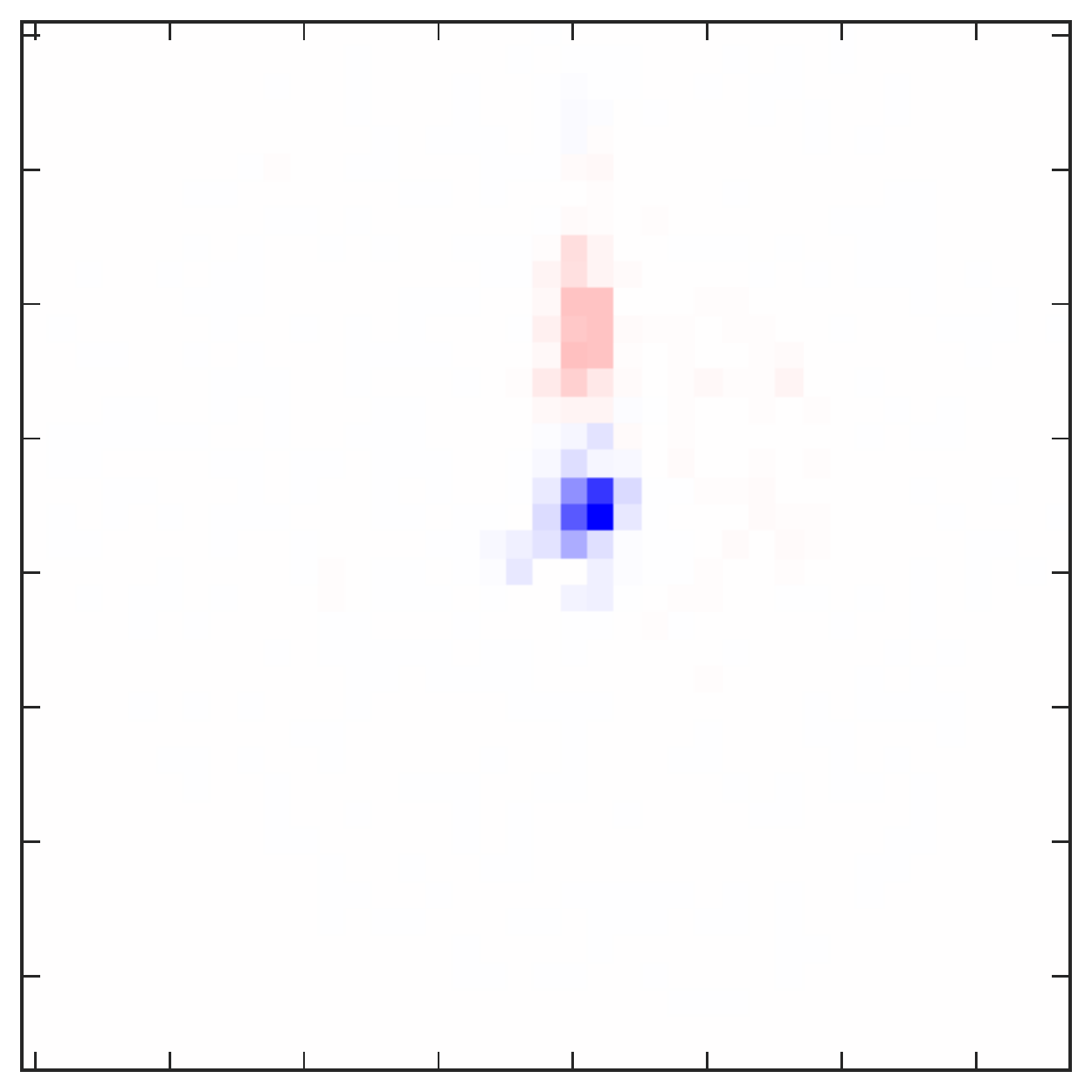} 
  \includegraphics[width=0.119\textwidth]{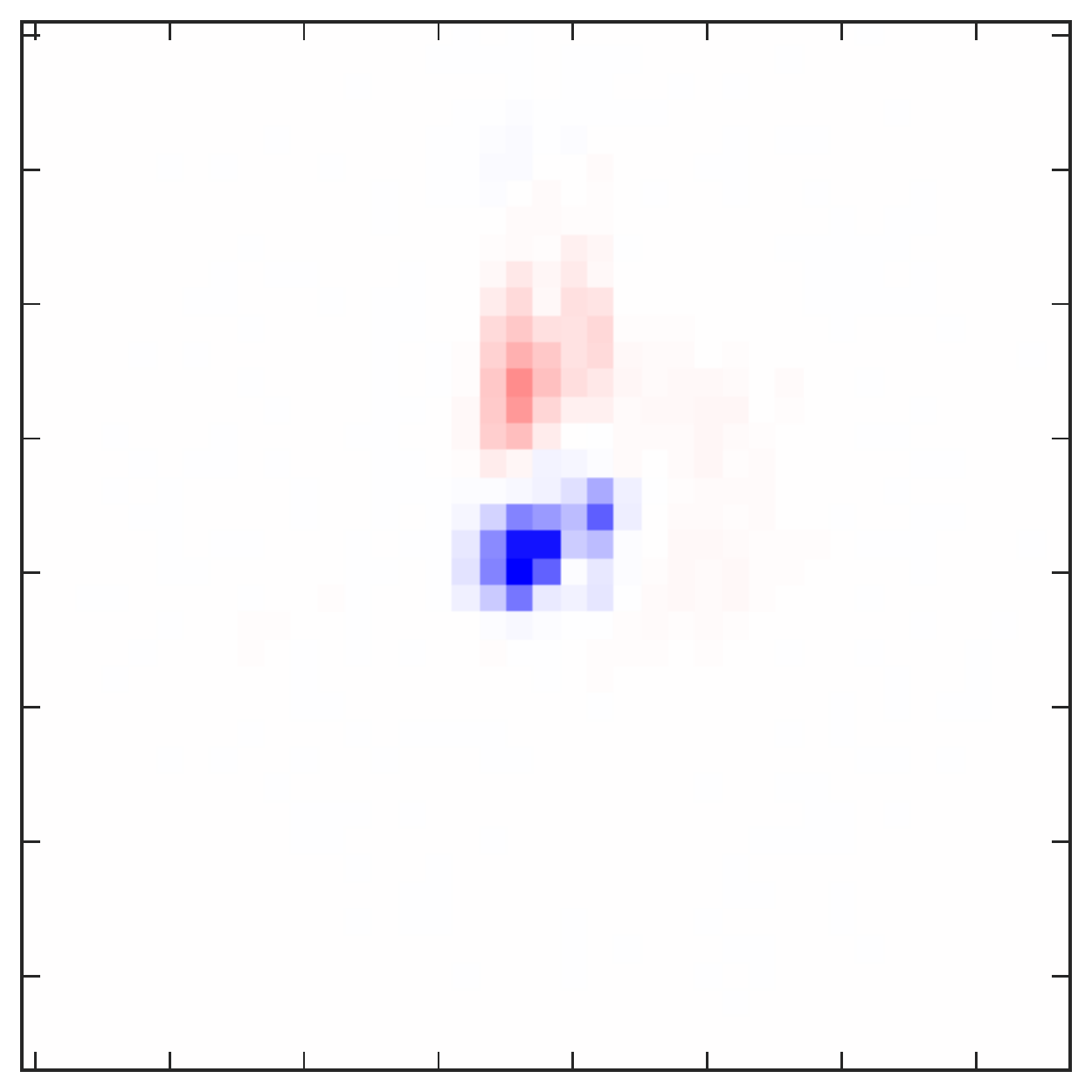} \\
  \includegraphics[width=0.119\textwidth]{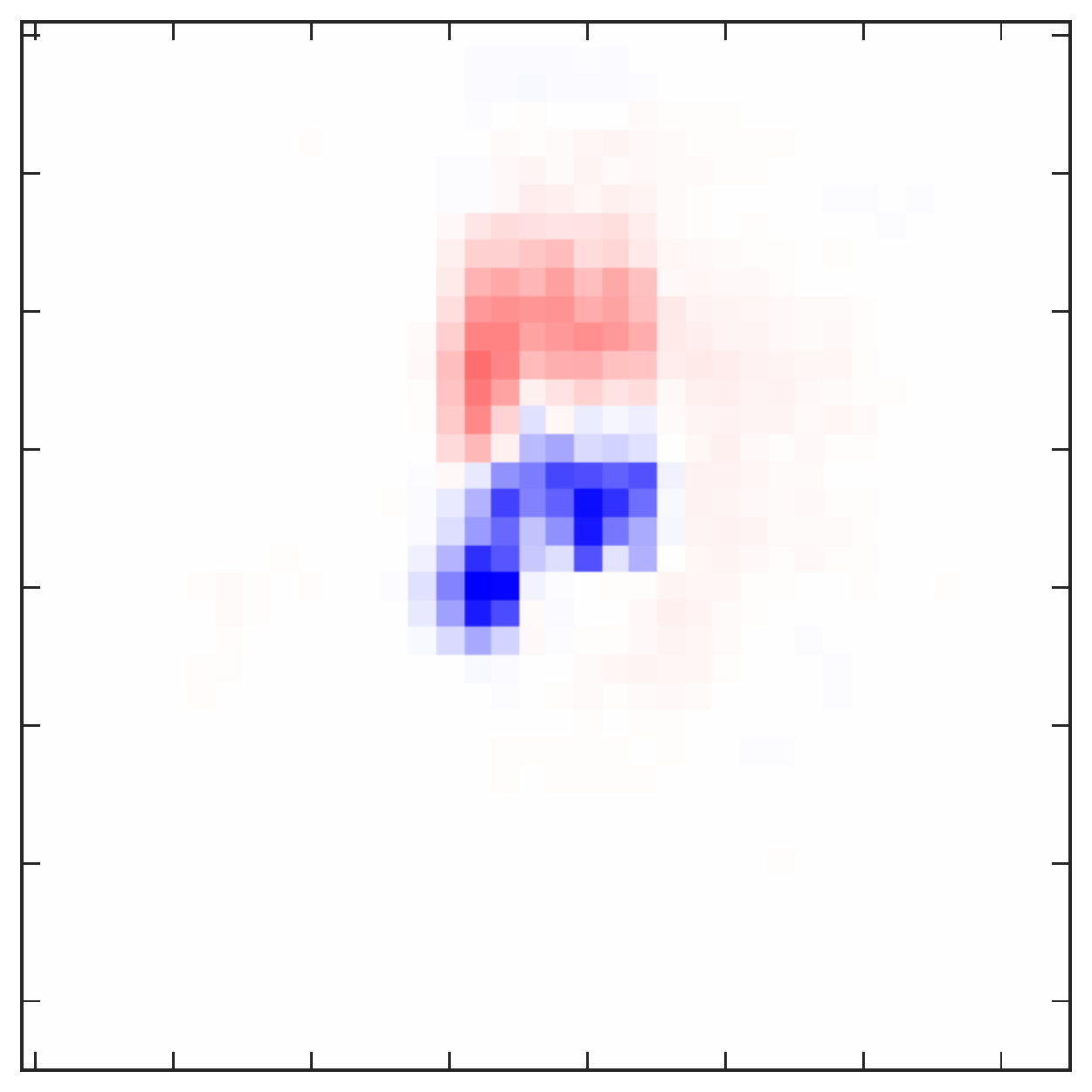}
  \includegraphics[width=0.119\textwidth]{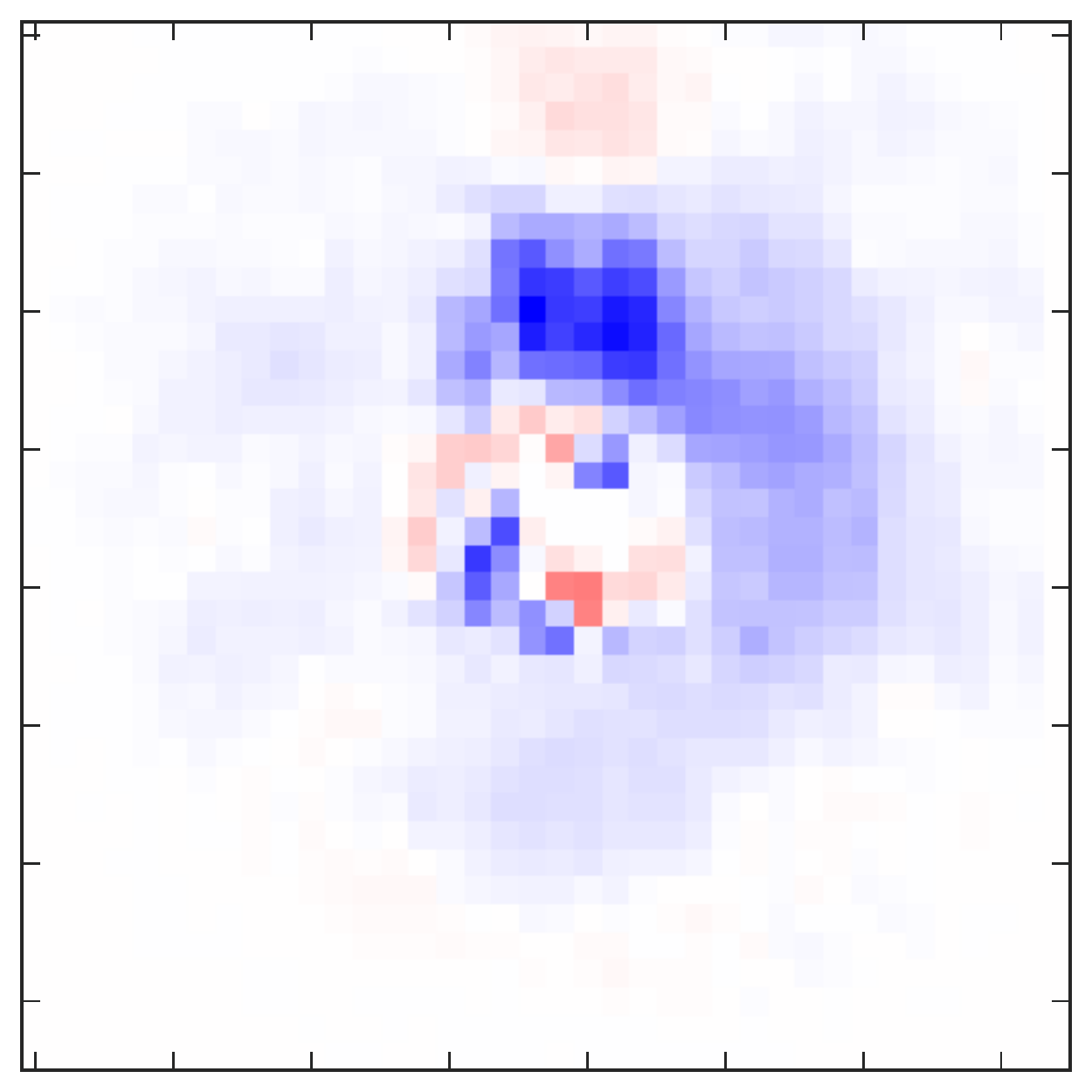}
  \includegraphics[width=0.119\textwidth]{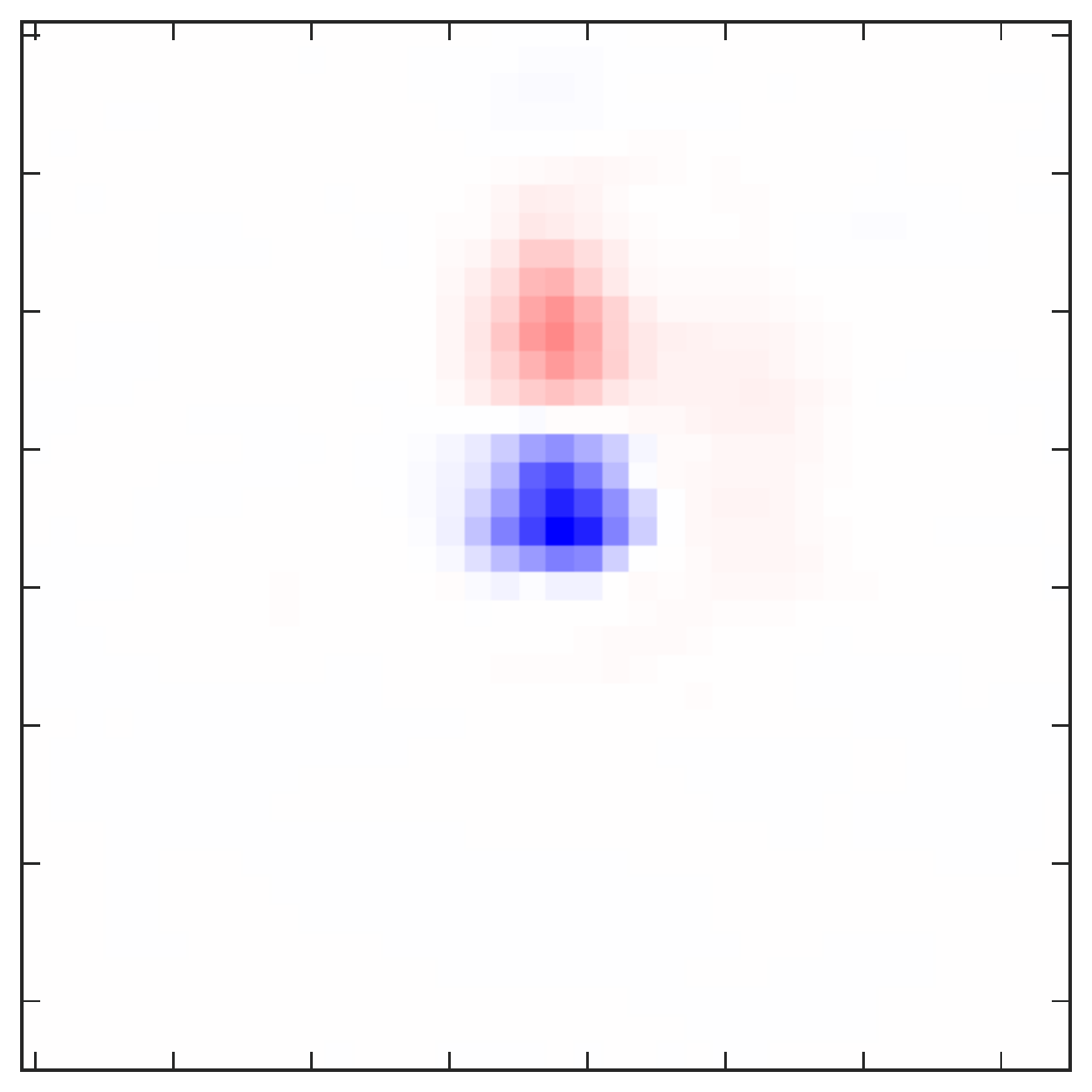}
  \includegraphics[width=0.119\textwidth]{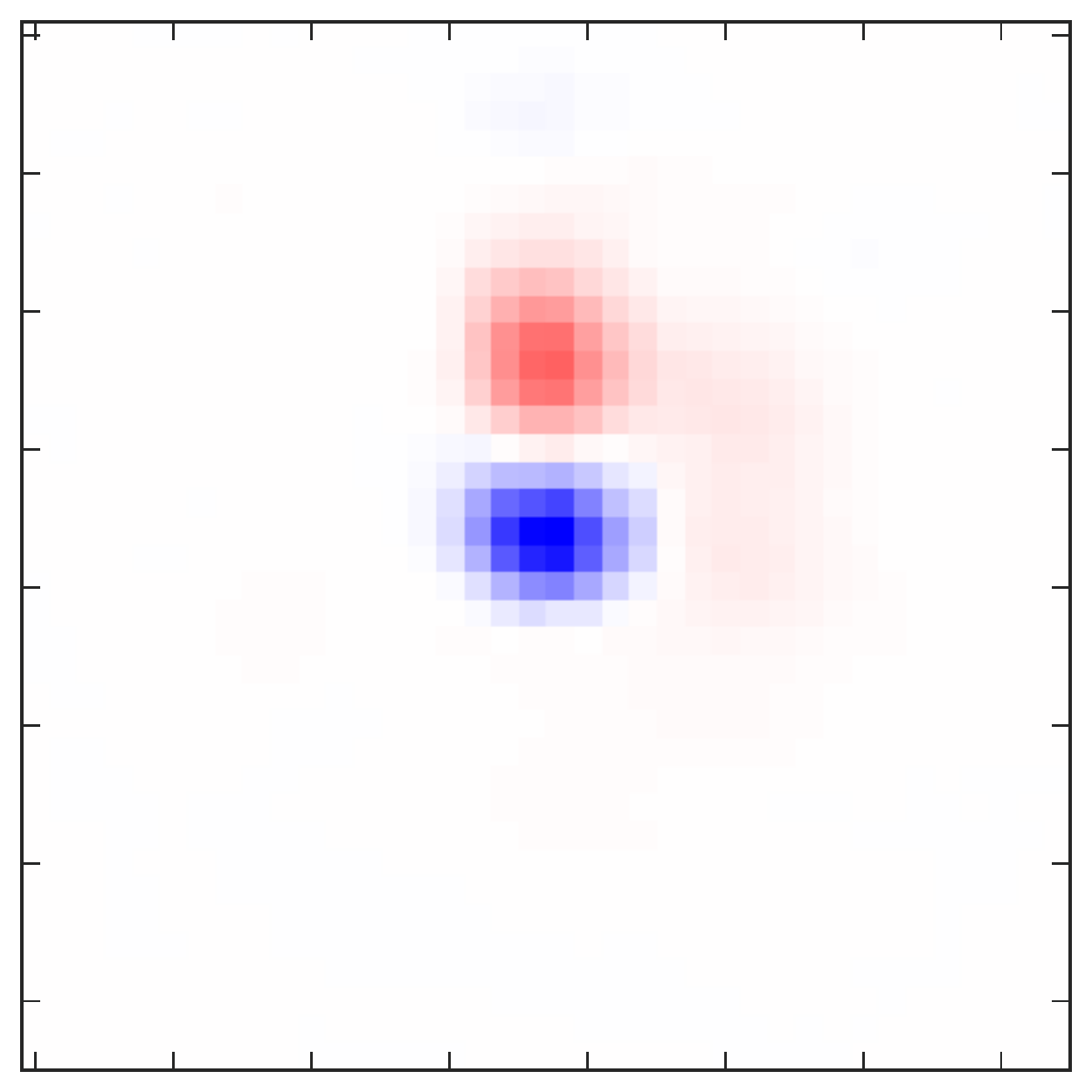}
  \includegraphics[width=0.119\textwidth]{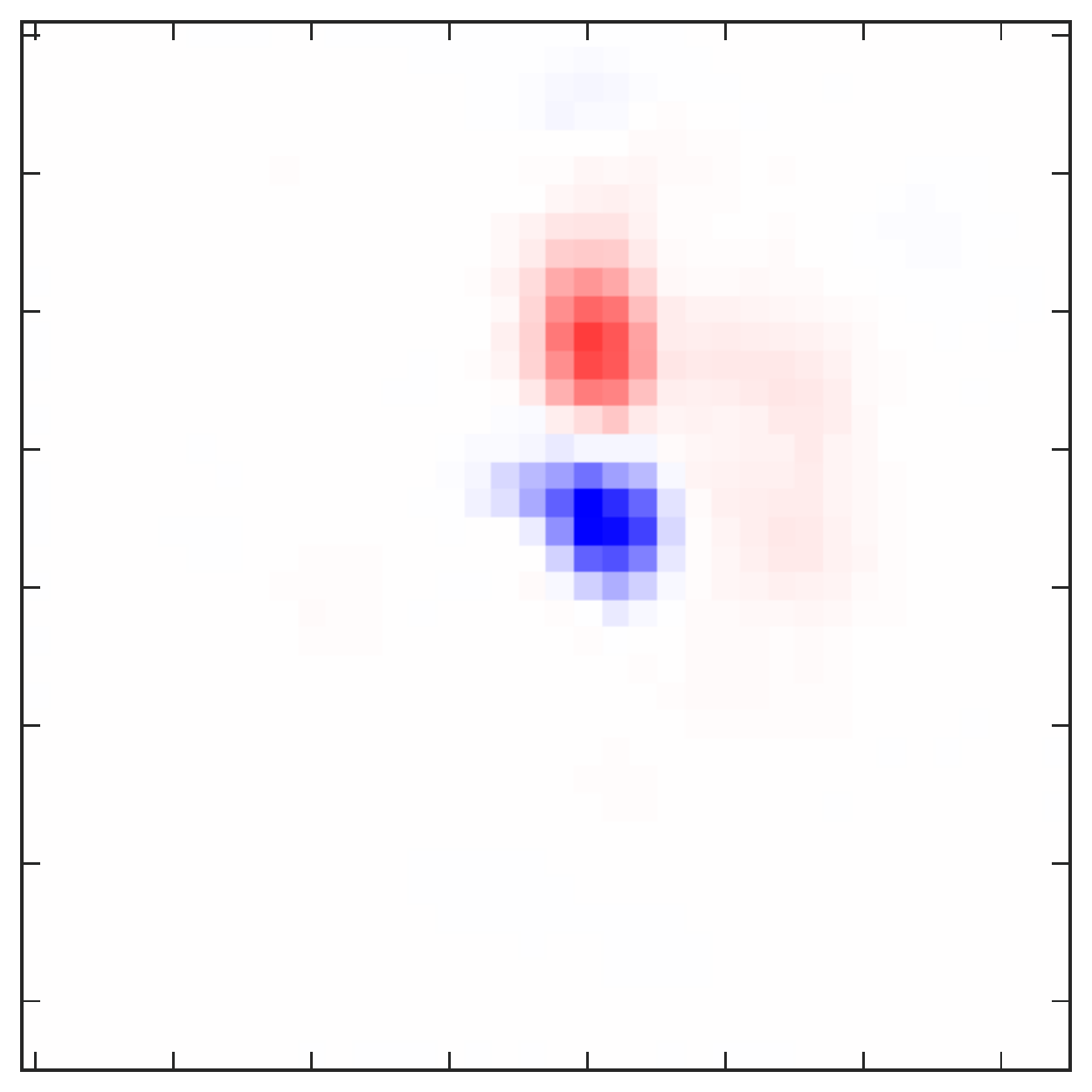}
  \includegraphics[width=0.119\textwidth]{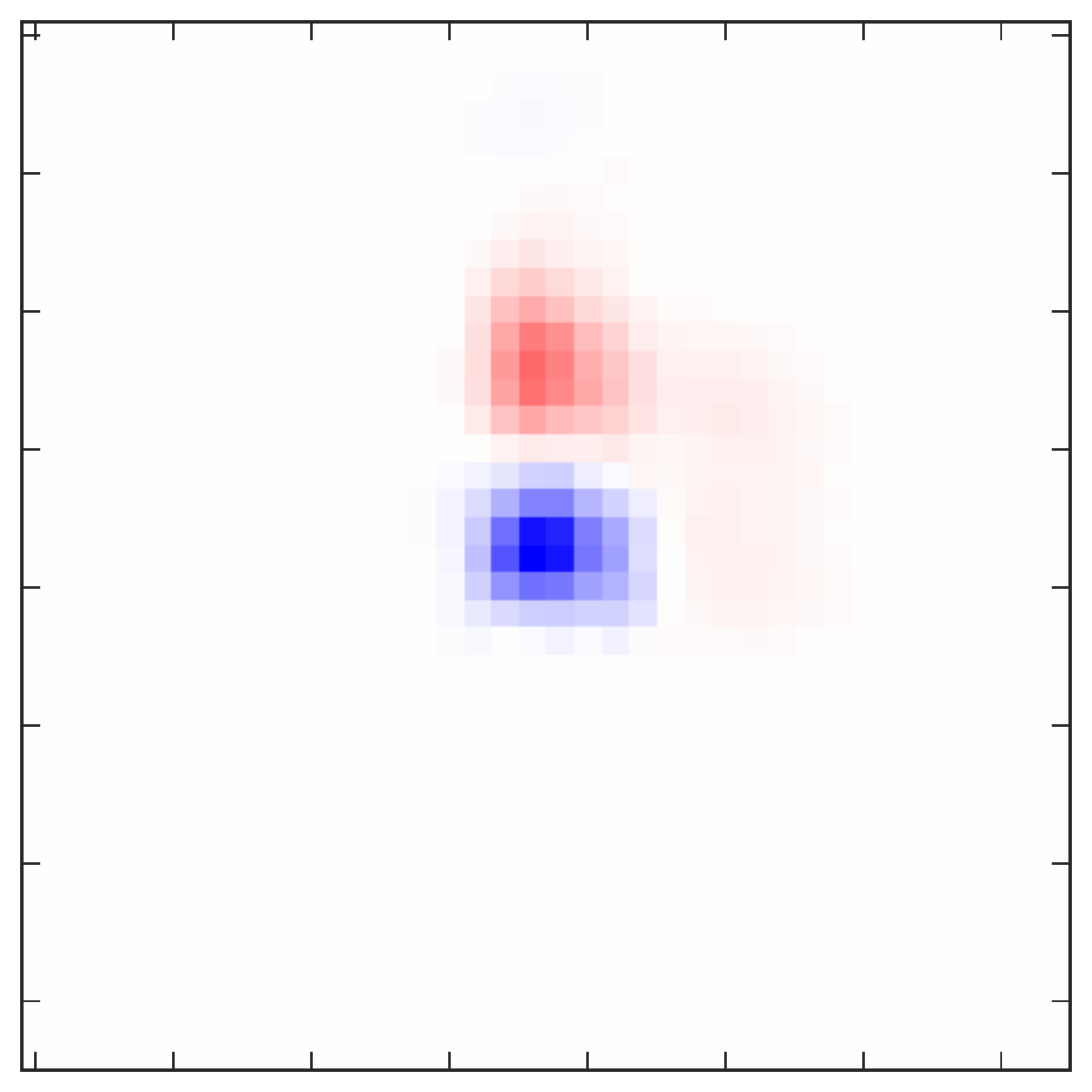} 
  \includegraphics[width=0.119\textwidth]{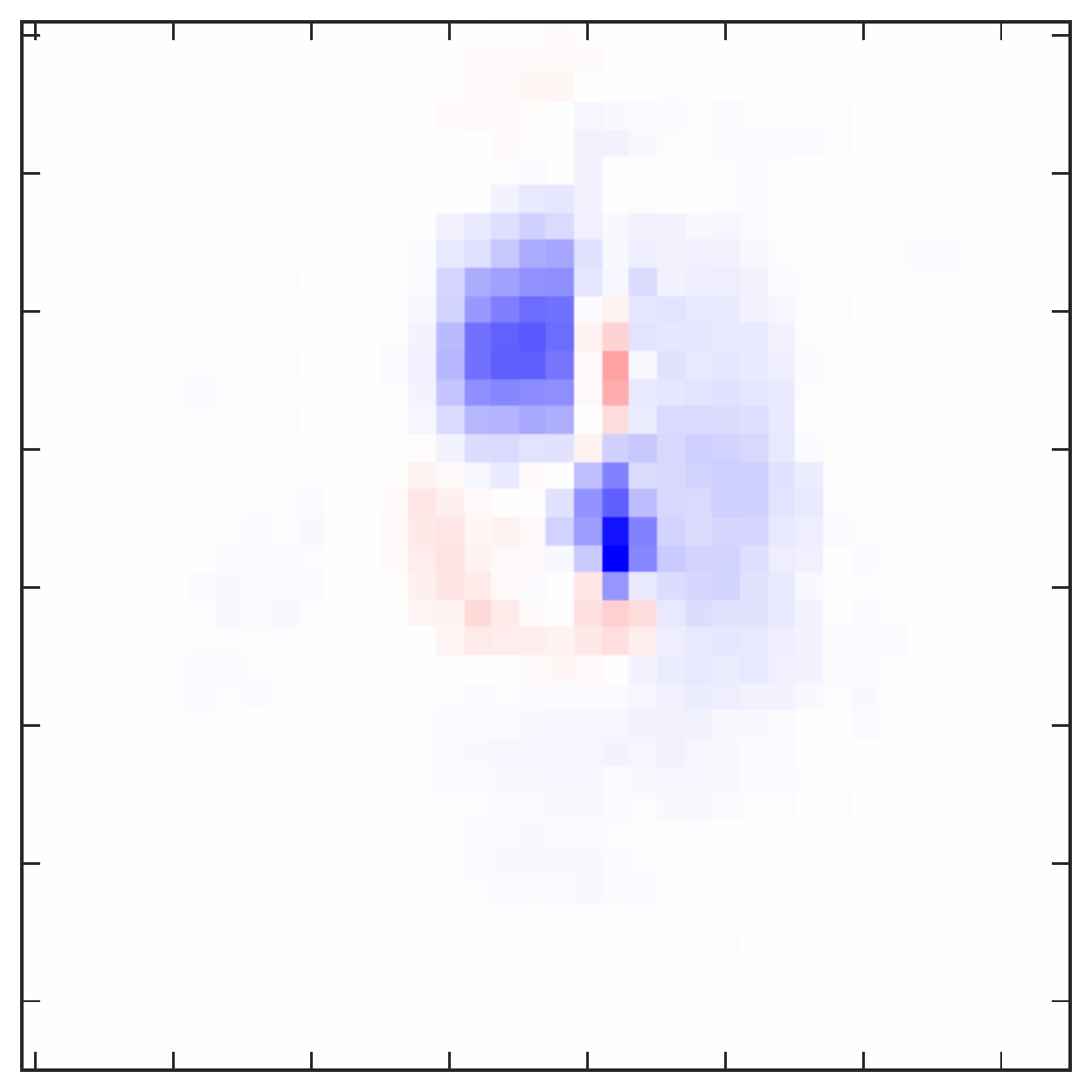} 
  \includegraphics[width=0.119\textwidth]{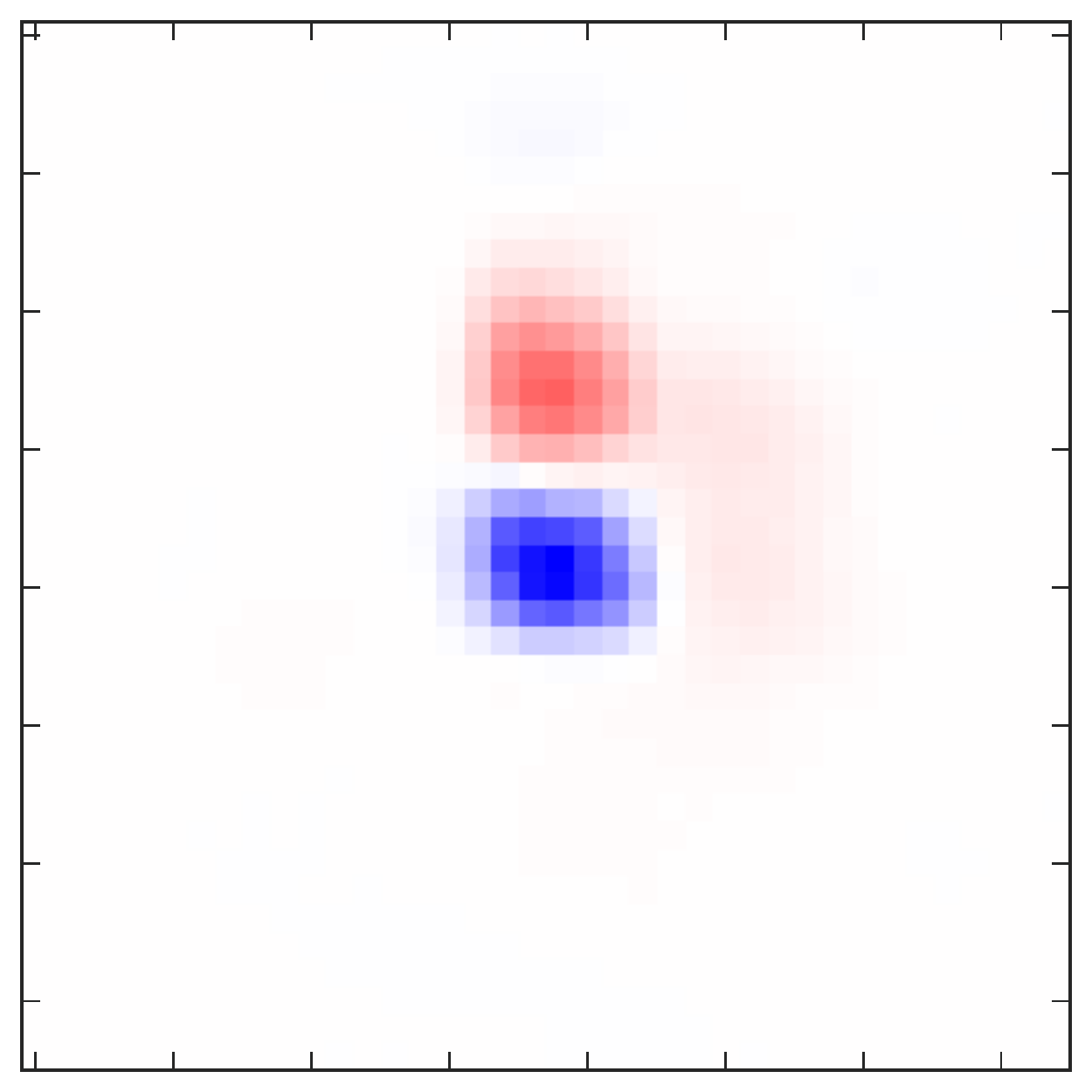} \\
  \includegraphics[width=0.119\textwidth]{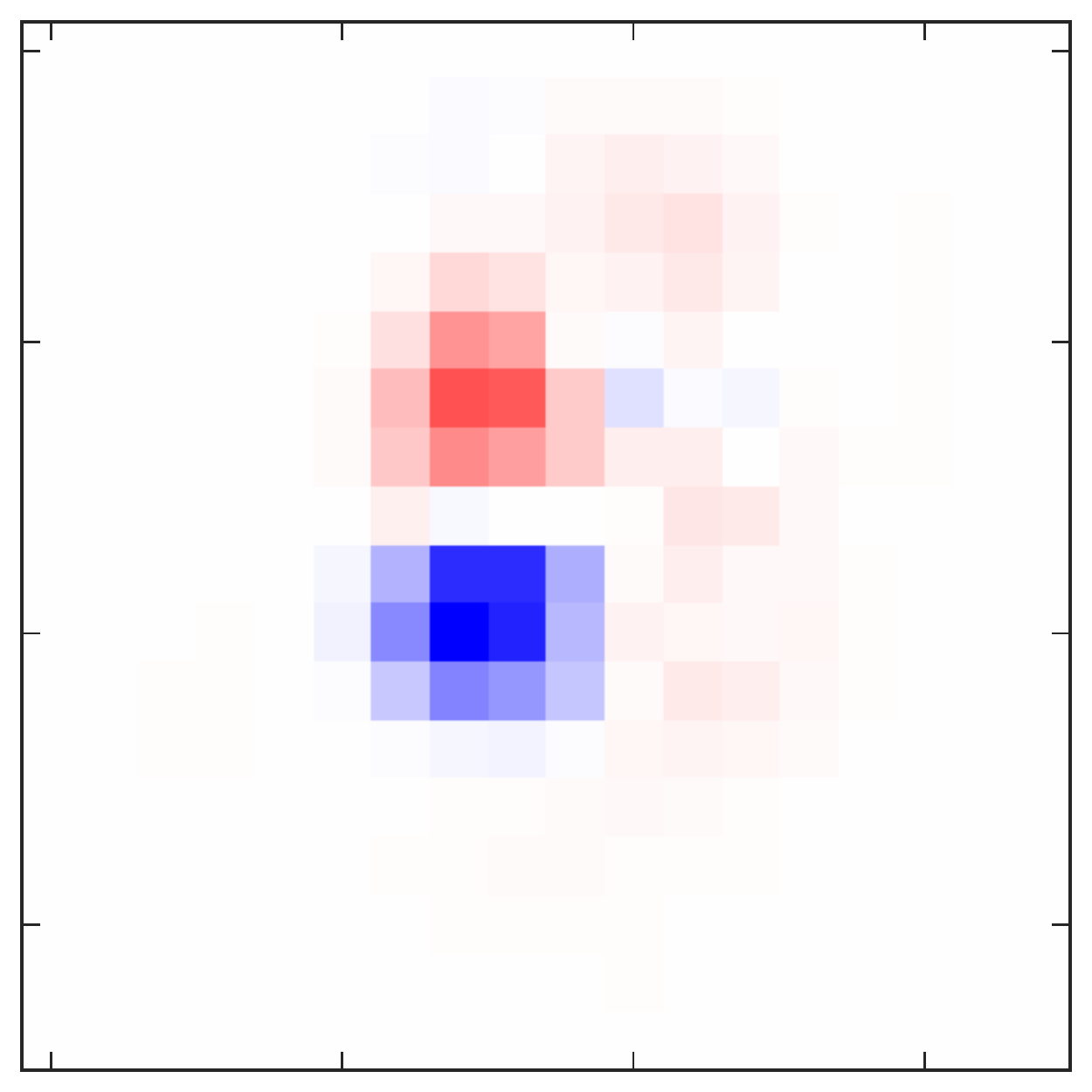}
  \includegraphics[width=0.119\textwidth]{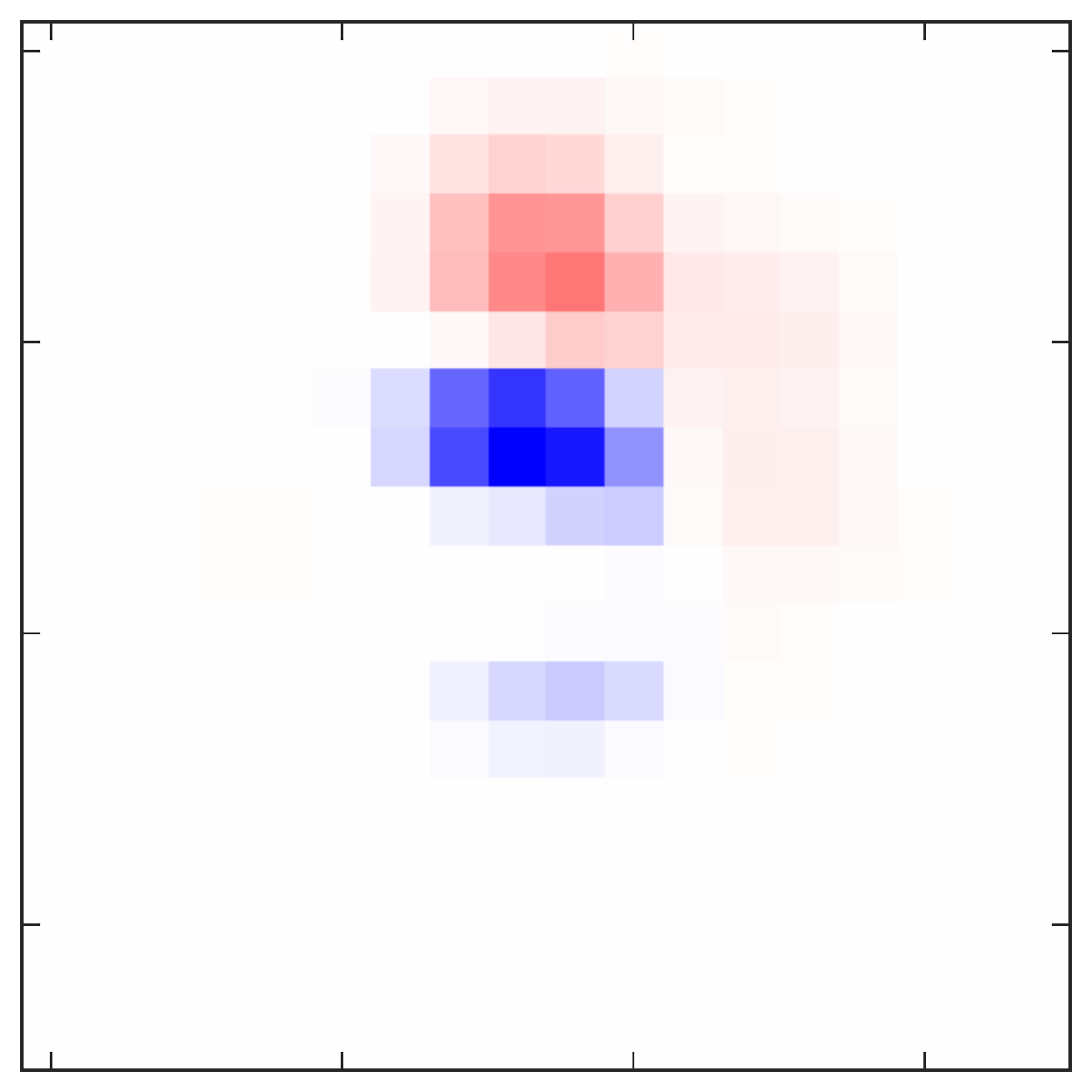}
  \includegraphics[width=0.119\textwidth]{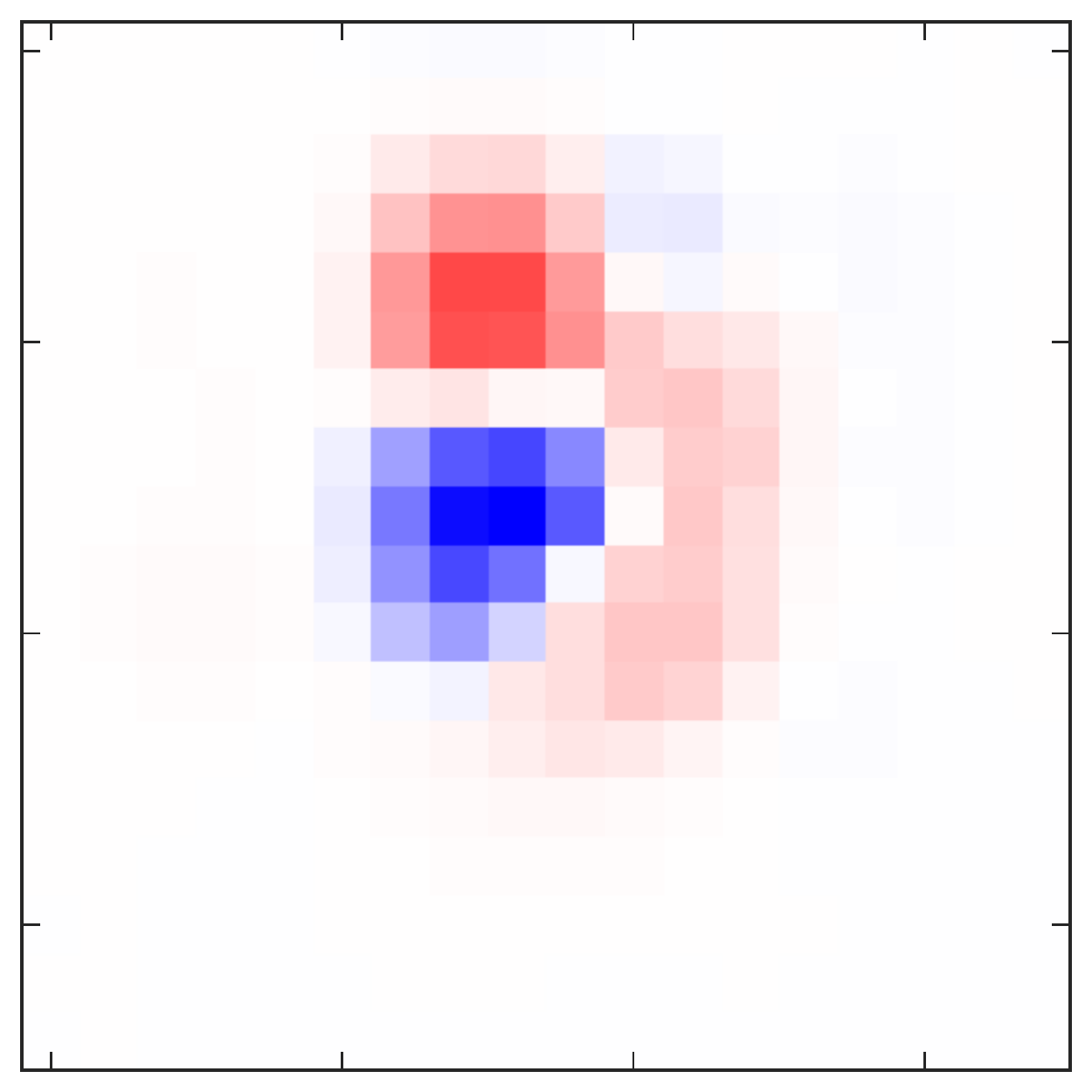}
  \includegraphics[width=0.119\textwidth]{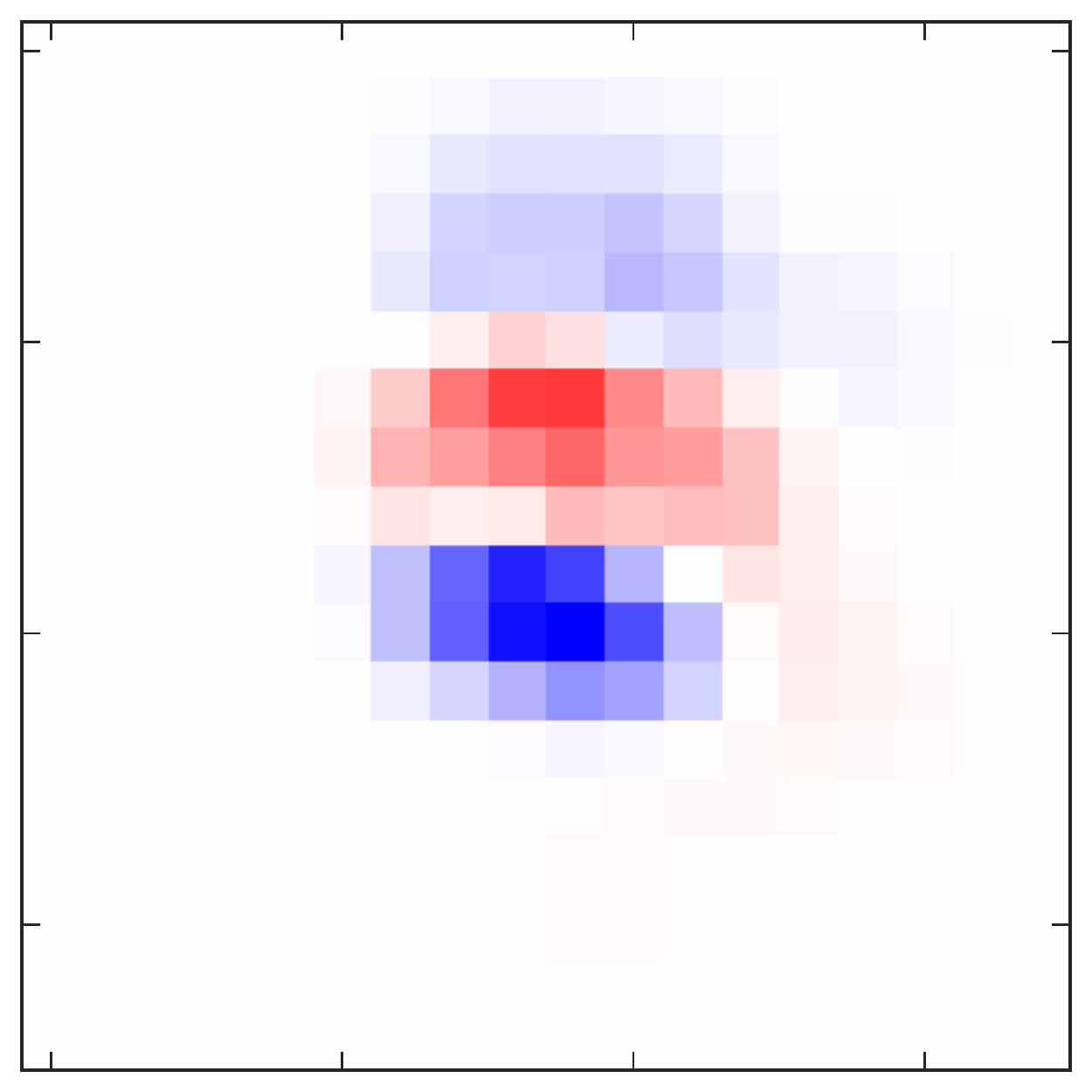}
  \includegraphics[width=0.119\textwidth]{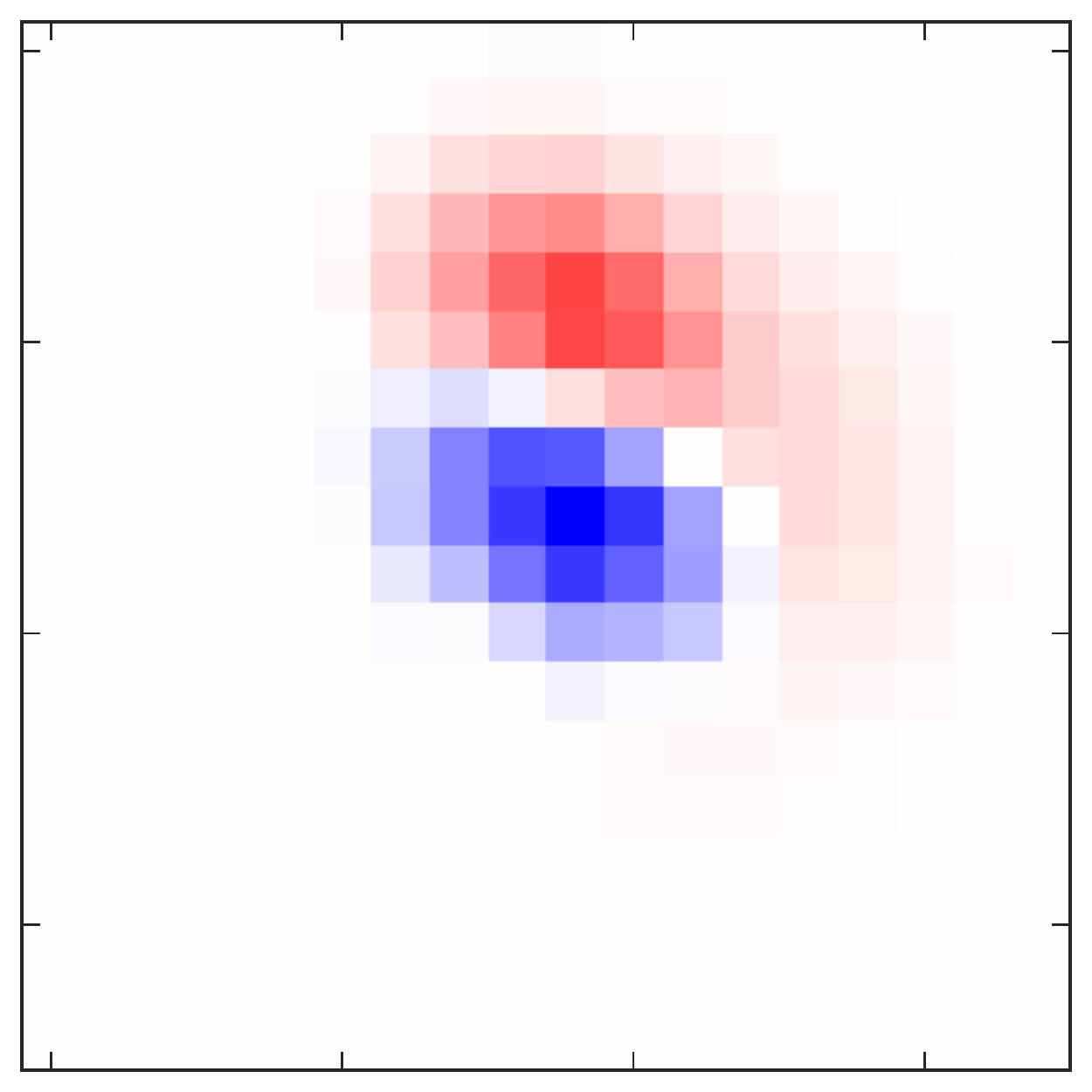}
  \includegraphics[width=0.119\textwidth]{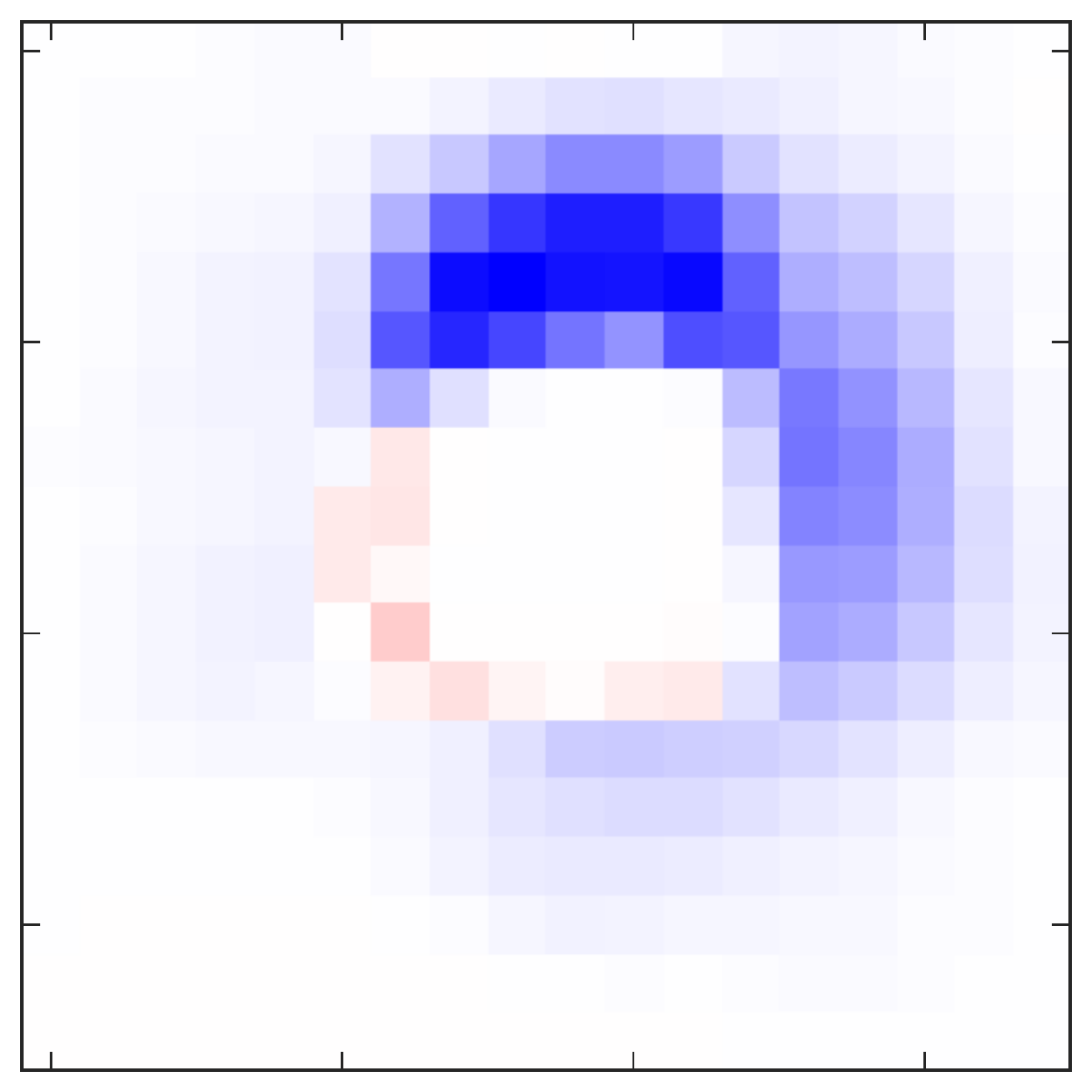} 
  \includegraphics[width=0.119\textwidth]{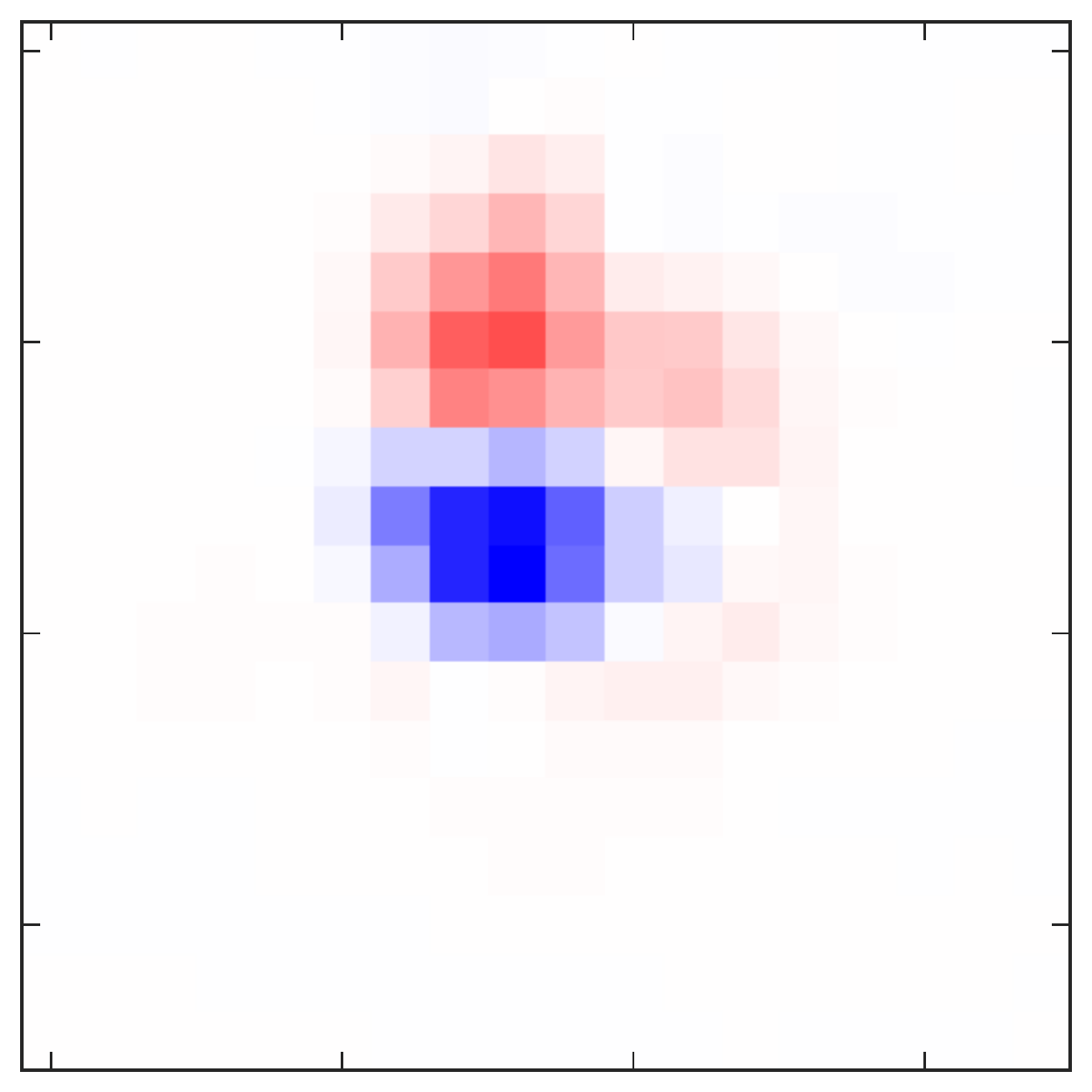} 
  \includegraphics[width=0.119\textwidth]{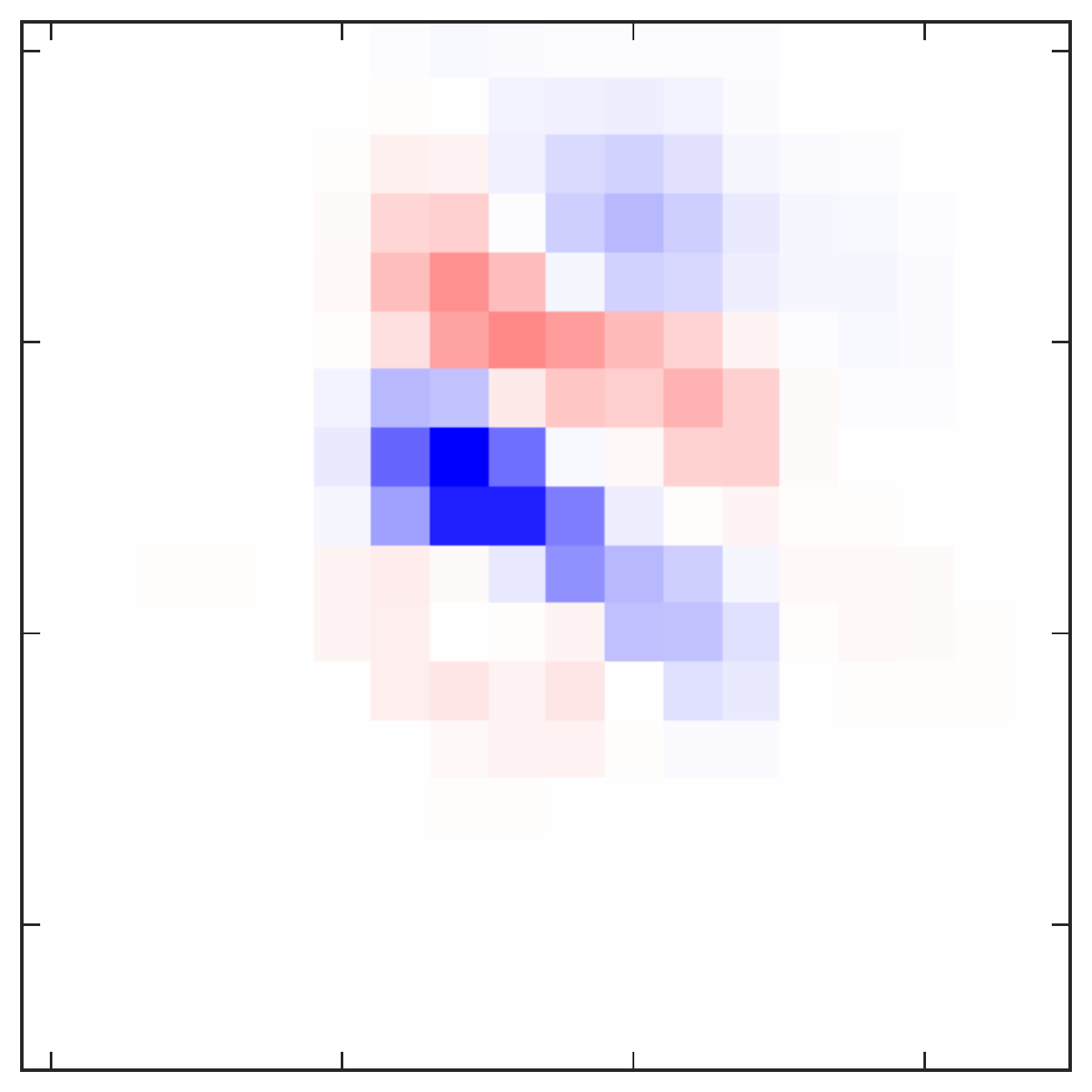} \\
  \includegraphics[width=0.119\textwidth]{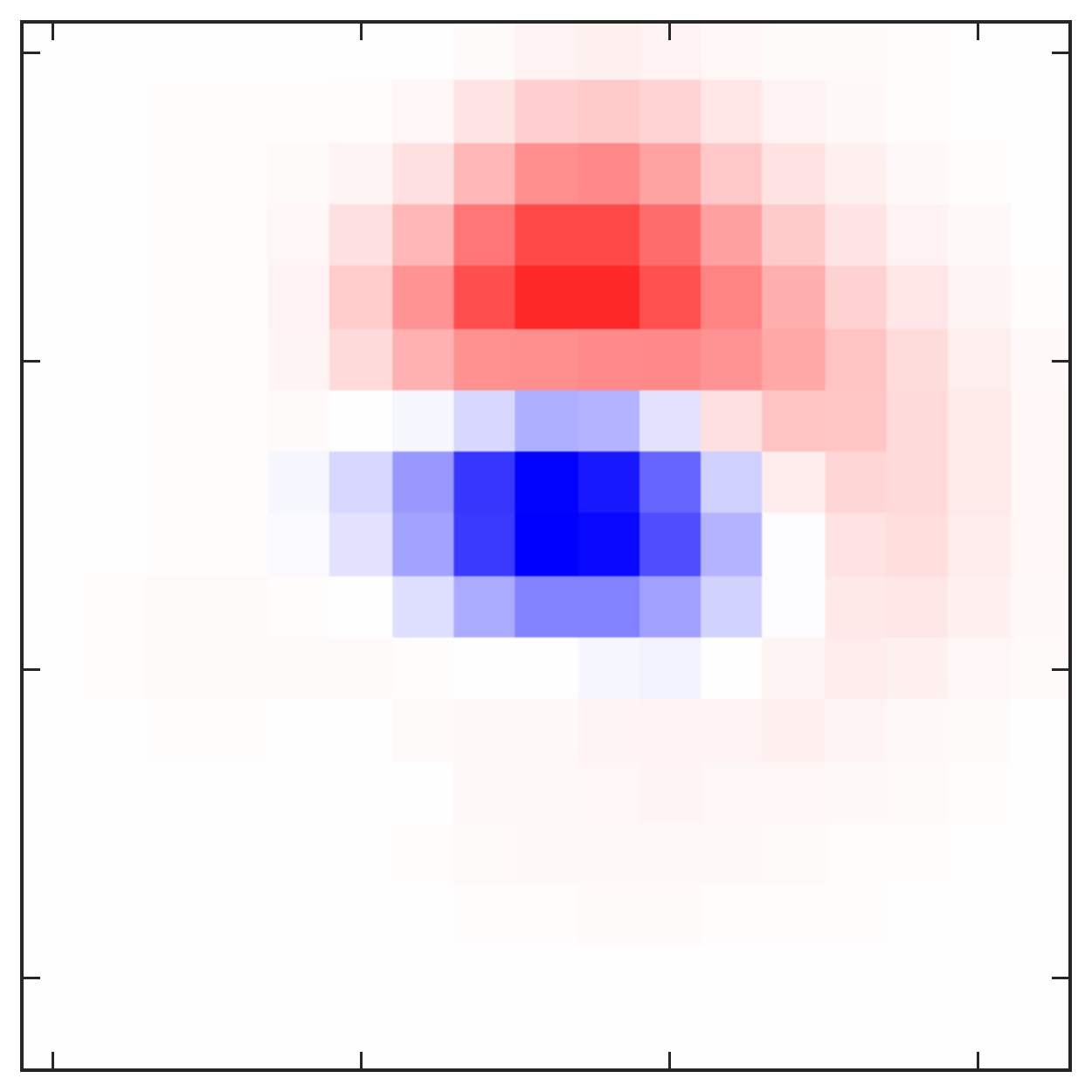}
  \includegraphics[width=0.119\textwidth]{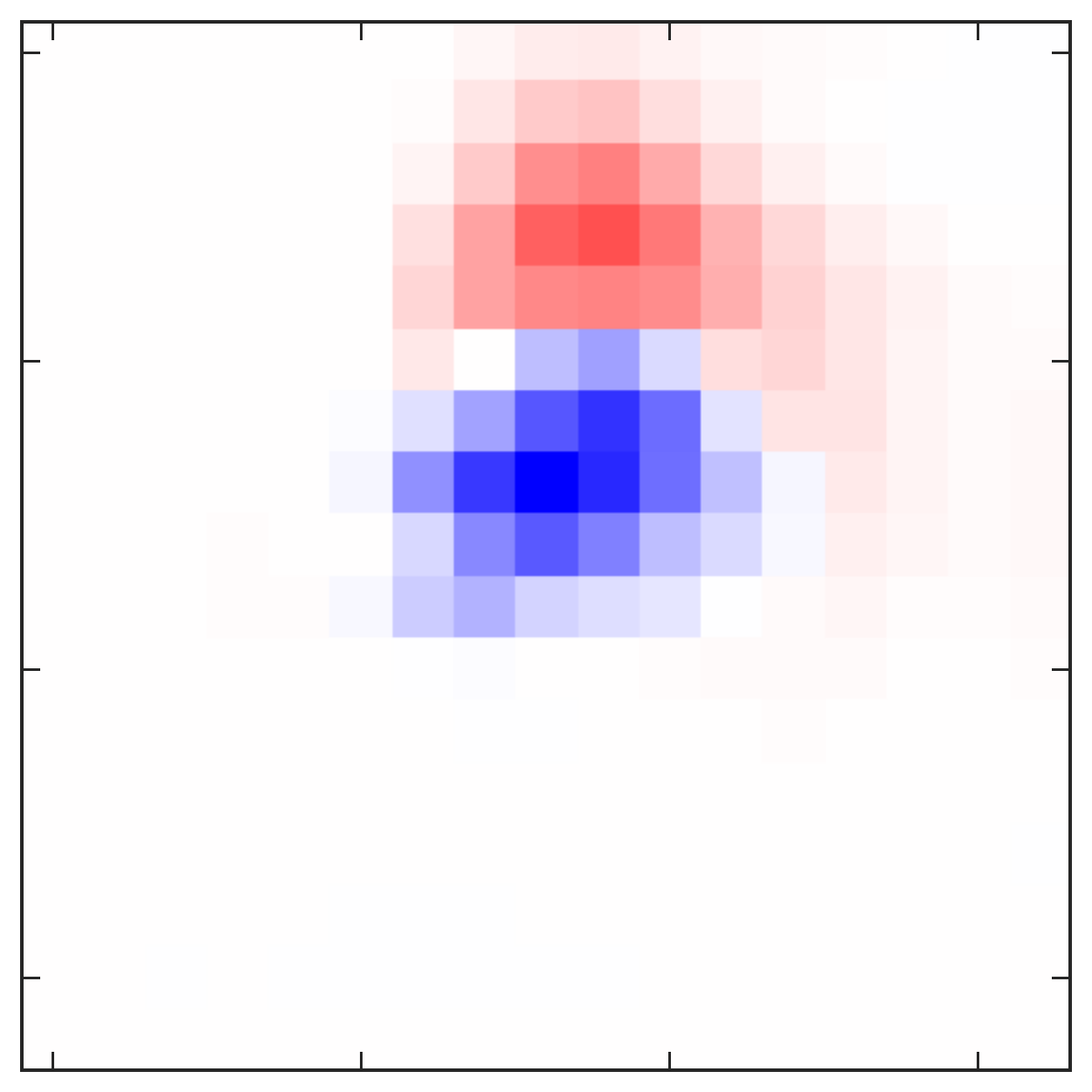}
  \includegraphics[width=0.119\textwidth]{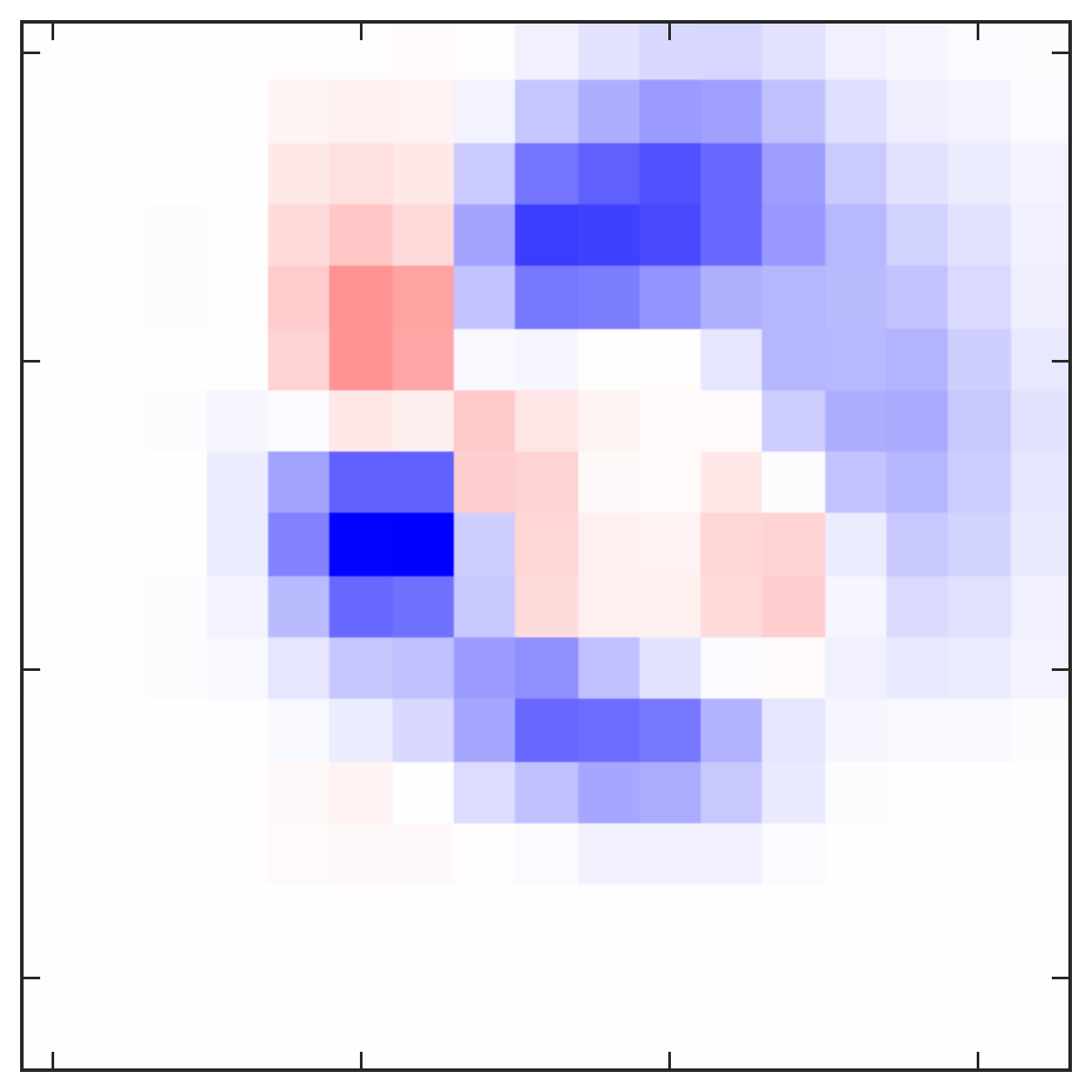}
  \includegraphics[width=0.119\textwidth]{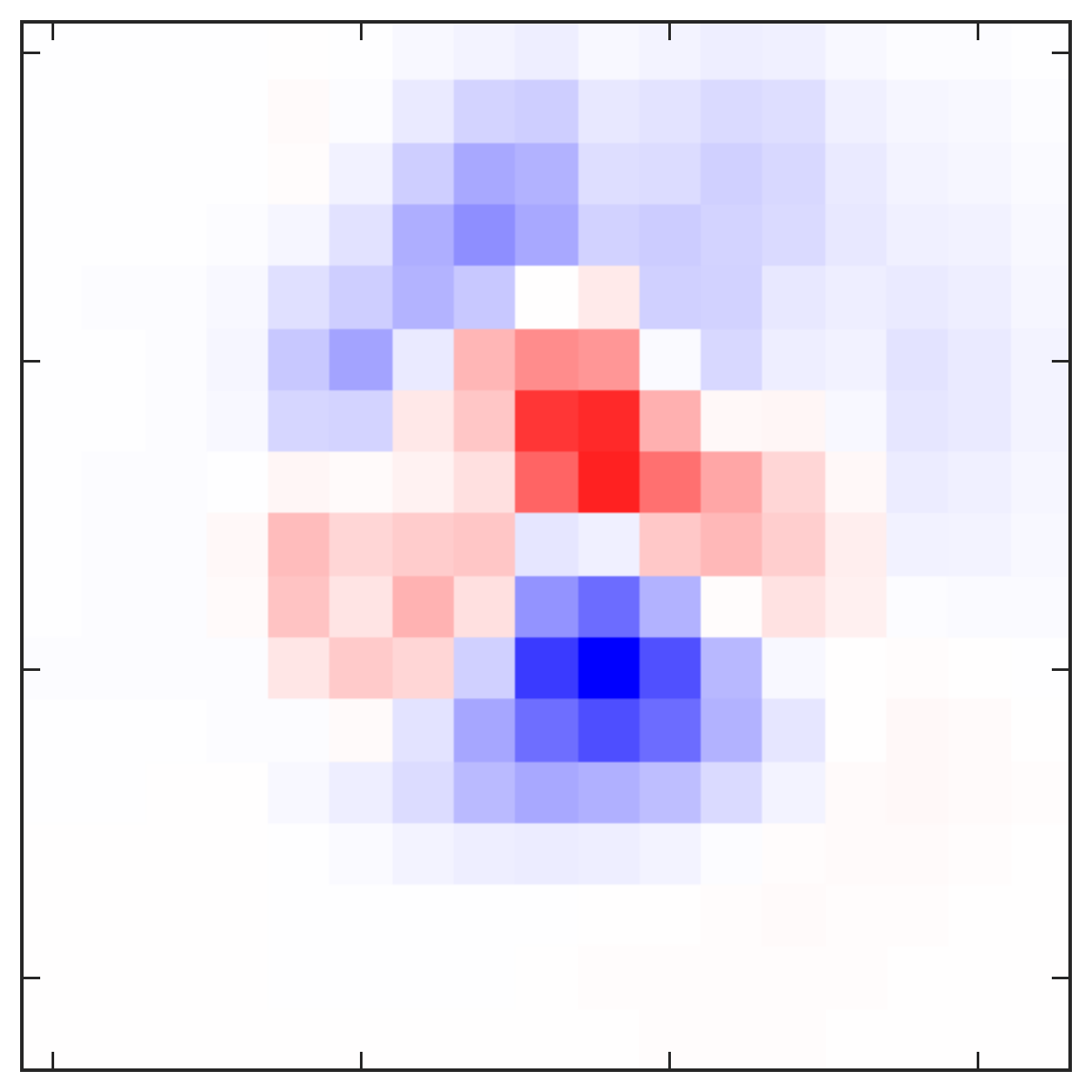}
  \includegraphics[width=0.119\textwidth]{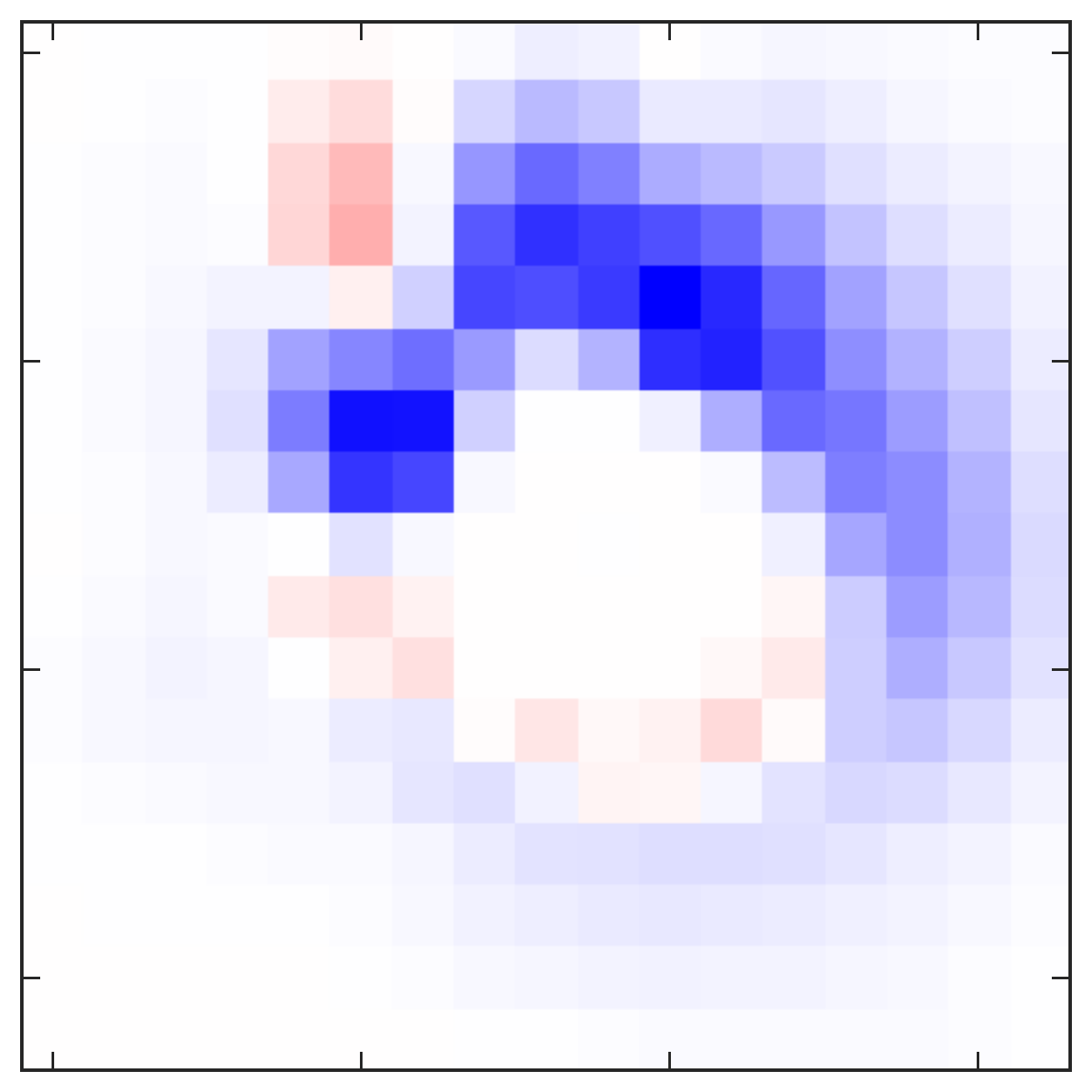}
  \includegraphics[width=0.119\textwidth]{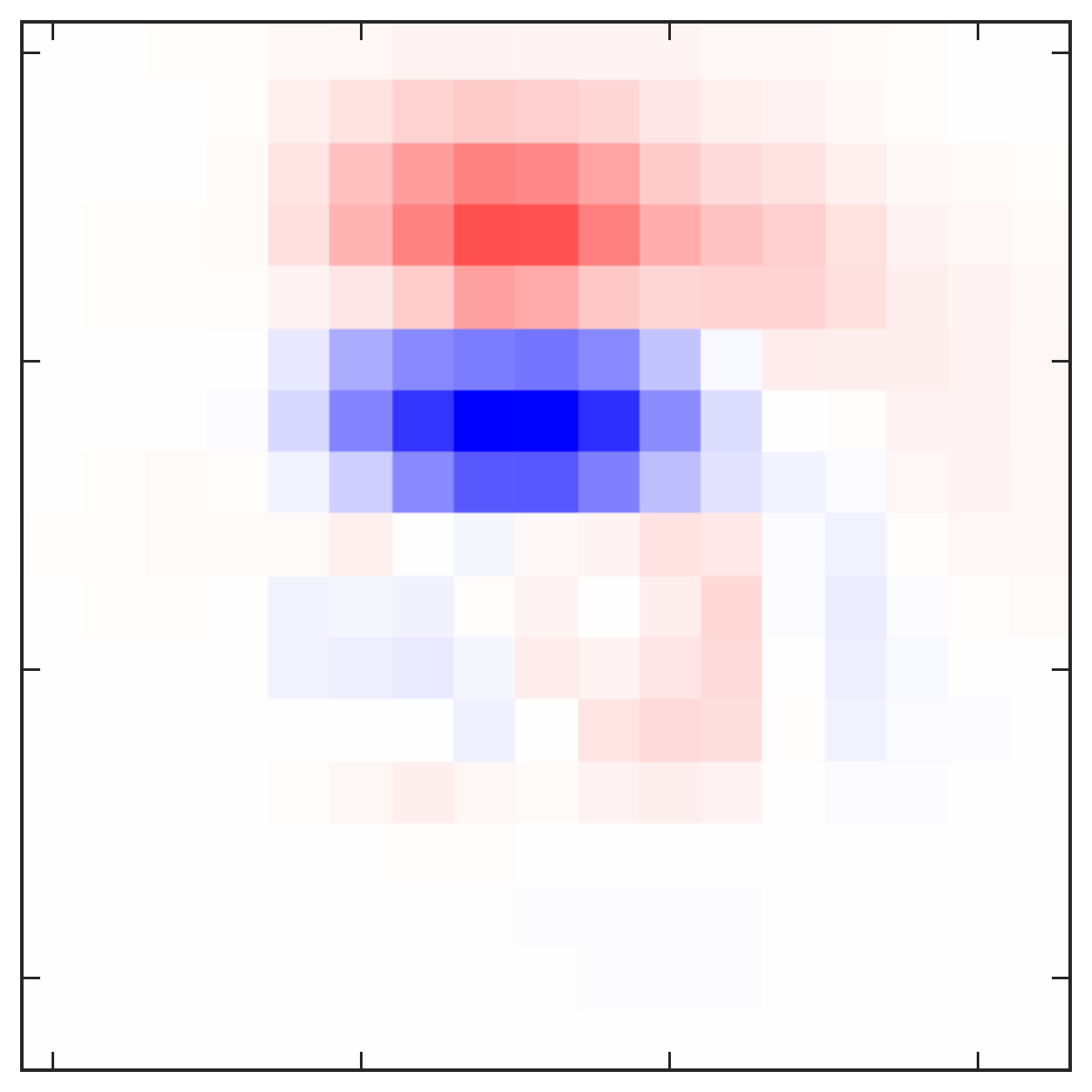}
  \includegraphics[width=0.119\textwidth]{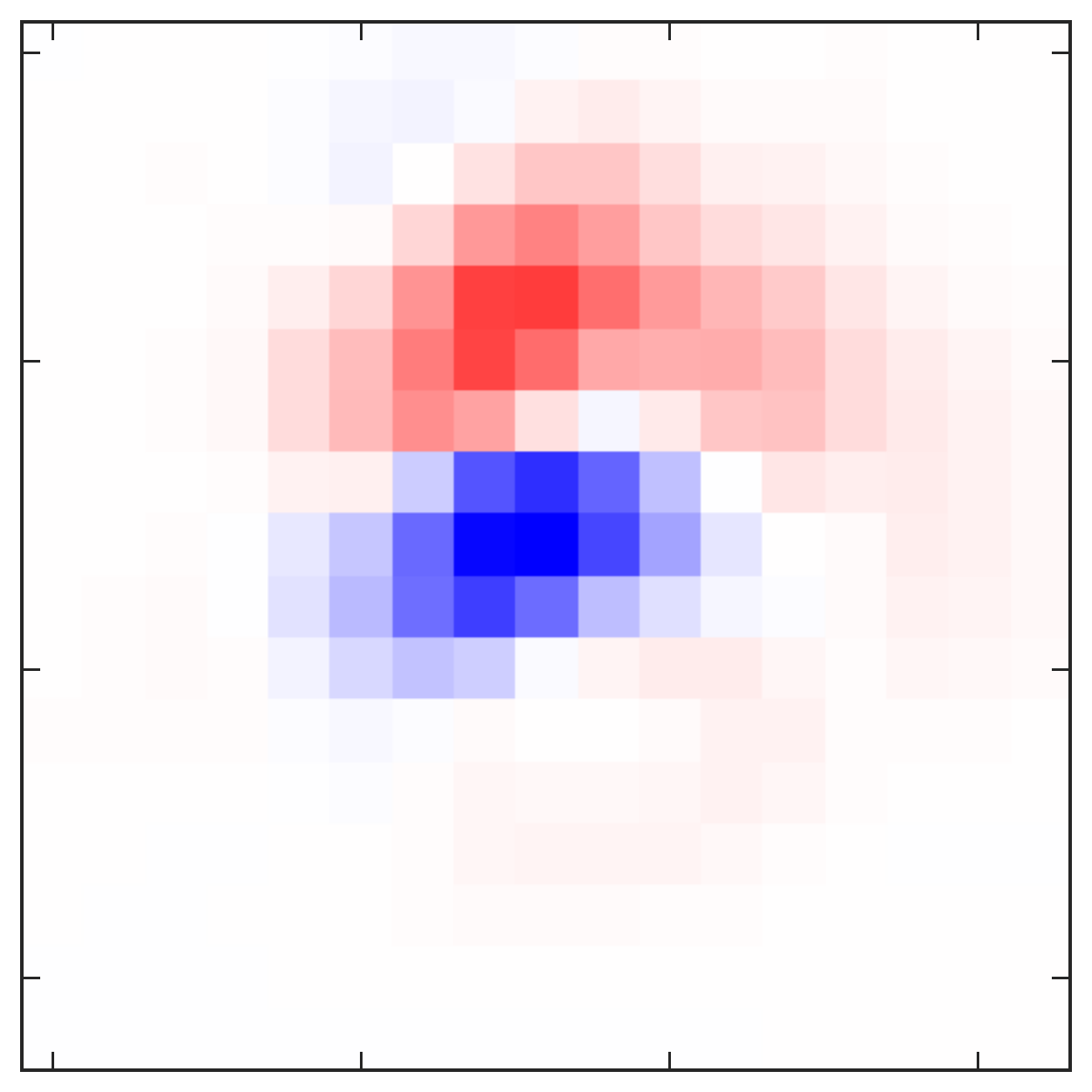}
  \includegraphics[width=0.119\textwidth]{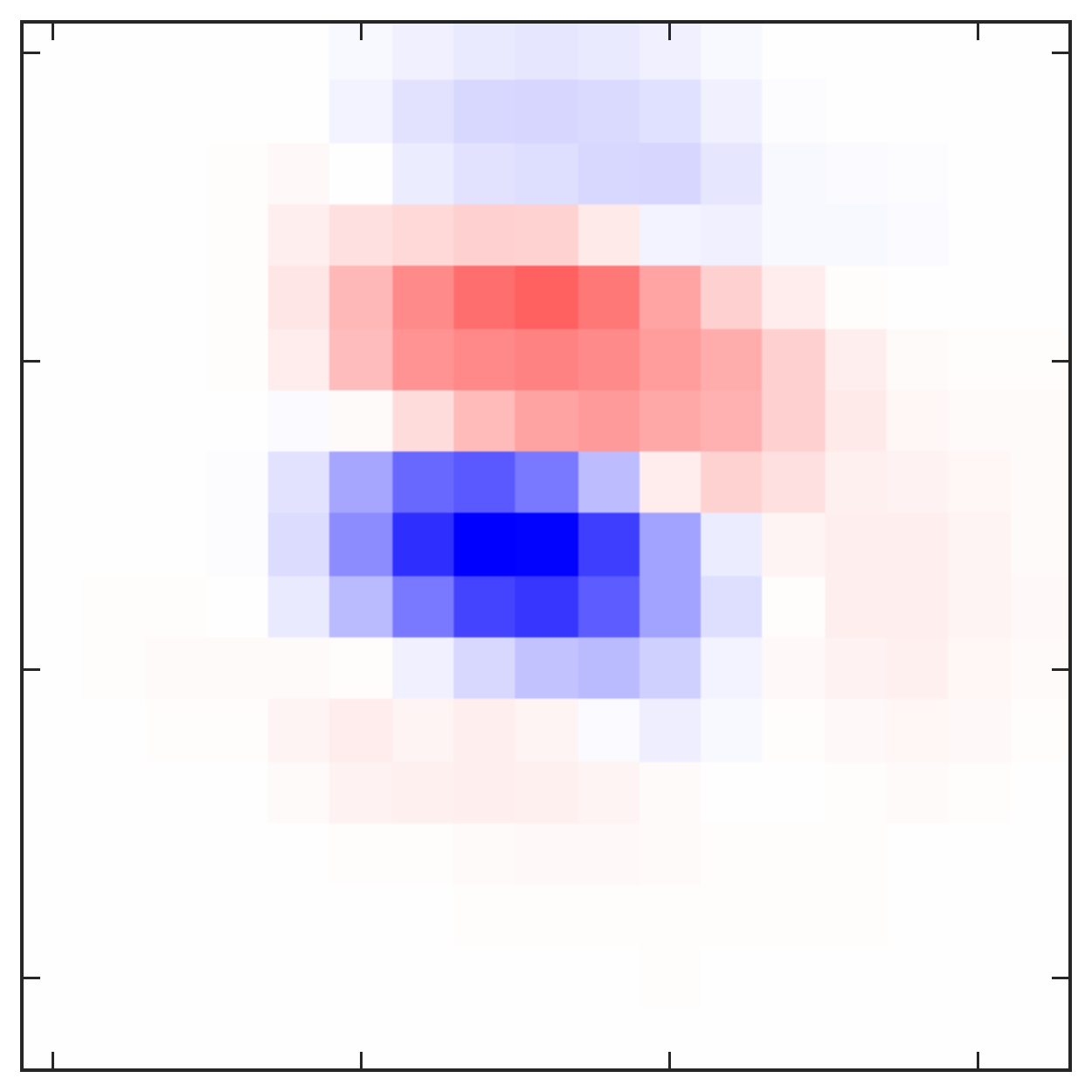}
  \caption[Averaged signal minus background for our default network
    and full pre-processing.]{Averaged signal minus background for our default network
    and full pre-processing. The rows correspond to ConvNet layers one
    to four. After two rows MaxPooling reduces the number of pixels by
    roughly a factor of four. The columns indicate the feature maps
    one to eight. Red areas indicate signal-like regions, blue areas
    indicate background-like regions.}
  \label{fig:conv_layers}
\end{figure}

Before we move to the performance study, we can get a feeling for what
is happening inside the trained ConvNet by looking at the output of
the different layers in the case of fully pre-processed images. In
Fig.~\ref{fig:conv_layers} we show the difference of the averaged
output for 100 signal and 100 background images. For each of those two
categories, we require a classifier output of at least 0.8. Each
row illustrates the output of a convolutional layer. Signal-like red
areas are typical for jet images originating from top decays; blue
areas are typical for backgrounds. The first layer seems to
consistently capture a well-separated second subjet, and some kernels
of the later layers seem to capture the third signal subjet in the
right half-plane. However, one should keep in mind that there is no
one-to-one correspondence between the location in feature maps of
later layers and the pixels in the input image.

On the left hand side of Fig.~\ref{fig:deep_layers_and_pcc} we show the same kind of intermediate
result for the two fully connected DNN layers. Each of the 64 linear
bars represents a node of the layer. We see that individual nodes are
quite distinctive for signal and background images. The fact that some
nodes are not discriminative indicates that in the interest of speed
the number of nodes could be reduced slightly. The output of the DNN
is essentially the same as the probabilities shown in the right panel of
Fig.~\ref{fig:arc_scan}, ignoring the central probability range
between 20\% and 80\%.

To see which pixels of the fully pre-processed $40 \times 40$ jet image have
an impact on the signal vs background label, we can correlate the
deviation of a pixel $x_{ij}$ from its mean value $\bar{x}_{ij}$ with
the deviation of the label $y$ from its mean value $\bar{y}$. A
properly normalized correlation function for a given set of combined
signal and background images can be defined as
\begin{align}
  r_{ij} = \frac{\sum_\text{images}\left(x_{ij} - \bar{x}_{ij}\right)
    \left(y - \bar{y}\right)}
  { \sqrt{ \sum_\text{images}\left(x_{ij} -
        \bar{x}_{ij}\right)^2} \sqrt{ \sum_\text{images}\left(y - \bar{y}\right)^2 }
  } \;.
\end{align}
It is usually referred to as the Pearson correlation coefficient. From
the definition we see that for a signal probability $y$ positive
values of $r_{ij}$ indicate signal-like patterns.  On the right hand side of
Fig.~\ref{fig:deep_layers_and_pcc} we show this correlation for our network
architecture.  A large energy deposition in the centre leads to
classification as background. A secondary energy deposition in the 12
o'clock position combined with additional energy deposits in the right
half-plane lead to a classification as signal. This is consistent with
our expectations after full pre-processing, shown in
Fig.~\ref{fig:averaged_images}.

\begin{figure}[t!]
  \begin{subfigure}{0.6\textwidth}
    \includegraphics[width=\textwidth]{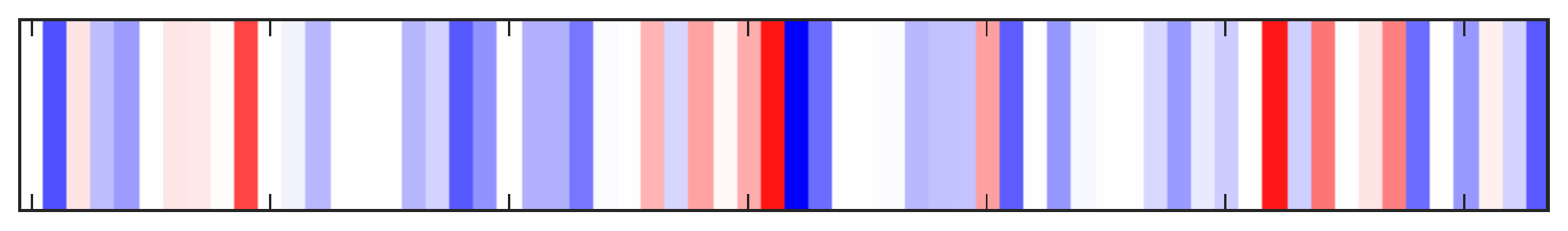}
    \includegraphics[width=\textwidth]{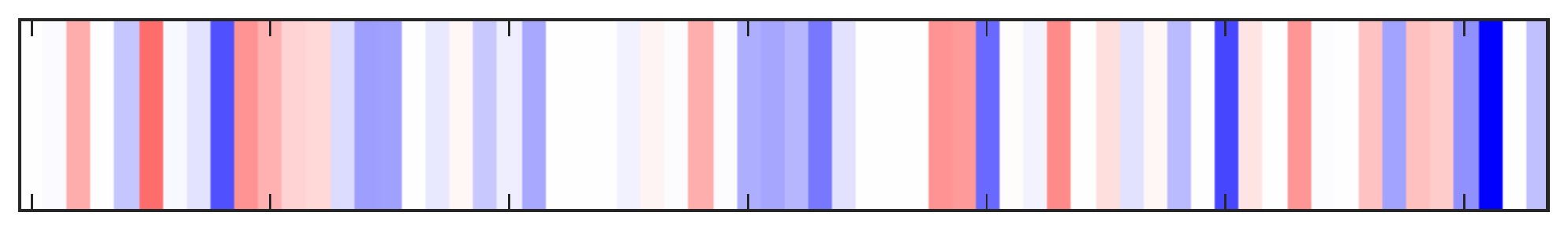}
    \includegraphics[width=\textwidth]{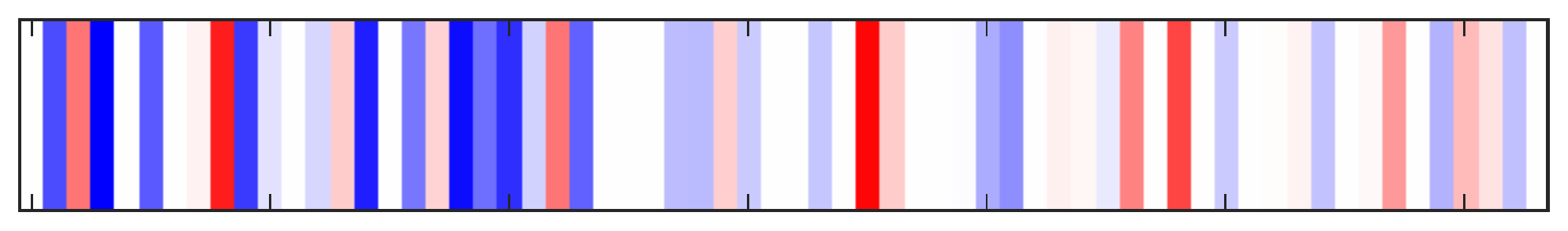}
  \end{subfigure}%
  ~
  \begin{subfigure}{0.4\textwidth}
    \includegraphics[width=\textwidth]{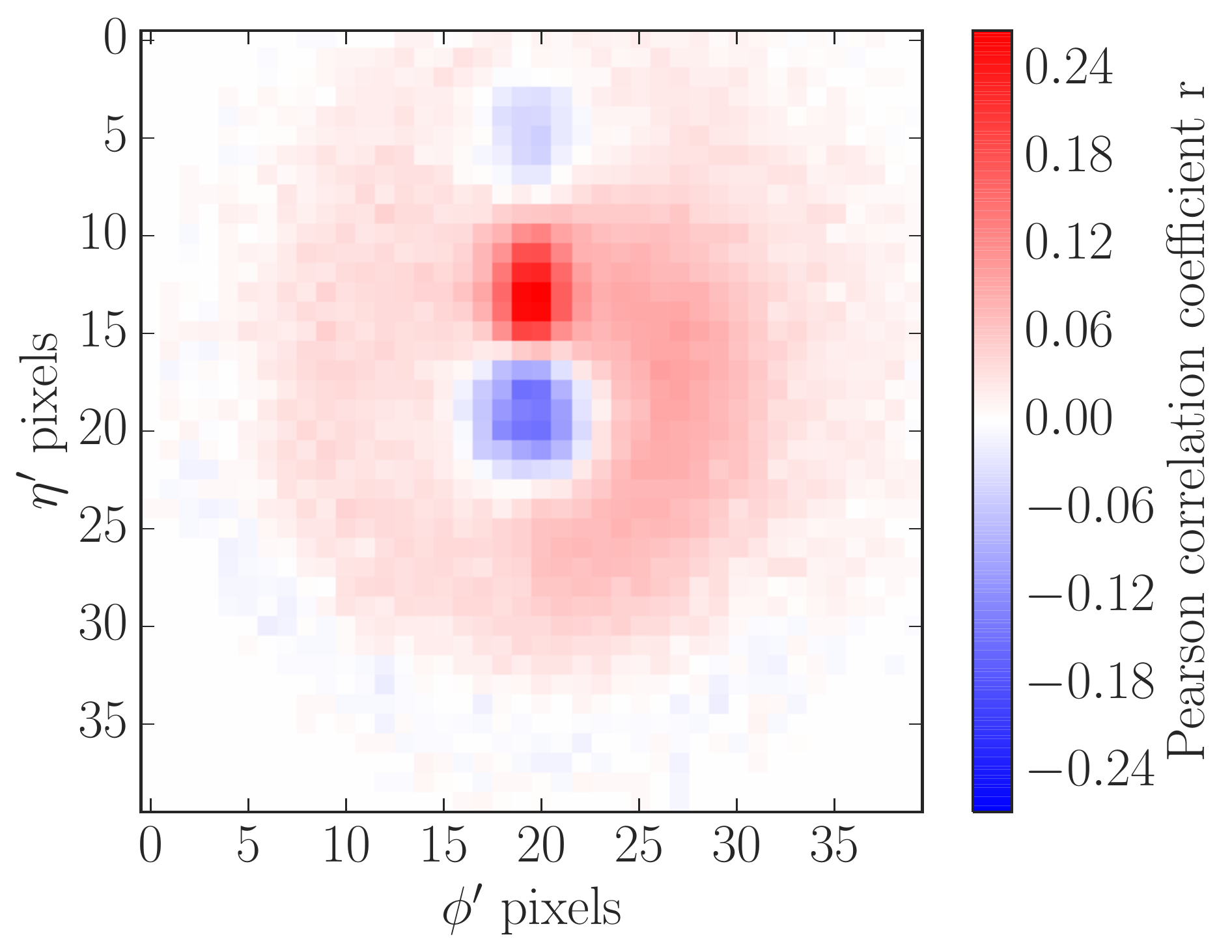}
   \end{subfigure}
 \caption[Averaged signal minus background for our default network
    and full pre-processing, and Pearson correlation coefficient for signal and background images.]{Left: Averaged signal minus background for our default network
    and full pre-processing. The rows show the three dense DNN
    layers. Red areas indicate signal-like regions, blue areas
    indicate background-like regions. Right: Pearson correlation coefficient for 10,000 signal and
    background images each. The corresponding jet image is illustrated
    in Fig.~\ref{fig:averaged_images}.  Red areas indicate signal-like
    regions, blue areas indicate background-like regions.}
  \label{fig:deep_layers_and_pcc}
\end{figure}

\subsection{Results and benchmarks}
\label{sec:results}
Given our optimized machine learning setup introduced in
Sec.~\ref{sec:machines} and the fact that we can understand its
workings and trust its outcome, we can now compare its performance
with state-of-the-art top taggers. The details of the signal and background
samples and jet images are discussed in Sec.~\ref{sec:analysis};
essentially, we attempt to separate a top decay inside a fat jet from
a QCD fat jet including fast detector simulation and for the
transverse momentum range $p_{T,\text{fat}} = 350~...~450$~GeV. Other
transverse momentum ranges for the fat jet can be targeted using the
same DNN method. 

Because we focus on a comparing the performance of the DNN approach
with the performance of standard multivariate top taggers we take our
Monte Carlo training and testing sample as a replacement of actual
data. This means that for our performance test
we do not have to include uncertainties in our \textsc{Pythia}
simulations compared to other Monte Carlo simulations and
data, see Ref.~\cite{Barnard:2016qma} for a study in this direction.

\begin{figure}[t]
  \begin{center}
  \includegraphics[width=0.46\textwidth]{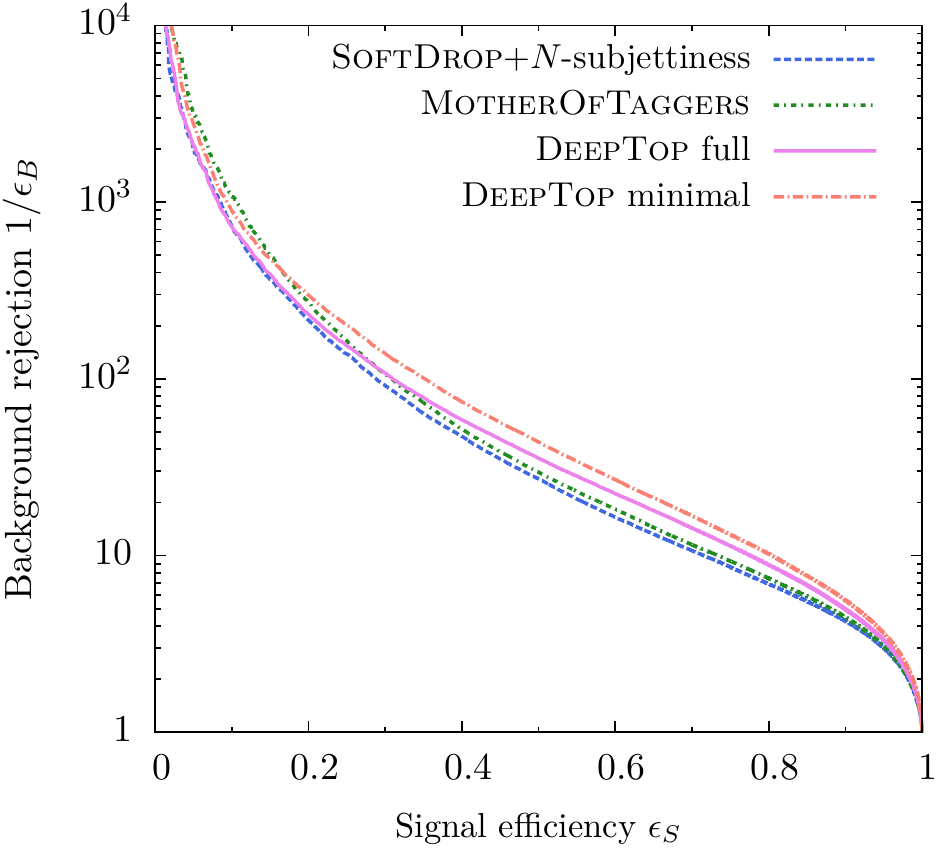}
  \end{center}
  \caption[ROC curves for \textsc{DeepTop} architectures compared with boosted decision trees.]{Performance of the neural network tagger compared to the QCD-based
    approaches \textsc{SoftDrop} plus $N$-subjettiness and including
    the \textsc{HEPTopTagger} variables.}
\label{fig:performance}
\end{figure}

\subsubsection{Performance of the network}
\label{sec:comp}

To benchmark the performance of our \textsc{DeepTop} DNN, we compare
its ROC curve with standard Boosted Decision Trees based on the C/A
jets using \textsc{SoftDrop} combined with $N$-subjettiness.  From
Fig.~\ref{fig:arc_scan} we know the spread of performance for the
different network architectures for fully pre-processed images.  In
Fig.~\ref{fig:performance} we see that minimal pre-processing actually
leads to slightly better results, because the combination or rotation
and cropping described in Sec.~\ref{sec:analysis} leads to a
small loss of information. Altogether, the band of different machine
learning results indicates how large the spread of performance will be
whenever for example binning issues in $p_{T,\text{fat}}$ are taken
into account, in which case we we would no longer be using the perfect
network for each fat jet.\medskip

For our BDT we use \textsc{GradientBoost} in the Python package
\textsc{sklearn}~\cite{sklearn} with 200 trees, a maximum depth of 2,
a learning rate of~0.1, and a sub-sampling fraction of $90\%$ for the
kinematic variables
\begin{align}
  \{\ m_\text{sd}, m_\text{fat}, \tau_2, \tau_3, \tau_2^\text{sd}, \tau_3^\text{sd}  \ \}
  \qqqquad \text{(\textsc{SoftDrop} + $N$-subjettiness)} \; ,
\end{align}
where $m_\text{fat}$ is the un-groomed mass of the fat jet.  This is
similar to standard experimental approaches for our transverse momentum range
$p_{T,\text{fat}} = 350~...~400$~GeV. In addition, we include the
\textsc{HEPTopTagger2} information from filtering combined with a mass
drop criterion,
\begin{align}
  \{\ m_\text{sd}, m_\text{fat}, m_\text{rec}, f_\text{rec}, \Delta R_\text{opt}, \tau_2, \tau_3, \tau_2^\text{sd}, \tau_3^\text{sd} \ \} 
  \qqqquad \text{(\textsc{MotherOfTaggers})} \; .
\label{eq:def_mother}
\end{align}
\medskip

In Fig.~\ref{fig:performance} we compare these two
QCD-based approaches with our best neural networks.  Firstly, we see
that both QCD-based BDT analyses and the two neural network setups are
close in performance. Indeed, adding \textsc{HEPTopTagger} information
slightly improves the \textsc{SoftDrop}+$N$-subjettiness setup,
reflecting the fact that our transverse momentum range is close to the
low-boost scenario where one should rely on the better-performing
\textsc{HEPTopTagger}. Second, we see that the difference between the
two pre-processing scenarios is in the same range as the difference
between the different approaches. Running the \textsc{DeepTop}
framework over signal samples with a 2-prong $W'$ decay to two jets
with $m_{W'} = m_t$ and over signal samples with a shifted value of
$m_t$ we have confirmed that the neural network setup learns both, the
number of decay subjets and the mass scale.\medskip

\begin{figure}[t]
  \includegraphics[width=0.49\textwidth]{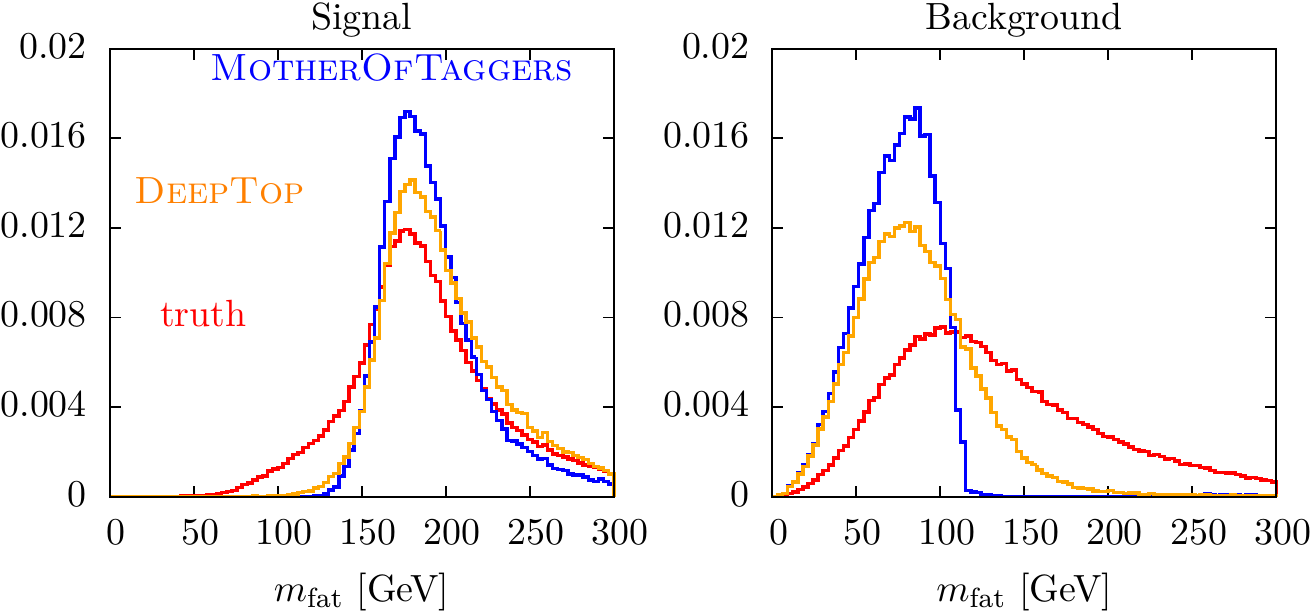}~
  \hspace{0.02\textwidth}
  \includegraphics[width=0.47\textwidth]{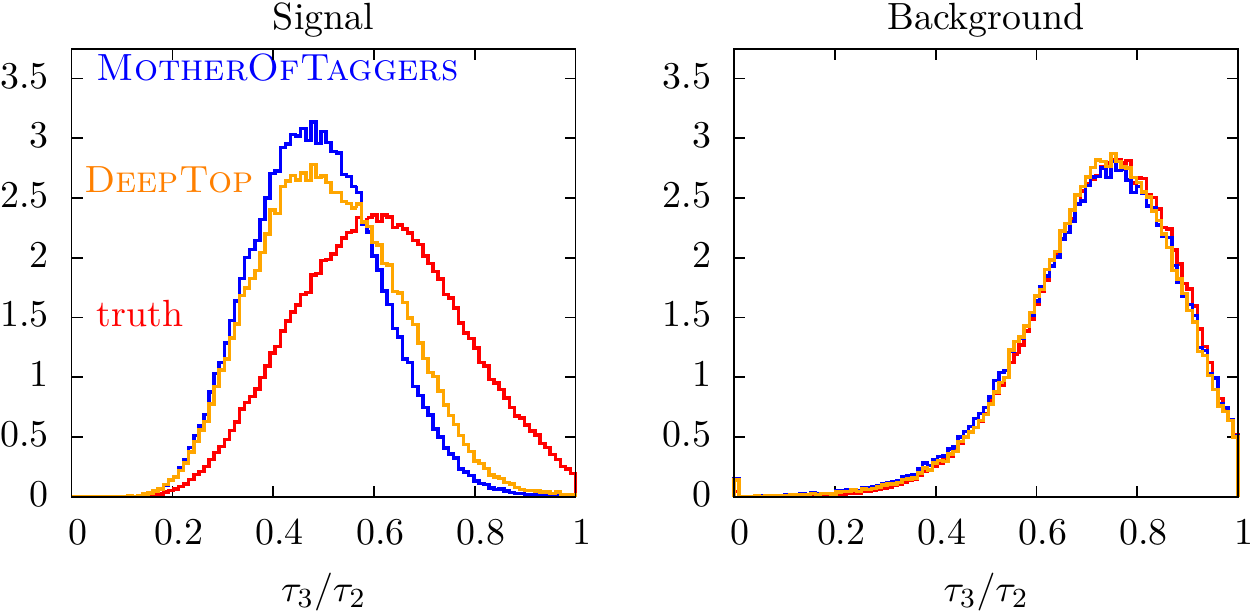}
  \caption[$m_\text{fat}$ and $\tau_3/\tau_2$ distributions.]{Kinematic observables $m_\text{fat}$ and $\tau_3/\tau_2$
    for events correctly determined to be signal or background by the
    \textsc{DeepTop} neutral network and by the
    \textsc{MotherOfTaggers} BDT, as well as Monte Carlo truth.}
  \label{fig:inputs}
\end{figure}

\subsubsection{What the network learns}

Following up on on the observation that the neural network and the
QCD-based taggers show similar performance in tagging a boosted top
decay inside a fat jet, we can check what kind of information
is used in this distinction.  Both for the DNN and for the
\textsc{MotherOfTaggers} BDT output we can study signal-like learned
patterns in actual signal events by cutting on the output label
$y$ corresponding to the 30\% most signal-like events shown on the right of Fig.~\ref{fig:arc_scan}.  Similarly, we
can select the 30\% most background-like events to test if the background
patterns are learned correctly. In addition, we can compare the
kinematic distributions in both cases to the Monte Carlo truth. In
Fig.~\ref{fig:inputs} we show the distributions for $m_\text{fat}$ and
$\tau_3/\tau_2$, both part the set of observables defined in
Eq.~\eqref{eq:def_mother}. 

We see that the DNN and BDT tagger indeed
learn essentially the same structures. The fact that the signal-like
features of those distributions are more pronounced 
than the Monte Carlo truth is linked to our stiff cut on $y$, which for the DNN and BDT tagger cases removes
events where the signal kinematic features are less pronounced. The
\textsc{MotherOfTaggers} curves for the signal are more peaked than
the \textsc{DeepTop} curves is due to the fact that the observables
are exactly the basis choice of the BDT, while for the neutral network
they are derived quantities. 

For our performance comparison of the QCD-based tagger approach and
the neutral network it is crucial that we understand what the
\textsc{DeepTop} network learns in terms of physics variables. The
relevant jet substructure observables differentiating between QCD jets
and top jets are those which which we evaluate in the
\textsc{MotherOfTaggers} BDT, Eq.~\eqref{eq:def_mother}.
To quantify which signal features the DNN and the BDT tagger have
correctly extracted we show observables for signal event correctly
identified as such, i.e. requiring events with a classifier response $y$ corresponding to the 30\% most signal like events.
As we can see from Fig.~\ref{fig:arc_scan} this cut value captures a
large fraction of correctly identified events. The same we also do for the 30\% most background like events
identified by each classifier.\medskip

\begin{figure}[t!]
  \includegraphics[width=0.49\textwidth]{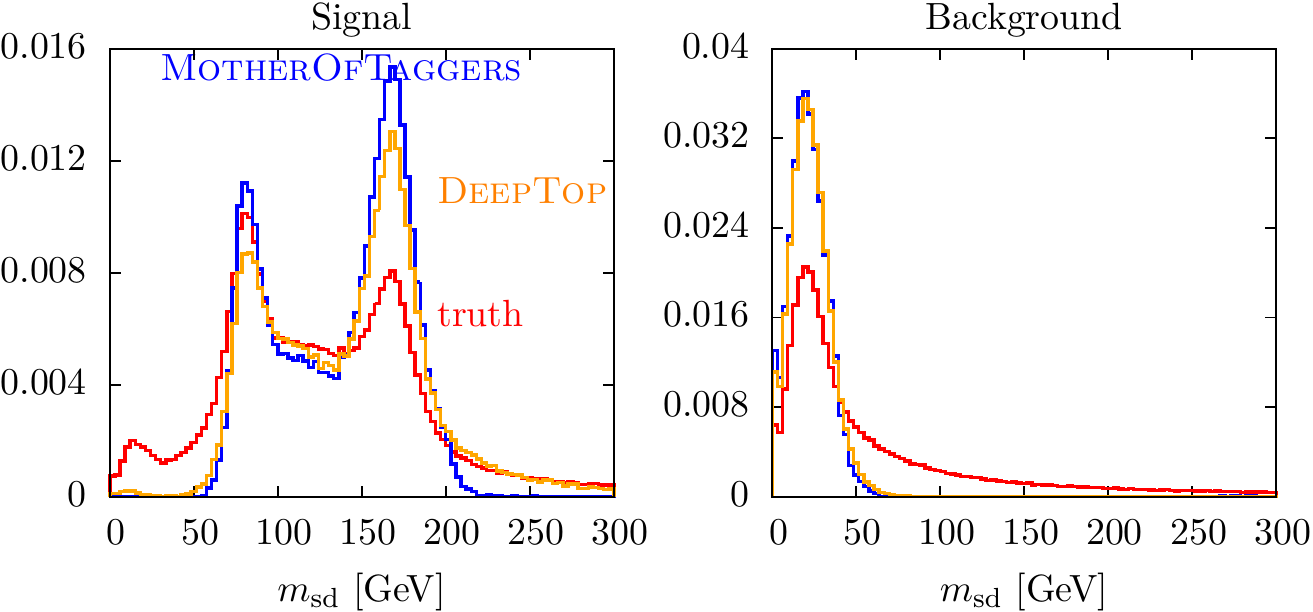}~
   \hspace{0.02\textwidth}
  \includegraphics[width=0.49\textwidth]{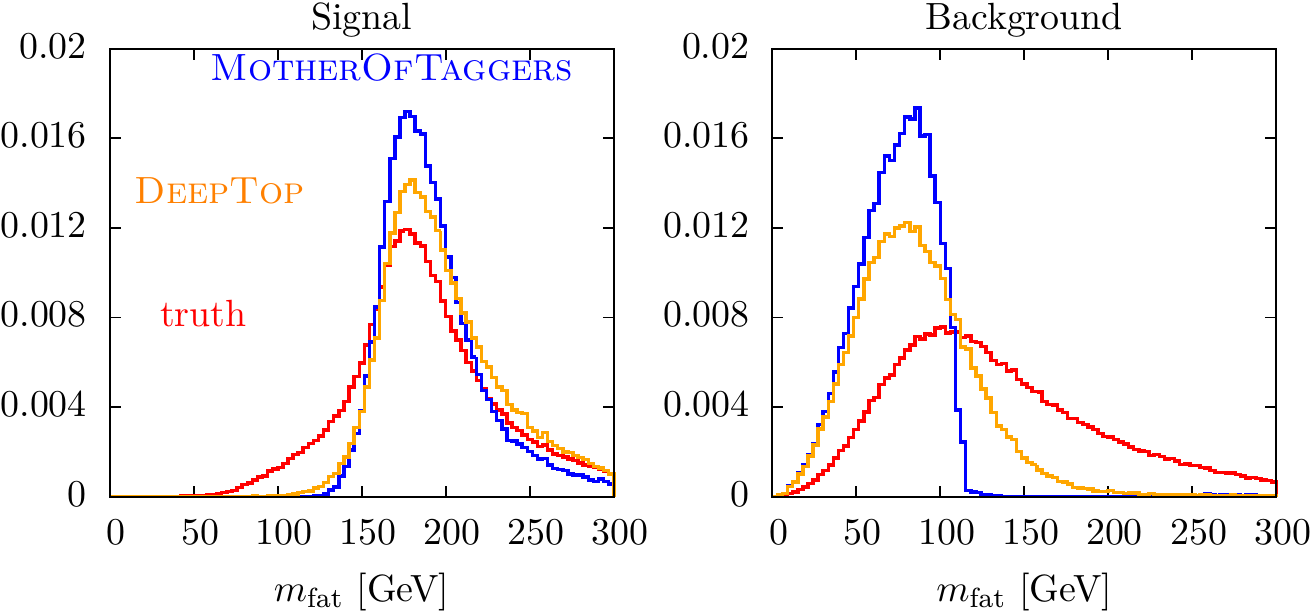} \\
  \includegraphics[width=0.49\textwidth]{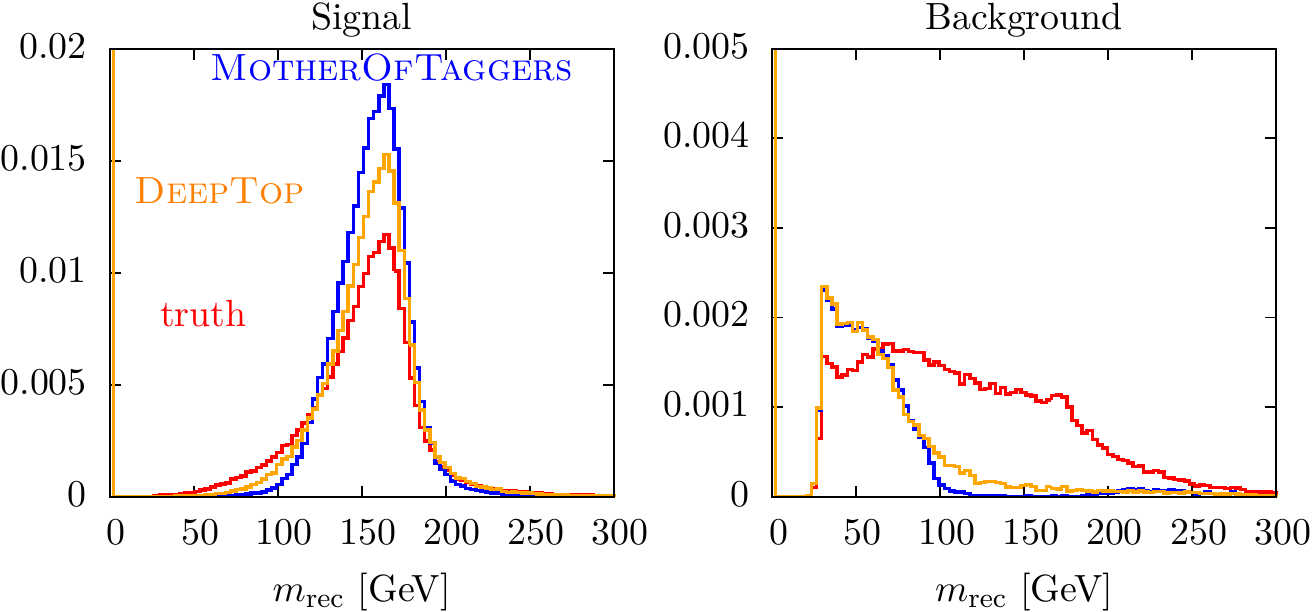}~
   \hspace{0.02\textwidth}
  \includegraphics[width=0.49\textwidth]{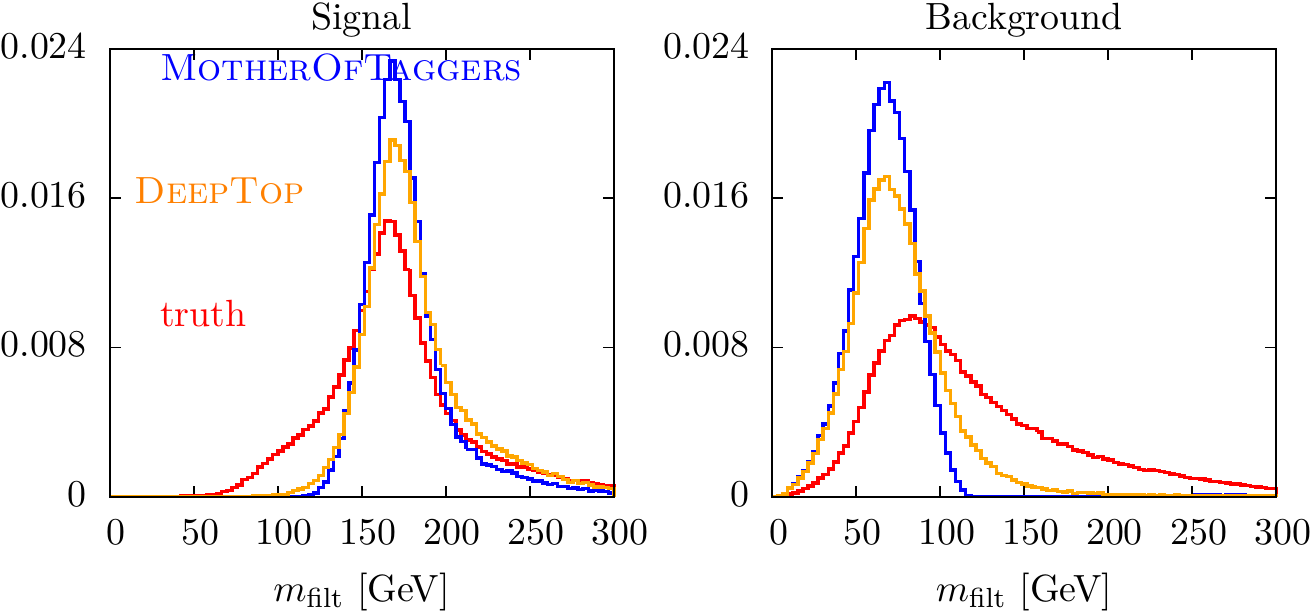} \\
  \includegraphics[width=0.49\textwidth]{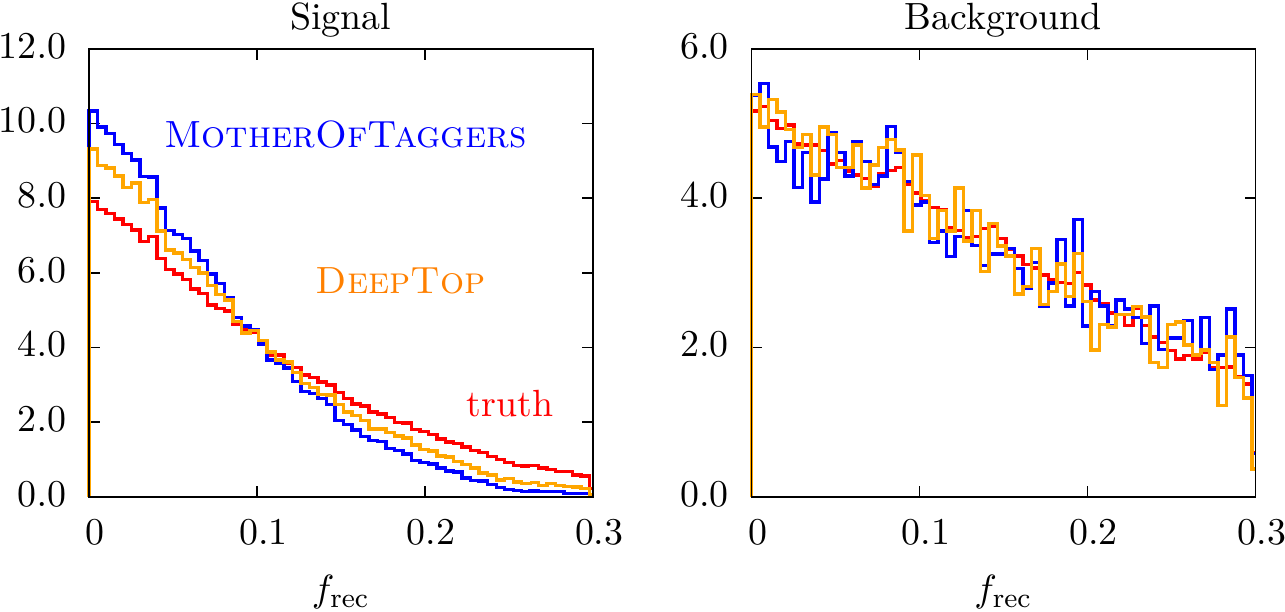}~
   \hspace{0.02\textwidth}
  \includegraphics[width=0.49\textwidth]{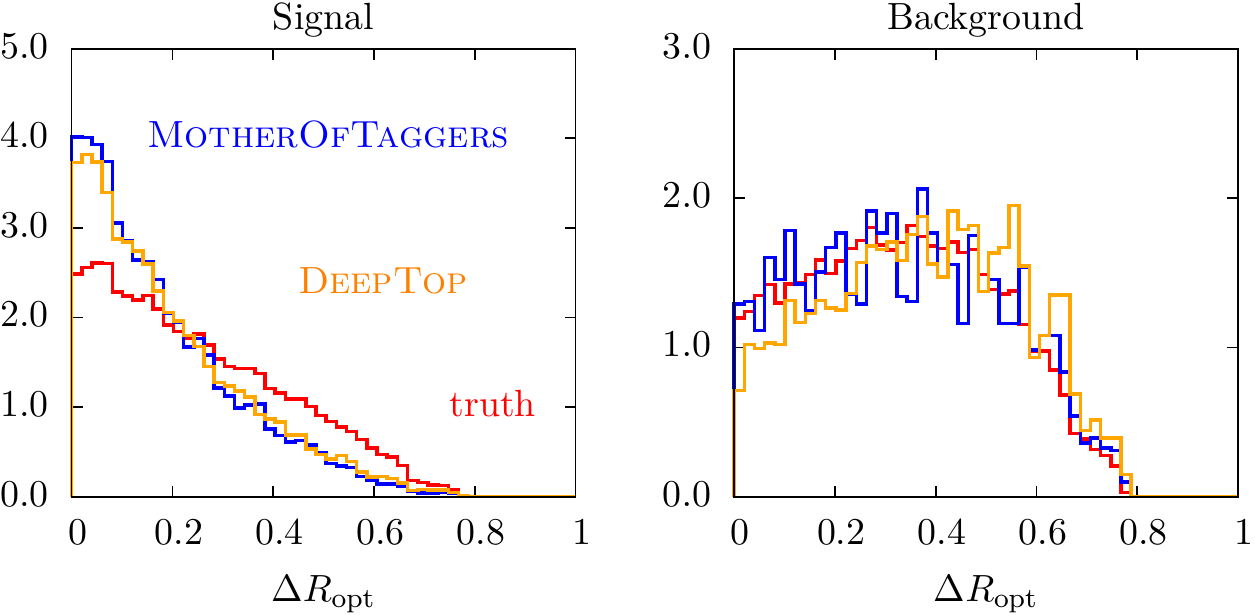} \\
  \includegraphics[width=0.49\textwidth]{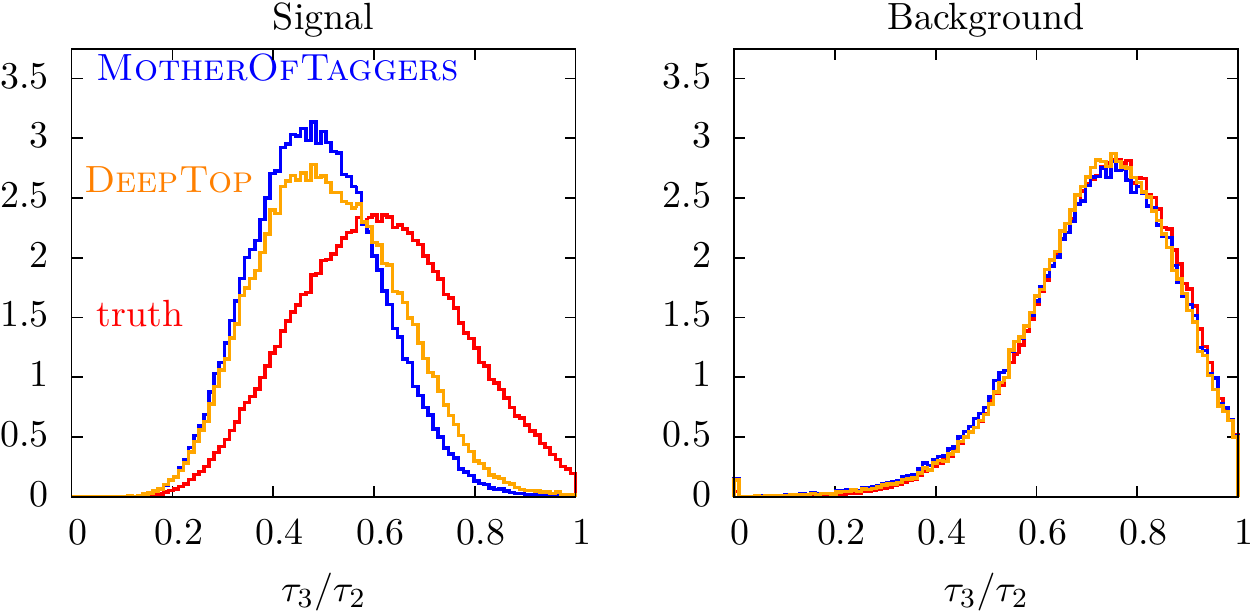}~
   \hspace{0.02\textwidth}
  \includegraphics[width=0.49\textwidth]{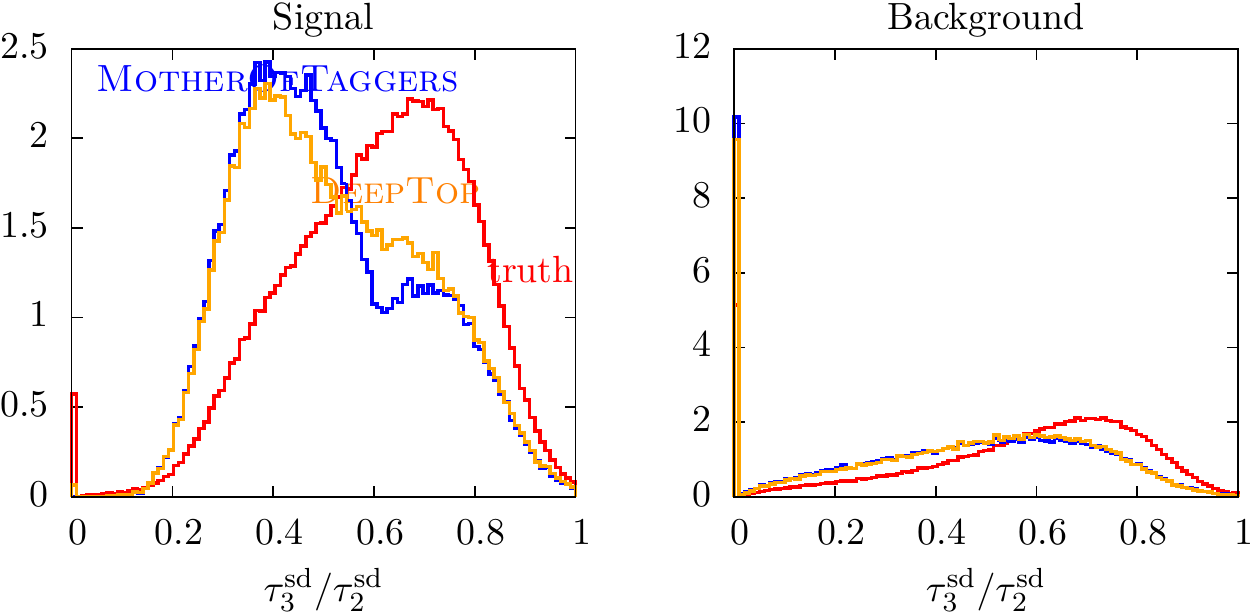}
  \caption[Groomed jet mass, $N$-subjettiness and HTT parameter distributions.]{Kinematic observables defined in Eq.\eqref{eq:def_mother}
    for events correctly determined to be signal or background by the
    \textsc{DeepTop} neutral network and by the
    \textsc{MotherOfTaggers} BDT, as well as Monte Carlo
    truth. Extended version of Fig.~\ref{fig:inputs}.}
\label{fig:physics}
\end{figure}

The upper two rows in Fig.~\ref{fig:physics} show the different mass
variables describing the fat jet. We see that the DNN and the BDT
tagger results are consistent, with a slightly better performance of
the BDT tagger for clear signal events. For the background the two
approaches deliver exactly the same performance. The deviation from
the true mass for the \textsc{HEPTopTagger} background performance is
explained by the fact that many events with no valid top candidate
return $m_\text{rec} = 0$.  Aside from generally comforting results we
observe a peculiarity: the \textsc{SoftDrop} mass identifies the
correct top mass in fewer that half of the correctly identified signal
events, while the fat jet mass $m_\text{fat}$ does correctly reproduce
the top mass. The reason why the \textsc{SoftDrop} mass is
nevertheless an excellent tool to identify top decays is that its
background distribution peaks at very low values, around $m_\text{sd}
\approx 20$~GeV. Even for $m_\text{sd} \approx m_W$ the hypothesis
test between top signal and QCD background can clearly identify a
massive particle decay.

In the third row we see that the \textsc{HEPTopTagger} $W$-to-top mass
ratio $f_\text{rec}$ only has little significance for the transverse
momentum range studied. For the optimalR variable $\Delta
R_\text{opt}$~\cite{Kasieczka:2015jma} the DNN and the BDT tagger again give consistent
results. Finally, for the $N$-subjettiness ratio $\tau_3/\tau_2$
before and after applying the \textsc{SoftDrop} criterion the results
are again consistent for the two tagging approaches.\medskip

Following up on the observation that \textsc{SoftDrop} shows excellent
performance as a hypothesis test, we show in Fig.~\ref{fig:momrec} the
reconstructed transverse momenta of the fat jet, or the top quark for
signal events. In the left panel we see that the transverse momentum
of the un-groomed fat jet reproduces our Monte-Carlo range
$p_{T,\text{fat}} = 350~...~450$~GeV.  Because the transverse momentum
distributions for the signal and background are very similar, the BDT
tagger and DNN curves agree very well with the correct behaviour. In
the right panel we see that the constituents identified by the
\textsc{SoftDrop} criterion have a significantly altered transverse
momentum spectrum. To measure the transverse momentum of the top quark
we therefore need to rely on a top identification with
\textsc{SoftDrop}, but a top reconstruction based on the (groomed) fat
jet properties.

\begin{figure}[t]
  \includegraphics[width=0.49\textwidth]{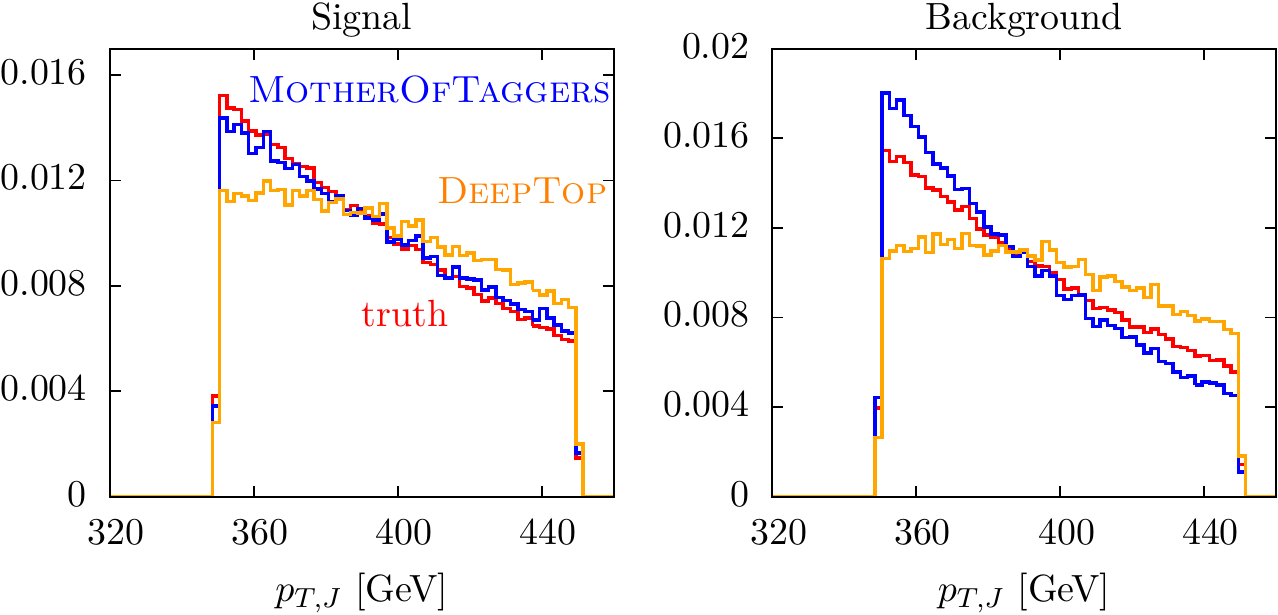}~
  \hspace{0.02\textwidth}
  \includegraphics[width=0.49\textwidth]{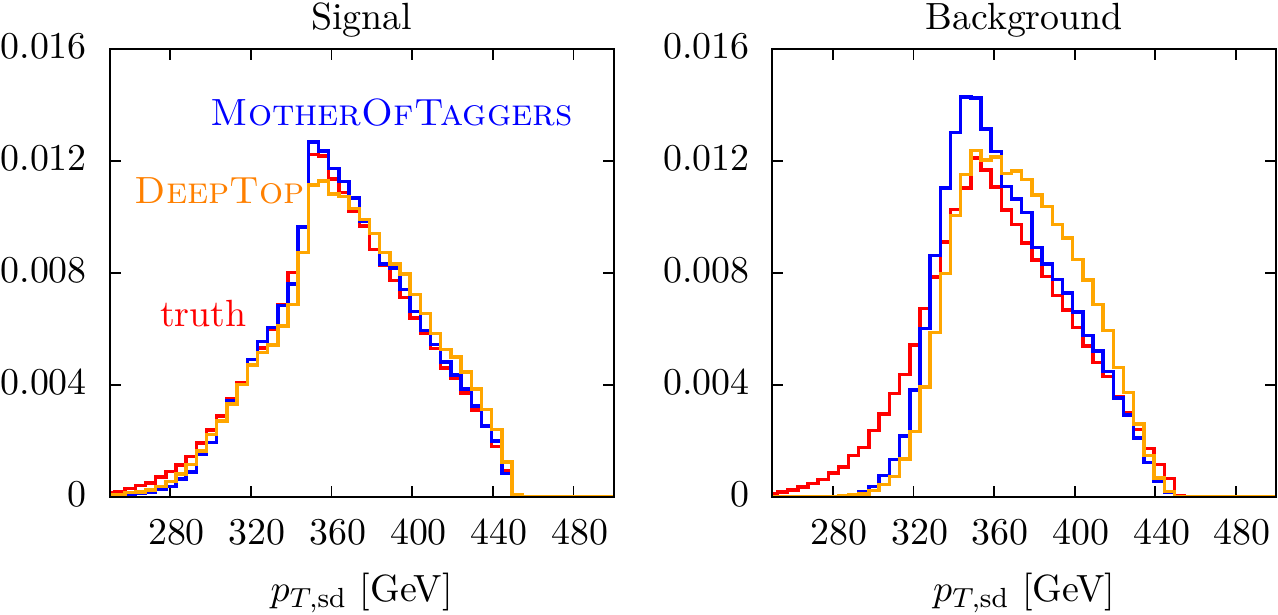}
  \caption[Fat jet transverse momentum distributions.]{Reconstructed transverse momenta for events correctly
    determined to be signal or background by the \textsc{DeepTop}
    neutral network and by the \textsc{MotherOfTaggers} BDT, as well
    as Monte Carlo truth.}
\label{fig:momrec}
\end{figure}

\subsubsection{Sensitivity to experimental effects}

Finally, a relevant question is to what degree the information used by
the neural network is dominated by low-$p_T$ effects. We can apply a
cutoff, for example including only pixels with a transverse energy
deposition $E_T > 5$~GeV. This is the typical energy scale where the
DNN performance starts to degrade. 
A key question for the tagging performance is the dependence on the
activation threshold. Fig.~(\ref{fig:rejection_vs_X}) shows the
impact of different thresholds on the pixel activation, i.e. $E_T$ used
both for training and testing the networks. Removing very soft
activity, below 3~GeV, only slightly degrades the network's
performance. Above 3~GeV the threshold leads to an approximately
linear decrease in background rejection with increasing
threshold.\medskip

A second, important experimental systematic uncertainty when working with
calorimeter images is the calorimeter energy scale (CES). We assess the
stability of our network by evaluating the performance on jet images
where the $E_T$ pixels are globally rescaled by $\pm 25\%$.
As shown in the right panel of Fig.~\ref{fig:rejection_vs_X}
this leads to a decline in the tagging performance of
approximately $10\%$ when reducing the CES
and $5\%$ when increasing the CES.

Next, we train a \textit{hardened} version of the network. It uses the
same architecture as our default, but during the training procedure
each image is randomly rescaled using a Gaussian distribution with a
mean of 1.0 and a width of 0.1. New random numbers are used from epoch
to epoch.  The resulting network has a similar performance as the
default and exhibits a further reduced sensitivity to changes in the
global CES. While other distortions of the image, such as non-uniform rescaling,
will need to be considered, the resilience of the network and our
ability to further harden it are very encouraging for experimental usage
where the mitigation and understanding of systematic uncertainties is
critical.

\begin{figure}[t]
  \includegraphics[width=0.45\textwidth]{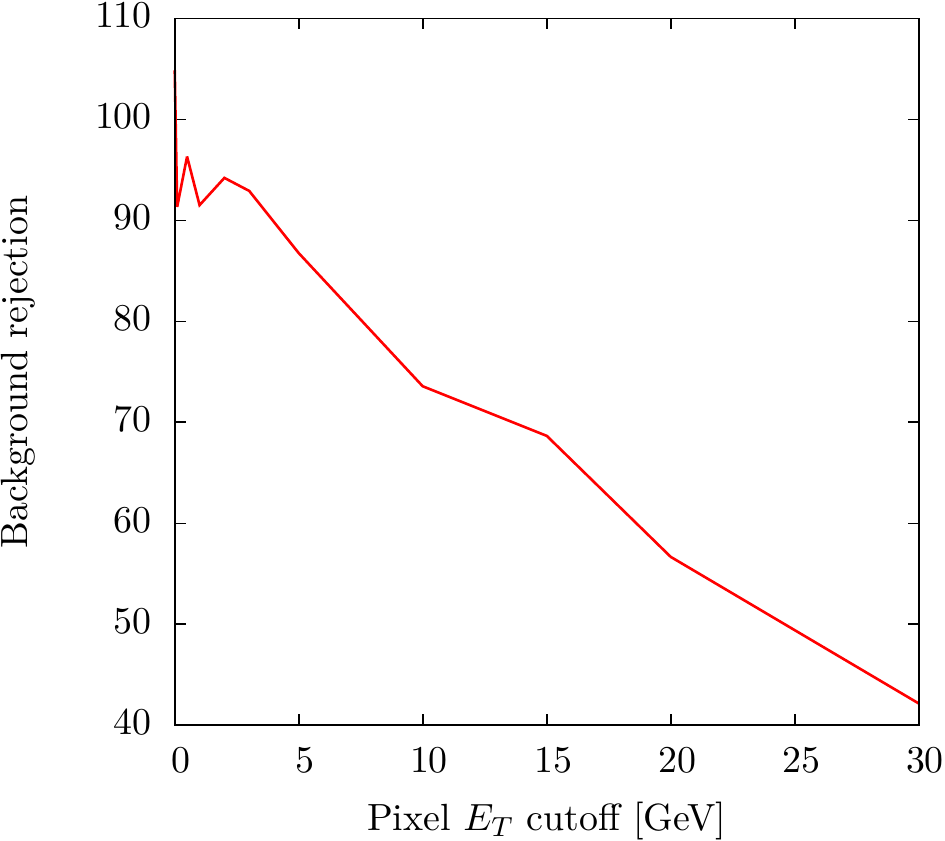}
  \hspace*{0.05\textwidth}
  \includegraphics[width=0.438\textwidth]{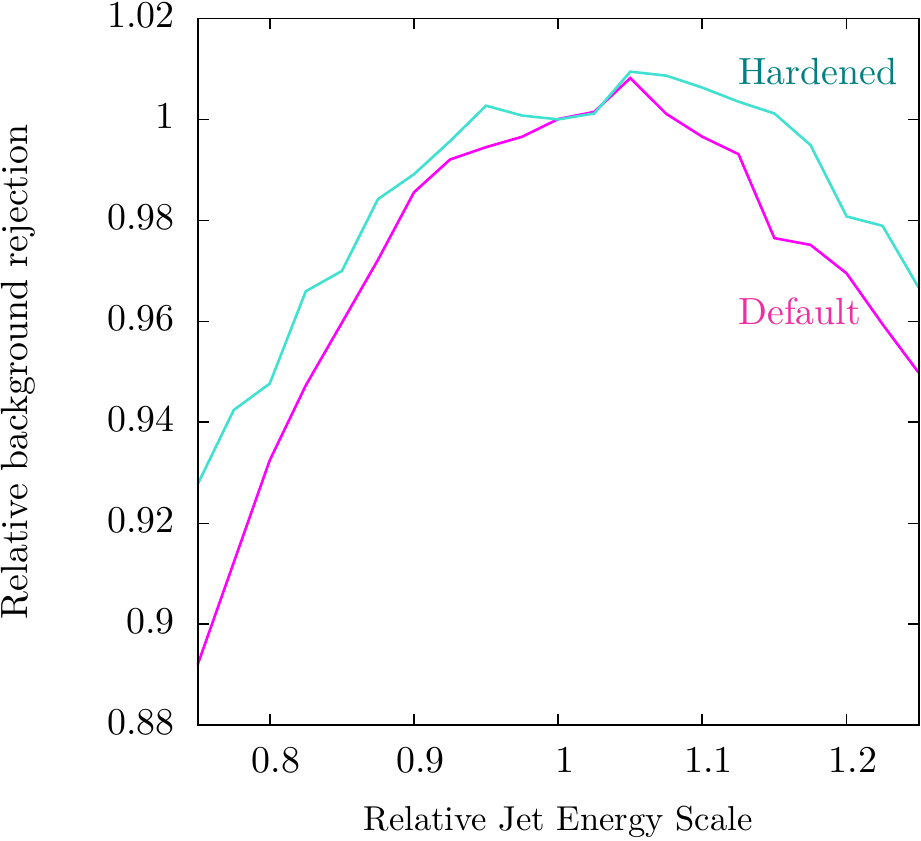}\\

  \caption[Background rejection versus pixel activation threshold and detector smearing.]{Left: Background rejection at a signal efficiency of 30~\%
    for different activation thresholds.  Right: Background rejection
    for the default and hardened training for different re-scalings of
    the jet images. The background rejection is evaluated at a signal
    efficiency of 30~\% and normalized to the rejection at nominal
    calorimeter energy scale.}
  \label{fig:rejection_vs_X}
\end{figure}

\subsection{Summary}
\label{sec:conc_ch6}
Like boosted decision trees before them, deep neural networks provide a powerful way of distinguishing between jets from heavy decaying objects, and jets from QCD background. This has already been shown for $W$-boson tagging, in this analysis we have shown that the same conclusions hold for top tagging as well. Using techniques from image processing, we trained a ConvNet on a Monte Carlo simulated sample of hadronically decaying tops and QCD dijets, and quantified its performance by testing the network on an independent testing sample. By benchmarking the network against the performance of other industry standard taggers, in terms of ROC curves, we showed that the network can offer comparable, and even superior background rejection to QCD inspired taggers. We also showed that preprocessing of the images is an optional, but not necessary step, and good network performance is observed even on minimally preprocessed images, albeit at the price of more sophisticated network architectures, and thus more intensive computational demands.

Interestingly, we showed that for some observables the network learns features that are not also captured by the BDTs, and vice versa. This shows that the network may be picking up on features of the jets that are not captured by currently used taggers and observables. We have also demonstrated that the network performance is fairly robust against degradation by detector level effects such as calorimeter deposit cutoffs and jet energy scale calibration, which is an encouraging sign that they may one day prove feasible in LHC experimental analyses, as well as phenomenological studies.

Still, there are some shortcomings. Our analysis, like others in this area, is based exclusively on Monte Carlo simulated data samples, for which the training labels are known for each jet. If these techniques are to actually prove useful for the experiments, it will have to be demonstrated that these networks can be trained on real hadron collider data, which, as well as bringing the extra complications of pileup, multiparton interactions and underlying event, does not come with `truth' labels on each event which can be used to evaluate the network performance. It thus remains to be seen whether or not these techniques will bear fruit in the long term. However, the impressive performance that we see here shows that these techniques certainly merit further study.

\newpage

\newpage
\section*{Summary and conclusions}
\addcontentsline{toc}{section}{Summary and conclusions}
This thesis has explored several aspects of top quark phenomenology at hadron and lepton colliders. In chapter 1, we reviewed the foundations of the Standard Model of particle physics, discussed some generalities about hadron collider physics and the main uncertainties that can hinder high theoretical precision there, and explored the properties of the top quark can that can be probed at hadron colliders through its various production mechanisms and decay observables, discussing in detail the state-of-the-art SM theory calculations where relevant. 

Chapter 2 focused on physics `Beyond the Standard Model'. After outlining some of the main flaws of the Standard Model and arguments for new physics (perhaps at the \tev scale); the hierarchy problem, gauge coupling unification and vacuum instability, we presented a few examples of popular new physics models constructed to address these flaws, and we briefly touched upon why they might be relevant for top quark phenomenology. We then moved from specific UV models to a more agnostic approach, by considering the Standard Model as the leading part of an effective theory, where heavy degrees of freedom have been integrated out, leaving behind a tower of higher-dimensional (that is, $D > 4$) operators; which can be studied perturbatively as an expansion in Wilson coefficient divided by UV cutoff $(\co{i}/\Lambda)^n$. 

By using a `bottom-up' approach, constructing the operators that are consistent with the SM gauge and global symmetries, we derived the only \D5 operator allowed by symmetry constraints. The terms that could generate effects at collider energy scales begin at \D6. Since there are many more operators allowed at this order, we did not reproduce the full derivation of the operator set, we just sketched some of its salient points, following the derivation of the `Warsaw basis'. Left with 64 non-redundant \D6 operators, we focused on those operators which are (at leading order in the EFT expansion) relevant for top quark physics, which amounted to calculating the effects of the operators on the observables introduced in the first chapter, and computing their numerical effects as a guide for how strongly they may be bounded with current measurements.

In chapter 3, we focused on confronting those operators with data. With the large production rates for top quark associated processes at the LHC and Tevatron, huge statistical samples of top quark data are now public, which make the precision scrutiny of the top quark sector of the Standard Model EFT a timely exercise. The large number of observables requires sophisticated fit machinery, which was achieved by adapting the {\sc{Professor}} software package, originally used for tuning of Monte Carlo generators, to BSM limit-setting code. We took various top quark production channels in turn, beginning with \ttbar production. We demonstrated that there is complementarity between LHC and Tevatron measurements, because they are dominated by different partonic subprocesses and so most sensitive to different operators. Differential measurements bring much more sensitivity than total rates alone, as they provide extra sensitivity to shape modifications by the new Lorentz structures of the \D6 operators that go beyond overall normalisation differences. Higher-order processes such as $\ttbar Z$ production bring sensitivity to operators that cannot otherwise be constrained, but the statistical uncertainties on the early measurements of these processes are weak and so the subsequent bounds are inconclusive. There is overlap in the operator set constrainable from single top production and from decay observables such as helicity fractions, and so combining these measurements gives stronger bounds. 

Some measurements remain in tension with the SM-only prediction, but there are no anomalies that are more significant than what would be na\"ively expected with this number of independent measurements.  We discussed in detail some of the validity issues inherent in the EFT formulation, such as the neglecting of operators of higher dimension than 6, and the potential for the fit to be excessively pulled by `overflow' bins in the differential distributions, for which there is no control over scales entering the fit. We concluded that the latter was not a problem in practice by performing the fit with and without those bins, and that for the former, in order to avoid unphysical effects such as negative cross-sections, it is best to keep in contributions from \ord{1/\Lambda^4} terms, even when \D8 operators are neglected. Though the fit is comprehensive, and the complementarity between different sets of measurements used allowed the system of operators considered to be easily overconstrained, we found that when our constraints mapped on to the parameter space of specific UV models, the final numerical bounds on the operators considered were in the end rather weak.

Chapter 4 was concerned with the prospects for improving this situation. We noted that the constraints on the \ttbar operators, which constituted the largest component of the fit, were dominated by observables reconstructed using `resolved' techniques in the low to medium $p_T$ region. While statistical uncertainties are smaller in this region, so is the sensitivity to the operators, whose interference typically scales as $\hat s/\Lambda^2$. High $p_T$ final states reconstructed by jet substructure methods thus have the potential to dramatically improve the constraints, especially as low statistics in the tails becomes less relevant over the LHC lifetime. 

Our analysis showed mixed results, however. While the constraints from a typical low $p_T$ analysis can be improved by up to 70\% when current systematics are improved upon and as we approach 3~\iab of data, the improvement from the boosted region is much milder when experimental error bars are reduced, indicating that SM (and SMEFT) theory uncertainties also have to be dramatically improved in order to exploit the full potential of boosted tagging performance. Given the timescales involved over the forecasted LHC lifetime, however, this does not seem an unrealistic expectation.

We also studied the role that proposed future lepton colliders could play in these improvements. By far the weakest constraints from our fit, and the least well-measured of the top quark couplings in general, were from the electroweak neutral vertices. The only direct handle on these from the LHC is from \ttz and $\ttbar \gamma$ production, and the improvements at 3~\iab are still modest. Lepton colliders are sensitive to the same operators, however, through the electroweak process $\ep \to \ttbar$. We found that orders of magnitude improvement over the current bounds was possible, even when marginalising over all operators in a global fit, which was unsurprising giving the improvement in precision on both the theory prediction and measurement sides. 

By benchmarking proposed scenarios at the 500 \gev ILC and 3 \tev CLIC machines, we saw that just as for the LHC, using as much information as possible (in this case running with high statistics at several incoming beam polarisations and making use of pseudo-observables such as forward-backward asymmetries in addition to total cross-sections) was crucial in maximising the sensitivity. Unlike for the LHC, however, we saw that the increasing the collider CM energy does not necessarily buy extra sensitivity, and as we move away from threshold and overall rates become smaller, sensitivity to the operators begins to degrade slightly, although in a full 4D fit the difference was not remarkable. We also highlighted the possibility of combining these bounds with measurements from LEP in order to lift a blind direction in the EFT parameter space, however, at this level of precision, the much more involved EFT loop corrections would have to also be considered in some detail for the numerical bounds to be completely trusted.  

In chapter 5, we returned again to hadron colliders, and discussed in detail what goes into the algorithms behind boosted taggers such as the {\sc{HepTopTagger}}, which we used in the boosted analysis of chapter 4. After summarising the current state-of-the-art in terms of taggers built from perturbative QCD, we moved on to concepts from machine learning; namely using so-called `deep learning' neural networks to improve the performance of boosted top reconstruction, by building an image out of the calorimeter plane in a hadronic top quark event, and training image classifier algorithms over top signal against QCD background. 

Quantifying its performance in terms of ROC curves (signal efficiency versus background rejection) we found comparable performance to well-established QCD based taggers, even in the presence of experimental degradation such as detector smearing and calorimeter energy cutoffs. The overarching question remains of whether these techniques will bear fruit when applied to real data and not just Monte Carlo simulation, but the robust performance of our network suggests that this is certainly a question that merits future investigation.

There are several other well-motivated directions for future work. For instance, in the global fit presented in chapter 3, a total of 12 parameters were constrained. However, in order to perform this fit, this parameter space was broken up into subsets of operators: 6 in \ttbar production, 3 in single top, and a further 3 in \ttz production. This factorisation, while easing the burden on the computational complexity of the fit, is not necessarily physically justified, since \ttz and other electroweak processes are sensitive to all 12 operators simultaneously, and because the \D6 operators also affect top quark decays, the division of the fit into production and decay observables is an imperfect approximation. This can be improved by only fitting to fiducial top measurements that are presented in terms of final state quantities and not `unfolded' to the level of tops. The experimental collaborations are beginning to favour presenting data in this way, and work on improving the capabilities of {\sc{TopFitter}} to harness this data is underway.

Improvements can be made on the theory side as well. All the constraints presented here are at leading order in the SMEFT, and the bounds on the Wilson coefficients can be interpreted as valid at the scale that they are probed. However, just as in QCD, when we truncate the perturbative expansion in $\co{i}/\Lambda^2$ we introduce a scale dependence of the \D6 Wilson coefficients. This can be modelled with an additional scale uncertainty propagated into each observable, or explicitly calculated through the RGEs for the \D6 operators, which are known at 1-loop. RGE improvement cannot completely capture the full NLO corrections, however, because the new operators will in general induce additional loop corrections that can substantially affect the shapes of differential distributions. This is also related to the imperfectness of our modelling of (N)NLO corrections with QCD $K$-factors, which inevitably misses out on some kinematic effects. On the other hand, a complete 1-loop EFT calculation of \ttbar production, for instance, is a formidable challenge, due to the sheer number of additional operators and diagrams involved, but certainly a worthwhile one, especially given the timescale of the future LHC programme.

Zooming out from the plethora of numerical results in this thesis, we can ask what general conclusions can be drawn? Firstly, though we often hear repeated that `top quark physics has entered a precision era' and that `the LHC is a top quark factory', it appears we still have some way to go before we saturate our understanding of the top quark's properties, and improvements in both experimental precision and theory understanding are both essential to this endeavour. Secondly, since at the time of writing there are no convincing hints of new resonant states from the current data and we are beginning to asymptote towards the maximum LHC reach for these states, it seems that precision understanding of the Standard Model, both as a full and an effective theory, will increasingly play a role in the hunt for new physics. It remains to be seen whether the \D6 extension of the SM will be the avenue that leads us to the next Standard Model, but its usefulness as a tool for collider phenomenology has shown us beyond question that it is an avenue worth pursuing.

\newpage
\null

\newpage
\appendix
\section{\D6 redefinitions of Standard Model input parameters}
As well as generating additional contributions to $S$-matrix elements through new Feynman rules, the \D6 operators also lead to modifications of the Standard Model Lagrangian parameters, which will propagate into observables. To see this, we note that an observable \op{} such as a decay width or cross-section can be written as a function of Lagrangian parameters $\mathcal{O}(\{\rho\})$, where $\{\rho\} = \{g,g^{\prime},v,y_t\}$ etc. These Lagrangian parameters bear a specific relation to the physical observables $\{obs\} =\{\alpha,G_F,m_Z,m_t \}$  so that the shift $\epsilon$ in \op{} due to a Wilson coefficient \co{i} has, in addition to an explicit dependence on \co{i}, an implicit dependence due to a modification of the relation between Lagrangian parameters $\{\rho\}$ and observables $\{obs\}$:
\begin{equation}
\mathcal{O}(\{\rho(obs)\}) \to \op{}+\epsilon = \mathcal{O}(\{\rho(obs,\co{i})\},\co{i} )
\end{equation}
In order to work consistently up to a given order in the EFT expansion, one must take care to ensure that the SM input parameters are appropriately renormalised as as function of \co{i} so that $obs$ are unchanged. The modifications to the relations $\rho(obs)$ is the subject of this section\footnote{To avoid unnecessary cluttering of notation, all \D6 Wilson coefficients are dimensional, i.e. $\co{i}/\Lambda^2 \to \co{i}$ in this section.}.

\subsection*{Higgs potential}
The Higgs potential receives a contribution from the operator \op{\varphi}:
\begin{equation}
V(\varphi) = \lambda\left(\varphi^\dagger\varphi-\frac{1}{2}v^2\right)^2 - \co{\varphi}(\varphi^\dagger\varphi)^3 .
\end{equation}
The shifted vacuum expectation value of the potential is at
\begin{equation}
\frac{1}{2}v_6^2 = \frac{1}{3\co{\varphi}} \left(\lambda - \lambda \sqrt{1-\frac{3\co{\varphi}v^2}{\lambda}}\right),
\end{equation}
This expression does not have a well-defined SM (\co{\varphi} = 0) limit, but this can be obtained by expanding to first order in \co{\varphi}:
\begin{equation}
\braket{ \varphi^\dagger\varphi } = \frac{1}{2}v_6^2 = \frac{1}{2}v^2\left(1+\frac{3\co{\varphi}v^2}{4\lambda} \right) ,
\end{equation}
so that the shift in the vev is linear in \co{\varphi} and vanishes in the SM limit.

\subsection*{Kinetic terms}
The scalar kinetic part of the SM is modified by the operators \op{\varphi\Box} and \op{\varphi D}
\begin{equation}
\lag{} = (D_\mu\varphi)^\dagger (D_\mu\varphi) + \co{\varphi\Box} (\varphi^\dagger\varphi)\Box(\varphi^\dagger\varphi)+\co{\varphi D}(\varphi^\dagger D_\mu \varphi)^*(\varphi^\dagger D^\mu \varphi) ,
\end{equation}
In the unitary gauge, we can write the field $\varphi$ as 
\begin{equation}
\varphi = \left( \begin{array}{c} -i (1+\kappa_\pm) G_\pm \\ \frac{1}{\sqrt{2}}(v_6+(1+\kappa_h) h + i(1+\kappa_0)G_0) \end{array} \right) ,
\end{equation}
where the coefficients $\kappa_h$, $\kappa_0$ and $\kappa_\pm$ can be chosen to ensure the kinetic terms will be canonically normalised. Upon expanding out the Lagrangian and keeping just the scalar-only terms (the gauge parts will be treated shortly), we see that the kinetic terms are given by:
\begin{equation}
\begin{split}
\lag{H} &= \frac{1}{2}(1+\kappa_h)^2\left(1+2v^2\left [\frac{\co{\varphi D}}{4} - \co{\varphi\Box}\right] \right)\partial_\mu h \partial^\mu h , \\
\lag{G0} &= \frac{1}{2}(1+\kappa_0)^2\left(1+2v^2 \frac{\co{\varphi D}}{4}\right) \partial_\mu G_0 \partial^\mu G_0 , \\
\lag{G \pm} &= \frac{1}{2}(1+\kappa_\pm)^2 \partial_\mu G_+ \partial^\mu G_-
\end{split}
\end{equation}
Canonical renormalisation of the kinetic terms then requires
\begin{equation}
\begin{split}
\kappa_h &= v^2\left(\co{\varphi\Box}-\frac{\co{\varphi D}}{4} \right), \\
\kappa_0 &= -v^2\frac{\co{\varphi D}}{4} , \\
\kappa_\pm &= 0.
\end{split}
\end{equation}
Combining the Higgs kinetic terms with the modified Higgs potential, the scalar Lagrangian reads
\begin{equation}
\begin{split}
\lag{} &= \frac{1}{2}(\partial_\mu h)^2 - \frac{\kappa_h}{v_6^2}\left[h^2(\partial_\mu h)^2+2vh(\partial_\mu h)^2\right] -\lambda v_6^2\left(1-\frac{3\co{\varphi}v^2}{2\lambda} + 2\kappa_h\right)h^2  \\
&-\lambda v_6^2 \left(1-\frac{5\co{\varphi}v^2}{2\lambda}+3\kappa_h \right)h^3 -\frac{1}{4}\lambda \left(1- \frac{15\co{\varphi}v^2}{2\lambda}+4\kappa_h\right)h^4 + \frac{3}{4}\co{\varphi}vh^5 +\frac{1}{8}\co{\varphi}h^6  ,
\end{split}
\end{equation}
so we see that there is a shift in the Higgs mass definition
\begin{equation}
m_h^2 = 2\lambda v_6^2\left(1-\frac{3\co{\varphi}v^2}{2\lambda} + 2\kappa_h\right) ,
\end{equation}
which does not correspond to a physical mass shift, as it can be absorbed into a renormalisation of the quartic coupling $\lambda(m_h)$. 

\subsection*{Gauge sector}
The \D6 operators induce redefinitions of the gauge fields and gauge couplings. The part of the \D6 Lagrangian relevant for this discussion is:
\begin{equation}
\begin{split}
\lag{} & \supset \co{\varphi G}(\varphi^\dagger\varphi)G^A_{\mu\nu}G^{A,\mu\nu} +  \co{\varphi W}(\varphi^\dagger\varphi)W^I_{\mu\nu}W^{I,\mu\nu} + \co{\varphi B}(\varphi^\dagger\varphi)B_{\mu\nu}B^{\mu\nu}  \\
&+  \co{\varphi WB}(\varphi^\dagger\tau^I\varphi)W^I_{\mu\nu}B^{\mu\nu} +  \co{G}f^{ABC}G^{A\nu}_\mu G^{B\nu}_\rho G^{C\rho}_\mu +  \co{W}\epsilon^{IJK}W^{I\nu}_\mu W^{J\nu}_\rho W^{K\rho}_\mu .
\end{split}
\end{equation}
After electroweak symmetry breaking, this becomes (combined with the \D4 Lagrangian)
\begin{equation}
\begin{split}
\lag{} & \supset -\frac{1}{2}W^+_{\mu\nu}W_-^{\mu\nu} -\frac{1}{4}W^3_{\mu\nu}W_3^{\mu\nu}-\frac{1}{4}B_{\mu\nu}B^{\mu\nu}-\frac{1}{4}G^A_{\mu\nu}G^{A,\mu\nu}+\frac{1}{2}v_6^2\co{\varphi G} G^A_{\mu\nu}G^{A,\mu\nu} \\
&+ \frac{1}{2}v_6^2\co{\varphi W} W^I_{\mu\nu}W^{I,\mu\nu} + \frac{1}{2}v_6^2\co{\varphi B} B_{\mu\nu}B^{\mu\nu} - \frac{1}{2}v_6^2\co{\varphi WB} W^3_{\mu\nu}B^{\mu\nu}.
\end{split}
\end{equation}
The gauge kinetic terms are no longer canonically normalised, and we have also induced kinetic mixing between $W^3$ and $B$. Beginning with the gluons, we can write the canonically normalised gluon field as $\mathcal{G}_\mu = (1+\kappa_G)^{\frac{1}{2}}G_\mu$, then the Lagrangian reads
\begin{equation}
\begin{split}
\lag{} & \supset - \frac{1}{4}(1-2v_6^2\co{\varphi G})G_{\mu\nu}G^{\mu\nu} \\
& = -\frac{1}{4}\mathcal{G}_{\mu\nu}\mathcal{G}^{\mu\nu} ,
\end{split}
\end{equation}
provided that
\be
\kappa_G  = -2 v_6^2 \co{\varphi G}.
\ee
The gluon field is rescaled 
\begin{equation}
G^A_\mu = \mathcal{G}^A_\mu (1+ \co{\varphi G} v^2), 
\end{equation}
so that the strong coupling constant must also be renormalised
\begin{equation}
\bar g_s = g_s (1+\co{\varphi G} v^2), 
\end{equation}
in order to keep the vector currents unchanged, i.e. $g_s G^A_\mu = \bar g_s \mathcal{G}^A_\mu$. The exercise can be repeated for the electroweak bosons 
\begin{equation}
\begin{split}
W^I_\mu &= \mathcal{W}^I_\mu(1+\co{\varphi W} v^2) \\ 
B^\mu &= \mathcal{B}_\mu(1+\co{\varphi B} v^2)
\end{split}
\end{equation}
where the renormalisation coefficients are $\kappa_W = -2v^2\co{\varphi W}$ and $\kappa_B = -2v^2\co{\varphi B}$, and the gauge couplings are renormalised to 
\begin{equation}
\begin{split}
\bar g &= g (1+\co{\varphi W} v^2), \\
\bar g^\prime &= g^\prime (1+\co{\varphi B} v^2).
\end{split}
\end{equation}
We must also take into account the kinetic mixing that has been induced. The Lagrangian for the kinetic terms of the $\mathcal{W}^3$ and $\mathcal{B}$ fields can be written as a non-diagonal matrix:
\begin{equation}
\lag{} = -\frac{1}{4} \left [\begin{array}{cc}  \mathcal{W}^3_{\mu\nu} & \mathcal{B}_{\mu\nu} \end{array} \right]  \left [\begin{array}{cc}  1 & v_6^2\co{\varphi WB} \\  v_6^2\co{\varphi WB} & 1 \end{array} \right] \left [\begin{array}{c}  \mathcal{W}^3_{\mu\nu} \\ \mathcal{B}_{\mu\nu} \end{array} \right] .
\end{equation}
Canonically normalising the $\mathcal{W}^3$ and $\mathcal{B}$ fields then just amounts to diagonalising this matrix, which can be done by rotating the fields
\begin{equation}
\begin{split}
\left [\begin{array}{c}  \bm{\mathcal{W}}^3_\mu \\ \bm{\mathcal{B}}_\mu \end{array} \right]  = \left [\begin{array}{cc}  1 & v_6^2\co{\varphi WB} \\  0 & 1 \end{array} \right] \left [\begin{array}{c}  \mathcal{W}^3_\mu \\ \mathcal{B}_\mu \end{array} \right] .
\end{split}
\end{equation}
The canonically normalised kinetic Lagrangian is then
\begin{equation}
\begin{split}
\lag{} = -\frac{1}{4} \bm{\mathcal{W}}^I_{\mu\nu} \bm{\mathcal{W}}^{\mu\nu}_I - \frac{1}{4} \bm{\mathcal{B}}_{\mu\nu} \bm{\mathcal{B}}^{\mu\nu} = -\frac{1}{2} \bm{\mathcal{W}}^+_{\mu\nu} \bm{\mathcal{W}}^{\mu\nu}_- -\frac{1}{4} \bm{\mathcal{W}}^3_{\mu\nu} \bm{\mathcal{W}}^{\mu\nu}_3 - \frac{1}{4} \bm{\mathcal{B}}_{\mu\nu} \bm{\mathcal{B}}^{\mu\nu}  ,
\end{split}
\end{equation}
where $\bm{\mathcal{W}}^{1,2}_\mu = \mathcal{W}^{1,2}_\mu$.

Turning now to the electroweak mass terms, we start with the Lagrangian
\begin{equation}
\lag{} = -\frac{1}{4}\bar{g}^2v_6^2\mathcal{W}_\mu^+\mathcal{W}^\mu_- + \frac{1}{8}v_6^2(\bar g\mathcal{W}_\mu^3-\bar g^\prime\mathcal{B}_\mu)^2+\frac{1}{16}v_6^2\co{\varphi D}(\bar g\mathcal{W}_\mu^3-\bar g^\prime\mathcal{B}_\mu)^2 ,
\end{equation}
so that the $W$ mass term can be straightforwardly read off as
\begin{equation}
M_W^2 = \frac{1}{4}\bar g^2 v_6^2.
\end{equation}
As for the $Z$ and photon, we can write the Lagrangian as
\begin{equation}
\lag{} = \frac{1}{2}\left(\frac{1}{4}v_6^2\left\{1+\frac{1}{2}v_6^2\co{\varphi D}\right\}\right) \left [\begin{array}{cc}  \mathcal{W}^3_{\mu\nu} & \mathcal{B}_{\mu\nu} \end{array} \right]  \left [\begin{array}{cc}  \bar g^2 & \bar g \bar g^\prime \\   \bar g \bar g^\prime & \bar{g}^{\prime 2} \end{array} \right] \left [\begin{array}{c}  \mathcal{W}^3_\mu \\ \mathcal{B}_\mu \end{array} \right]
\end{equation}
Diagonalising the mass matrix gives us the updated expression
\begin{equation}
\left[ \begin{array}{c}  \bm{\mathcal{Z}}_\mu \\ \bm{\mathcal{A}}_\mu \end{array} \right] = \left[ \begin{array}{cc} \cos\bar\theta_W & -\sin\bar\theta_W \\ \sin\bar\theta_W & \cos\bar\theta_W \end{array} \right] \left[ \begin{array}{c}  \bm{\mathcal{W}}^3_\mu \\ \bm{\mathcal{B}}_\mu \end{array} \right] ,
\end{equation}
where 
\begin{equation}
\begin{split}
\cos\bar\theta_W &= \frac{\bar g}{\sqrt{\bar g^2 +\bar g^{\prime 2}}} \left[1 + v^2 \frac{\bar g}{\bar g^\prime}\frac{\bar g^{\prime 2}}{\bar g^2 +\bar g^{\prime 2}} \co{\varphi WB} \right] \\
\sin\bar\theta_W &= \frac{\bar g^\prime}{\sqrt{\bar g^2 +\bar g^{\prime 2}}} \left[1 - v^2 \frac{\bar g^\prime}{\bar g}\frac{\bar g^2}{\bar g^2 +\bar g^{\prime 2}} \co{\varphi WB} \right]  .
\end{split}
\end{equation}
Then the $Z$ mass is given by 
\begin{equation}
M_Z^2 = \frac{v_6^2}{4}(\bar g^2+ \bar g^{\prime 2})+\frac{1}{8}v_6^4\co{\varphi D}(\bar{g}^2+\bar{g}^{\prime 2})+\frac{1}{2}v_6^4\bar{g}\bar{g}^\prime\co{\varphi WB},
\end{equation}
while the photon remains massless as required. Substituting the renormalised gauge fields into the covariant derivative operator gives the expression
\begin{equation}
D_\mu = \partial_\mu + i\frac{g^\prime}{\sqrt{2}}[ \bm{\mathcal{W}}^+_\mu T^+ +  \bm{\mathcal{W}}^-_\mu T^- ] + i \bar g_Z[T^3 - (\sin^2\bar\theta+\kappa)Q] \bm{\mathcal{Z}}_\mu+i\bar e Q \bm{\mathcal{A}}_\mu ,
\end{equation}
where, as usual, the generators $T^\pm = T_1 \mp iT_2$ and the electric charge is $Q = T_3 + Y$. The effective neutral couplings are
\begin{equation}
\begin{split}
\bar e &= \frac{\bar{g} \bar{g^\prime}}{\sqrt{\bar{g}^2+\bar{g}^{\prime 2}}}\left[1-\frac{\bar{g}\bar{g^\prime}}{g^2+g^{\prime 2}}v_6^2\co{\varphi WB} \right] = \bar g \sin\bar\theta - \frac{1}{2}\cos\bar\theta \bar g v_6^2 \co{\varphi WB} \\
\bar g_Z &= \sqrt{\bar{g}^2+\bar{g}^{\prime 2}}+\frac{\bar{g}\bar{g^\prime}}{g^2+g^{\prime 2}}v_6^2\co{\varphi WB} = \frac{\bar e}{\sin\bar\theta\cos\bar\theta}\left[1+\frac{\bar{g}^2+\bar{g}^{\prime 2}}{2\bar{g}\bar{g^\prime}} v_6^2 \co{\varphi WB}\right] \\
\sin^2\bar\theta &= \frac{g^{\prime 2}}{\bar{g}^2+\bar{g}^{\prime 2}} + \frac{\bar{g}\bar{g^\prime}(\bar{g}^2-\bar{g}^{\prime 2})}{(\bar{g}^2+\bar{g}^{\prime 2})^2}v_6^2 \co{\varphi WB} .
\end{split}
\end{equation}

\subsection*{Yukawa sector}
The Yukawa sector will be modified by the operators of type $\psi^2\varphi^3$. The Yukawa Lagrangian for the unbroken theory now reads 
\begin{equation}
\begin{split}
\lag{Yukawa} &=  - (\varphi^\dagger \bar d_s [y_d]_{st} Q_{t} + \tilde \varphi^\dagger \bar u_s [y_u]_{st} Q_{t} + \varphi^\dagger \bar e_s [y_e]_{st} L_t + h.c.) \\
&+ (\co{d\varphi}(\varphi^\dagger\varphi)\varphi^\dagger \bar{d}_s Q_s + \co{u\varphi}(\varphi^\dagger\varphi) \varphi^\dagger \bar{u}_s Q_s +  \co{u\varphi}(\varphi^\dagger\varphi) \varphi^\dagger \bar{e}_s L_s + h.c.) ,\\
\end{split}
\end{equation}
which in the unbroken theory leads to the fermion mass matrices
\begin{equation}
[M_\psi]_{rs} = \frac{v_6}{\sqrt{2}}\left([y_\psi]_{rs} -\frac{1}{2}v^2[\co{\psi\varphi}]_{rs}\right), \quad \text{where} \quad \psi = u,d,e 
\end{equation}
and to the Higgs fermion couplings
\begin{equation}
\begin{split}
[\kappa_\psi]_{rs} &= \frac{1}{\sqrt{2}}[y_\psi]_{rs} (1+\kappa_h) - \frac{3}{2\sqrt{2}}v^2\co{\psi\varphi} \\
&=  \frac{1}{\sqrt{2}}[M_\psi]_{rs} (1+\kappa_h) - \frac{v^2}{\sqrt{2}}v^2\co{\psi\varphi} , \quad \text{where} \quad \psi = u,d,e  \\
\end{split}
\end{equation}
which, unlike in the SM, are not simply proportional to the fermion mass matrices. Furthermore, because the fermion mass matrices and Yukawa matrices have different RGEs, they are not simultaneously diagonalisable, so Higgs-fermion couplings will be no longer flavour diagonal.

\subsection*{Fermi sector}
The Fermi coupling constant is measured from the transition rate for $\mu^- \to e^- \bar\nu_e \nu_\mu$, and this in turn defines the value of the electroweak scale $v$. In the SM alone, this is described by the effective operator
\begin{equation}
\lag{\mathnormal{G_F}}= \frac{4 G_F}{\sqrt{2}} (\bar{\nu}_\mu\gamma^\mu P_L \mu) (\bar e \gamma_\mu P_L \nu_e).
\end{equation}
The Fermi constant $G_F$ will also receive corrections from \D6 operators, leading to the new value:
\begin{equation}
\sqrt{2}G_F = \frac{1}{v_6^2} - \frac{1}{2}\left( \co[2112]{ll}+\co[1221]{ll} \right) + \left(\co[(3)11]{\varphi l}+\co[(3)22]{\varphi l}\right) .
\end{equation}
Although this looks like a physical shift, in fact it can be combined with the modified expression for the Higgs mass to define renormalised values for the Higgs self-coupling and vev:
\begin{equation}
\begin{split}
\lambda &= \frac{3\sqrt{2}\co{\varphi}}{4G_F} +\frac{m_h^2}{4}\left( \co[2112]{ll}+\co[1221]{ll}- 2(\co[(3)11]{\varphi l}+\co[(3)22]{\varphi l}) \right) + \frac{1}{\sqrt{2}}m_h^2G_F(1-2\kappa_h) \\
v_6 &= \frac{1}{(\sqrt{2}G_F)^{\frac{1}{2}}} +\frac{1}{2(\sqrt{2}G_F)^{\frac{3}{2}}}\left(\co[(3)11]{\varphi l}+\co[(3)22]{\varphi l}-\frac{1}{2}\left(\co[2112]{ll}+\co[1221]{ll} \right)\right)   .
\end{split}
\end{equation}
Likewise, we can rearrange the expressions for the observables $\alpha_{em} (\equiv e^2/4\pi)$ and $M_Z$ to obtain the renormalised \U1Y and \su2L couplings
\begin{equation}
\begin{split}
\bar g^\prime &= g^\prime + \frac{v_6^2(4\co{\varphi WB}g+\co{\varphi D}g^\prime)(-4M_Z^2 + v_6^2(g^2-g^{\prime 2}+16\pi\alpha_{em}))}{32(M_Z^2-4\pi v_6^2\alpha_{em})} \\
\bar g &= g - \frac{2v_6^2(4\co{\varphi WB}g+\co{\varphi D}g^\prime)(4M_Z^4+M_Z^2v_6^2)}{M_Z^2(M_Z^2-4\pi v_6^2\alpha_{em})} \\
&\times \left [\frac{v_6^4((g^2-g^{\prime 2}-20\pi\alpha_{em})+\pi\alpha_{em}(-3g^2+3g^{\prime 2}+16\pi\alpha_{em}))}{(4M_Z^2+v_6^2(g^2-g^{\prime 2}))^2}\right]
\end{split}
\end{equation}
where the SM coupling constants are as usual given by
\begin{equation}
\begin{split}
g^\prime &= \frac{\sqrt{2}}{v_6}\left( M_Z^2 -\sqrt{M_Z^4-4\pi\alpha_{em}M_Z^2v_6^2} \right)^{\frac{1}{2}} \\
g &= \frac{\sqrt{2}}{v_6}\left( M_Z^2 +\sqrt{M_Z^4-4\pi\alpha_{em}M_Z^2v_6^2} \right)^{\frac{1}{2}} .\\
\end{split}
\end{equation}
This completes the finite renormalisation of the Standard Model Lagrangian due to the effects of operators of dimension \D6. The modifications to all other derived parameters can be obtained from the relations presented here.

\newpage

\newpage
\phantomsection
\addcontentsline{toc}{section}{\listfigurename}
\listoffigures

\newpage
\phantomsection
\addcontentsline{toc}{section}{\listtablename}
\listoftables 

\newpage
\bibliographystyle{utphys}
\phantomsection
\addcontentsline{toc}{section}{References}
\bibliography{thesis}

\end{document}